\providecommand{\N}[1]{\ensuremath{\tilde{\chi}^o_#1}}

\providecommand{\GeV}{\,\,{\rm{GeV}}}
\providecommand{\MeV}{\,\,{\rm{MeV}}}
\providecommand{\TeV}{\,\,{\rm{TeV}}}
\providecommand{\fb}{\,\,{\rm{fb}}}
\providecommand{\cm}{\,\,{\rm{cm}}}
\providecommand{\herwig}{{\tt HERWIG}}
\providecommand{\Max}{{\tt{max}}\ }
\providecommand{\max}{{\tt{max}}\ }
\providecommand{\Deltam}{M_{-}}
\providecommand{\etal}{{\emph{et.al}}}
\providecommand{\Poincare}{Poincar$\acute{\rm{e}}$}

\documentclass[11pt]{ociamthesis}
\pdfoutput=1

\usepackage{amssymb,epsfig}
\usepackage{amsmath,amssymb,amsfonts} 
\usepackage{graphicx}
\usepackage{feynmf}
\usepackage{color}
\usepackage{ifthen}
\usepackage{slashed}

\usepackage{hyperref}  

\newboolean{found}

\title{Mass Determination \\ of New Particle States}
\author{Mario A Serna Jr}
\college{Wadham College}
\degree{Doctor of Philosophy}
\degreedate{Trinity 2008}
\renewcommand{\crest}{\beltcrest}

\newcommand{\Mvariable}[1]{}

\begin{document}

\baselineskip=21pt plus1pt

\setcounter{secnumdepth}{2}
\setcounter{tocdepth}{1}

\setlength{\unitlength}{1mm}

\maketitle

\begin{dedication}
This thesis is dedicated to\\
 my wife\\
 for joyfully supporting me and our daughter while I studied.\\
\end{dedication}

\newpage

\centerline{\Large \bf{ Mass Determination of New Particle States}}

\vspace{0.3in}
\centerline{\large Mario Andres Serna Jr}
\centerline{ Wadham College}

\normalsize
\vspace{0.3in}
\centerline{Thesis submitted for the degree of Doctor of Philosophy}
\centerline{Trinity Term 2008}

\vspace{0.3in}

\centerline{\Large{ \bf{Abstract}}}

\vspace{0.3in}
\normalsize
\parbox[c]{5.5in}{
We study theoretical and experimental facets of mass determination of new particle states.
Assuming supersymmetry, we update the quark and lepton mass matrices at the grand unification scale accounting for threshold corrections enhanced by large ratios of the vacuum expectation value of the two supersymmetric Higgs fields $v_u/ v_d \equiv \tan \beta$.
From the hypothesis that
quark and lepton masses satisfy a classic set of relationships suggested in some Grand Unified Theories (GUTs), we predict $\tan \beta$ needs to be large, and the gluino's soft mass needs to have the opposite sign to the wino's soft mass.  Existing tools to measure the phase of the gluino's mass at upcoming hadron colliders require model-independent, kinematic techniques to determine the
masses of the new supersymmetric particle states.
The mass determination is made difficult because supersymmetry is likely to have a dark-matter particle which will be invisible to the detector, and because the reference frame and energy of the parton collisions are unknown at hadron colliders.
We discuss the current techniques to determine the mass of invisible particles.
We review the transverse mass kinematic variable $M_{T2}$ and the use of invariant-mass edges to find relationships between masses.
Next, we introduce a new technique to add additional constraints between the masses of new particle states using $M_{T2}$ at
different stages in  a symmetric decay chain.
These new relationships further constrain the mass differences between new particle states, but still leave the absolute mass weakly determined.
Next, we introduce the constrained mass variables $M_{2C,LB}$, $M_{2C,UB}$, $M_{3C,LB}$, $M_{3C,UB}$ to provide event-by-event lower-bounds and upper-bounds to the mass scale given mass differences.
We demonstrate mass scale determination in realistic case studies of supersymmetry models by fitting ideal distributions to simulated data.
We conclude that the techniques introduced in this thesis have precision and accuracy that rival or exceed the best known techniques for invisible-particle mass-determination at hadron colliders.
}

\newpage

\newpage

\centerline{\Large \bf{Declaration}}
\normalsize

\vspace{0.5in}

\parbox[c]{6in}{This thesis is the result of my own work, except where reference is made to the work of others, and has not been submitted for other qualification to this or any other university.}

\vspace{0.5in}
\begin{flushright}
Mario Andres Serna Jr
\end{flushright}

\newpage

\begin{acknowledgements}
I would like to thank my supervisor Graham Ross
for his patient guidance as I explored many topics
about which I have wanted to learn for a long time.
For the many hours of
teaching, proofreading, computer time and conversations they have provided me,
I owe many thanks to my fellow students
Tanya Elliot, Lara Anderson, Babiker
Hassanain, Ivo de Medeiros Varzielas, Simon
Wilshin,  Dean Horton, and the friendly department post-docs  James Gray, Yang-Hui He, and David Maybury.
I also need to thank Laura Serna, Alan Barr, Chris Lester, John March Russell, Dan Tovey, and Giulia Zanderighi for many helpful comments on my published papers which were forged into this thesis.
Next I need to thank Tilman Plehn, Fabio Maltoni and Tim Stelzer for providing us
online access to MadGraph and MadEvent tools.
I would also like to thank  Alex Pinder, Tom Melina, and Elco Makker, with whom I've worked on fourth-year projects this past year, for questions and conversations that have helped my work on this thesis.  I also would like to thank my mentors over the years: Scotty Johnson, Leemon Baird, Iyad Dajani, Richard Cook, Alan Guth, Lisa Randall, Dave Cardimona, Sanjay Krishna, Scott Dudley, and Kevin Cahill, and the many others who have offered me valuable advice that I lack the space to mention here.
Finally I'd like to thank my family and family-in-law for their support and engaging distractions during these past few years while I've been developing this thesis.

I also acknowledge support from the United States Air Force Institute of Technology.
The views expressed in this thesis are those of the author and
do not reflect the official policy or position of the United States Air Force, Department of Defense, or the US Government.
\end{acknowledgements}

\begin{romanpages}
 \tableofcontents
 \listoffigures
\end{romanpages}

\baselineskip=21pt plus1pt



\chapter{Introduction}



In the mid-seventeenth century,
a group of Oxford natural philosophers, including
Robert Boyle and Robert Hooke,  argued for the inclusion of experiments
in the natural philosopher's toolkit as a means of falsifying theories \cite{Boyle:1686ab}\@.
This essentially marks the beginning of modern physics, which is rooted in an interplay between creating theoretical models,
developing experimental techniques, making observations, and falsifying theories\footnote{There are many philosophy of physics subtleties about how physics progress is made \cite{wilczek:2004428ab} or how theories are falsified \cite{dagostini:2004ab} which we leave for future philosophers.
Nevertheless, acknowledging the philosophical interplay between experiment and scientific
theory is an important foundational concept for a physics thesis.}.  This thesis is concerned with mass determination of new particle states in these first two stages: we present a new theoretical observation leading to predictions about particle masses and their relationships, and we develop new experimental analysis techniques to extract the masses of new states produced in hadron collider experiments.   The remaining steps of the cycle will follow in the next several years: New high-energy observations will begin soon at the Large Hadron Collider (LHC)\footnote{This does not mean that the LHC will observe new particles, but that the LHC will perform experiments with unexplored energies and luminocities that will constrain theories.}, and history will record which theories were falsified and which (if any) were chosen by nature.

No accepted theory of fundamental particle physics claims to be `true',  rather theories claim to make successful predictions within a given domain of validity for experiments performed within a range of accuracy and precision.
The Standard Model with an extension to include neutrino masses is the best current `minimal' model.  Its predictions agree with all the highest-energy laboratory-based observations ($1.96 \TeV$ at the Tevatron) to within the best attainable accuracy and precision (an integrated luminosity ${\mathcal{L}}$ of about $\int dt {\mathcal{L}} \approx 4 \fb^{-1}$)
 \cite{FermiLuminocity} \footnote{Grossly speaking in particle physics, the domain of validity is given in terms of the energy of the collision, and the precision of the experiment is
 dictated by the integrated luminosity $\int dt {\mathcal{L}}$.
 Multiplying the $\int dt {\mathcal{L}}$ with the cross-section for a process gives the number of events of that process one expects to occur.  The larger the integrated luminosity, the more sensitive one is potentially to processes with small cross sections.}.
The Standard Model's agreement with collider data requires agreement with quantum corrections \cite[Ch 10]{PDBook2008}.  The success of the Standard Model will soon be challenged.

At about the time this thesis is submitted, the Large Hadron Collider (LHC) will
 (hopefully) begin taking data at a substantially larger collision energy of $10 \TeV$ and soon after at $14 \TeV$ \cite{AtlasTDR}.  The high luminosity of this collider ($10^{34} \cm^{-2} \sec^{-1}$) \cite{AtlasTDR} enables us to potentially measure processes with much greater precision (hopefully about $\int dt {\mathcal{L}}=300 \fb^{-1}$ after three years).
The LHC holds promise to test many candidate theories of fundamental particle physics that make testable claims at these new energy and precision frontiers.

The thesis is arranged in two parts.  The first regards
theoretical determination of masses of new particle states, and the second regards
experimental determination of masses at hadron colliders.
Each part begins with a review of important developments
relevant to placing the new results of this thesis in context.
The content is aimed at recording enough details to enable
future beginning graduate students to follow the motivations and, with the help of the
cited references, reproduce the work.
A full pedagogical explanation of quantum field theory, gauge theories, supersymmetry, grand unified theories (GUTs)
and renormalization would require several thousand pages reproducing easily accessed lessons found in current textbooks.
Instead we present summaries of key features and refer the reader to the author's favored textbooks or review articles on the subjects for refining details.

Part I of this thesis contains Chapters \ref{ChapterPastCurrentMassPredictions}-\ref{ChapterUnificationAndFermionMassStructure}.
Part II of this thesis contains Chapters \ref{ChapterMassDeterminationReview}-\ref{ChapterM3C}.

Chapter \ref{ChapterPastCurrentMassPredictions}
outlines theoretical approaches to mass determination that motivate
the principles behind our own work.
Exact symmetry, broken symmetry, and fine-tuning of radiative corrections
form the pillars of
past successes in predicting mass properties of new particle states.  By mass properties, we mean both the magnitude and any CP violating phase that cannot be absorbed into redefinition of the respective field operators.  We highlight a few examples of the powers of each pillar through the discoveries of the past century.  We continue with a discussion of what the future likely holds for predicting masses of new particle states.  The observation of dark-matter  suggests that nature has a stable, massive, neutral particle that has not yet been discovered.  What is the underlying theoretical origin of this dark matter?  The answer on which we focus is supersymmetry.
SUSY, short for supersymmetry, relates the couplings of current fermions and bosons
to the couplings of new bosons and fermions known as superpartners.  The masses of the new SUSY particles reflect the origin of supersymmetry breaking.
Many theories provide plausible origins of supersymmetry breaking but are beyond the scope of this thesis.
We review selected elements of SUSY related to why one might believe it has something to do with nature and to key elements needed for the predictions in the following chapter.

Chapter \ref{ChapterUnificationAndFermionMassStructure}, which is drawn from work published by the author and his supervisor in \cite{Ross:2007az}, presents a series of arguments using supersymmetry, unification, and precision measurements of low-energy observables to suggest some mass relationships and phases of the yet undiscovered superpartner masses\footnote{The phase of a Majorana fermion mass contributes to the degree to which a particle and its antiparticle have distinguishable interactions.  The conventions is to remove the phase from the mass by redefining the fields such as to make the mass positive.  This redefinition sometimes transfers the CP violating phase to the particle's interactions.}.
In the early 1980s,  Georgi and Jarlskog \cite{Georgi:1979df} observed that the masses of the quarks and leptons satisfied several properties at the grand unification scale. Because these masses cover six orders of magnitude, these relationships are very surprising.  We discover that with updated experimental observations,
the Georgi Jarlskog mass relationships  can now only be satisfied for a specific class of quantum corrections enhanced by  the ratio of the vacuum expectation value of the two supersymmetric Higgs $v_u/v_d \equiv \tan \beta$.  We predict that $\tan \beta$ be large $(\gtrsim 20)$ and the gluino mass has the sign opposite to the sign of the wino mass.

Chapter \ref{ChapterMassDeterminationReview} reviews the existing toolbox of experimental techniques to determine masses and phases of masses.
If the model is known,
then determining
the masses and phases can be done by fitting a complete simulation of the model to the LHC observations.
However, to discover properties of an unknown model, one would like to
study mass determination tools that are mostly model-independent\footnote{By model-independent, we mean techniques that do not rely on the cross section's magnitude and apply generally to broad classes of models.}.
Astrophysical dark-matter observations suggests the lightest new particle states be stable and leave the detector unnoticed.
The dark-matter's stability suggests the new class of particles be produced  in pairs.
The pair-produced final-state invisible particles, if produced, would lead to large missing transverse momentum.
Properties of hadron colliders make the task of mass determination of the dark-matter particles more difficult because we do not know the rest frame or energy of the initial parton collision.
We review techniques based on combining kinematic edges, and others based on assuming enough on-shell states in one or two events such that the masses can be numerically solved.
We continue with the transverse mass $M_{T}$ which forms a lower-bound on the mass of the new particle which decays into a particle with missing transverse momentum, and was used to measure the mass of the $W$.  The transverse mass $M_T$ has a symmetry under boosts along the beam line.  Next we review $M_{T2}$ \cite{Lester:1999tx}\cite{Barr:2003rg} which generalizes this to the case where there are two particles produced where each of them decays to invisible particle states.

Chapter \ref{ChapterMT2onCascadeDecays} begins the original contributions of this thesis towards more accurate and robust mass determination techniques in the presence of invisible new particle states.
We begin by introducing novel ways to use the kinematic variable $M_{T2}$ to discover new constraints among the masses of new states. We also discuss the relationship of $M_{T2}$ to a recent new kinematic variable called $M_{CT}$.  This work was published by the author in Ref.~\cite{Serna:2008zk}.

Chapter \ref{ChapterM2Cdirect} introduces a new kinematic variable $M_{2C}(M_{-})$.  Most model-independent mass determination approaches succeed in constraining the mass difference $M_{-}=M_Y-M_N$ between a new states $Y$ and the dark-matter particle $N$, but leave the mass scale $M_{+}=M_Y+M_N$ more poorly determined.  $M_{2C}(M_{-})$ assume the mass difference, and then provides an event-by-event lower bound on the mass scale. The end-point of the  $M_{2C}$ distribution gives the mass $M_Y$ which is equivalent to the mass scale if $M_{-}$ is known.   In this chapter we also discover a symmetry of the $M_{2C}$ distribution which for direct $Y$ pair production, makes the shape of the distribution entirely independent of the unknown collision energy or rest frame.  Fitting the shape of the distribution improves our accuracy and precision considerably.
We perform some initial estimates of the performance with several simplifying assumptions and find that with $250$ signal events we are able to determine $M_N$ with a precision and accuracy of $\pm 6$ GeV for models with $M_Y+M_N/(M_Y-M_N) \approx 3$.  This chapters results were published in Ref~\cite{Ross:2007rm}.

Chapter \ref{ChapterM2CwUTM}, which is based on work published by
the author with Alan Barr and Graham Ross in Ref.~\cite{Barr:2008ba}, extends the $M_{2C}$ kinematic variable in two ways.
First we discover that in the presence of large upstream transverse momentum (UTM), that we are able to bound the mass scale from above.
This upper bound is referred to as $M_{2C,UB}$.
Here we perform a more realistic case study of the performance including backgrounds, combinatorics, detector acceptance, missing transverse momentum cuts and energy resolution.  The case study uses data provided by Dr Alan Barr created with \herwig\ Monte Carlo generator \cite{Corcella:2002jc,Moretti:2002eu,Marchesini:1991ch} to simulate the experimental data.
The author wrote Mathematica codes that predict the shape of the distributions built from properties one can measure with the detector.
We find the mass $M_N$ by fitting to the lower-bound distribution $M_{2C}$ and the upper-bound distribution $M_{2C,UB}$ distribution shapes.  Our simulation indicates that with $700$ events and all anticipated effects taken into consideration that we are able to measure the $M_N$ to $\pm 4$ GeV for models with $M_Y+M_N/(M_Y-M_N) \approx 3$.
This indicates that the method described is as good as, if not better than, the other known kinematic mass determination techniques.

Chapter \ref{ChapterM3C}, which is based on work published by the author with Alan Barr and Alex Pinder in Ref.~\cite{Barr:2008hv}, extends the constrained mass variables to include two mass differences in the variable $M_{3C}$.
We discuss the properties of the $M_{3C}$ distribution.
We observe that although the technique is more sensitive to energy resolution errors, we are still able to determine both the mass scale and the two mass differences as good if not better than other known kinematic mass determination techniques.  For SPS 1a we forecast determining the LSP mass with $\pm 2.4$ GeV with about $3600$ events.

Chapter \ref{ChapterConclusionsThesis} concludes the thesis.  We predict the wino's mass will have the opposite sign of the gluino's mass.  We develop new techniques to measure the mass of dark-matter particles produced at the LHC.  Our techniques work with only two or three new particle states, and have precision and accuracy as good or better than other known kinematic mass determination techniques.


To facilitate identifying
the original ideas contained in this thesis for which the author is responsible, we list them here explicitly along with the location in the thesis which elaborates on them:
 \begin{itemize}
   \item {\bf Chapter \ref{ChapterUnificationAndFermionMassStructure}} The updated values of the strong coupling constant, top quark mass, and strange quark mass lead to quantitative disagreement with the Georgi-Jarslkog mass relationships unless one uses $\tan \beta$ enhanced threshold corrections and fix the gluino's mass to be opposite sign of the wino's mass published in Ref~\cite{Ross:2007az}.
   \item {\bf Chapter \ref{ChapterMT2onCascadeDecays}} The relationship between Tovey's $M_{CT}$ and Lester, Barr, and Summer's $M_{T2}$ variables published in Ref~\cite{Serna:2008zk}.
   \item {\bf Chapter \ref{ChapterMT2onCascadeDecays}} Using $M_{T2}$ along intermediate stages of a symmetric cascade decay to discover new constraints among the masses published in Ref~\cite{Serna:2008zk}.
   \item {\bf{ Chapter \ref{ChapterM2Cdirect} and \ref{ChapterM2CwUTM}}} The definition and use of the variable $M_{2C}$  to determine mass of dark-matter-like new particle states published in Ref~\cite{Ross:2007rm} and \cite{Barr:2008ba}.
   \item {\bf{Chapter \ref{ChapterM2CwUTM}}} The ability to obtain an event-by-event upper bound on the mass scale given the mass difference when the mass difference is known and the event has large upstream transverse momentum published in Ref~\cite{Barr:2008ba}.
   \item {\bf{Chapter \ref{ChapterM3C}}} Complications and advantages to the logical extension variable $M_{3C}$ and their use as a distribution which is  published in Ref~\cite{Barr:2008hv}.
   \item A set of Mathematica Monte Carlo simulations, and $M_{T2}$ calculators used to test the above contributions for $M_{2C}$, $M_{3C}$ and some C++ codes for $M_{2C}$ and $M_{2C,UB}$ which has not yet been published.  The author will be happy to share any codes related to this thesis upon request.
 \end{itemize}

\chapter{Mass Determination in the Standard Model and Supersymmetry}
\label{ChapterPastCurrentMassPredictions}

\section*{Chapter Overview}

This chapter highlights three pillars of past successful new-particle state prediction and mass determination, and it shows how these pillars are precedents for the toolbox and concepts employed in Chapter \ref{ChapterUnificationAndFermionMassStructure}'s contributions.
The three pillars on which rest most demonstrated successful theoretical new-particle-state predictions and mass determinations in particle physics are:
(i) Symmetries, (ii)  Broken symmetries, and (iii)  Fine Tuning of Radiative Corrections.  We use the narrative of these pillars to review and introduce the
Standard Model and supersymmetry.

The chapter is organized as follows. Section \ref{SecPastMassDetermination} gives a few historical examples of these pillars: the positron's mass from Lorentz symmetry and from fine tuning, the $\Omega^{-}$'s mass from an explicitly broken flavor symmetry, the charm quark's mass prediction from fine tuning, and
the $W^\pm$ and $Z^o$ masses from  broken $SU(2)_L \times U(1)_Y$ gauge symmetries of the Standard Model.  These historical examples give us confidence that these three pillars hold predictive power and that future employments of these techniques may predict new particle states and their masses.
Section \ref{SecDarkMatterEvidence} introduces astrophysical observation of dark-matter which suggests that nature has a stable, massive, neutral particle that has not yet been discovered.
Section \ref{SecSupersymmetryReviewAndIntro} introduces key features of Supersymmetry, a model with promising potential to provide a dark-matter particle and simultaneously address many other issues in particle physics.  We discuss reasons why Supersymmetry may describe nature, observational hints like the top-quark mass, anomaly in the muon's magnetic moment $(g-2)_\mu$, and gauge coupling unification.  We also review the classic $SU(5)$ Georgi-Jarlskog mass relations and $\tan \beta$-enhanced SUSY threshold corrections.
Chapter \ref{ChapterUnificationAndFermionMassStructure} will discuss how updated low-energy data and the Georgi-Jarlskog mass relationships may provide a window on predicting mass relationships of the low-energy supersymmetry.

\section{Past Mass Determination and Discovery of New Particle States}
\label{SecPastMassDetermination}

\subsection{Unbroken Lorentz Symmetry: Positron Mass}
\label{SubSecPositronMass}
Our first example uses Lorentz symmetry to predict the existence of and mass of the positron.
Lorentz symmetry refers to the invariance of physics to the choice of inertial frame of reference.  The Lorentz group, which transforms vectors from their expression in one inertial frame to another, is generated by the matrix $M^{\alpha \beta}$.  For a spin-1 object (single index 4-vector), the group elements, parameterized by a antisymmetric $4 \times 4$ matrix $\theta_{\mu \nu}$, are given by
 \begin{equation}
 \Lambda (\theta ) = \exp \left( i M^{\mu \nu} \theta_{\mu \nu} \right)
 \end{equation}
where the antisymmetric matrix of generators $M^{\mu \nu}$ satisfies
 \begin{equation}
   [ M^{\mu \nu}, M^{\rho \sigma} ] = i ( g^{\mu \rho} M^{\nu \sigma} -  g^{\nu \rho} M^{\mu \sigma} - g^{\mu \sigma} M^{\nu \rho} +  g^{\nu \sigma} M^{\mu \rho} ) \label{EqLorentzGroup}
 \end{equation}
and $g^{\mu \nu} = {\rm{diag}}(1,-1,-1,-1)$ is the Lorentz metric.  We emphasize that
each entry of $M^{\mu \nu}$ is an operator (generator), where each entry of $\theta_{\mu \nu}$ is a number.
The generators of the \Poincare\ symmetry are the generators of the Lorentz symmetry
supplemented with generators for
space-time translations $P_\mu$ satisfying $[P_\mu , P_\nu ] = 0 $ and
  \begin{equation}
  [ M_{\mu \nu}, P_\rho ] = -i( g_{\mu \rho} P_\nu -  g_{\nu \rho} P_\mu).
  \label{EqMGeneratorsBoosts}
  \end{equation}
Supersymmetry, introduced later in this thesis, is a generalization of the \Poincare\ symmetry.

In 1928 designing a theory with Lorentz symmetry was a major goal in early quantum mechanics. As Dirac described in his original paper, two problems persisted, (1) preservation of probability in quantum mechanics requires a first order linear equation for time evolution and (2) the presence of positive and negative energy solutions.  The Klein-Gordon equation is
 \begin{equation}
  ( D_\mu D^\mu - m^2 ) \phi = 0
 \end{equation}
where $D_\mu = \partial_\mu - i e A_\mu$.  It is invariant under Lorentz transformations  but suffers from both problems:
  the equation is second order in $\partial_t$, and it has solutions proportional to $\exp (i \omega t)$ and $\exp( -i \omega t)$.

Dirac's 1928 paper on the Dirac equation \cite{Dirac:1928hu} claims only to solve problem (1). Because problem (2) was not solved, Dirac claimed ``The resulting theory is therefore still only an approximation".  However, the paper shows how to do Lorentz invariant quantum mechanics of spin $1/2$ fields.
Although in different notation, Dirac's paper showed that if one has a set of four matrices $\gamma^\mu$ that satisfy
  \begin{equation}
   \gamma^\mu \gamma^\nu + \gamma^\nu \gamma^\mu = 2 g^{\mu \nu}
  \end{equation}
then the system of equations
 \begin{equation}
 ( i \gamma^\mu D_\mu - m) \psi = 0
 \end{equation}
transforms covariantly under Lorentz transformations if
we define a new class of transforms $\Lambda_{1/2} = \exp ( i \theta_{\mu \nu} (M_{1/2})^{\mu \nu})$ where the group is generated by
\begin{equation}
 (M_{1/2})^{\mu \nu} = \frac{i}{2} (\gamma^\mu \gamma^\nu - \gamma^\nu \gamma^\mu).
 \label{EqGeneratorSpinHalf}
\end{equation}
The new generators $M_{1/2}$ satisfy Eq(\ref{EqLorentzGroup}) so they form a representation of the Lorentz group specific to spin $1/2$. The Dirac field transforms as  $\psi' = \Lambda_{1/2} \psi$
and the $\gamma^\mu$ matrices transform as $\Lambda^\mu_{\ \nu} \gamma^\nu = \Lambda^{-1}_{1/2} \gamma^\mu \Lambda_{1/2}$.

Dirac interprets that the negative-energy solutions to the equations
will behave as if they have the opposite charge but same mass in the presence of a background electromagnetic field.  The formation of a relativistic quantum mechanics which still possesses negative energy solutions suggests that this alternative solution with negative energy and the opposite charge and same mass may be a prediction of relativistic quantum mechanics.  Indeed, Anderson observed the negative energy version of the electron in 1933 \cite{PhysRev.43.491} \footnote{Anderson's first paper makes no reference to Dirac.  However, he does introduce the term positron and suggests renaming the electron the negatron.}.   The positron has the opposite charge of the electron but the same mass.

Today the Dirac equation is interpreted in terms of quantum field theory where all energies are always considered positive.  With hindsight we see that Dirac's motivation was partially wrong, and the Klein-Gordon equation provides just as good an argument for antiparticles.

\subsection{Renormalization, Fine Tuning and the Positron}
\label{SubSecFineTuningPositron}

The next example is not the historical origin of the positron's prediction,
but could have been
if physicists in the early 1900s understood the world with today's effective field theory tools.
The electron's self-energy poses another problem which effective quantum field theory and
the existence of the positron solve \cite{NimaFineTune}.

If we divide up the electron's charge into two pieces\footnote{The choice of two pieces is arbitrary and simplifies the calculations to illuminate the fine-tuning concept trying to be communicated.}, we ask how much
energy is stored in the system in the process of bringing the two halves together
from infinity?
At what distance apart will the
`self energy' of the electron be equal to the mass energy of the electron?
This is approximately the classical electron radius\footnote{The classical electron radius is $4 \times$ this value.}, and the answer is around ten times bigger than atomic nuclei at around $r_e = 7 \times 10^{-14}$ m.  Electron-electron scattering can probe closer than this for collisions with $\sqrt{s} \approx 2$ MeV.
Also electrons are emitted from nuclei during $\beta$ decay suggesting the electron must be smaller than the nucleus.
We now break the mass up into two quantities: the bare mass $m_{e\,o}$ and the `self-energy' mass $\delta m$ with the observed mass equalling $m_e = m_{e\,o} + \delta m_e$.
Phrasing self energy in terms of a cut-off one finds
 \begin{equation}
   \delta m_e \approx  \frac{\Lambda}{4}
 \end{equation}
and the cut off $\Lambda$ indicates the energy needed to probe the minimum distance within which two pieces of the electrons mass and charge must be contained.
At the Plank scale, this requires a cancelation between the bare mass $m_{e,o}$ and the
self energy $\delta m_e$ to more than $22$ decimal places to end up with the observed mass $m_e=0.511 \MeV$!
Fine tuning is where large cancelations between terms are needed to give an observable quantity.
This large cancelation could be perceived as an indication of missing physics below the
scale of about $2 \MeV$ where we would have had a cancelation of the same size as the observable quantity $m_e$.

The effective quantum field theory used to describe electromagnetism introduces the positron with the same mass as the electron.
The positron acts to moderate this ``self-energy".
In QFT, the self energy of the electron is partly due to electron interaction with electron, but also partly due to electron interaction with a virtual positron.
This is because the two interaction vertices $x$ and $y$ are integrated over $\int d^4 x d^4 y$ so the interaction
The resulting self energy, in terms of a cut off $\Lambda$, is
 \begin{equation}
  \frac{\delta m_e}{m_e} \approx  \frac{\alpha}{4 \pi} \ln \frac{\Lambda^2}{m_e^2}
 \end{equation}
where $\alpha=e^2/4 \pi$.
Weisskopf and Furry were the first to discover that quantum field theory with both positrons and electrons leads to only a logarithmic divergence  \cite{CERNCourierWeisskopfSpecial,PhysRev.56.72}.
Now we see that taking the cut off $\Lambda$ to the Plank scale only gives a $6 \%$ correction.
There is no longer a cancelation of two large terms, an issue solved by introducing new physics below the scale at which the low-energy effective theory became fine tuned.

\subsection{Broken Flavor Symmetry: $\Omega^{-}$ Mass}
\label{SubSecOmegaMinusMass}

The next example brings us to the early 1960s when Murray Gell-mann \cite{GellMann:1961ky} and Yuval Ne'eman \cite{Ne'eman:1961cd}
\footnote{Yuval Ne'eman, like the author of this thesis, was a member of the military while studying for his PhD \cite{NeemanRembered}.} were both studying broken $SU(3)$ flavor symmetry as a way to understand the zoo of particles
being discovered in the 1950s and 1960s.  Their study led to the theoretical determination of mass of the $\Omega^{-}$ baryon before it was observed.

To place the $SU(3)$ flavor symmetry in context, we begin first with the $SU(2)$ isospin symmetry.  Isospin symmetry postulates that protons $p$ and neutrons $n$ are indistinguishable if one ignores electromagnetism.  Therefore, the equations and forces governing $p$ and $n$ interactions should have a symmetry on rotating the proton's complex field $p$ into the neutron's complex field $n$ with an $SU(2)$ rotation.
The symmetry is explicitly broken by the electromagnetic force and to a lesser extent the quark masses, and a special direction is singled out.
We can label the eigenvalues along this broken direction.
The eigenvalues of $I_3$ isospin generator label the states along the broken isospin axis:
$I_3=1/2$ denotes a proton, and $I_3=-1/2$ denotes a neutron.  Today we can trace the isospin symmetry to the presence of up and down quarks with nearly the same mass.  Isospin symmetry is broken both by electromagnetism and by the up and down quark mass difference.

Next we wish to understand how $SU(3)_f$ symmetry predicted the mass of the $\Omega^{-}$. The $SU(3)$ flavor symmetry is an extension of isospin symmetry.  It can be understood in modern language as the existence of three light quarks.
The symmetry is broken because the strange quark is considerably more massive ( $\approx 103 \pm 12 \GeV$ ) than the up and down quark ( $m_u \approx 2.7 \pm 0.5$ GeV, and $m_d=5.7 \pm 0.5$ GeV)
\footnote{Quark masses are very difficult to define because they are not free particles.  Here we quote the current quark masses
at an $\overline{MS}$ renormalization scheme scale of $\mu=2 \GeV$ as fit to observations in Chapter \ref{ChapterUnificationAndFermionMassStructure}.}.
The group $SU(3)$ has isospin $SU(2)$ as a supgroup so $I_3$ remains a good quantum number; in addition $SU(3)$ states are also specified by the hypercharge $Y$.
Quarks and anti-quarks are given the following charge assignments ($I_3,Y$): $u=(1/2,1/3)$, $d=(-1/2,1/3)$, $s=(0,-2/3)$, $\bar u =(-1/2,-1/3)$, $\bar d=(1/2,-1/3)$, $\bar s=(0,2/3)$.
Representations of $SU(3)$ are formed by tensors of the fundamental its conjugate with either symmetrized or antisymmetrized indices and with traces between upper and lower indices removed.
Representations are named by the number of independent components that tensor possesses and shown by bold numbers like {\bf{3}}, {\bf{8}}, {\bf{10}}, {\bf{27}} etc.

\begin{figure}
\centerline{\includegraphics[width=4in]{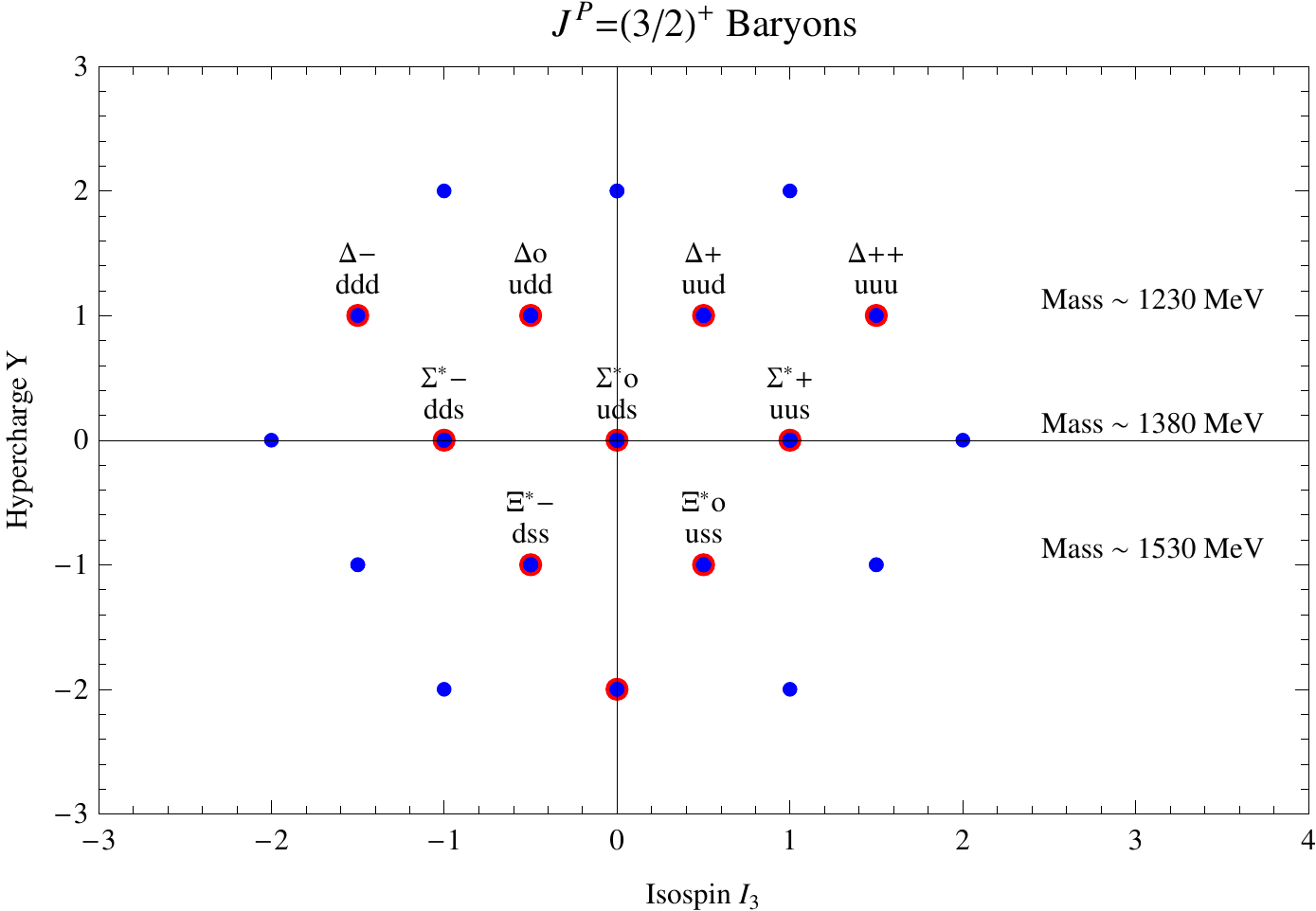}}
\caption{\label{FigOmengaMinusDiscovery} The states of the $SU(3)$ {\bf{10}} of $J^P=3/2^+$ baryons in 1962.  Also shown is the $SU(3)$ {\bf{27}}.}
\end{figure}

Gell-mann and Ne'eman did not know what representation of $SU(3)$ was the correct one to describe the many baryons and mesons being discovered; to describe the spin $3/2$ light baryons, they were each considering the {\bf{10}} and the {\bf{27}}.
The representation {\bf{10}} is formed by $B^{abc}$ where $a$,$b$, and $c$ are indexes that run over $u$, $d$, and $s$, and where the tensor $B^{abc}$ is symmetric on interchanges of the three indices.  The states are displayed as the red dots in Fig.~\ref{FigOmengaMinusDiscovery} where the conserved quantum numbers of $I_3$ is plotted against hypercharge $Y$.  The {\bf{27}} is given by a tensor $B^{ab}_{\ \ cd}$ where $(ab)$ is symmetrized, $(cd)$ is symmetrized, and the $9$ traces $B^{ac}_{\ \ cd}=0$ are removed.  The {\bf{27}} is shown by the smaller blue dots.
The particles and the observed masses as of 1962 are shown in the Fig~\ref{FigOmengaMinusDiscovery}.

In July 1962, both Gell-mann and Ne'eman went to the 11$^{\rm{th}}$ International Conference on High-Energy Physics at {CERN}.
Ne'eman learned from
Sulamith and Gerson Goldhaber (a husband and wife pair)
that $K^{+}$(u$\bar{\rm{s}}$,$I_3=+1/2$,$Y=1$)  vs $N$ (ddu,$I_3=-1/2$,$Y=1$) scattering did not lead to a resonance at $(I_3=0,Y=2)$ with a mass range near $1000$ MeV as one would expect if the pattern followed was that of the {\bf{27}} \cite{Fraser:1998ih}.
Gell-mann had also learned of the Goldhabers' negative result, now known as the
Goldhaber gap. During the discussion after the next talk at the conference, Gell-mann and Ne'eman both planned on announcing their observation; Gell-mann was called on first, and announced that one should
discover the $\Omega^{-}$ with a mass near $1680$ GeV.
The $\Omega^{-}$ was first seen in 1964 at Brookhaven \cite{PhysRevLett.12.204} with a mass of $M_{\Omega^-}=1686 \pm 12$ MeV. Amazingly the spin of $\Omega^{-}$ was not experimentally verified until 2006 \cite{Aubert:2006dc}.

The broken $SU(3)_F$ flavor symmetry successfully predicted and explained the masses of the $\Omega^{-}$ and many other baryons and mesons.  The general formula
for the masses in terms of the broken symmetry was developed both by Gell-mann and by
Okubo and is known as the Gell-mann-Okubo mass formula \cite{Okubo:1961jc}.

Despite this success (and many others) the quark-model theoretical description used here is still approximate; the strong force is only being parameterized.  The quarks here classify the hadrons, but the masses of the quarks do not add up to the mass of the hadron.  Quark masses that are thought of as being about $1/3$ the mass of the baryon are called ``constituent quark masses".  These are not the quark masses that are most often referenced in particle physics.
A consistent definition of quark mass and explanation of how these quarks masses relate to hadron masses will wait for the discovery of color $SU(3)_c$.  In $SU(3)_c$ the masses of the quarks are defined in terms of Chiral Perturbation Theory ($\chi$PT) and are called ``current" quark masses.   We will be using current quark masses as a basis for the arguments in Chapter \ref{ChapterUnificationAndFermionMassStructure}.

Another important precedent set here is that of arranging the observed particles in a representation of a larger group that is broken.  This is the idea behind both the Standard Model and the Grand Unified Theories to be discussed later.  The particle content and the forces are arranged into representations of a group.  When the group is broken, the particles distinguish themselves as those we more readily observe.


\subsection{Charm Quark Mass Prediction and Fine Tuning of Radiative Corrections}
\label{SubSecGIM}

In the non-relativistic quark model, the neutral $K^o$ is a bound state of $d$ and $\bar{s}$ and  $\bar{K}^o$ is a bound state of $s$ and $\bar d$.
The Hamilontian for this system is given by
\footnote{In this Hamilontian, we're neglecting the {CP} violating features.}
\begin{equation}
 H_{K,\bar K} =  \left( \begin{matrix} K^o & \bar{K}^o \end{matrix} \right)
 \left( \begin{matrix} M^2_o & \delta^2 \cr \delta^2 & M^2_o \end{matrix} \right)
 \left( \begin{matrix} K^o \cr \bar{K}^o \end{matrix} \right).
\end{equation}
A non-zero coupling $\delta$ between the two states leads to two distinct
mass eigenstates.
For $\delta << M_o$, the two mass eigenstates have almost equal mixtures of $K^o$ and $\bar{K}^o$ and
are called $K_1$ and $K_2$.
Experimentally the mass splitting between the two eigenstates is
$M_{K_2}-M_{K_1} = 3.5 \times 10^{-6}$ eV.

During the 1960s, a combination of chiral four-Fermi interactions with the approximate chiral symmetry $SU(3)_{L} \times SU(3)_{R}$ of the three lightest quarks was proving successful at describing many hadronic phenomena; however, it predicted
$\delta$ was non zero.
Let's see why. The effective weak interaction Lagrangian that was successful in atomic decays and Kaon decays \cite{WeinbergQFT}\cite{ChengLi} was given by
 \begin{equation}
  {\mathcal{L}} = \frac{2\,G_F}{\sqrt{2}} J^\mu J^\dag_\mu \ \ \ {\rm{where}} \ \ \
   J^\mu  =  J^\mu_{(L)} + J^\mu_{(H)}
 \end{equation}
and
 \begin{eqnarray}
   J_{(L)}^\mu & = & \bar{\nu_e} \gamma^\nu \frac{1}{2}(1-\gamma^5) e + \bar{\nu_\mu} \gamma^\nu \frac{1}{2}(1-\gamma^5) \mu \\
   J_{(H)}^\mu & = & \bar{u} \gamma^\nu \frac{1}{2}(1-\gamma^5)\, d\, \cos \theta_c +
   \bar{u} \gamma^\nu \frac{1}{2}(1-\gamma^5)\, s\, \sin \theta_c.
 \end{eqnarray}
and $\theta_c$ is the Cabibbo angle.
Using a spontaneously broken $SU(3)_L \times SU(3)_R$ \cite{Gaillard:1974hs} one can calculate the loops connecting $u \bar{s}$ to $\bar{u} s$.  These loops are responsible for the Kaon mass splitting and give
 \begin{equation}
 M_{K_2} - M_{K_1} = \frac{\delta^2}{M_K} \approx \frac{\cos^2 \theta_c \sin \theta_c }{4 \pi^2} f^2_K M_K G_F (G_F \Lambda^2) \label{EqNaiveMassSplitting}
 \end{equation}
where $\Lambda$ is the cut off on the divergent loop and $f_K \approx 180 \GeV$ \cite{Lee:1968ks}\cite{Glashow:1970gm}\cite{Gaillard:1974hs}\cite{WeinbergQFT}.  Using this relation, the cut-off cannot be above $2$ GeV
without exceeding the observed Kaon mass splitting.
There are higher order contributions, each with higher powers of $G_F \Lambda^2$,
that may cancel to give the correct answer.
Indeed Glashow, Ilioupoulos, and Maiani (GIM)
\cite{Glashow:1970gm} observe
in a footnote ``Of course there is no reason that one cannot exclude \emph{a priori}
the possibility of a cancelation in the sum of the relevant perturbation expansion in the limit $\Lambda \rightarrow \infty$".  The need for an extremely low cut off was a problem of fine tuning and naturalness with respect to radiative corrections.

This solution to the unnaturally low scale of $\Lambda$ suggested by the mass splitting $M_{K_2}-M_{K_1}$ holds foreshadowing for supersymmetry
\footnote{I have learned from Alan Barr that Maiani also sees the GIM mechanism as a precedent in favor of supersymmetry.}.
GIM proposed a new broken quark symmetry $SU(4)$ which required the existence of
the charm quark, and also discuss its manifestation in terms of a massive intermediate vector boson $W^+$, $W^-$.  They introduce a unitary matrix $U$, which will later
be known as the $CKM$ matrix.
We place group the quarks into groups: $d^i={d,s, \ldots}$ and $u^i = {u,c,\ldots}$; then the matrix $U_{ij}$ links up-quark $i$ to down-quark $j$ to the charged current
 \begin{equation}
  (J_{(H)})^\mu = \bar{u}^i \gamma^\mu \frac{1}{2} (1-\gamma^5) d^j U_{ij}.
 \end{equation}
The coupling of the Kaon to the neutral current formed by the pair of $W^+$ with $W^-$.  In the limit of exact $SU(4)$ (all quark masses equal) the coupling of the Kaon to the neutral current is proportional to ${\mathcal{A}} \propto (\sum_{j=u,c,\ldots} U_{dj} U^\dag_{j s}) = \delta_{ds} = 0$.  The coupling $\delta^2$ between $K^o$ and $\bar{K}^o$ is proportional to ${\mathcal{A}}^2$.
This means that in the limit of $SU(4)$ quark symmetry, there would be no coupling to enable $K_1$ and $K_2$ mass splitting.

However, the observed $K_1$ $K_2$ mass splitting is non-zero, and $SU(4)$ is not an exact symmetry; it is broken by the quark masses.  The mass splitting is dominated by the mass of the new quark $m_c$.  GIM placed a limit on $m_c \lessapprox 3 \GeV$.
One might think of the proposed $SU(4)$ symmetry becoming approximately valid above scale $\Lambda \approx m_c$.  The new physics (in this case the charm quark) was therefore predicted and found to lie at the scale where fine tuning could be avoided.

\subsection{The Standard Model: Broken Gauge Symmetry: $W^\pm$, $Z^o$ Bosons Mass}
\label{SubSecStandardModel}

The Standard Model begins with a gauge symmetry $SU(3)_c \times SU(2)_L \times U(1)_Y$ that gets spontaneously broken to $SU(3)_c \times U(1)_{EM}$. For a more complete pedagogical introduction, the author
 has found useful Refs.~\cite{Borodulin:1995xd,Srednicki:2007qs,Ramond:1999vh} and the
 PDG \cite[Ch 10]{PDBook2006}.

The field content of the Standard Model with its modern extensions for our future use is given in Table \ref{TableSMFields}. The $i$ indexes the three generations.
The fermions are all represented as 2-component Weyl left-handed transforming states\footnote{\label{FootnoteLorentzNotation}
The $(2,0)$ is the projection of Eq(\ref{EqGeneratorSpinHalf}) onto left handed states with $P_L=1/2(1-\gamma^5)$. In a Weyl basis, Eq(\ref{EqGeneratorSpinHalf})
is block diagonal so one can use just the two components that survive $P_L$. If
$e^c$ transforms as a $(2,0)$ then  $i \sigma^2 (e^c)^*$ transforms as $(0,2)$.}.
The gauge bosons do not transform covariantly under the gauge group, rather they transform as connections \footnote{\label{ConnectionTransform}The gauge fields transform as connections under gauge transformations: $W'_\mu=U W_\mu U^{-1} + (i/g) U \partial_\mu U^{-1}$.}.
\begin{table}
  \centering
  \begin{tabular}{|c|c|c|c|c|}
    \hline
  Field    & Lorentz$^{\rm{\ref{FootnoteLorentzNotation}}}$ & $SU(3)_c$ & $SU(2)_L$ & $U(1)_Y$ \\   \hline
  $(L)^i$   & $(2,0)$ & $1$     &  $2$      & $-1$ \\
  $(e^c)^i$ & $(2,0)$ & $1$     &  $1$      & $2$ \\
  $(\nu^c)^i$ & $(2,0)$ & $1$     &  $1$      & $0$ \\
  $(Q)^i$   & $(2,0)$ & $3$     &  $2$      & $1/3$ \\
  $(u^c)^i$ & $(2,0)$ & $\bar 3$     &  $1$      & $-4/3$ \\
  $(d^c)^i$ & $(2,0)$ & $\bar 3$     &  $1$      & $2/3$ \\
  $H$   &    $1$        &  $1$     &  $2$      & $+1$ \\
  $B_\mu$   &    $4$        & $1$     &  $1$      & $0^{^\ddagger}$ \\
  $W_\mu$   &    $4$        & $1$     &  $3^{^\ddagger}$      & $0$ \\
  $G_\mu$   &    $4$        & $8^{^\ddagger}$     &  $1$      & $0$ \\
    \hline
  \end{tabular}
  \caption{Standard Model field's transformation properties. $^\ddagger$ Indicates the field does not transform covariantly but rather transform as a connection$^{\rm{\ref{ConnectionTransform}}}$. }\label{TableSMFields}
\end{table}
In 1967 \cite{Weinberg:1967tq}, Weinberg set the stage with a theory of leptons
which consisted of
the left-handed leptons which form $SU(2)$ doublets $L^i$, the right-handed charged leptons which form $SU(2)$ singlets $(e^c)^i$, and the Higgs fields $H$ which is an $SU(2)$ doublet.
Oscar Greenberg first proposed 3-color internal charges of $SU(3)_c$ in 1964 \cite{Greenberg:1964pe}\footnote{Greenberg, like the author, also has ties to the US Air Force. Greenberg served as a Lieutenant in the {USAF} from 1957 to 1959. A discussion of the history of $SU(3)_c$ can be found in Ref.~\cite{Greenberg:2008fs}.}. It was not until Gross and Wilczek and Politzer discovered asymptotic
freedom in the mid 1970s that $SU(3)_c$ was being taken seriously as a theory of the strong nuclear force \cite{Gross:1973ju}\cite{Politzer:1973fx}.
The hypercharge in the SM is
not the same hypercharge as in the previous subsection \footnote{In the SM, the $d$ and $s$ (both left-handed or right-handed) have the same hypercharge, but in flavor $SU(3)_f$ (\ref{SubSecOmegaMinusMass}) they have different hypercharge. }.  The $U(1)_Y$ hypercharge assignments $Y$ are designed to satisfy $Q=\sigma^3/2 + Y/2$ where $Q$ is the electric charge operator and $\sigma^3 / 2$ is with respect to the $SU(2)_L$ doublets (if it has a charge under $SU(2)_L$ otherwise it is $0$).

The gauge fields have the standard Lagrangian
\begin{equation}
  {\mathcal{L}}_{W,B,G} = \frac{-1}{4}  B_{\mu \nu} B^{\mu \nu}
 + \frac{-1}{2} {\rm{Tr}} W_{\mu \nu} W^{\mu \nu}
 + \frac{-1}{2} {\rm{Tr}} G_{\mu \nu} G^{\mu \nu}
\end{equation}
where $B_{\mu\nu} = \frac{i}{g'} [D_\mu , D_\nu]$ when $D_\mu$ is the covariant derivative with respect to $U(1)$: $D_\mu = \partial_\mu - i g' B_\mu$.
Likewise $W_{\mu\nu} =  \frac{i}{g} [D_\mu , D_\nu]$ with $D_\mu = \partial_\mu - i g \frac{\sigma^a}{2} W^a_\mu$ and $G_{\mu\nu} =  \frac{i}{g_3} [D_\mu , D_\nu]$ with $D_\mu = \partial_\mu - i g_3 \frac{\lambda^a}{2} G^a_\mu$ with $\sigma/2$ and $\lambda/2$  the generators of $SU(2)$ and $SU(3)$ gauge symmetry respectively. The $1/g$ factors in these definitions give the conventional normalization of the field strength tensors.  At this stage, $SU(2)$ gauge symmetry prevents terms in the Lagrangian like $\frac{1}{2} M^2_W W_\mu W^\mu$ which would give $W$ a mass.

The leptons and quarks acquire mass through the Yukawa sector given by
 \begin{equation}
   {\mathcal{L}}_Y = Y^e_{ij} L^i (i\sigma^2 H^*) (e^c)^j + Y^d_{ij} Q^i (i\sigma^2 H^*) (d^c)^j + Y^u_{ij} Q^i H (u^c)^j + h.c. \label{EqYukawaSM}
 \end{equation}
where we have suppressed all but the flavor indices \footnote{Because both
$H$ and $L$ transform as ${\textbf{2}}$ their contraction to form an invariant is done like $L^a H^b \epsilon_{ab}$ where $\epsilon_{ab}$ is the antisymmetric tensor.}.
Because $H$ and $i \sigma_2 H^*$ both transform as a $\textbf{2}$ (as opposed to a $\overline{\textbf{2}}$), these can be used to couple a single Higgs field to both up-like and down-like quarks and leptons.
If neutrinos have a Dirac mass
then Eq(\ref{EqYukawaSM}) will also have a term $Y^{\nu}_{ij} L^i H (\nu^c)^j$.  If the right-handed neutrinos have a Majorana mass, there will also be a term $M_{R_{jk}} (\nu^c)^j (\nu^c)^k$.

With these preliminaries, we can describe how the $W$ boson's mass was predicted.
The story relies on the Higgs mechanism that enables a theory with an exact gauge symmetry to give mass to gauge bosons in such a way that the gauge symmetry is preserved, although hidden.
The Higgs sector Lagrangian is
\begin{equation}
 {\mathcal{L}}_H = D_\mu H^\dag D^\mu H - \mu^2 H^\dag H - \lambda (H^\dag H)^2
\end{equation}
where the covariant derivative coupling to $H$ is given by
$D_\mu = \partial_\mu - i g \frac{\sigma^a}{2} W^a_\mu - i \frac{g'}{2} B_\mu$.
The gauge symmetry is spontaneously broken if $\mu^2<0$, in which case $H$ develops a vacuum expectation value (VEV) which by choice of gauge can be chosen to be along $\langle H \rangle = (0 , v/\sqrt{2})$.
The gauge boson's receive an effective mass due to the coupling between the vacuum state of $H$ and the fluctuations of $W$ and $B$:
\begin{equation}
 \langle {\mathcal{L}}_H \rangle = \frac{g^2 v^2}{8} \left( (W^1)_\mu^2
+(W^2)_\mu^2  \right) + \frac{v^2}{8} \left(g' B_\mu + g (W^3)_\mu \right)^2.
\end{equation}
From this expression, one can deduce  $W^1_\mu$ and $W^2_\mu$ have a mass of $M^2_W=g^2 v^2 / 4$, and the linear combination $Z_\mu=(g^2+g'^2)^{-1/2}(g' B_\mu + g (W^3)_\mu)$ has a mass
$M^2_Z=(g^2+g'^2) v^2 /4$.  The massless photon $A_\mu$ is given by the orthogonal combination
$A_\mu=(g^2+g'^2)^{-1/2}(g B_\mu - g' (W^3)_\mu)$.  The weak mixing angle $\theta_W$ is given by $\sin \theta_W = g'(g^2+g'^2)^{-1/2}$, and electric charge coupling $e$ is given by $e= g \sin \theta_W$.

Before $W^\pm$ or $Z^o$ were observed, the value of $v$ and $\sin^2 \theta_W$ could be extracted from the rate of neutral current weak processes, and through left right asymmetries in weak process experiments \cite{LlewellynSmith:1981yk}.  At tree-level, for momentum much less than $M_W$, the four-Fermi interaction can be compared to predict $M_W^2   = \sqrt{2} e^2 / (8 G_F  \sin^2 \theta_W)$.  By 1983 when the $W$ boson was first observed, the predicted mass including quantum corrections was given by $M_W = 82 \pm 2.4$ GeV to be compared
with the {UA1} Collaboration's measurement $M_W=81 \pm 5$ GeV \cite{Arnison:1983rp}.
A few months later the $Z^o$ boson was observed \cite{Arnison:1983mk}.
Details of the $W$ boson's experimental mass determination will be discussed in Section \ref{SecWMassDetermination} as it is relevant for this thesis's contributions in
Chapters \ref{ChapterMT2onCascadeDecays} - \ref{ChapterM3C}.

\begin{figure}
\centerline{\includegraphics[width=2in]{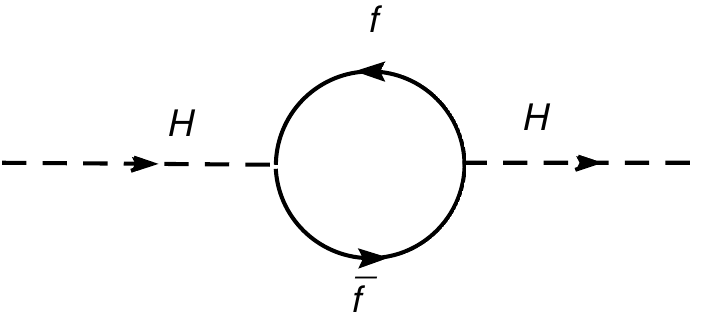}\ \includegraphics[width=2in]{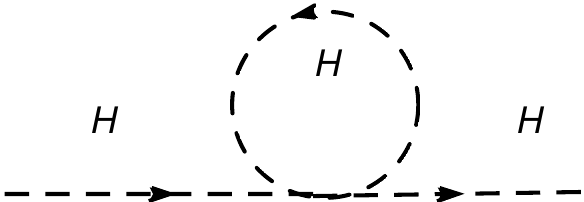}}
\centerline{(a) \ \ \ \ \ \  \  \ \  \ \ \ \ \ \ \ \ \ \ \ \ \ \ \ \ \ \ \ \ \ \ \ (b)}
\caption{\label{FigHiggsSelfEnergy} One Loop Contributions to the Higgs mass
self energy. (a) Fermion loop.   (b) Higgs loop.}
\end{figure}
The Standard Model described in here has one serious difficulty regarding
fine tuning of the Higgs mass radiative corrections.
The Higgs field has two contributions to the mass self-energy shown in
Fig.~\ref{FigHiggsSelfEnergy}.
The physical mass $m_H$ is given by
\begin{equation}
 m^2_H \approx m_{o,H}^2  - \sum_{f=u,d,e}\frac{{\rm{Tr}}(Y^f (Y^f)^\dag)}{8 \pi^2} \Lambda^2 +  \frac{\lambda}{2 \pi^2} \Lambda^2
\end{equation}
where $m_{o,H}$ is the bare Higgs mass, $\Lambda$ is the cut off energy scale, $Y^u$ $Y^d$ $Y^e$ are the three Yukawa coupling matrices, third term is from the Higgs loop.
Assuming the physical Higgs mass $m_H$ is near the electo-weak scale ($\approx 100 \GeV$)\footnote{We use $100$ GeV as a general electroweak scale.  The current fits to Higgs radiative corrections suggest $m_H=95 \pm 35$ GeV \cite[Ch10]{PDBook2008}, but the direct search limits require $m_H \gtrsim  115$ GeV.} and $\Lambda$ is at the Plank scale means that these terms need to cancel to some $34$ orders of magnitude.
If we make the cut off at $\Lambda \approx 1$ TeV the corrections from either
Higgs-fermion loop\footnote{We use the fermion loop because the Higgs loop depends linearly on the
currently unknown quartic Higgs self coupling $\lambda$.} are equal to a physical Higgs mass of $m_H = 100$ \GeV.
Once again there is no reason that we cannot exclude
\emph{a priori} the possibility of a cancelation between terms of this magnitude.  However, arguing that such a cancelation  is unnatural has successfully predicted the charm quark mass in Sec \ref{SubSecGIM}.  Arguing such fine-tuning is unnatural
in the Higgs sector suggests new physics should be present below around $1$ TeV.


\section{Dark Matter: Evidence for New, Massive, Invisible Particle States}
\label{SecDarkMatterEvidence}

Astronomical observations also point to a something missing in the Standard Model.
Astronomers see evidence for `dark matter' in galaxy rotation curves, gravitational lensing, the cosmic microwave background (CMB), colliding galaxy clusters, large-scale structure, and high red-shift supernova.
Some recent detailed reviews of astrophysical evidence for dark-matter
can be found in  Bertone \etal\cite{Bertone:2004pz} and Baer and Tata \cite{Baer:2008uu}.
We discuss the evidence for dark matter below.

\begin{itemize}
\item {\textbf{Galactic Rotation Curves}}
\end{itemize}
Observations of the velocity of stars normal to the galactic plane of the Milky Way led Jan Oort \cite{Oortz1932} in 1932 to observe the need for `dark matter'.
The 1970s saw the beginning of using doppler shifts of galactic centers versus the periphery.
Using a few ideal assumptions, we can estimate
how the velocity should trend with changing $r$.  The $21$ cm HI line allows rotation curves of galaxies to be measured far outside the visible disk of a galaxy \cite{2001ARA&A..39..137S}
To get a general estimate of how to interpret the observations, we assume spherical symmetry and circular orbits.
If the mass responsible for the motion of a star at a radius $r$ from the galactic center is dominated by a central mass $M$ contained in $R \ll r$, then the tangential velocity is $v_T \approx \sqrt{G_N \, M / r}$ where $G_N$ is Newton's constant.
If instead the star's motion is governed by mass distributed with a uniform density $\rho$ then $v_T= r\,\sqrt{4\, G_N\, \rho\, \pi / 3}$.
Using these two extremes, we can find when the mass of the galaxy has been mostly bounded by seeking the distance $r$ where the velocity begins to fall like $\sqrt{r}$.
The observations \cite{2001ARA&A..39..137S,2004IAUS..220..211R} show that for no $r$ does $v_T$ fall as $\sqrt{1/r}$.  Typically $v_T$ begins to rise linearly with $r$ and then for some $r>R_o$ it stabilizes at $v_T = \rm{const.}$.  In our galaxy the constant $v_T$ is approximately $220 {\rm{km}} / {\rm{sec}}$.
The flat $v_T$ vs $r$ outside the optical disk of the galaxy implies $\rho(r) \propto r^{-2}$ for the dark matter.  The density profile of the dark matter near the center is still in dispute.
The rotation curves of the galaxies is one of the most compelling pieces of evidence that
a non-absorbing, non-luminous, source of gravitation permeates galaxies and extends far beyond the visible disk of galaxies.

\begin{itemize}
\item \textbf{Galaxy clusters}
\end{itemize}
In 1937 Zwicky \cite{Zwicky1937} published research showing that galaxies within clusters of galaxies were moving with velocities such that they could not have been bound together without large amounts of `dark matter'.
Because dark matter is found near galaxies, a modified theory of newtonian gravity (MOND)
\cite{Bottema:2002zv} 
has been shown to agree with the galactic rotation curves of galaxies.
A recent observation of two clusters of galaxies that passed through each other, known as the bullet cluster, shows that
the dark matter is separated from the luminous matter of the galaxies\cite{Clowe:2006eq}.
The dark matter is measured through the gravitational lensing effect on the shape of the background galaxies.
The visible `pressureless' stars and galaxies pass right through each other during the collision of the two galaxy clusters.
The dark matter lags behind the visible matter.
This separation indicates that the excess gravitation observed in galaxies and galaxy clusters cannot be accounted for by a modification of the gravitation of the visible sector, but requires a separate `dark' gravitational source not equal to the luminous matter that can lag behind and induce the observed gravitational lensing.
The bullet cluster observation is a more direct, empirical observation of dark matter.

The dark matter seen in the rotation curves and galaxy clusters could not be due to $H$ gas (the most likely baryonic candidate) because it would have been observed in the $21$ cm $HI$ observations and is bounded to compose no more than $1 \%$ of the mass of the system \cite{Spergel:1996hw}.
Baryonic dark matter through MAssive Compact Halo Objects (MACHOs) is bounded to be $\lesssim 40\%$ of the total dark matter \cite[Ch22]{PDBook2008}

\begin{itemize}
\item{\textbf{Not Baryonic: The Anisotropy of the Cosmic Microwave Background and Nucleosythesis}}
\end{itemize}

The anisotropy of the cosmic microwave background radiation provides an strong constraint on both the total energy density of the universe, but also on the dark-matter and the baryonic part of the dark matter.
Here is a short explanation of why.

The anisotropy provides a snapshot of the density fluctuations in the primordial plasma at the time the universe cools enough so that the protons finally `recombine' with the electrons to form neutral hydrogen.  The initial density fluctuations imparted by inflation follow a simple power law.
The universe's equation-of-state shifts from radiation dominated to matter dominated before the time of recombination.  Before this transition the matter-photon-baryon system oscillates between gravitational collapse and radiation pressure.  After matter domination, the dark matter stops oscillating and only collapses, but the baryons and photons remain linked and oscillating.
The peaks are caused by the number of times the baryon-photon fluid collapses and rebounds before recombination.

The first peak of the CMB anisotropy power spectrum requires the total energy density of the universe to be very close to the critical density $\Omega=1$.
This first peak is the scale where the photon-baryon plasma is collapsing following a dark-matter gravitational well, reaches its peak density, but doesn't have time to rebound under pressure before recombination.
This attaches a length scale to the angular scale observed in the CMB anisotropy in the sky.  The red-shift attaches a distance scale to the long side of an isosceles triangle.   We can determine the geometry from the angles subtended at vertex of this isosceles triangle with all three sides of fixed geodesic length. The angle is larger for such a triangle on a sphere than on a flat surface.
By comparing the red-shift, angular scale, and length scale allows one to measure the total spatial curvature of space-time \cite{Hartle:2003yu} to be nearly the critical density $\Omega =1$.

We exclude baryonic matter as being the dark matter both from direct searches of $H$ and MACHOs and from cosmological measurement of $\Omega_M$ vs $\Omega_b$.
$\Omega_M$ is the fraction of the critical density that needs to obey a cold-matter equation of state and $\Omega_b$ is the fraction of the critical density that is composed of baryons.  $\Omega_\Lambda$ is the fraction of the critical density with a vacuum-like equation of state (dark energy).  The anisotropy of the Cosmic Microwave Background (CMB) provides a constraint the reduced baryon density ($h^2 \Omega_b$) and reduced matter density ($h^2 \Omega_b$).

As the dark-matter collapses, the baryon-photon oscillates within these wells.
The baryons however have more inertia than the photons and therefore `baryon-load'
the oscillation.  The relative heights of the destructive and constructive interference positions indicate how much gravitational collapse due to dark matter occur during the time of a baryon-photon oscillation cycle. The relative heights of different peaks measure the baryon loading.
Therefore, the third peak provides data about the dark matter density as opposed to the photon-baryon density.

Using these concepts, the WMAP results give estimates of the baryonic dark-matter:  $\Omega_b h^2 = 0.0224 \pm 0.0009$ and the dark-matter plus baryonic matter $\Omega_M h^2 = 0.135 \pm 0.008$ \cite{Bertone:2004pz}.  The CMB value for $\Omega_b$ agrees with the big-bang nucleousythesis (BBN) value
 $0.018 < \Omega_b h^2 < 0.023$.  The dark-matter abundance is found to be $\Omega_M = 0.25$ using $h=0.73 \pm 0.03$ \cite[Ch21]{PDBook2008}.  Combining the CMB results with supernova observations with baryon acoustic oscillations (BAO) compared to the galaxy distribution on large scales all lead to a three-way intersection on $\Omega_M$ vs $\Omega_\Lambda$ that give $\Omega_M=0.245 \pm 0.028$, $\Omega_\Lambda=0.757 \pm 0.021$, $\Omega_b = 0.042 \pm 0.002$ \cite{Frieman:2008sn}.
This is called the concordance model or the $\Lambda$CDM model.

Because all these tests confirm $\Omega_b << \Omega_M$, we see that the
cosmological observations confirm the non-baryonic dark matter observed in galactic rotation curves.

\begin{itemize}
\item{\textbf{Dark Matter Direct Detection and Experimental Anomalies}}
\end{itemize}

\begin{figure}
\centerline{\includegraphics{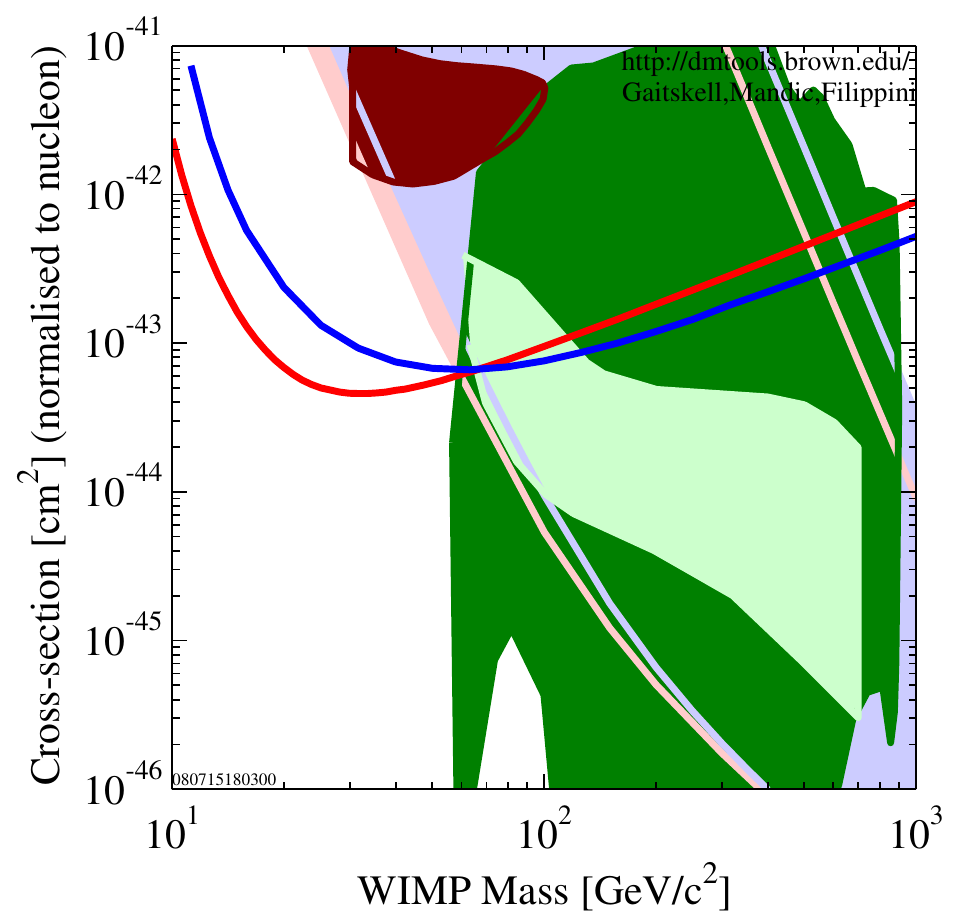}}
\centerline{\includegraphics{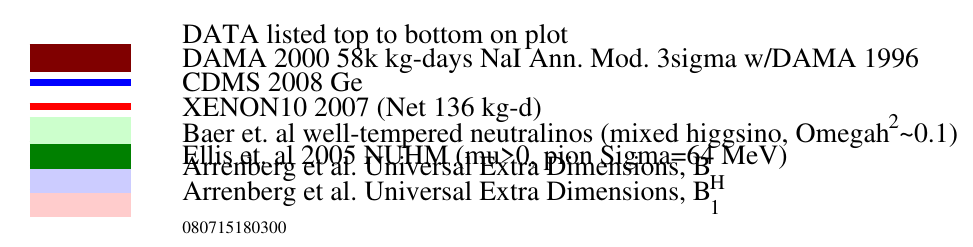}}
\caption{\label{FigDMDirectSearch}Direct dark matter searches showing the limits from the experiments CDMS, XENON10, and the DAMA signal.
Also shown are sample dark-matter cross sections and masses predicted for several supersymmetric models and universal extra dimension models. Figures generated using \newline \tt{http://dmtools.berkeley.edu/limitplots/}.}
\end{figure}

First we assume a local halo density of the dark matter of $0.3 \GeV / \cm^3$ as extracted from $\Omega_M$ and galaxy simulations.
For a given mass of dark matter particle, one can now find a number density.  The motion of the earth through this background number density creates a flux.  Cryogenic experiments shielded deep underground with high purity crystals search for interactions of this dark matter flux with nucleons.
Using the lack of a signal they place bounds on the cross section as a function of the mass of the dark matter particle which can be seen in Fig.~\ref{FigDMDirectSearch}.
This figure shows direct dark-matter limits from the experiments CDMS (2008)\cite{Ahmed:2008eu}, XEONON10 (2007)\cite{Angle:2007uj}, and the DAMA anomaly (1999) \cite{Bernabei:1998td}\cite{Bernabei:2008yi}.
Also shown are sample  dark-matter cross sections and masses predicted for several supersymmetric models \cite{Baer:2006te}\cite{Ellis:2005mb} and universal extra dimension models \cite{Arrenberg:2008wy}.  This shows that the direct searches have not excluded supersymmetry as a viable source for the dark matter.

Although direct searches have no confirmed positive results, there are two anomalies of which we should be aware.
The first is an excess in gamma rays observed in the EGRET experiment
\cite{deBoer:2005tm} which points to dark matter particle annihilation with the dark-matter particles mass between $50 \GeV < M_{N} < 100 \GeV$.
The uncertainties in the background gamma ray sources could also explain this excess.
The second anomaly is an annual variation observed in the DAMA experiment
\cite{Bernabei:1998td}.
The annual variation could reflect the annual variation
 of the  dark matter flux as the earth orbits the sun.  The faster and slower flux of dark-matter particles triggering the process would manifest as an annual variation.  Unfortunately,
CDMS did not confirm DAMA's initial results \cite{Akerib:2004fq}\cite{Ahmed:2008eu}.  This year the DAMA/LIBRA experiment
released new results which claim to confirm their earlier result\cite{Bernabei:2008yi} with an $8\,\sigma$ confidence.  Until another group's experiment confirms the DAMA result, the claims will be approached cautiously.

\begin{itemize}
\item{\textbf{Dark matter Candidates}}
\end{itemize}
There are many models that provide possible dark-matter candidates.
To name a few possibilities that have been considered we list: $R$-parity conserving supersymmetry,
Universal Extra Dimensions (UED), axions, degenerate massive neutrinos \cite{Tegmark:2005cy},
stable black holes.
In Chapter \ref{ChapterUnificationAndFermionMassStructure} we focus on Supersymmetry as the model framework for explaining the dark-matter, but our
results in Chapters \ref{ChapterM2Cdirect} - \ref{ChapterM3C} apply to any model where new particle states are pair produced and decay to two semi-stable dark matter particles that escape the detector unnoticed.

The pair-produced nature of dark-matter particles is relatively common trait of models with dark matter.  For example the UED models \cite{ArkaniHamed:1998rs,Appelquist:2000nn,Servant:2002aq} can also pair produced dark-matter particles at a collider. The lightest Kaluza-Klein particle (LKP) is also a dark-matter candidate.  In fact UED and SUSY have very similar hadron collider signatures \cite{Battaglia:2005zf}.

\section{Supersymmetry: Predicting New Particle States}
\label{SecSupersymmetryReviewAndIntro}

Supersymmetry will be the theoretical frame-work for new-particle states
on which this thesis focuses.  Supersymmetry has proven a very popular and
powerful framework with many successes and unexpected benefits:
Supersymmetry provides a natural dark-matter candidate.
As a fledgling speculative theory, supersymmetry showed a preference for a heavy top-quark mass.
In a close analogy with the {GIM} mechanism, Supersymmetry's minimal phenomenological implementation eliminates a fine-tuning problem
associated with the Higgs boson in the Standard Model.
Supersymmetry illuminates a coupling constant symmetry (Grand Unification) among the three non-graviational forces at an energy scale around $2 \times 10^{16} \GeV$.
Supersymmetry is the only extension of \Poincare\ symmetry discussed in \ref{SubSecPositronMass} allowed with a graded algebra \cite{Haag:1974qh}.
It successfully eliminates the
tachyon in String Theory through the GSO projection \cite{Gliozzi:1976qd}.
Last, it is a candidate explanation for the $3\,\sigma$ deviation
of the muons magnetic moment known as the $(g-2)_\mu$ anomaly \cite{Stockinger:2006zn}.

These successes are exciting because they
follow many of the precedents and clues described earlier in this chapter that have successfully
predicted new-particle states in the past.
SUSY, short for supersymmetry, relates the couplings of the Standard-Model  fermions and bosons
to the couplings of new bosons and fermions known as superpartners.
SUSY is based on an extension of the Poincare-symmetry (Sec \ref{SubSecPositronMass}).
In the limit of exact Supersymmetry, the Higgs self-energy problem (Sec \ref{SubSecStandardModel}) vanishes.
In analogy to the GIM mechanism (Sec \ref{SubSecGIM}), the masses of the new SUSY particles reflect the breaking of supersymmetry.
There are many theories providing the origin of supersymmetry breaking which are beyond the scope of this thesis.  The belief that nature is described by Supersymmetry follows from how SUSY connects to the past successes in predicting new-particle states and their masses from symmetries, broken-symmetries, and fine-tuning arguments.

Excellent introductions and detailed textbooks on Supersymmetry exist, and there is no point to reproduce these textbooks here.
Srednicki provides a very comprehensible yet compact introduction to supersymmetry via superspace at the end of Ref~\cite{Srednicki:2007qs}.  Reviews of SUSY
that have proved useful in developing this thesis are Refs~\cite{Martin:1997ns,Baer:2006rs,Gates:1983nr,Wess:1992cp,Cahill:1999aq,Aitchison:2005cf}.
In this introduction, we wish to highlight a few simple parallels to past successes in new-particle state predictions.
We review supersymmetric radiative corrections to the coupling constants and to the Yukawa couplings.
These radiative corrections can be summarized in
the renormalization group equations (RGE).  The RGE have surprising predictions for Grand Unification of the non-gravitational forces and also show evidence of a quark lepton mass relations
suggested by Georgi and Jarlskog which we refer to as mass matrix unification.
The mass matrix unification predictions can be affected by potentially large corrections enhanced by the ratio of the vacuum expectation value of the two supersymmetric Higgs particles $v_u / v_d = \tan \beta$.
Together, these renormalization group equations and the large $\tan \beta$ enhanced corrections provide the basis for the new contributions this thesis presents in Chapter \ref{ChapterUnificationAndFermionMassStructure}.

\subsection{Supersymmetry, Radiative Effects, and the Top Quark Mass}


Supersymmetry extends the \Poincare\ symmetry used in \ref{SubSecPositronMass}.
The super-\Poincare\ algebra involves a graded Lie algebra which has generators that anticommute as well as generators that commute.
One way of understanding Supersymmetry is to extend four space-time coordinates to include two complex Grassmann coordinates which transform as two-component spinors.
This extended space is called superspace.
In the same way that $P_\mu$ generates space-time translations, we introduce one supercharge\footnote{We will only consider theories with one supercharge.} $Q$ that generates translations in $\theta$ and $\bar{\theta}$.
The graded super-\Poincare\ algebra includes the \Poincare\ supplemented by
 \begin{eqnarray}
  \left[ Q^\dag_{\dot{a}} , P^\mu \right] = \left[ Q_a , P^\mu \right]  &  =  & 0 \label{EqSuperQConserved} \\
  \left\{ Q_a , Q_b \right\} = \left\{ Q^\dag_{\dot{a}} , Q^\dag_{\dot{b}} \right\} & = & 0 \label{EqQQAnticommute}\\
  \left[ Q_a , M^{\mu \nu} \right]& = &
  \frac{i}{2} ( \bar{\sigma}^\mu {\sigma}^\nu -  \bar{\sigma}^\nu {\sigma}^\mu)_{{a}}^{\ {c}} Q^\dag_{{c}} \label{EqSuperQTransform} \\
  \left[ Q^\dag_{\dot{a}} , M^{\mu \nu} \right]  & = &
  \frac{i}{2} ( \sigma^\mu \bar{\sigma}^\nu -  \sigma^\nu \bar{\sigma}^\mu)_{\dot{a}}^{\ \dot{c}} Q^\dag_{\dot{c}} \label{EqSuperQTransform2}
  \\
  \left\{ Q_a , Q^\dag_{\dot{a}} \right\} & = &  2\, \sigma^\mu_{a \dot{a}} P_\mu. \label{EqSuperQQTranslate}
 \end{eqnarray}
where $M$ are the boost generators satisfying Eq(\ref{EqMGeneratorsBoosts}),
$\{ , \}$ indicate anticommutation, $a, b, \dot{a}, \dot{b}$ are spinor indices, $\sigma_\mu=(1,\vec{\sigma})$ and $\bar{\sigma}_\mu = (1,-\vec{\sigma})$.
Eq(\ref{EqSuperQConserved}) indicates htat the charge $Q$ is conserved by space-time translations.  Eq(\ref{EqQQAnticommute}) indicates that no more than states of two different spins can be connected by the action of a supercharge. Eq(\ref{EqSuperQTransform}-\ref{EqSuperQTransform2}) indicate that
$Q$ and $Q^\dag$ transform as left and right handed spinors respectively.
Eq(\ref{EqSuperQQTranslate}) indicates that two supercharge generators can generate a space-time translation.

In relativistic QFT, we begin with fields $\phi(x)$ that are functions of space-time coordinates $x^\mu$ and require the Lagrangian to be invariant under a representation of the \Poincare\ group acting on the space-time and the fields.  To form supersymmetric theories, we begin with superfields $\hat{\Phi}(x,\theta,\bar{\theta})$ that are functions of space-time coordinates $x^\mu$ and the anticommuting Grassmann coordinates $\theta_a$ and $\bar{\theta}_{\dot{a}}$ and require the Lagrangian to be invariant under a representation of the
super-\Poincare\ symmetry acting on the superspace and the superfields.  The procedure described here is very tedious:
defining a representation of the super-\Poincare\
 algebra, and formulating a Lagrangian that is invariant under actions of the group
 involves many iterations of trial and error. Several
short-cuts have been discovered to form supersymmetric theories very quickly.
These shortcuts involve studying properties of superfields.

Supersymmetric theories can be expressed as ordinary relativistic QFT by expressing the superfield $\hat{\Phi}(x,\theta,\bar{\theta})$ in terms of space-time fields like $\phi(x)$ and $\psi(x)$.  The superfields, which we denote with hats, can be expanded as a Taylor series in $\theta$ and $\bar \theta$.  Because $\theta_a^2=\bar{\theta}_{\dot{a}}^{\,2}=0$, the superfield expansions consist of a finite number of space-time fields (independent of $\theta$ or $\bar \theta$); some of which transform as scalars, and some transform as spinors, vectors, or higher level tensors.
A supermultiplet is the set of fields of different spin interconnected because they are part of the same superfield.
If supersymmetry were not broken, then these fields of different spin would be indistinguishable.  The fields of a supermultiplet share the same quantum numbers (including mass) except spin.

Different members of a supermultiplet are connected by the action of the supercharge operator $Q$ or $Q^\dag$: roughly speaking $Q | \rm{boson} \rangle = | \rm{fermion} \rangle$ and $Q | \rm{fermion} \rangle = | \rm{boson} \rangle$.
Because $Q_a^2=(Q_{\dot{a}}^\dag)^2=0$ a supermultiplet only has fields of two different spins.
Also because it is a symmetry transformation, the fields of different spin within a supermultiplet need to have equal numbers of degrees of freedom \footnote{There are auxiliary fields in supermultiplets that, while not dynamical, preserve the degrees of freedom when virtual states go off mass shell.}.
A simple type of superfield is a chiral superfield.  A chiral superfield consists of a complex scalar field and a chiral fermion field each with $2$ degrees of freedom.  Another type of superfield is the vector superfield $\hat{V} = \hat{V}^\dag$ which consists of a vector field and a Weyl fermion field.  Magically vector superfields have natural gauge transformations, the spin-1 fields transform as connections under
gauge transformation whereas the superpartner spin-1/2 field transforms covariantly  in the adjoint of the gauge transformation\footnote{\label{ConnectionTransform2} Although this seems very unsymmetric, Wess and Bagger\cite{Wess:1992cp} show a supersymmetric differential geometry with tetrads that illuminate the magic of how the spin-1 fields transform as connections but the superpartners transform covariantly in the adjoint representation.}.

The shortcuts to form Lagrangians invariant under supersymmetry transformations
are based on three observations:  (1) the products of several superfields is again a superfield (2) the term in an expansion of a superfield (or product of superfields)
proportional to $\theta_a \theta_b \epsilon^{ab}$ is invariant under supersymmetry transformations up to a total derivative (called an $F$-term), and (3) the term in an expansion of a superfield (or product of superfields)
proportional to $\theta_a \theta_b \epsilon^{ab} \bar{\theta}_{\dot{a}} \bar{\theta}_{\dot{b}} \epsilon^{\dot{a}\dot{b}}$ is invariant under supersymmetry transformations up to a total derivative (called a $D$-term).

These observations have made constructing theories invariant under supersymmetry a relatively painless procedure:  the $D$-term of $\hat{\Psi}^\dag e^{-\hat{V}} \hat{\Psi}$ provides supersymmetricly invariant kinetic terms.
A superpotential $\hat{W}$ governs the Yukawa interactions among chiral superfields.
The $F$-term of $\hat{W}$ gives the supersymmetricly invariant interaction Lagrangian.
The $F$-term of $\hat{W}$ can be found with the following shortcuts: If we take all the chiral superfields in $\hat{W}$ to be enumerated by $i$ in
$\hat{\Psi}_i$ then the interactions follow from two simple calculations: The scalar potential is given by $\mathcal{L} \supset  = - (\sum_j |\partial \hat{W} / \partial \hat{\Psi}_j|^2)$
where $\hat{\Psi}_j$ is a the superfield and all the superfields inside $()$ are replaced with their scalar part of their chiral supermultiplet.
The fermion interactions with the scalars are given by
$\mathcal{L} \supset - \sum_{i,j} (\partial^2 \hat{W} / \partial \hat{\Psi}_i \partial \hat{\Psi}_j ) \psi_i \psi_j$.
where the superfields in $()$ are replaced with the scalar part of the chiral supermultiplet and $\psi_i$ and $\psi_j$ are the 2-component Weyl fermions that are part of their chiral supermultiplet $\hat{\Psi}_j$.  Superpotential terms must be gauge invariant just as one would expect for terms in the Lagrangian and must be holomorphic function of the superfields\footnote{By holomorphic we mean the superpotential can only be formed from unconjugated superfields $\hat{\Psi}$ and not the conjugate of superfields like $\hat{\Psi}^*$.}.
Another way to express the interaction Lagrangian is by ${\mathcal{L}_W}= \int d \theta^2 \hat{W} + \int d \bar{\theta}^2 W^\dag$ where the integrals pick out the $F$ terms of the superpotential $\hat{W}$.

In addition to supersymmetry preserving terms, we also need to add `soft' terms which parameterize the breaking of supersymmetry. `Soft' refers to only SUSY breaking terms which do not spoil the fine-tuning solution discussed below.

\begin{table}
\footnotesize
  \centering
  \begin{tabular}{|c|c|c|c|c|}
    \hline
  Field    & Lorentz & $SU(3)_c$ & $SU(2)_L$ & $U(1)_Y$ \\   \hline
  $(L)^i$   & $(2,0)$ & $1$     &  $2$      & $-1$ \\
  $(\tilde{L})^i$   & $0$ & $1$     &  $2$      & $-1$ \\ \hline
  $(e^c)^i$ & $(2,0)$ & $1$     &  $1$      & $2$ \\
  $(\tilde{e}^c)^i$ & $0$ & $1$     &  $1$      & $2$ \\ \hline
  $(\nu^c)^i$ & $(2,0)$ & $1$     &  $1$      & $0$ \\
  $(\tilde{\nu}^c)^i$ & $0$ & $1$     &  $1$      & $0$ \\ \hline
  $(Q)^i$   & $(2,0)$ & $3$     &  $2$      & $1/3$ \\
  $(\tilde{Q})^i$   & $0$ & $3$     &  $2$      & $1/3$ \\ \hline
  $(u^c)^i$ & $(2,0)$ & $\bar 3$     &  $1$      & $-4/3$ \\
  $(\tilde{u}^c)^i$ & $0$ & $\bar 3$     &  $1$      & $-4/3$ \\ \hline
  $(d^c)^i$ & $(2,0)$ & $\bar 3$     &  $1$      & $2/3$ \\
  $(\tilde{d}^c)^i$ & $0$ & $\bar 3$     &  $1$      & $2/3$ \\ \hline
  $H_u$   &    $1$        &  $1$     &  $2$      & $+1$ \\
  $\tilde{H}_u$   &    $(2,0)$        &  $1$     &  $2$      & $+1$ \\ \hline
  $H_d$   &    $1$        &  $1$     &  $2$      & $-1$ \\
  $\tilde{H}_d$   &    $(2,0)$        &  $1$     &  $2$      & $-1$ \\ \hline
  $B_\mu$   &    $4$        & $1$     &  $1$      & $0^\ddagger$ \\
  $\tilde{B}$   &    $(2,0)$        & $1$     &  $1$      & $0$ \\ \hline
  $W_\mu$   &    $4$        & $1$     &  $3^\ddagger$      & $0$ \\
  $\tilde{W}$   &    $(2,0)$        & $1$     &  $3$      & $0$ \\ \hline
  $g_\mu$   &    $4$        & $8^\ddagger$     &  $1$      & $0$ \\
  $\tilde{g}$   &    $(2,0)$        & $8$     &  $1$      & $0$ \\
    \hline
  \end{tabular}
\normalsize
  \caption{Minimial Supersymmetric Standard Model (MSSM) field's transformation properties. Fields grouped together are part of the same supermultiplet. $^\ddagger$ Indicated fields transform as a connection as opposed to covariantly$^{\rm{\ref{ConnectionTransform2}}}$. }\label{TableMSSMFields}
\end{table}
To form the minimal supersymmetric version of the Standard Model,
known as the MSSM,
we need to identify the Standard Model fields with supermultiplets.  The resulting list of fields is given in Table \ref{TableMSSMFields}.
Supersymmetry cannot be preserved in the Yukawa sector with just one Higgs field because the superpotential which will lead to the $F$-terms in the theory must be holomorphic so we cannot include both $\hat{H}$ and $\hat{H}^*$ superfields in the same superpotential; instead the MSSM has two Higgs fields $H_u = (h^+_u, h^o_u)$ and $H_d=(h^o_d , h^{-}_d)$.  The
neutral component of each Higgs will acquire a vacuum expectation value (VEV): $\langle h_u^o \rangle = v_u$ and $\langle h_d^o \rangle = v_d$.
The parameter $\tan \beta=v_u /v_d$ is ratio of the VEV of the two Higgs fields.

The structure of the remaining terms can be understood by studying a field like a right-handed up quark.
The $(u^c)$ transforms as a $\textbf{3}$ under $SU(3)$ so its superpartner must be
also transform as a $\textbf{3}$ but have spin $0$ or $1$. No Standard Model candidate exists that fits either option.  A spin-$1$ superpartner is excluded because the fermion component of a vector superfield transforms as a connection in the adjoint of the gauge group; not the fundamental representation like a quark.   If $(u^c)$ is part of a chiral superfield, then there is an undiscovered spin-0 partner.  Thus the $(u^c)$ is part of a chiral supermultiplet with a scalar partner called a right-handed squark $\tilde{u}^c$ \footnote{The right-handed refers to which fermion it is a partner with.  The field is a Lorentz scalar.}.

The remaining supersymmetric partner states being predicted in Table \ref{TableMSSMFields} can be deduced following similar arguments.  The superpartners are named after their SM counterparts with preceding `s' indicating it is a scalar superpartner of a fermion or affixing `ino' to the end indicating it is a fermionic partner of a boson: for example selectron, smuon, stop-quark, Higgsino, photino, gluino, etc.

\begin{figure}
\centerline{\includegraphics[width=2in]{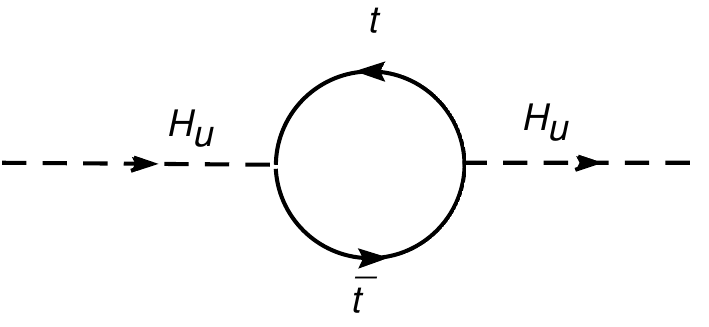}\ \includegraphics[width=2in]{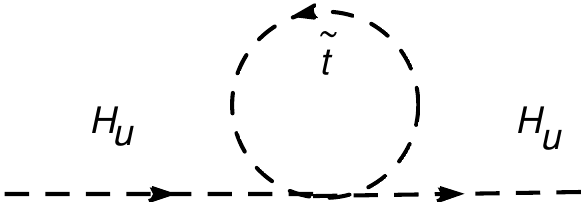}}
\centerline{(a) \ \ \ \ \ \  \  \ \  \ \ \ \ \ \ \ \ \ \ \ \ \ \ \ \ \ \ \ \ \ \ \ (b)}
\caption{\label{FigSUSYHiggsSelfEnergy} One Loop Contributions to the supersymmetric Higgs $H_u$ mass
self energy. (a) top loop.   (b) stop loop.}
\end{figure}
Supersymmetry solves the fine-tuning problem of the Higgs self energy.
The superpotential describing the Yukawa sector of the MSSM is given by
 \begin{equation}
   \hat{W} = Y^e_{ij} \hat{L}^i \hat{H}_d (\hat{e}^c)^j + Y^d_{ij} \hat{Q}^i \hat{H}_d (\hat{d}^c)^j + Y^u_{ij} \hat{Q}^i \hat{H}_u (\hat{u}^c)^j + \mu \hat{H}_u \hat{H}_d
 \end{equation}
where the fields with hats like $\hat{L}$, $\hat{H}_{u,d}$, $\hat{Q}$, etc. are all superfields.
In the limit of exact supersymmetry the fine-tuning problem is eliminated because the resulting potential and interactions lead to a
cancelation of the quadratic divergences between the fermion and scalar loops in Fig~\ref{FigSUSYHiggsSelfEnergy}.
The self-energy of the neutral Higgs $H_u=(h_u^+,h_u^o)$ is now
approximately given by
\begin{equation}
 m^2_{h^o_u} \approx |\mu_{o}|^2  - \frac{{\rm{Tr}}(Y^u (Y^u)^\dag)}{8 \pi^2} \Lambda^2 +  \frac{{\rm{Tr}}(Y^u (Y^u)^\dag)}{8 \pi^2} \Lambda^2
\end{equation}
where $|\mu_o|^2$ is the modulus squared of the bare parameter in the superpotential which must be positive, the second term comes from the fermion loop and therefore has a minus sign, and the third term comes from the scalar loop. Both divergent loops follow from the MSSM superpotential:
the first loop term follows from the fermion coupling $\partial^2 W / \partial t \partial t^c  =  y_t h^o_u t t^c$ and the second loop term follows from the scalar potential $|\partial W / \partial t^c|^2 = y_t^2 |h^o_u|^2 |\tilde{t}|^2$ and $|\partial W / \partial t|^2 = y_t^2 |h^o_u|^2 |\tilde{t^c}|^2$ where we assume the top-quark dominates the process.
Exact supersymmetry ensures these two quadratically divergent loops cancel.
However two issues remain and share a common solution:  supersymmetry is not exact, and the Higgs mass squared must go negative to trigger spontaneous symmetry breaking (SSB).
In the effective theory well above the scale where all superpartners are energetically accessible the cancelation dominates.  The fine-tuning arguments in Sec.~\ref{SubSecStandardModel} suggest that if $\mu_o \approx 100 \GeV$ \footnote{Again we choose $100$ GeV as a generic electroweak scale.} this cancelation should dominate above about $1$ TeV.  In an effective theory between the scale of the top-quark mass and the stop-quark mass only the fermion loop (Fig~\ref{FigSUSYHiggsSelfEnergy} a) will contribute significantly. In this energy-scale region we neglect the scalar loop (Fig~\ref{FigSUSYHiggsSelfEnergy} b).
With only the fermion loop contributing significantly and if $y_t$ is large enough then the fermion loop will overpower $\mu^2_o$ then the mass squared $m^2_{h^o_u}$ can be driven negative.

In this way the need for SSB without fine tuning in the MSSM prefers
a large top-quark mass and the existence of heavier stop scalar states.  Assuming $\mu_o \approx 100 \GeV$ predicts $\tilde{t}$ and $\tilde{t}^c$ below around $1$ TeV\footnote{As fine tuning is an aesthetic argument, there is a wide range of opinions on the tolerable amount of fine tuning acceptable.  There is also a wide range of values for $\mu$ that are tolerable.}.  This phenomena for triggering SSB is known as radiative electroweak symmetry breaking (REWSB). We have shown a very coarse approach to understanding the major features; details can be found in a recent review \cite{Ibanez:2007pf} or any of the supersymmetry textbooks listed above.

The detailed REWSB \cite{Ibanez:1982fr} technology was developed in the early 1980s, and predicted $M_t \approx 55-200$ GeV \cite{Inoue:1982pi}\cite{AlvarezGaume:1983gj}; a prediction far out of the general expectation of $M_t \approx 15-40$ GeV of the early 1980s
\footnote{Raby \cite{Raby:1980uk} Glashow \cite{Glashow:1980xq} and others \cite{Mahanthappa:1979ek} \cite{Yanagida:1979gs} all made top-quark mass predictions in the range of $15 - 40 \GeV$.} and closer to the observed value of $M_t=170.9 \pm 1.9$ GeV.  If supersymmetric particles are observed at the LHC, the large top-quark mass  may be looked at as the first confirmed prediction of supersymmetry.

The MSSM provides another independent reason to prefer a large top quark mass.
The top Yukawa coupling's radiative corrections are dominated by the difference between terms proportional to $g_3^2$ and $y_{t}^2$.  If the ratio of these two terms is fixed, then $y_{t}$ will remain fixed \cite{Pendleton:1980as}.
For the standard model this gives a top quark mass around $120$ GeV.
However for the MSSM, assuming $\alpha_3=g_3^2/4\pi=0.12$ then one finds a top quark mass around $180$ GeV assuming a moderate $\tan \beta$ \cite{Bardeen:1993rv}.
Our observations of the top quark mass very near this fixed point again points to supersymmetry as a theory describing physics at scalers above the electroweak scales.

\subsection{Dark Matter and SUSY}

Supersymmetry is broken in nature. The masses of the superpartners reflect this breaking.
Let's assume the superpartner masses are at a scale that avoids bounds set from current searches yet still solve the fine-tuning problem.
Even in this case there are still problems that need to be resolved \footnote{There
is also a flavor changing neutral current (FCNC) problem not discussed here.}.
There are many couplings allowed by the charge assignments
displayed in Table~\ref{TableMSSMFields} that would immediately lead to unobserved phenomena.  For example the superpotential could contain superfield interactions $\hat{u}^c \hat{u}^c \hat{d}^c$ or $\hat{Q} \hat{L} \hat{d}^c$ or $\hat{L} \hat{L} \hat{e}^c$ or $\kappa \hat{L} \hat{H}_u$ where $\kappa$ is a mass scale.  Each of these interactions is invariant under the $SU(3)_c \times SU(2)_L \times U(1)_Y$ charges listed in Table \ref{TableMSSMFields}.  These couplings, if allowed with order $1$ coefficients, would violate the universality of the four-Fermi decay, lead to rapid proton decay, lepton and baryon number violation, etc.

These couplings can be avoided in several ways.
We can require a global baryon number or
lepton number $U(1)$ on the superpotential; in the Standard Model
these were accidental symmetries.
However, successful baryogenesis requires baryon number violation so imposing it
directly is only an approximation.
Another option is to impose an additional discrete symmetry
on the Lagrangian; a common choice is $R$-parity
 \begin{equation}
  R=(-1)^{2j+3B+L}
 \end{equation}
where $j$ is the spin of the particle.  This gives the Standard Model particles $R=1$ and the superpartners $R=-1$.  Each interaction of the MSSM Lagrangian conserves R-parity.  The specific choice of how to remove these interactions is more relevant for GUT model building.  The different choices lead to different predictions for proton decay lifetime.

The $R$-Parity, which is needed to effectively avoid these unobserved interactions at tree level,
has the unexpected benefit of also making stable the lightest supersymmetric particle (LSP).
A stable massive particle that is non-baryonic is exactly what is needed to provide the dark matter observed in the galactic rotation curves of Sec.~\ref{SecDarkMatterEvidence}.

\subsection{Renormalization Group and the Discovered Supersymmetry Symmetries}

\subsubsection{Unification of $SU(3)_c \times SU(2)_L \times U(1)_Y$ coupling constants}

Up until now, the divergent loops have been treated with a {UV} cut off.  Renormalization of non-abelian gauge theories is more easily done using dimensional regularization where the dimensions of space-time are taken to be  $d=4-2\epsilon$.  The dimensionless coupling constants $g$ pick up a dimensionfull coefficient $g \,\mu^{-\epsilon}$ where $\mu$ is an arbitrary energy scale.
The divergent term in loops diagrams is now proportional to $1/\epsilon$, and the counter-terms can be chosen to cancel these divergent parts of these results.
By comparison with observable quantities, all the parameters in the theory are measured assuming a choice of $\mu$.
The couplings with one choice of $\mu$ can be related to an alternative choice of $\mu$ by means of a set of differential equation known as the renormalization group equations (RGE).
The choice of $\mu$ is similar to the choice of the zero of potential energy; in principle any choice will do, but in practice some choices are easier than others.
Weinberg shows how this arbitrary scale can be related to the typical energy scale of a process \cite{ChengLi}.

The renormalization group at one loop for the coupling constant $g$ of an $SU(N)$ gauge theory coupled to fermions and scalars is
\begin{equation}
  \frac{\partial }{ \partial \log \mu} g =  \frac{g^3}{16 \pi^2} \left[ \frac{11}{3} C(G) - \frac{2}{3} n_F S(R_F) -\frac{1}{3} n_S S(R_S) \right] \label{EqRGCouplingRunningGeneral}
\end{equation}
where $n_F$ is the number of 2-component fermions, $n_S$ is the number of complex scalars,
 $S(R_{F,S})$ is the Dynkin index for the representation of the fermions or scalars respectively.
Applying this to our gauge groups in the MSSM
$C(G)=N$ for $SU(N)$ and $C(G)=0$ for $U(1)$.
For the fundamental $SU(N)$ we have $S(R)=1/2$; for the adjoint $S(R)=N$; for $U(1)$ we have $S(R)$ equal to the $\sum (Y/2)^2$ over all the scalars (which accounts for the number of scalars).
When the $SU(3)_c \times SU(2)_L \times U(1)_Y$ is embedded in a larger group like $SU(5)$, the $U(1)_Y$ coupling $g'$ is rescaled to the normalization appropriate to the $SU(5)$ generator that becomes hypercharge.  This rescaling causes us to work with $g_1 = \sqrt{5/3} g'$ \footnote{Eq(\ref{EqRGCouplingRunningGeneral}) is for $g'$ and one must substitute the definition of $g_1$ to arrive at Eq(\ref{EqRGGaugeCouplings}).}.

\begin{figure}
\centerline{\includegraphics[width=3in]{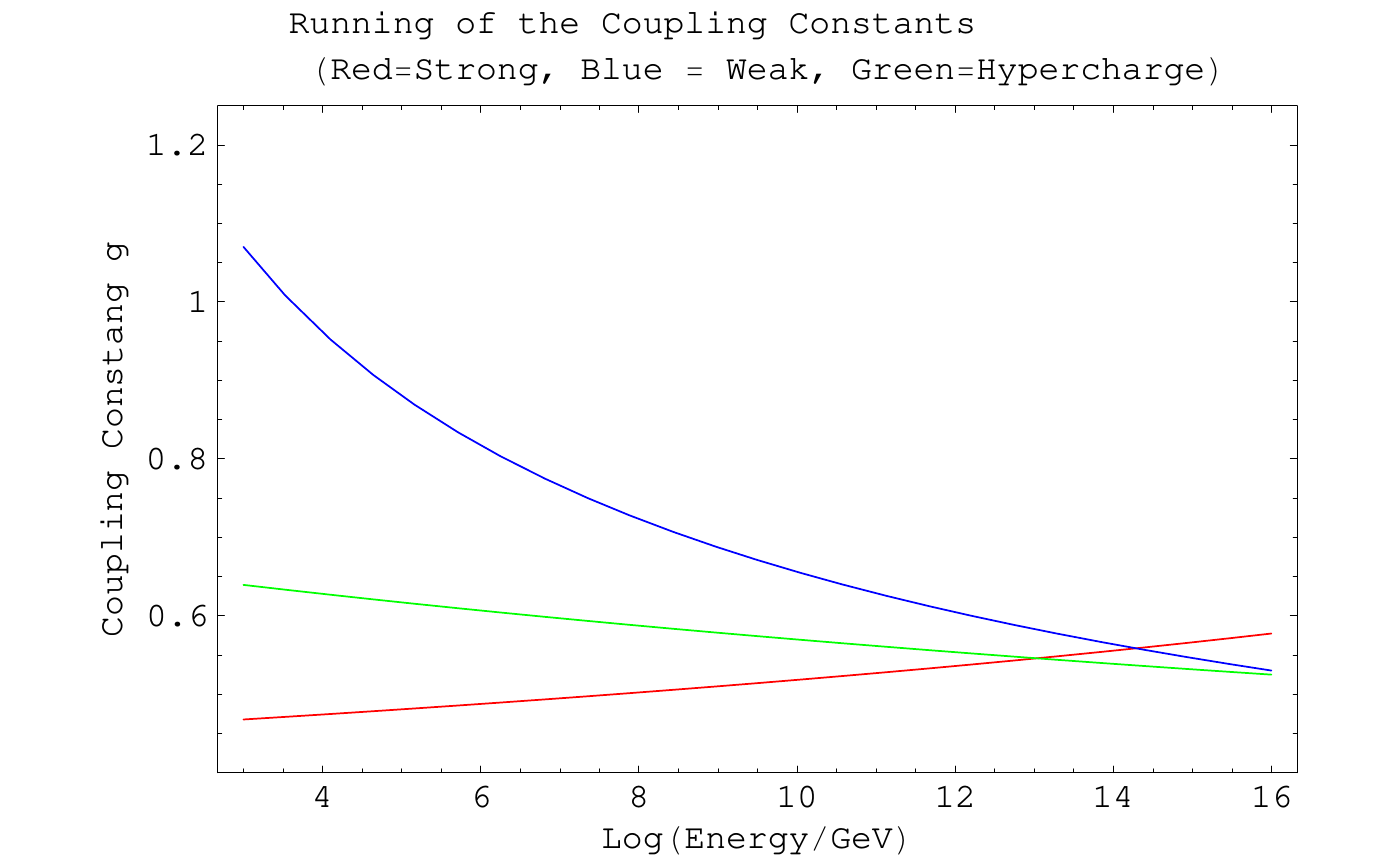}
\includegraphics[width=3in]{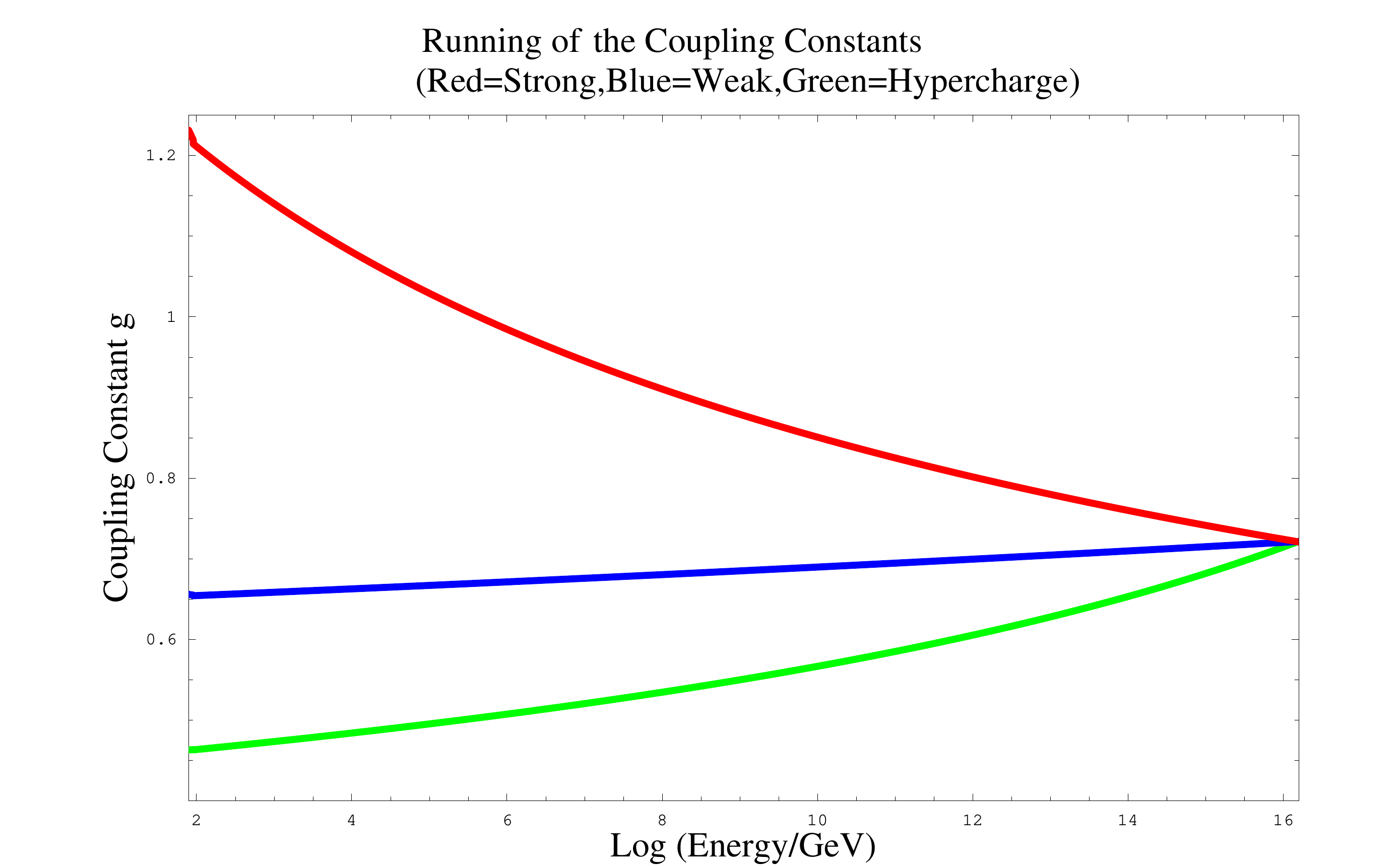}}
\caption{\label{FigGaugeCouplingUnification} Gauge couplings for the three non-gravitational forces as as a function of energy scale for the (left) Standard Model and (right) MSSM.}
\end{figure}
Applying this general formula to both the Standard Model (SM) and the MSSM
leads to
 \begin{eqnarray}
 \frac{\partial}{\partial \log \mu} g_i &=& \frac{g_i^3}{16 \pi^2} b_i \label{EqRGGaugeCouplings} \\
 {\rm{SM}}  \begin{cases}
         b_1 =& n_G\,4/3  + n_H\,1/10  \\
         b_2 =& -22/3 + n_G\,4/3  + n_H\,1/6  \\
         b_3 =& -11 + n_G\,4/3
       \end{cases}
       \ & & \ \
        {\rm{MSSM}} = \begin{cases}
         b_1 =& n_G\,2  + n_H\,3/10  \\
         b_2 =& -6 +  n_G\,2 +  n_H\,1/2  \\
         b_3 =& -9+ n_G\,2
       \end{cases} \nonumber
 \end{eqnarray}
where $n_G=3$ is the number of generations and $n_H$ is the number of Higgs doublets ($n_H=1$ in SM and $n_H=2$ in MSSM). A miracle is shown in Fig.~\ref{FigGaugeCouplingUnification}.
The year 1981 saw a flurry of papers from Dimopoulos, Ibanez, Georgi, Raby, Ross, Sakai and Wilczek who were detailing the consequences of this miracle \cite{PhysRevD.24.1681}\cite{Ibanez:1981yh}\cite{Dimopoulos:1981zb}\cite{Sakai:1981gr}.
There is one degree of freedom in terms of where to place an effective supersymmetry scale $M_S$ where Standard Model RG running turns into MSSM RG running.  At one-loop order, unification requires $M_S\approx M_Z$; at two-loop order $200 \GeV < M_S < 1 $ TeV \footnote{This range comes from a recent study \cite{deBoer:2003xm} which assumes $\alpha_S(M_Z) = 0.122$.  If we assumes $\alpha_S(M_Z)=0.119$, then we find $2 \TeV < M_S < 6 $ TeV.  Current PDG \cite[Ch 10]{PDBook2008} SM global fits give $\alpha_S = 0.1216 \pm 0.0017$.}.

The MSSM was not designed for this purpose, but the particle spectrum gives this result
effortlessly\footnote{Very close coupling constant unification can also occur in non-supersymmetric models. The Standard Model with six Higgs doublets is one such example \cite{Willenbrock:2003ca}, but the unification occurs at too low a scale $\approx 10^{14}$ GeV.  GUT-scale gauge-bosons lead to proton decay.  Such a low scale proton decay at a rate in contradiction with current experimental bounds.}. A symmetry among the coupling of the three forces is discovered through the RG equations.  If the coupling unify, they may all originate from a common grand-unified force that is spontaneously broken at $\mu \approx 2 \times 10^{16}$ GeV.

There are several possibilities for SUSY GUT theories:
$SU(5)$ \cite{Georgi:1974sy}
 or $SO(10)$ or $SU(4) \times SU(2)_L \times SU(2)_R$ \cite{Pati:1974yy} to list just a few.
$SU(5)$, although the minimal version is now excluded experimentally, is the prototypical example with which we work.
A classic review of GUT theories can be found in \cite{Ross:1985ai}

\subsubsection{Georgi-Jarlskog Factors}

It is truly miraculous that the three coupling constants unify (to within current experimental errors) with two-loop running when adjusted to an $SU(5)$ grand unified gauge group and when the SUSY scale is placed in a region where the fine-tuning arguments suggest new-particle states should exist.  Let's now follow the unification of forces arguments to the next level.
Above the unification scale, there is no longer a distinction between $SU(3)_c$ and $U(1)_Y$.
If color and hypercharge are indistinguishable, what distinguishes an electron from a down-quark?
The Yukawa couplings, which give rise to the quark and lepton masses, are also
functions of the scale $\mu$ and RG equations relate the low-energy values to their values at the GUT scale.
Appendix \ref{AppendixRGRunningDetails} gives details of the RG procedure used in this thesis to take the measured low-energy parameters and use the RG equations to relate them to the predictions at the GUT scale.  Do the mass parameters also unify?


With much more crude estimates for the quark masses, strong force, and without the knowledge of the value of the top quark mass Georgi and Jarlskog (GJ)
\cite{Georgi:1979df} noticed that at
the GUT scale the masses satisfied the approximate relations
\footnote{To the best of my knowledge, the $b=\tau$ relations were first noticed by Buras \etal \cite{Buras:1977yy}.}:
 \begin{equation}
  m_\tau \approx m_b \ \ \ \  m_\mu \approx 3 m_s \ \ \  \ 3 m_e \approx m_d.
 \label{EqGJMassRelations}
 \end{equation}
This is a very non-trivial result.
The masses of the quarks and charged leptons span more than $5$ orders of magnitude.
The factor $3$ is coincidentally equal to the number of colors.
At the scale of the $Z^o$-boson's mass $\mu=M_Z$ the ratios look like $m_\mu \approx 2 m_s$, and $1.6 m_\tau \approx m_b$ so the factor of three is quite miraculous.
Using this surprising observation, GJ constructed a model where this relation followed from an $SU(5)$ theory with the second generation coupled to a Higgs in a different representation.

In the $SU(5)$ model the fermions are arranged into a  $\bf{\bar{5}}$ $(\psi^a_i)$
and a $\bf{10}$ $(\psi_{ab\,i})$ where $a,b,c,..$ are the $SU(5)$ indexes and $i,j,..$ are the family indexes.
The particle assignments are
 \begin{eqnarray}
  \psi_a^i &=& \left( \begin{matrix} d^c & d^c & d^c & \nu & e \end{matrix} \right)^i \\
  \psi^{ab\,j} &=& \frac{1}{\sqrt{2}} \left(
\begin{array}{lllll}
 0 & u^c & -u^c & -u &
   -d \\
 -u^c & 0 & u^c & -u &
   -d \\
 u^c & -u^c & 0 & -u &
   -d \\
 u & u & u & 0 & -e^c
   \\
 d & d & d & e^c & 0
\end{array}
\right)^j
 \end{eqnarray}
where $^c$ indicates the conjugate field.
There are also a $\bf{\bar{5}}$ Higgs fields $(H_d)_{a}$
and a $\bf{5}$ Higgs field $(H_u)^a$.
The key to getting the mass relations hypothesized in
Eq(\ref{EqGJMassRelations}) is coupling only the second
generation to a $\bf{45}$ Higgs \footnote{In tensor notation the {\bf{45}} representation is given by $(H_{d,45})^c_{ab}$ where $ab$ are antisymmetric and
the five traces $(H_{d,45})^a_{ab}=0$ are removed.}.
The VEVs of the Higgs fields are $H_u = (0,0,0,0,v_u)$, $H_d=(0,0,0,0,v_d)$
and $(H_{45,d})^a_{b5}=v_{45} (\delta^a_b-4 \delta^{a4} \delta_{b4})$.
The coupling to matter that gives mass to the down-like states is
\begin{equation}
W_{Y\,d} = Y_{d\,5\,ij} \psi^{ab\,i} \psi_{a}^j (H_d)_b +
   Y_{d\,45\,ij} \psi^{ab\,i} \psi^j_c (H_{d,45})_{ab}^{c}
   \label{EqWdownLikeGJ}
\end{equation}
and that give mass to the up-like states
\begin{equation}
  W_{Y\,u} =  Y_{5u\,ij} \psi^{ab\,i} \psi^{cd\,j} (H_u)^e \epsilon_{abcde}.
\end{equation}
Georgi and Jarlskog do not concern themselves with relating the neutrino masses to the up-quark masses, so we will focus on the predictions for the down-like masses.
The six masses of both the down-like quarks and the charged leptons may now be satisfied by
arranging for
 \begin{equation}
 Y_{d\,45} = \left( \begin{matrix} 0 & 0 & 0 \cr 0 & C & 0 \cr 0 & 0 & 0 \end{matrix} \right)  \ \ \ Y_{d\,5} = \left( \begin{matrix} 0 & A & 0 \cr A & 0 & 0 \cr 0 & 0 & B \end{matrix} \right)
 \end{equation}
and fitting the three parameters $A$, $B$, and $C$. The fitting will create the hierarchy $B >> C >> A$. Coupling these Yukawa matricies to the Eq(\ref{EqWdownLikeGJ}) gives the factor of $-3 C$ for the (2,2) entry of the leptons mass matrix relative to the $(2,2)$ entry of the down-like quark mass matrix.
Because $B >> C >> A$, the equality of the (3,3) entry leads to $m_b / m_\tau \approx 1$.
The (2,2) entry dominates the mass of the second generation so $m_\mu / m_s \approx 3$.
The determinant of the resulting mass matrix is independent of $C$ so at the GUT scale the product of the charged lepton masses is predicted to equal the product of the down-like quark masses.

These results have been generalized to other GUT models
like the Pati-Salam model \footnote{In Pati-Salam the $\bf{45}$'s VEV
is such that one has a factor of $3$ and not $-3$ for the charged leptons vs the down-like quark Yukawa coupling.}.
Family symmetries have been used to arrange the general structure shown here
\cite{deMedeirosVarzielas:2005Earlier}.  The continued validity of the Georgi-Jarlskog mass relations is one of the novel contributions of the thesis presented in Chapter \ref{ChapterUnificationAndFermionMassStructure}.

\subsection{$\tan \beta$ Enhanced Threshold Effects}

The Appelquist Carazzone \cite{Appelquist:1974tg} decoupling theorem indicates that particle states heavier than the energy scales being considered can be integrated out and decoupled from the low-energy effective theory.  A good review of working with effective theories and decoupling relations is found in Pich \cite{Pich:1998xt}.  The parameters we measure are in some cases in an effective theory of $SU(3)_c \times U(1)_{EM}$; in other cases we measure the parameters with global fits to the Standard Model.  At the energy scale of the sparticles, we need to match onto the MSSM effective theory.  Finally at the GUT scale we need to match onto the GUT effective theory.

As a general rule, matching conditions are needed to maintain the order of accuracy of the results. If we are using one-loop RG running, we can use trivial matching conditions at the interface of the two effective theories.
If we are using two-loop RG running, we should use one-loop matching conditions at the boundaries.   This is to maintain the expected order of accuracy and precision of the results.   There is an important exception to this general rule relevant to SUSY theories with large $\tan \beta$.

At tree level the VEV of $H_u$ gives mass to the up-like states (t,c,u) and the neutrinos.  At tree-level the VEV of $H_d$ gives mass to the down-like states (b, s, d, $\tau$, $\mu$, e).  However `soft' interactions which break supersymmetry allow the VEV of $H_u$ to couple to down-like Yukawa couplings through loop diagrams.  Two such soft terms are the trilinear couplings ${\mathcal{L}} \supset y_t A^t H_u \tilde{t} \tilde{t}^c$ where $A_t$ is a mass parameter and the gluino mass ${\mathcal{L}} \supset  \tilde g \tilde g M_3$ where $M_3$ is the gluino's soft mass parameter.

The matching conditions for two effective theories are deduced by expressing a common observable in terms of the two effective theories.
For example the Pole mass $M_b$  of the bottom quark \footnote{If it existed as a free state.}
would be expressed as
\begin{equation*}
M_b  \approx
 \parbox{1.8in}{\includegraphics[width=1.8in]{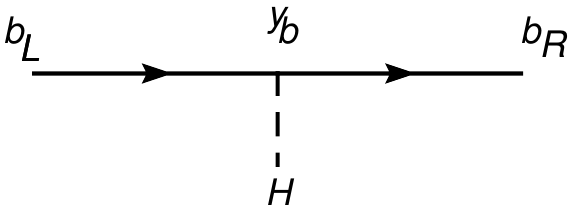}}
 \ + \
 \parbox{1.8in}{\includegraphics[width=1.8in]{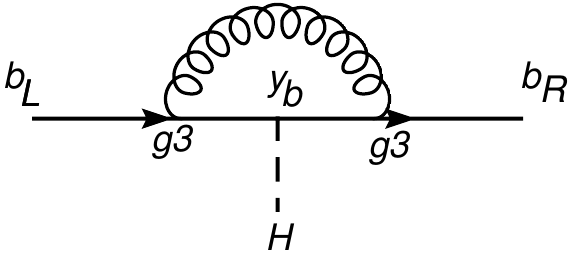}}
 + \ldots
 \end{equation*}
where $H$ takes on the vacuum expectation value, $y_b(\mu)$ and $g_3(\mu)$ are the QFT parameters which depends on an unphysical choice of scale $\mu$.
The same observable expressed in MSSM involves new diagrams
 \begin{equation*}
 M_b \approx
 \parbox{1.8in}{\includegraphics[width=1.8in]{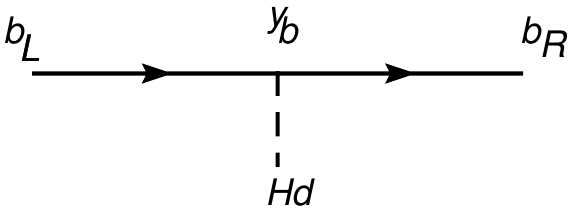}}
 + \parbox{1.8in}{\includegraphics[width=1.8in]{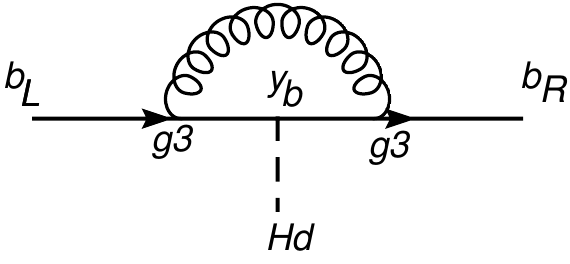}}
 \end{equation*}
 \begin{equation*}
 +  \parbox{1.8in}{\includegraphics[width=1.8in]{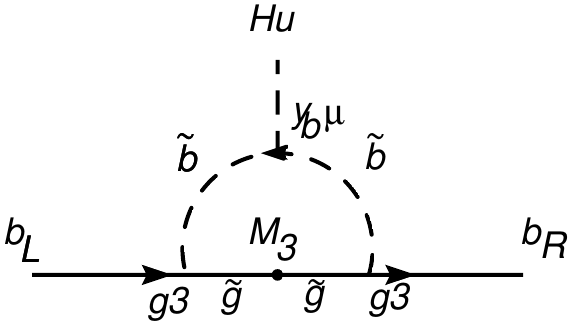}}
 +
 \parbox{1.8in}{\includegraphics[width=1.8in]{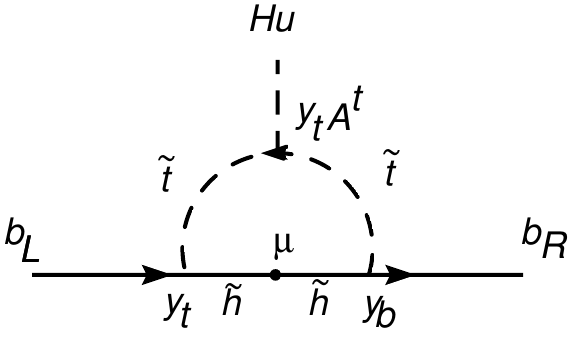}}
 + \ldots
 \end{equation*}
The parameters $y_b^{SM} (\mu), g_3^{SM}(\mu)$ are not equal to the parameters
$y_b^{MSSM} (\mu), g_3^{MSSM}(\mu)$.
By expressing $y_b^{MSSM} = y_b^{SM} + \delta y_b$ and likewise for $g_3^{MSSM} = g_d^{SM} + \delta g_3$, we find many common graphs which cancel.
We are left with an expression for $\delta y_b$ equal to the graphs not common between the two effective theories.
\begin{equation}
 -(\delta y_b) v_d =    \parbox{1.8in}{\includegraphics[width=1.8in]{GluinoLoop}}
 +
 \parbox{1.8in}{\includegraphics[width=1.8in]{StopLoop}}
 + \ldots
\end{equation}
The two graphs in this correction are proportional to the VEV of $\langle h^o_u \rangle=v_u$.  However the $y_b$ Yukawa coupling is the ratio of $m_b$ to $\langle h^o_d \rangle=v_d$.
This makes the correction $\delta y_b / y_b$
due to the two loops shown proportional to $v_u/v_d = \tan \beta$. If $\tan \beta$ were small $\lesssim 5$, then the loop result times the $\tan \beta$ would remain small and the effect would be only relevant at two-loop accuracy.  However when $\tan \beta \gtrsim 10$ then the factor of $\tan \beta$ makes the contribution an order of magnitude bigger and the effect can be of the same size as the one-loop running itself.

These $\tan \beta$ enhanced SUSY threshold corrections can have a large effect on the GUT-scale parameters.  More precise observations of the low-energy parameters have driven the  Georgi-Jarlskog mass relations out of quantitative agreement. However there is a class of $\tan \beta$ enhanced corrections that can bring the relations back into quantitative agreement. Chapter \ref{ChapterUnificationAndFermionMassStructure} of this thesis makes predictions for properties of the SUSY mass spectrum by updating the GUT-scale parameters to the new low-energy observations and considering properties of $\tan \beta$ enhanced SUSY threshold corrections needed to maintain the quantitative agreement of the Georgi Jarlskog mass relations.

%
%
%
%

\section*{Chapter Summary}

In this chapter we have introduced the ingredients of the Standard Model and its supersymmetric extension and given examples of how symmetries, broken symmetries, and fine-tuning arguments have successfully predicted the mass of the positron, the $\Omega^-$, the charm quark, and the $W^\pm$ and $Z^o$ bosons.  We have introduced astrophysical
evidence that indicate a significant fraction of the mass-energy density of the universe is in a particle type yet to be discovered.  We have introduced supersymmetry as a plausible framework for solving the fine-tuning of the Higgs self energy, for
explaining the top-quarks large mass, and for providing a dark-matter particle.  In addition we have discussed how SUSY predicts gauge coupling unification, and a framework for mass-matrix unification.  Last we have introduced potentially large corrections to the RG running of the mass matrices. 


\chapter{Predictions from Unification and Fermion Mass Structure}

\label{ChapterUnificationAndFermionMassStructure}

\providecommand{\Mvariable}[1]{ERROR}
\providecommand{\etal}{\textit{et.al.}}
 \providecommand{\foundit}{\setboolean{found}{true}}
 \providecommand{\MeV}{{\rm{MeV}}}
 \providecommand{\GeV}{{\rm{GeV}}}
 \providecommand{\eV}{{\rm{eV}}}
 \providecommand{\imag}[1]{\,i\,}
 \providecommand{\tanbloopfactor}{}

\section*{Chapter Overview}

\begin{figure}
 \centerline{\includegraphics[width=3in]{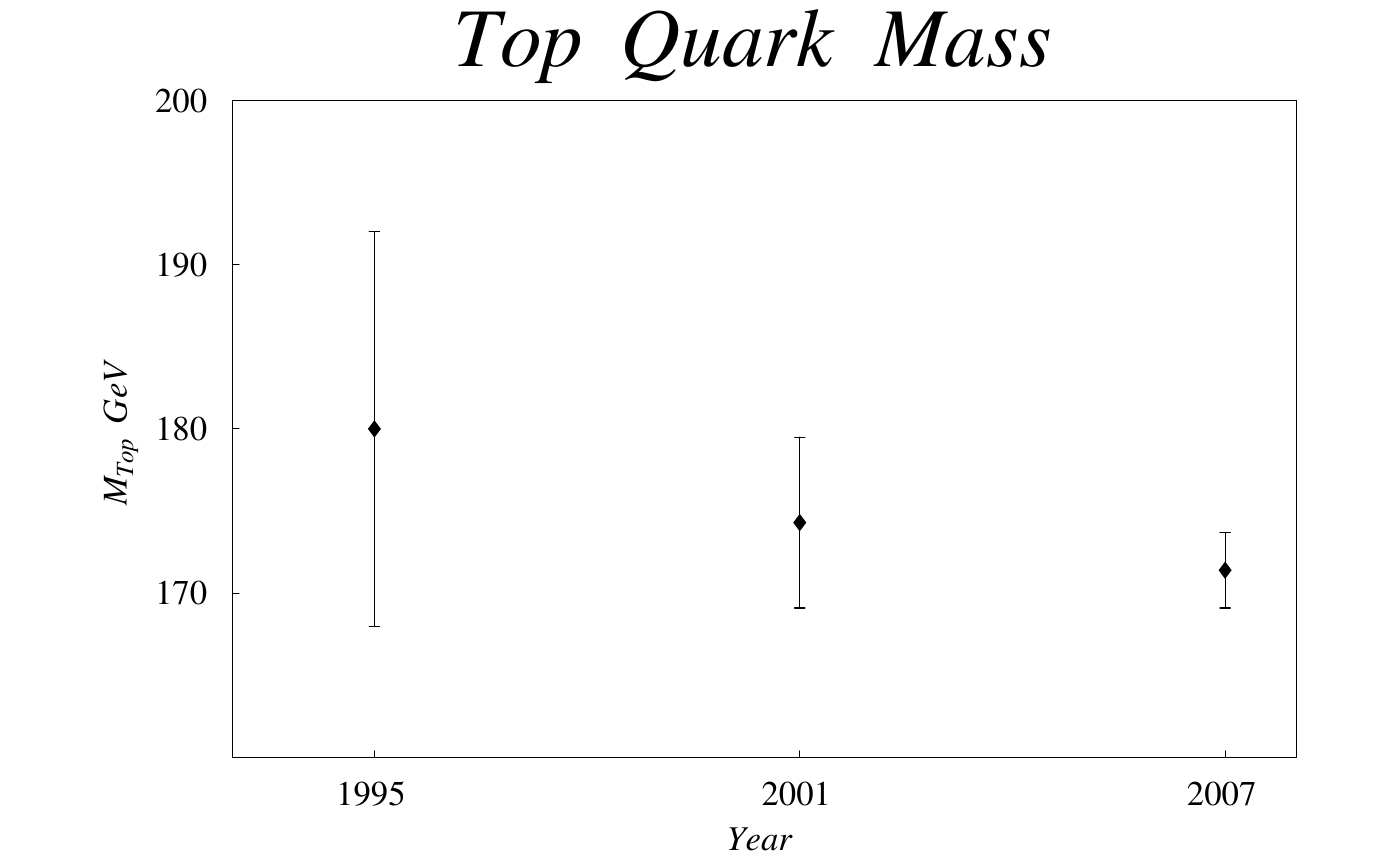}
  \includegraphics[width=3in]{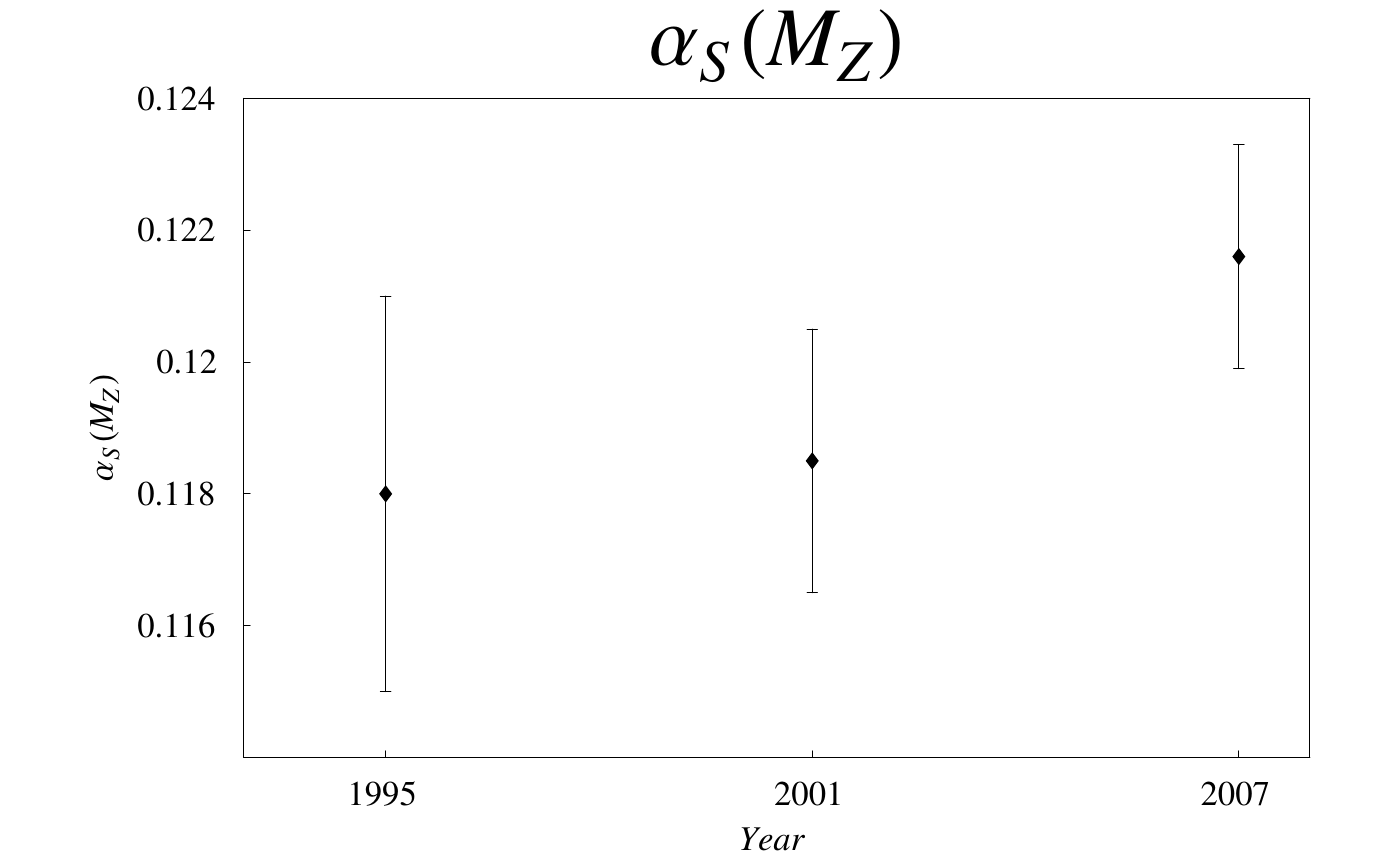}}
  \centerline{ \includegraphics[width=3in]{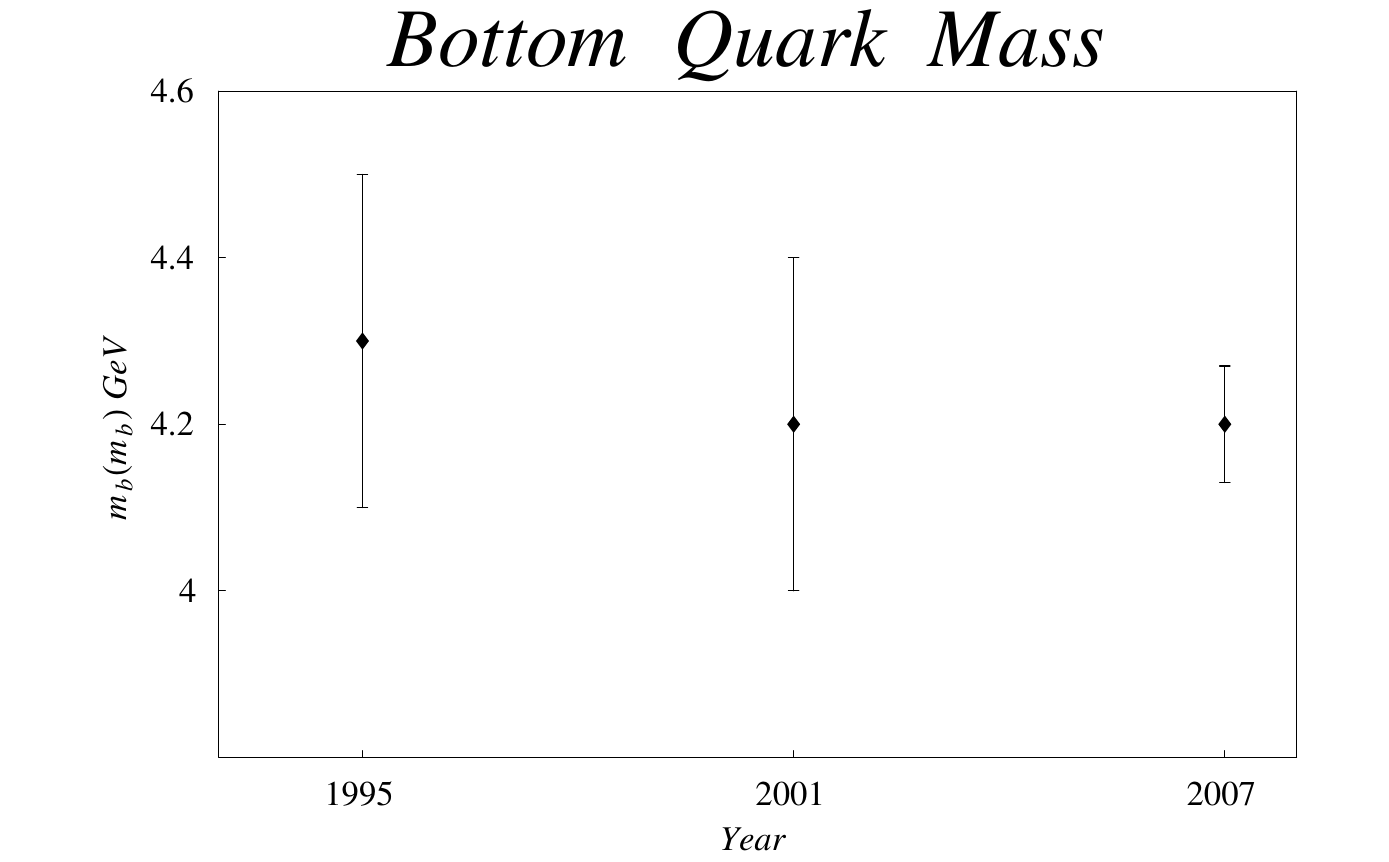}}
  \caption{\label{FigUpdatesToRunningDominantTerms}Updates to the top-quark mass, strong coupling constant, and bottom-quark mass are responsible for the quantitative stress of the classic GUT relation for $y_b/y_\tau$.}
\end{figure}

\begin{figure}
  \centerline{\includegraphics[width=3in]{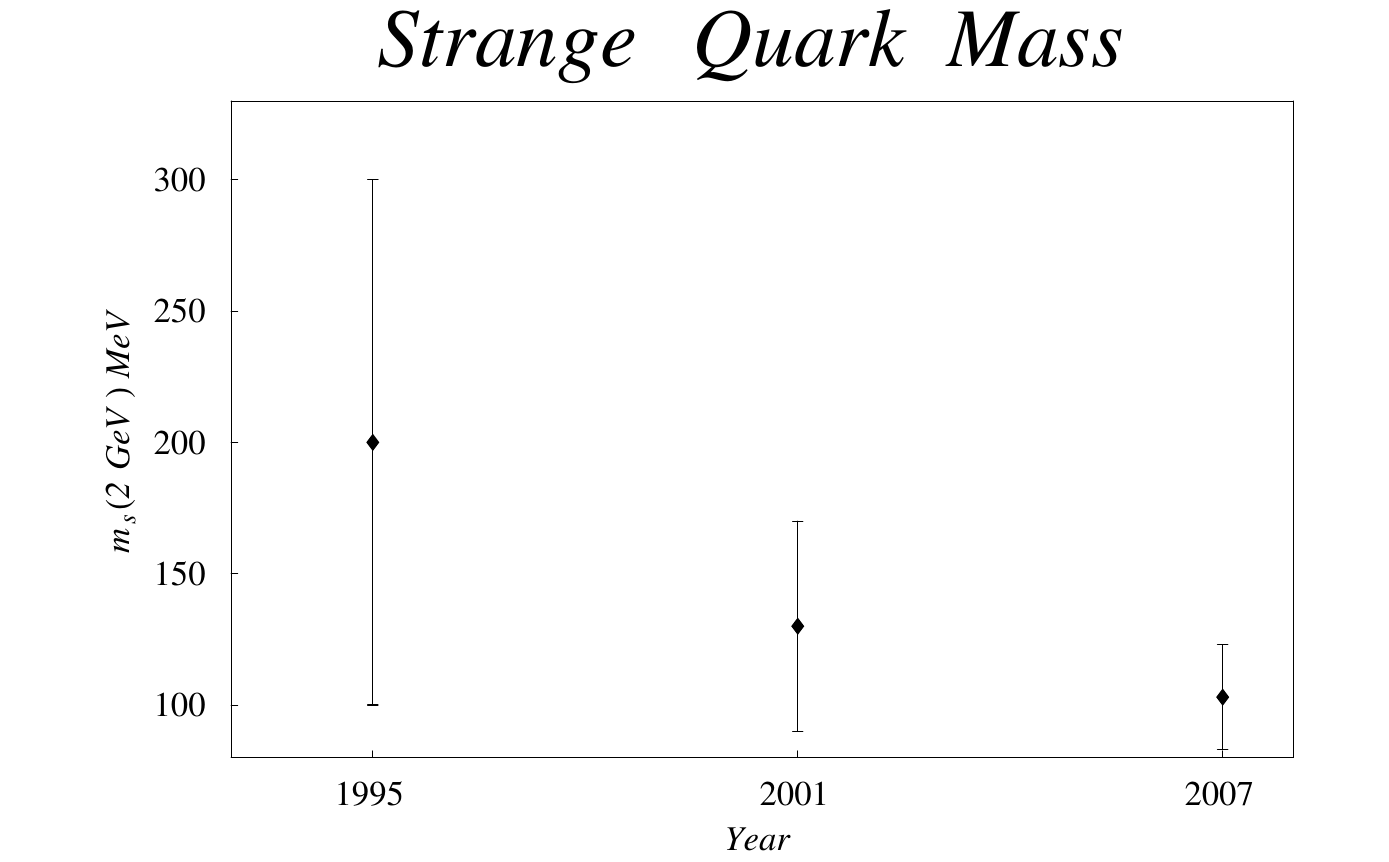} \includegraphics[width=3in]{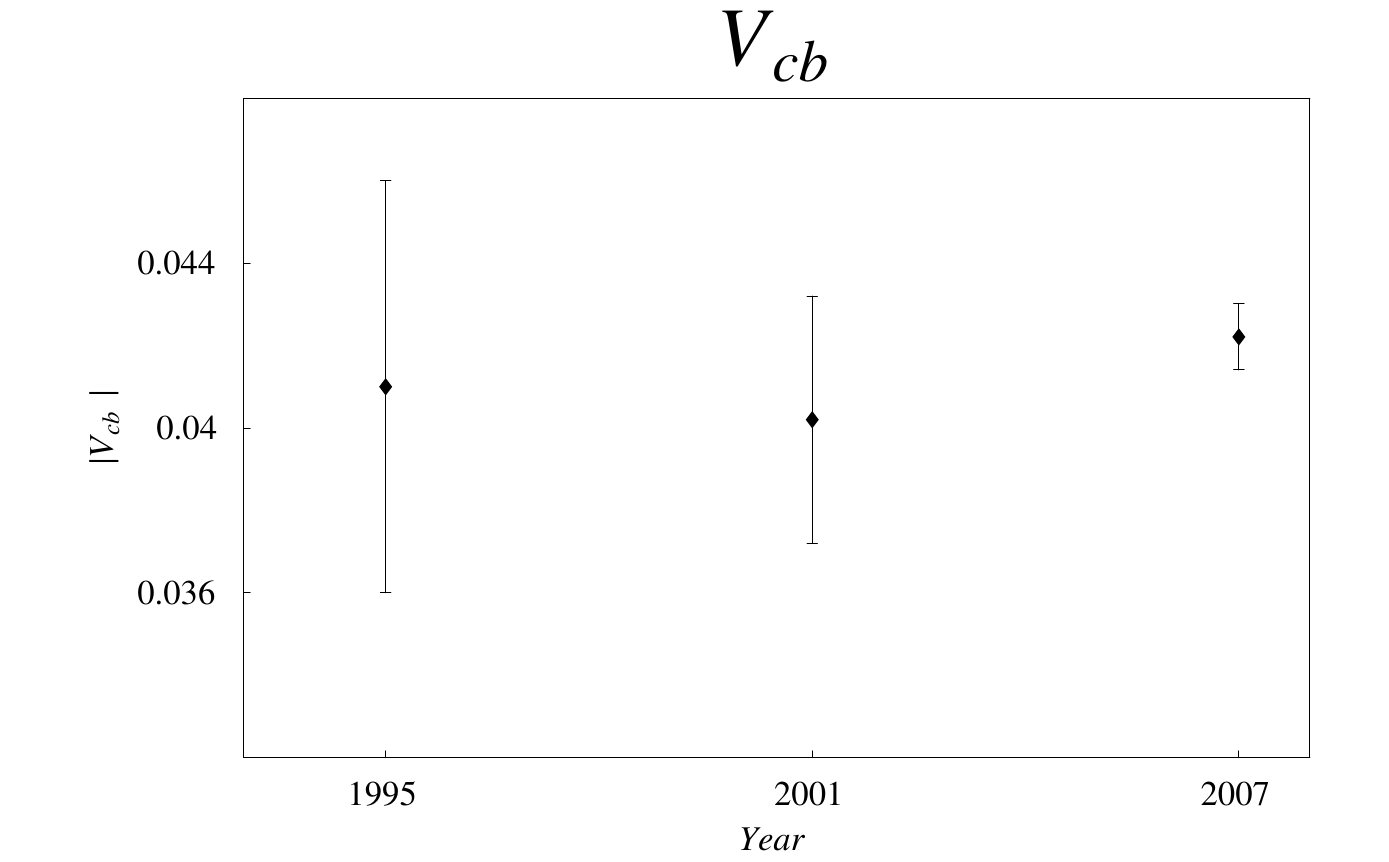}}
  \caption{\label{FigUpdatedObservablesThatAffectGJSuccess} Updates to
  the strange-quark mass and $V_{cb}$ are responsible for the quantitative stress of the Georgi-Jarlskog mass relations and the need to update values from Ref.~\cite{Roberts:2001zy}.}
\end{figure}

Grand Unified Theories predict relationships between the GUT-scale quark and
lepton masses. Using new data in the context of the MSSM, we update the
values and uncertainties of the masses and mixing angles for the three
generations at the GUT scale. We also update fits to hierarchical patterns
in the GUT-scale Yukawa matrices. The new data shows that not all the classic
GUT-scale mass relationships remain in quantitative agreement at small to
moderate $\tan \beta $. However, at large $\tan \beta $, these discrepancies
can be eliminated by finite, $\tan \beta $-enhanced, radiative, threshold
corrections if the gluino mass has the opposite sign to the wino mass.
This chapter is based on work first published by the author and his supervisor in Ref.~\cite{Ross:2007az}.

Explaining the origin of fermion masses and
mixings remains one of the most important
goals in our attempts to go beyond the
Standard Model. In this, one very promising
possibility is that there is an underlying
stage of unification relating the couplings
responsible for the fermion masses. However
we are hindered by the fact that the measured
masses and mixings do not directly give the
structure of the underlying Lagrangian both
because the data is insufficient
unambiguously to reconstruct the full fermion
mass matrices and because radiative
corrections can obscure the underlying
structure. In this chapter we will address
both these points in the context of the MSSM.

We first present an analysis of the measured mass and mixing angles
continued to the GUT scale.
The analysis updates Ref \cite{Roberts:2001zy}
using the
precise measurements of fermion masses and mixing angles from the
b-factories and the updated top-quark mass from CDF and D0. The resulting
data at the GUT scale allows us to look for underlying patterns which may
suggest a unified origin. We also explore the sensitivity of these patterns
to $\tan \beta $-enhanced, radiative threshold corrections.

We next proceed to extract the underlying Yukawa coupling matrices for the
quarks and leptons. There are two difficulties in this. The first is that
the data cannot, without some assumptions, determine all elements of these
matrices. The second is that the Yukawa coupling matrices are basis
dependent.
We choose to work in a basis in which the mass matrices are
hierarchical in structure with the off-diagonal elements small relative to
the appropriate combinations of on-diagonal matrix elements.
Appendix \ref{SecCKMParams}, Eq(\ref{EqYukawaOrderOfMagnitude}) defines this basis more precisely.
This is the
basis we think is most likely to display the structure of the underlying
theory, for example that of a spontaneously broken family symmetry in which
the hierarchical structure is ordered by the (small) order parameter
breaking the symmetry. With this structure to leading order the observed
masses and mixing angles determine the mass matrix elements on and above the
diagonal, and our analysis determines these entries, again allowing for
significant $\tan \beta $ enhanced radiative corrections. The resulting form
of the mass matrices provides the \textquotedblleft data\textquotedblright\
for developing models of fermion masses such as those based on a broken
family symmetry.

\section{Supersymmetric Thresholds and GUT-Scale Mass Relations}

\begin{table}[tb]
\footnotesize
\centerline{
\begin{tabular}{|c|c|p{1.8in}|}
 \hline
 Low-Energy Parameter &
  Value(Uncertainty in last digit(s)) &
  Notes and Reference \cr
  \hline
 $m_u(\mu_L)/m_d(\mu_L)$
   &   0.45(15)
   & PDB  Estimation \cite{PDBook2006} \cr
 $m_s(\mu_L)/m_d(\mu_L)$
   & 19.5(1.5)
   & PDB  Estimation \cite{PDBook2006} \cr
 $m_u(\mu_L)+ m_d(\mu_L)$
 & $\left[ 8.8(3.0),\ 7.6(1.6) \right]$ MeV  &
 PDB, Quark Masses, pg 15 \cite{PDBook2006}. (
 Non-lattice, Lattice )
 \cr
 $Q = \sqrt{\frac{m_s^2 - (m_d + m_u)^2/4}{m_d^2 - m_u^2}}$
   & 22.8(4)
    & Martemyanov and Sopov
\cite{Martemyanov:2005bt} \cr
 $m_s(\mu_L)$
 & $\left[ 103 (20)\, , 95 (20) \right]$ MeV & PDB, Quark Masses, pg 15
\cite{PDBook2006}. [Non-lattice, lattice] \cr
   $m_u(\mu_L)$ & 3(1)\,   MeV & PDB, Quark
Masses, pg 15 \cite{PDBook2006}.
Non-lattice.\cr
 $m_d(\mu_L)$
   & 6.0(1.5)\, MeV &
    PDB, Quark Masses, pg 15 \cite{PDBook2006}. Non-lattice.  \cr
  $m_c(m_c)$ & 1.24(09) GeV & PDB, Quark Masses, pg 16
  \cite{PDBook2006}. Non-lattice.\cr
 $m_b(m_b)$ &
4.20(07) \, GeV & PDB, Quark Masses, pg 16,19
\cite{PDBook2006}. Non-lattice. \cr
 $M_{t}$ & 170.9 (1.9)\,{\rm{GeV}} &  CDF \& D0 \cite{Group:2007bx} Pole Mass   \cr
  $(M_e,M_\mu,M_\tau)$ &
($0.511(15)$,$\ 105.6(3.1)$,$\ 1777(53)$ )
MeV & $3\%$ uncertainty from neglecting $Y^e$
thresholds.  \cr $A$ Wolfenstein parameter &
0.818(17)  &  PDB Ch 11 Eq.~11.25
\cite{PDBook2006}  \cr $\overline{\rho}$
Wolfenstein parameter & 0.221(64)   &  PDB Ch
11 Eq.~11.25 \cite{PDBook2006}  \cr $\lambda$
Wolfenstein parameter &   0.2272(10)     &
PDB Ch 11 Eq.~11.25 \cite{PDBook2006}  \cr
$\overline{\eta}$ Wolfenstein parameter &
0.340(45) &  PDB Ch 11 Eq.~11.25
\cite{PDBook2006}  \cr $|V_{CKM}|$ & $\left(
\begin{matrix} 0.97383(24) & 0.2272(10) & 0.00396(09) \cr
0.2271(10) & 0.97296(24) & 0.04221(80) \cr
0.00814(64) & 0.04161(78) & 0.999100(34)
\end{matrix} \right)$ &   PDB Ch 11 Eq.~11.26
\cite{PDBook2006}  \cr $\sin 2\beta$ from CKM
& 0.687(32) & PDB Ch 11 Eq.~11.19
\cite{PDBook2006} \cr Jarlskog Invariant &
 $3.08(18) \times 10^{-5}$ &  PDB Ch 11
 Eq.~11.26 \cite{PDBook2006}\cr
\hline
 $v_{Higgs}(M_Z)$ & $246.221(20)$ GeV &
 Uncertainty expanded. \cite{PDBook2006}  \cr
 ( $\alpha_{EM}^{-1}(M_Z)$, $\alpha_s(M_Z)$,
 $\sin^2 \theta_W(M_Z)$ ) &  ($\ 127.904(19)$, $\ 0.1216(17)$, $\ 0.23122(15)$) & PDB Sec 10.6
 \cite{PDBook2006} \cr
\hline
\end{tabular}}
\normalsize
\caption{Low-energy observables. Masses in lower-case $m$ are $\overline{MS}$
running masses. Capital $M$ indicates pole mass. The light quark's ($u$,$d$,$%
s$) mass are specified at a scale $\protect\mu _{L}=2\ \mathrm{GeV}$. $%
V_{CKM}$ are the Standard Model's best fit values. }
\label{TableParametersToFitTo}
\end{table}

The data set used is summarized in Table
\ref{TableParametersToFitTo}. Since the fit
of reference \cite{Roberts:2001zy} (RRRV) to
the Yukawa texture was done, the measurement
of the Standard-Model parameters has improved
considerably.
Figs~\ref{FigUpdatesToRunningDominantTerms} and \ref{FigUpdatedObservablesThatAffectGJSuccess} highlight a few of the
changes in the data since: The top-quark
mass has gone from $M_t=174.3 \pm 5$ GeV to
$M_t=170.9 \pm 1.9$ GeV.
In 2000 the Particle Data Book reported $m_{b}(m_{b})=4.2\pm 0.2$ GeV \cite%
{PDBook2000} which has improved to $m_{b}(m_{b})=4.2\pm 0.07$ GeV today. In
addition each higher order QCD correction pushes down the value of $%
m_{b}(M_{Z})$ at the scale of the $Z$ bosons mass. In 1998 $%
m_{b}(M_{Z})=3.0\pm 0.2$ GeV \cite{Fusaoka:1998vc} and today it is $%
m_{b}(M_{Z})=2.87\pm 0.06$ GeV \cite{Baer:2002ek}. The most significant
shift in the data relevant to the RRRV fit is a downward revision to the
strange-quark mass at the scale $\mu _{L}=2$ GeV from $m_{s}(\mu
_{L})\approx 120\pm 50$ MeV \cite{PDBook2000} to today's value $m_{s}(\mu
_{L})=103\pm 20$ MeV. We also know the CKM unitarity triangle parameters
better today than six years ago. For example, in 2000 the Particle Data book
reported $\sin 2\beta =0.79\pm 0.4$ \cite{PDBook2000} which is improved to $%
\sin 2\beta =0.69\pm 0.032$ in 2006 \cite{PDBook2006}. Figures \ref{FigUpdatesToRunningDominantTerms} and  \ref{FigUpdatedObservablesThatAffectGJSuccess}
show these updates visually.
The $\sin 2\beta $
value is about $1.2\,\sigma $ off from a global fit to all the CKM data \cite%
{Lubicz:2007zv}, our fits generally lock onto the global-fit data and
exhibit a $1\,\sigma$ tension for $\sin 2\beta$. Together, the improved CKM
matrix observations add stronger constraints to the textures compared to
data from several years ago.

We first consider the determination of the fundamental mass parameters at
the GUT scale in order simply to compare to GUT predictions. The starting
point for the light-quark masses at low scale is given by the $\chi ^{2}$
fit to the data of Table \ref{TableParametersToFitTo}
\begin{equation}
m_{u}(\mu _{L})=2.7\pm 0.5\ \mathrm{MeV}\ \ m_{d}(\mu _{L})=5.3\pm 0.5\
\mathrm{MeV}\ \ m_{s}(\mu _{L})=103\pm 12\ \mathrm{MeV}.
\label{EqLightQuarkMassConsistent}
\end{equation}%
Using these as input we determine the values
of the mass parameters at the GUT scale for
various choices of $\tan \beta$ but not
including possible $\tan \beta $ enhanced
threshold corrections.  We do this using
numerical solutions to the RG equations. The
one-loop and two-loop RG equations for the
gauge couplings and the Yukawa couplings in
the Standard Model and in
the MSSM that we use in this study come from a number of sources \cite{Fusaoka:1998vc}\cite{Chankowski:2001mx}\cite{Ramond:1999vh}\cite{Barger:1992ac} and are detailed in
Appendix \ref{AppendixRGRunningDetails}.
The results are given in the
first five columns of Table \ref{table2}.
These can readily be compared to expectations
in various Grand Unified models. The classic
prediction of $SU(5)$ with third generation
down-quark and charged-lepton masses given by
the coupling
$B\;\overline{5}_{f}.10_{f}.5_{H}$\footnote{$\overline{5}_{f}$, $10_{f}$ refer
to the $SU(5)$ representations making up a
family of quarks and leptons while $5_{H}$ is
a five dimensional representation of Higgs
scalars.} is $m_{b}(M_{X})/m_{\tau}(M_{X})=1$
\cite{Buras:1977yy}.  This ratio is given in
Table \ref{table2} where it may be seen that
the value agrees at a special low $\tan
\beta$ value but for large $\tan \beta $ it
is some $25\%$ smaller than the GUT
prediction\footnote{We'd like to thank Ilja
Dorsner for pointing out that the $\tan
\beta$ dependence of $m_{b}/m_{\tau}(M_X)$ is
more flat than in previous studies (e.g.
ref.~\cite{Barr:2002mw}).  This change is
mostly due to the higher effective SUSY scale
$M_S$, the higher value of $\alpha_s(M_Z)$
found in global standard model fits, and
smaller top-quark mass $M_t$.}. A similar
relation between the strange quark and the
muon is untenable and to describe the masses
consistently in $SU(5)$ Georgi and Jarlskog
\cite{Georgi:1979df} proposed that the second
generation masses should come instead from
the coupling
$C\;\overline{5}_{f}.10_{f}.45_{H}$ leading
instead to the relation 3$m_{s}(M_{X})/m_{\mu
}(M_{X})=1.$ As may be seen from Table
\ref{table2} in
all cases this ratio is approximately
$0.69(8)$. The prediction of Georgi and
Jarlskog
for the lightest generation masses follows from the relation $%
Det(M^{d})/Det(M^{l})=1$. This results from the form of their mass matrix
which is given by\footnote{%
The remaining mass matrix elements may be non-zero provided they do not
contribute significantly to the determinant.}
\begin{equation}
M^{d}=\left(
\begin{array}{ccc}
0 & A^{\prime } &  \\
A & C &  \\
&  & B%
\end{array}%
\right) ,\;M^{l}=\left(
\begin{array}{ccc}
0 & A^{\prime } &  \\
A & -3C &  \\
&  & B%
\end{array}%
\right)   \label{GeorgiJarlskog}
\end{equation}%
in which there is a $(1,1)$ texture zero\footnote{%
Below we discuss an independent reason for having a $(1,1)$ texture zero.}
and the determinant is given by the product of the $(3,3)$, $(1,2)$ and
$(2,1)$ elements. If the $(1,2)$ and $(2,1)$ elements are also given by $%
\overline{5}_{f}.10_{f}.5_{H}$ couplings they will be the same in the
down-quark and charged-lepton mass matrices giving rise to the equality of
the determinants. The form of Eq(\ref{GeorgiJarlskog}) may be arranged by
imposing additional continuous or discrete symmetries. One may see from
Table \ref{table2} that the actual value of the ratio of the determinants is
quite far from unity disagreeing with the Georgi Jarlskog relation.

In summary the latest data on fermion masses,
while qualitatively in agreement with the
simple GUT relations, has significant
quantitative discrepancies. However the
analysis has not, so far, included the SUSY
threshold corrections which substantially
affect the GUT mass relations at large $\tan
\beta$ \cite{Diaz-Cruz:2000mn}.
\begin{table}[tbh]
\centerline{
\begin{tabular}{|c|c|c|c|c||c|c|}
 \hline
 Parameters & \multicolumn{6}{|c||}{Input SUSY Parameters }  \\
 \hline
 $\tan \beta$ & $1.3$ & $10$ & $ 38$ &  $ 50$ & $38$ & $38$ \cr
 $\gamma_b$ & $0$ & $0$ & $ 0$ &  $0$ & $-0.22$ & $+0.22$ \cr
 $\gamma_d$ & $0$ & $0$ & $ 0$ &  $0$ & $-0.21$ & $+0.21$ \cr
 $\gamma_t$ & $0$ & $0$ & $ 0$ &  $0$ & $0$ &  $-0.44$ \cr
 \hline
 Parameters &
 \multicolumn{6}{|c|}{Corresponding GUT-Scale Parameters
 with
 Propagated Uncertainty}  \cr
 \hline
 $y^t(M_X)$ & $6^{+1}_{-5}$ & $0.48(2)$ & $0.49(2)$ & $0.51(3)$               	 & $0.51(2)$ & $0.51(2)$ \cr
  $y^b(M_X)$ & $0.0113^{+0.0002}_{-0.01}$ & $0.051(2)$ & $0.23(1)$ & $0.37(2)$ 	 & $0.34(3)$ & $0.34(3)$ \cr
  $y^\tau(M_X)$ & $0.0114(3)$ &  $0.070(3)$  & $0.32(2)$ & $0.51(4)$          	 & $0.34(2)$ & $0.34(2)$   \cr
  $(m_u/m_c)(M_X)$ & $0.0027(6)$ & $0.0027(6)$ & $0.0027(6)$ & $0.0027(6)$     	 & $0.0026(6)$ & $0.0026(6)$ \cr
  $(m_d/m_s)(M_X)$ & $0.051(7)$ & $0.051(7)$ & $0.051(7)$ & $0.051(7)$        	 & $0.051(7)$  & $0.051(7)$   \cr
  $(m_e/m_\mu)(M_X)$ & $0.0048(2)$ & $0.0048(2)$ & $0.0048(2)$ & $0.0048(2)$  	 & $0.0048(2)$ & $0.0048(2)$ \cr
  $(m_c/m_t)(M_X)$ & $0.0009^{+0.001}_{-0.00006}$  & $0.0025(2)$ & $0.0024(2)$ & $0.0023(2)$  	 & $0.0023(2)$ & $0.0023(2)$ \cr
  $(m_s/m_b)(M_X)$  & $0.014(4)$ & $0.019(2)$ & $0.017(2)$ & $0.016(2)$       	 & $0.018(2)$ & $0.010(2)$ \cr
  $(m_\mu / m_\tau)(M_X)$ & $0.059(2)$  & $0.059(2)$ & $0.054(2)$ & $0.050(2)$      	 & $0.054(2)$ & $0.054(2)$ \cr
  $A(M_X)$ & $0.56^{+0.34}_{-0.01}$ & $0.77(2)$ & $0.75(2)$ & $0.72(2)$       	 & $0.73(3)$  & $0.46(3)$ \cr
  $\lambda(M_X)$ & $0.227(1)$ & $0.227(1)$ & $0.227(1)$ & $0.227(1)$          	 & $0.227(1)$ & $0.227(1)$ \cr
  $\bar{\rho}(M_X)$ & $0.22(6)$ & $0.22(6)$ & $0.22(6)$ &  $0.22(6)$          	 & $0.22(6)$ & $0.22(6)$  \cr
  $\bar{\eta}(M_X)$ & $0.33(4)$ & $0.33(4)$ & $0.33(4)$ & $0.33(4)$           	 & $0.33(4)$  & $0.33(4)$ \cr
  $J(M_X)\,  \times 10^{-5} $ & $1.4^{+2.2}_{-0.2}$  & $2.6(4)$ &  $2.5(4)$ & $2.3(4)$  	 & $2.3(4)$ & $1.0(2)$ \cr
  \hline
  Parameters &
 \multicolumn{6}{|c|}{Comparison with GUT Mass Ratios}  \cr
  \hline
  $(m_b/m_\tau)(M_X)$
     & $1.00^{+0.04}_{-0.4}$
     & $0.73(3)$
     & $0.73(3)$
     & $0.73(4)$       	 & $1.00(4)$ & $1.00(4)$ \cr
  $({3 m_s / m_\mu})(M_X)$
       & $0.70^{+0.8}_{-0.05}$
       & $0.69(8)$
       & $0.69(8)$
       & $0.69(8)$     	 & $0.9(1)$ & $0.6(1)$ \cr
  $({m_d / 3\,m_e})(M_X)$
      & $0.82(7)$
      & $0.83(7)$
      & $0.83(7)$
      & $0.83(7)$      	 & $1.05(8)$ & $0.68(6)$ \cr
  $(\frac{\det Y^d}{\det Y^e})(M_X)$
       & $0.57^{+0.08}_{-0.26}$
       & $0.42(7)$
       & $0.42(7)$
       & $0.42(7)$    	 & $0.92(14)$ & $0.39(7)$ \cr
  \hline
 \end{tabular}}
\caption{The mass parameters continued to the GUT-scale $M_X$ for various
values of $\tan \protect\beta $ and threshold corrections $\protect\gamma %
_{t,b,d}$. These are calculated with the 2-loop gauge coupling and 2-loop
Yukawa coupling RG equations assuming an effective SUSY scale $M_{S}=500$
GeV.}
\label{table2}
\end{table}
A catalog of the full SUSY threshold
corrections is given in \cite{Pierce:1996zz}.
The particular finite SUSY thresholds
discussed in this letter do not decouple as
the superpartners become massive. We follow
the approximation described in Blazek, Raby,
and Pokorski (BRP) for threshold corrections
to the CKM elements and down-like mass
eigenstates \cite{Blazek:1995nv}. The finite
threshold corrections to $Y^{e}$ and $Y^{u}$
and are generally about 3\% or smaller
\begin{equation}
\delta Y^{u},\ \delta Y^{d}\lesssim 0.03
\end{equation}%
and will be neglected in our study. The logarithmic threshold corrections
are approximated by using the Standard-Model RG equations from $M_Z$ to an
effective SUSY scale $M_S$.

The finite, $\tan \beta$-enhanced $Y^{d}$ SUSY threshold corrections are
dominated by the a sbottom-gluino loop, a stop-higgsino loop, and a
stop-chargino loop. Integrating out the SUSY particles at a scale $M_{S}$
leaves the matching condition at that scale for the Standard-Model Yukawa
couplings:
\begin{eqnarray}
\delta m_{sch}\,Y^{u\,SM} &=&\sin \beta \ \,Y^{u} \\
\delta m_{sch}\,Y^{d\,SM} &=&\cos \beta \ \,U_{L}^{d\dag }\,\left( 1+%
\tanbloopfactor {\Gamma}^{d}+\tanbloopfactor V_{CKM}^{\dag }\,{\Gamma }%
^{u}\,V_{CKM}\right) \,Y^{d}_{\mathrm{diag}}\,U_{R}^{d} \\
Y^{e\,SM} &=&\cos \beta \,\ Y^{e}.
\end{eqnarray}%
All the parameters on the right-hand side
take on their MSSM values in the
$\overline{DR}$ scheme. The factor $\delta
m_{sch}$ converts the quark running masses
from $\overline{MS}$ to $\overline{DR}$
scheme. Details about this scheme conversion
are listed in Appendix \ref{SecRGEMSSM}. The $\beta$ corresponds to the ratio
of the two Higgs VEVs $v_u / v_d=\tan \beta$.
The $U$ matrices decompose the MSSM Yukawa
couplings at the scale $M_{S}$:
$Y^{u}=U_{L}^{u\dag
}Y_{\mathrm{diag}}^{u}U_{R}^{u}$ and
$Y^{d}=U_{L}^{d\dag
}Y_{\mathrm{diag}}^{d}U_{R}^{d}$. The
matrices $Y_{\mathrm{diag}}^{u}$ and
$Y_{\mathrm{diag}}^{d}$ are diagonal and
correspond to the mass eigenstates divided by
the appropriate VEV at the scale $M_{S}$. The
CKM matrix is given by
$V_{CKM}=U_{L}^{u}U_{L}^{d\dag }$. The
left-hand side involves the Standard-Model
Yukawa couplings. The matrices $\Gamma ^{u}$
and $\Gamma ^{d}$ encode the SUSY threshold
corrections.

If the squarks are diagonalized in flavor space by the same rotations that
diagonalize the quarks, the matrices $\Gamma ^{u}$ and $\Gamma ^{d}$ are
diagonal: $\Gamma ^{d}=\mathrm{diag}(\gamma _{d},\gamma _{d},\gamma _{b}),$ $%
\ \Gamma ^{u}=\mathrm{diag}(\gamma _{u},\gamma _{u},\gamma _{t})$. In
general the squarks are not diagonalized by the same rotations as the quarks
but provided the relative mixing angles are reasonably small the corrections
to flavour conserving masses, which are our primary concern here, will be
second order in these mixing angles. We will assume $\Gamma ^{u}$ and $%
\Gamma ^{d}$ are diagonal in what follows.

Approximations for $\Gamma^{u}$ and $\Gamma
^{d}$ based on the mass
insertion approximation are found in \cite{Carena:1999py}\cite{Carena:2002es}%
\cite{Tobe:2003bc}:
\begin{eqnarray}
\gamma _{t} &\approx &y_{t}^{2}\,\mu \,A^{t}\,\frac{\tan \beta }{16\pi ^{2}}%
I_{3}(m_{\tilde{t}_{1}}^{2},m_{\tilde{t}_{2}}^{2},\mu ^{2})\ \ \sim \ \
y_{t}^{2}\,\frac{\tan \beta }{32\pi ^{2}}\frac{\mu \,A^{t}\,}{m_{\tilde{t}%
}^{2}}  \label{Eqgammat} \\
\gamma _{u} &\approx &-g_{2}^{2}\,M_{2}\,\mu \,\frac{\tan \beta }{16\pi ^{2}}%
I_{3}(m_{\chi _{1}}^{2},m_{\chi _{2}}^{2},m_{\tilde{u}}^{2})\ \ \sim \ \ 0
\label{Eqgammau} \\
\gamma _{b} &\approx &\frac{8}{3}\,g_{3}^{2}\,\frac{\tan \beta }{16\pi ^{2}}%
\,M_{3}\,\mu \,I_{3}(m_{\tilde{b}_{1}}^{2},m_{\tilde{b}_{2}}^{2},{M_{3}}%
^{2})\ \ \sim \ \ \frac{4}{3}\,g_{3}^{2}\,\frac{\tan \beta }{16\pi ^{2}}\,%
\frac{\mu \,M_{3}}{m_{\tilde{b}}^{2}}  \label{Eqgammab} \\
\gamma _{d} &\approx &\frac{8}{3}g_{3}^{2}\frac{\tan \beta }{16\pi ^{2}}%
M_{3}\,\mu \,I_{3}(m_{\tilde{d}_{1}}^{2},m_{\tilde{d}_{2}}^{2},{M_{3}}^{2})\
\ \sim \ \ \frac{4}{3}\,g_{3}^{2}\,\frac{\tan \beta }{16\pi ^{2}}\,\frac{\mu
\,M_{3}}{m_{\tilde{d}}^{2}}  \label{Eqgammad}
\end{eqnarray}%
where $I_{3}$ is given by
\begin{equation}
I_{3}(a^{2},b^{2},c^{2})=\frac{a^{2}b^{2}\log \frac{a^{2}}{b^{2}}%
+b^{2}c^{2}\log \frac{b^{2}}{c^{2}}+c^{2}a^{2}\log \frac{c^{2}}{a^{2}}}{%
(a^{2}-b^{2})(b^{2}-c^{2})(a^{2}-c^{2})}.
\end{equation}%
In these expressions $\tilde{q}$ refers to
superpartner of $q$. $\chi ^{j}$ indicate
chargino mass eigenstates. $\mu $ is the
coefficient to the $H^{u}$ $H^{d}$
interaction in the superpotential.
$M_{1},M_{2},M_{3}$ are the gaugino soft
breaking terms. $A^{t}$ refers to the soft
top-quark trilinear coupling. The mass
insertion approximation breaks down if there
is large mixing between the mass eigenstates
of the stop or the sbottom. The right-most
expressions in
Eqs(\ref{Eqgammat},\ref{Eqgammab},\ref{Eqgammad})
assume the relevant squark mass eigenstates
are nearly degenerate and heavier than
$M_{3}$ and $\mu$.
  These
 expressions ( eqs
 \ref{Eqgammat} - \ref{Eqgammad}) provide an
 approximate mapping from a supersymmetric
 spectra to the $\gamma_i$ parameters through which we parameterize
 the threshold corrections; however, with the
 exception of Column A of Table \ref{Table4},
 we do not
 specify a SUSY spectra but directly
 parameterize the thresholds corrections through
 $\gamma_i$.

The separation between $\gamma_b$ and
$\gamma_d$ is set by the lack of degeneracy
of the down-like squarks. If the squark
masses for the first two generations are not
degenerate, then there will be a
corresponding separation between the (1,1)
and (2,2) entries of  $\Gamma^d$ and
$\Gamma^u$. If the sparticle spectra is
designed to have a large $A^t$ and a light
stop, $\gamma_t$ can be enhanced and dominate
over $\gamma_b$. Because the charm Yukawa
coupling is so small, the scharm-higgsino
loop is negligible, and $\gamma_u$ follows
from a chargino squark loop and is also
generally small with values around $0.02$
because of the smaller $g_2$ coupling. In our
work, we approximate $\Gamma^u_{22} \sim
\Gamma^u_{11} \sim 0$. The only substantial
correction to the first and second
generations is given by $\gamma_d$
\cite{Diaz-Cruz:2000mn}.

As described in BRP, the threshold corrections leave $|V_{us}|$ and $%
|V_{ub}/V_{cb}|$ unchanged to a good approximation. Threshold corrections in
$\Gamma ^{u}$ do affect the $V_{ub}$ and $V_{cb}$ at the scale $M_{S}$
giving
\begin{equation}
\frac{V_{ub}^{SM}-V_{ub}^{MSSM}}{V_{ub}^{MSSM}}\backsimeq \frac{%
V_{cb}^{SM}-V_{cb}^{MSSM}}{V_{cb}^{MSSM}}\backsimeq -\left( \gamma
_{t}-\gamma _{u}\right) .
\end{equation}
The threshold corrections for the down-quark
masses are given  approximately by
\begin{eqnarray*}
m_{d} & \backsimeq & m_{d}^{0}\, (1+\gamma _{d}+\gamma _{u})^{-1} \\
m_{s} & \backsimeq &m_{s}^{0}\, (1+\gamma _{d}+\gamma _{u})^{-1} \\
m_{b} & \backsimeq &m_{b}^{0}\, (1+\gamma _{b}+\gamma_{t})^{-1}
\end{eqnarray*}%
where the superscript $0$ denotes the mass without threshold corrections.
Not shown are the nonlinear effects which arise through the RG equations
when the bottom Yukawa coupling is changed by threshold effects. These are
properly included in our final results obtained by numerically solving the
RG equations.

Due to our assumption that the squark masses for the first two generations
are degenerate, the combination of the GUT relations given by $\left( \det
M^{l}/\det M^{d}\right) \left( 3\,m_{s}/m_{\mu }\right) ^{2}\left(
m_{b}/m_{\tau }\right) =1$\ is unaffected up to nonlinear effects. Thus we
cannot simultaneously fit all three GUT relations through the threshold
corrections. A best fit requires the threshold effects given by
\begin{eqnarray}
\gamma _{b}+\gamma _{t} &\approx &-0.22\pm 0.02  \label{threshold} \\
\gamma _{d}+\gamma _{u} &\approx &-0.21\pm 0.02.  \label{threshold2}
\end{eqnarray}%
giving the results shown in the penultimate column of Table \ref{table2},
just consistent with the GUT predictions. The question is whether these
threshold effects are of a reasonable magnitude and, if so, what are the
implications for the SUSY\ spectra which determine the $\gamma _{i}?$ From
Eqs(\ref{Eqgammab},\ref{Eqgammad}), at $\tan \beta =38$ we have
\begin{equation*}
\frac{\mu \,M_{3}}{m_{\tilde{b}}^{2}}\sim -0.5,\;\ \ \ \frac{m_{\tilde{b}%
}^{2}}{m_{\tilde{d}}^{2}}\sim 1.0
\end{equation*}

The current observation of the muon's $(g-2)_{\mu }$ is $3.4\,\sigma $ \cite%
{Hagiwara:2006jt} away from the
Standard-Model prediction. If SUSY is to
explain the observed deviation, one needs
$\tan \beta >8$ \cite{Everett:2001tq} and
$\mu M_{2}>0$ \cite{Stockinger:2006zn}. With
this sign we must have $\mu M_{3}$ negative
and the $\widetilde{d},$ $\widetilde{s}$
squarks only lightly split from the
$\widetilde{b}$ squarks. $M_{3}$ negative is
characteristic of anomaly mediated SUSY\
breaking\cite{Randall:1998uk} and is discussed in \cite{Hall:1993gn}\cite{Komine:2001rm}%
\cite{Tobe:2003bc}\cite{Pallis:2003aw}.
 Although we have deduced
$M_3<0$ from the approximate
Eqs(\ref{Eqgammab},\ref{Eqgammad}), the
correlation persists in the near exact
expression found in Eq(23) of Ref
\cite{Blazek:1995nv}.
Adjusting to different squark splitting can
occur in various schemes \cite{Ramage:2003pf}.
However the squark splitting can readily be
adjusted without spoiling the fit because, up
to nonlinear effects, the solution only
requires the constraints implied by
Eq(\ref{threshold}), so we may make $\gamma
_{b}>\gamma _{d}$ and hence make
$m_{\tilde{b}}^{2}<m_{\tilde{d}}^{2}$ by
allowing for a small positive value for
$\gamma _{t}.$ In this case $A^{t}$ must be
positive.

It is of interest also to consider  the
threshold effects in the case that $\mu
M_{3}$ is positive. This is illustrated in
the last column of Table \ref{table2} in
which we have reversed the sign of $\gamma
_{d},$ consistent with positive $\mu M_{3}$ ,
and chosen $\gamma _{b}\simeq \gamma _{d}$ as
is expected for similar down squark masses.
The value of $\gamma _{t}$ is chosen to keep
the equality between $m_{b}$ and $m_{\tau }.$
One may see that the other GUT relations are
not satisfied, being driven further away by
the threshold corrections. Reducing the
magnitude of $\gamma _{b}$ and $\gamma _{d}$
reduces the discrepancy somewhat but still
limited by the deviation found in the
no-threshold case (the fourth column of Table
\ref{table2}).

At $\tan \beta $ near $50$ the non-linear
effects are large and $b-\tau $ unification
requires $\gamma _{b}+\gamma _{t}\sim -0.1$
to $-0.15.$ In this case it is possible to
have $t-b-\tau $ unification of the Yukawa
couplings. For $\mu >0,M_{3}>0$, the
\textquotedblleft Just-so\textquotedblright\
Split-Higgs solution of references
\cite{King:2000vp,Blazek:2001sb,Blazek:2002ta,Auto:2003ys}
can achieve this while satisfying both
$b\rightarrow s\ \gamma $ and $(g-2)_{\mu }$
constraints but only with large $\gamma _{b}$
and $\gamma _{t}$ and a large cancellation in
$\gamma _{b}+\gamma _{t}$. In this case, as
in the example given above, the threshold
corrections drive the masses further from the
mass relations for the first and second
generations because $\mu \,M_{3}>0$. It
\textit{is} possble to have $t-b-\tau $
unification with $\mu \,M_{3}<0$,  satisfying
the $b\rightarrow s\ \gamma $ and $(g-2)_{\mu
}$ constraints in which the GUT predictions
for the first and second generation of quarks
is acceptable. Examples include Non-Universal
Gaugino Mediation \cite{Balazs:2003mm} and
AMSB; both have some very heavy sparticle masses ( $\gtrsim 4$ TeV) \cite%
{Tobe:2003bc}. Minimal AMSB with a light sparticle spectra( $\lesssim 1$
TeV), while satisfying $(g-2)_{\mu }$ and $b\rightarrow s\ \gamma $
constraints, requires $\tan \beta $ less than about $30$ \cite%
{Stockinger:2006zn}.

\section{Updated fits to Yukawa matrices}

We turn now to the second part of our study in which we update previous fits
to the Yukawa matrices responsible for quark and lepton masses. As discussed
above we choose to work in a basis in which the mass matrices are
hierarchical with the off-diagonal elements small relative to the
appropriate combinations of on-diagonal matrix elements defined in Eq(\ref{EqYukawaOrderOfMagnitude}). This is the basis
we think is most likely to display the structure of the underlying theory,
for example that of a spontaneously broken family symmetry, in which the
hierarchical structure is ordered by the (small) order parameter breaking
the symmetry. With this structure to leading order in the ratio of light to
heavy quarks the observed masses and mixing angles determine the mass matrix
elements on and above the diagonal provided the elements below the diagonal
are not anomalously large. This is the case for matrices that are nearly
symmetrical or for nearly Hermitian as is the case in models based on an $%
SO(10)$ GUT.

\begin{table}[tp]
\centerline{
\begin{tabular}{|c|c|c|c|c|c|c|c|}
\hline
 Parameter & 2001 RRRV & Fit A0 &Fit B0 & Fit A1 & Fit B1  & Fit A2  & Fit B2 \cr
 \hline
 $\tan \beta$
          &   Small
          & $1.3$
          & $1.3$
          & $38$
          & $38$
          & $38$
          & $38$
          \cr
 $a'$
         &  ${\mathcal{O}}(1)$
         &  $0$
         &  $0$
         &  $0$
         &  $0$
         &  $-2.0$
         &  $-2.0$ \cr
 $\epsilon_u$
         & $0.05$
         & $0.030(1)$
         & $0.030(1)$
         & $0.0491(16)$          & $0.0491(15)$          & $0.0493(16)$          & $0.0493(14)$            \cr
 $\epsilon_d$
         & $0.15(1)$
         & $0.117(4)$
         & $0.117(4)$
         & $0.134(7)$
         & $0.134(7)$
         & $0.132(7)$
         & $0.132(7)$ \cr
 $|b'|$
         & $1.0$
         & $1.75(20)$
         & $1.75(21)$
         & $1.05(12)$
         & $1.05(13)$
         & $1.04(12)$
         & $1.04(13)$\cr
 ${\rm{arg}}(b')$
         & $90^o$
         & $+\,93(16)^o$
         & $-\,93(13)^o$
         & $+\,91(16)^o$
         & $-\,91(13)^o$
         & $+\,93(16)^o$
         & $-\,93(13)^o$\cr
 $a$
         & $1.31(14)$
         & $2.05(14)$
         & $2.05(14)$
         & $2.16(23)$
         & $2.16(24)$
         & $1.92(21)$
         & $1.92(22)$ \cr
 $b$
         & $1.50(10)$
         & $1.92(14)$
         & $1.92(15)$
         & $1.66(13)$
         & $1.66(13)$
         & $1.70(13)$
         & $1.70(13)$ \cr
 $|c|$
         & $0.40(2)$
         & $0.85(13)$
         & $2.30(20)$
         & $0.78(15)$          & $2.12(36)$
         & $0.83(17)$
         & $2.19(38)$\cr
 ${\rm{arg}}(c)$
         & ${-\,24(3)^o}$
         & ${-\,39(18)^o}$
         & ${-\,61(14)^o}$
         & ${-\,43(14)^o}$          & ${-\,59(13)^o}$          & ${-\,37(25)^o}$          & ${-\,60(13)^o}$          \cr
 \hline
 \end{tabular}}
\caption{ Results of a $\protect\chi^2$ fit of eqs(\protect\ref{EqYuTexture},%
\protect\ref{EqYdTexture}) to the data in Table \protect\ref{table2} in
the absence of threshold corrections. We set $a^{\prime }$ as indicated and
set $c^{\prime}=d^{\prime}=d=0$ and $f=f^{\prime}=1$ at fixed values. }
\label{Table3}
\end{table}

For convenience we fit to symmetric Yukawa coupling matrices but, as
stressed above, this is not a critical assumption as the data is insensitive
to the off-diagonal elements below the diagonal and the quality of the fit
is not changed if, for example, we use Hermitian forms.
For comparison Appendix \ref{SecCKMParams} gives the observables in terms of
Yukawa matrix entries following a general hierarchical texture.
We parameterize a
set of general, symmetric Yukawa matrices as:
\begin{eqnarray}
Y^{u}(M_{X}) &=&y_{33}^{u}\left(
\begin{matrix}
d^{\prime }\epsilon _{u}^{4} & b^{\prime }\,\epsilon _{u}^{3} & c^{\prime
}\,\epsilon _{u}^{3}\cr b^{\prime }\,\epsilon _{u}^{3} & f^{\prime
}\,\epsilon _{u}^{2} & a^{\prime }\,\epsilon _{u}^{2}\cr c^{\prime
}\,\epsilon _{u}^{3} & a^{\prime }\,\epsilon _{u}^{2} & 1%
\end{matrix}%
\right) ,  \label{EqYuTexture} \\
Y^{d}(M_{X}) &=&y_{33}^{d}\left(
\begin{matrix}
d\,\epsilon _{d}^{4} & b\,\epsilon _{d}^{3} & c\,\epsilon _{d}^{3}\cr %
b\,\epsilon _{d}^{3} & f\,\epsilon _{d}^{2} & a\,\epsilon _{d}^{2}\cr %
c\,\epsilon _{d}^{3} & a\,\epsilon _{d}^{2} & 1%
\end{matrix}%
\right) .  \label{EqYdTexture}
\end{eqnarray}%
Although not shown, we always choose lepton Yukawa couplings at $M_{X}$
consistent with the low-energy lepton masses. Notice that the $f$
coefficient and $\epsilon _{d}$ are redundant (likewise in $Y^{u}$). We
include $f$ to be able to discuss the phase of the (2,2) term. We write all
the entries in terms of $\epsilon $ so that our coefficients will be ${%
\mathcal{O}}(1)$. We will always select our best $\epsilon $ parameters such
that $|f|=1$.

RRRV noted that all solutions, to leading order in the small expansion
parameters, only depend on two phases $\phi _{1}$ and $\phi _{2} $ given by
\begin{eqnarray}
\phi _{1} &=&(\phi _{b}^{\prime }-\phi _{f}^{\prime })-(\phi _{b}-\phi _{f})
\\
\phi _{2} &=&(\phi _{c}-\phi _{a})-(\phi _{b}-\phi _{f})
\end{eqnarray}%
where $\phi_x$ is the phase of parameter $x$.
For this reason it is sufficient to consider
only $b^{\prime}$ and $c$ as complex with all
other parameters real.

As mentioned above the data favours a texture zero in the $(1,1)$ position.
With a symmetric form for the mass matrix for the first two families, this
leads to the phenomenologically successful Gatto Sartori Tonin \cite%
{Gatto:1968ss} relation
\begin{equation}
V_{us}(M_{X})\approx \left\vert b\epsilon _{d}-|b^{\prime }|e^{i\,\phi
_{b^{\prime }}}\epsilon _{u}\right\vert \approx \left\vert \sqrt{(\frac{m_{d}%
}{m_{s}})_{0}}-\sqrt{(\frac{m_{u}}{m_{c}})_{0}}e^{i\,\phi _{1}}\right\vert .
\label{EqVusInTermsOfMasses}
\end{equation}%
This relation gives an excellent fit to $V_{us}$ with $\phi _{1}\approx
\,\pm \,90^{o}$, and to preserve it we take $d,$ $d^{\prime }$ to be zero in
our fits. As discussed above, in $SU(5)$ this texture zero leads to the GUT
relation $Det(M^{d})/Det(M^{l})=1$ which, with threshold corrections, is in
good agreement with experiment. In the case that $c$ is small it was shown
in RRRV that $\phi _{1}$ is to a good approximation the CP violating phase $%
\delta $ in the Wolfenstein parameterization.
A non-zero $c$ is necessary to avoid the
relation $V_{ub}/V_{cb}=\sqrt{m_{u}/m_{c}}$
and with the improvement in the data, it is
now necessary to have $c$ larger than was
found in RRRV \footnote{As shown in
ref.~\cite{Matsuda:2006xa}, it is possible,
in a basis with large off-diagonal entries,
to have an Hermitian pattern with the (1,1)
and (1,3) zero provided one carefully
orchestrates cancelations among $Y^{u}$ and
$Y^{d}$ parameters.  We find this approach
requires a strange-quark mass near its upper
limit.}. As a result the contribution to CP\
violation coming from $\phi _{2}$ is at least
$30\%$. The sign ambiguity in $\phi _{1}$
gives rise to an ambiguity in $c$ with the
positive sign corresponding to the larger
value of $c$ seen in Tables \ref{Table3} and
\ref{Table4}.

\begin{table}
\small
\centerline{
\begin{tabular}{|c|c|c|c|c|c|}
\hline
 Parameter & A & B & C  & B2 & C2 \cr
 \hline
 $\tan \beta$
          & $30$
          & $38$
          & $38$
          & $38$
          & $38$\cr
 $\gamma_b$ & $0.20$
            & $-0.22$
            & $+0.22$
            & $-0.22$
            & $+0.22$ \cr
 $\gamma_t$ & $-0.03$
            & $0$
            & $-0.44$
            & $0$
            & $-0.44$\cr
 $\gamma_d$ & $0.20$
            & $-0.21$
            & $+0.21$
            & $-0.21$
            & $+0.21$\cr
 $a'$
         &  $0$
         &  $0$
         &  $0$
         &  $-2$
         &  $-2$\cr
 \hline
 $\epsilon_u$
         & $0.0495(17)$
         & $0.0483(16)$
         & $0.0483(18)$
         & $0.0485(17)$
         & $0.0485(18)$\cr
 $\epsilon_d$
         & $0.131(7)$
         & $0.128(7)$
         & $0.102(9)$
         & $0.127(7)$
         & $0.101(9)$ \cr
 $|b'|$
         & $1.04(12)$
         & $1.07(12)$
         & $1.07(11)$
         & $1.05(12)$
         & $1.06(10)$\cr
 ${\rm{arg}}(b')$
         & ${\,90(12)^o}$
         & ${\,91(12)^o}$
         & ${\,93(12)^o}$
         & ${\,95(12)^o}$
         & ${\,95(12)^o}$\cr
 $a$
         & $2.17(24)$
         & $2.27(26)$
         & $2.30(42)$
         & $2.03(24)$
         & $1.89(35)$ \cr
 $b$
         & $1.69(13)$
         & $1.73(13)$
         & $2.21(18)$
         & $1.74(10)$
         & $2.26(20)$\cr
 $|c|$
         & $0.80(16)$
         & $0.86(17)$
         & $1.09(33)$
         & $0.81(17)$
         & $1.10(35)$
         \cr
 ${\rm{arg}}(c)$
         & ${-\,41(18)^o}$
         & ${-\,42(19)^o}$
         & ${-\,41(14)^o}$
         & ${-\,53(10)^o}$
         & ${-\,41(12)^o}$
         \cr
 $ Y^u_{33}$
       & $0.48(2)$
       & $0.51(2)$
       & $0.51(2)$
       & $0.51(2)$
       & $0.51(2)$\cr
 $ Y^d_{33}$
       & $0.15(1)$
       & $0.34(3)$
       & $0.34(3)$
       & $0.34(3)$
       & $0.34(3)$\cr
 $ Y^e_{33}$
       & $0.23(1)$
       & $0.34(2)$
       & $0.34(2)$
       & $0.34(2)$
       & $0.34(2)$ \cr
       \hline
  $(m_b/m_\tau)(M_X)$
     & $0.67(4)$
     & $1.00(4)$
     & $1.00(4)$
     & $1.00(4)$
     & $1.00(4)$ \cr
  $({3 m_s / m_\mu})(M_X)$
       & $0.60(3)$
       & $0.9(1)$
       & $0.6(1)$
       & $0.9(1)$
       & $0.6(1)$  \cr
  $({m_d / 3\,m_e})(M_X)$
      & $0.71(7)$
      & $1.04(8)$
      & $0.68(6)$
      & $1.04(8)$
      & $0.68(6)$ \cr
 $ \left|\frac{\det Y^d(M_X)}{\det Y^e(M_X)} \right|$
       & $0.3(1)$
       & $0.92(14)$
       & $0.4(1)$
       & $0.92(14)$
       & $0.4(1)$\cr
 \hline
 \end{tabular}}
\normalsize
\caption{ A $\protect\chi^2$ fit of Eqs(\protect\ref{EqYuTexture},\protect
\ref{EqYdTexture}) including the SUSY threshold effects parameterized by the
specified $\protect\gamma_i$.}
\label{Table4}
\end{table}

Table \ref{Table3} shows results from a $\chi ^{2}$ fit of Eqs(\ref%
{EqYuTexture},\ref{EqYdTexture}) to the
data in Table \ref{table2} in the absence of
threshold corrections. The error, indicated
by the term in brackets, represent the widest
axis of the $1\sigma $ error ellipse in
parameter space. The fits labeled `A' have
phases such that we have the smaller
magnitude solution of $|c|$, and fits labeled
`B' have phases such that we have the larger
magnitude solution of $|c|$. As discussed
above, it is not possible unambiguously to
determine the relative contributions of the
off-diagonal elements of the up and down
Yukawa matrices to the mixing angles. In the
fit A2 and B2 we illustrate the uncertainty
associated with this ambiguity, allowing for
$O(1)$ coefficients $a^{\prime }$. In all the
examples in Table \ref{Table3}, the mass
ratios, and Wolfenstein parameters are
essentially the same as in Table
\ref{table2}.

The effects of the large $\tan \beta $
threshold corrections are shown in Table
\ref{Table4}. The threshold corrections
depend on the details of the SUSY\ spectrum,
and we have displayed the effects
corresponding to a variety of choices for
this spectrum. Column A corresponds to a
\textquotedblleft standard\textquotedblright\
SUGRA fit - the benchmark Snowmass Points and
Slopes (SPS) spectra 1b of
ref(\cite{Allanach:2002nj}). Because the
spectra SPS 1b has large stop and sbottom
squark mixing angles, the approximations
given in Eqns(\ref{Eqgammat}-\ref{Eqgammad})
break down, and the value for the correction
$\gamma_i$ in Column A need to be calculated
with the more complete expressions in BRP
\cite{Blazek:1995nv}
.
In the column A fit and
the next two fits in columns B and C, we set
$a^{\prime }$ and $c^{\prime }$ to zero.
Column B corresponds to the fit given in the
penultimate column of Table \ref{table2}
which agrees very well with the simple GUT
predictions. It is characterized by the
\textquotedblleft
anomaly-like\textquotedblright\ spectrum with
$M_{3}$ negative. Column C examines the
$M_{3}$ positive case while maintaining the
GUT\ prediction for the third generation
$m_{b}=m_{\tau }.$ It corresponds to the
\textquotedblleft Just-so\textquotedblright\
Split-Higgs solution. In the fits A, B and C
the value of the parameter $a$ is
significantly larger than that found in RRRV.
This causes problems for models based on
non-Abelian family symmetries, and it is of
interest to try to reduce $a$ by allowing
$a^{\prime },$ $b^{\prime }$ and $c^{\prime
}$ to vary  while remaining
${\mathcal{O}}(1)$ parameters. Doing this for
the fits B and C leads to the fits B2 and C2
given in Table \ref{Table4} where it may be
seen that the extent to which $a$ can be
reduced is quite limited. Adjusting to this
is a challenge for the broken family-symmetry
models.

Although we have included the finite corrections to match the MSSM theory onto the Standard Model at an effective SUSY scale $M_S=500$ GeV, we have not included finite corrections from matching onto a specific GUT model.  Precise threshold corrections cannot be rigorously calculated without a specific GUT model.  Here we only estimate the order of magnitude of corrections to the mass relations in Table \ref{table2} from matching the MSSM values onto a GUT model at the GUT scale.  The $\tan \beta$ enhanced corrections in Eq(\ref{Eqgammat}-\ref{Eqgammad}) arise from soft SUSY breaking interactions and are suppressed by factors of $M_{SUSY}/M_{GUT}$ in the high-scale matching.  Allowing for ${\mathcal{O}}(1)$ splitting of the mass ratios of the heavy states, one obtains corrections to $y^b/y^\tau$ (likewise for the lighter generations) of ${\mathcal{O}}(\frac{g^2}{(4\pi)^2})$ from the $X$ and $Y$ gauge bosons and $ {\mathcal{O}}(\frac{y_b^2}{(4 \pi)^2})$ from colored Higgs states. Because we have a different Higgs representations for different generations, these threshold correction will be different for correcting the $3 m_s / m_\mu$ relation than the $m_b / m_\tau$ relation.  These factors can be enhanced in the case there are multiple Higgs representation.  For an $SU(5)$ SUSY GUT these corrections are of the order of $2\,\%$.  Planck scale suppressed operators can also induce corrections to both the unification scale \cite{Hill:1983xh} and may have significant effects on the masses of the lighter generations \cite{Ellis:1979fg}.  In the case that the Yukawa texture is given by a broken family symmetry in terms of an expansion parameter $\epsilon$, one expects model dependent corrections of order $\epsilon$ which may be significant.

\section*{Chapter Summary}

In summary, in the light of the significant
improvement in the measurement of fermion
mass parameters, we have analyzed the
possibility that the fermion mass structure
results from an underlying supersymmetric GUT
at a very high-scale mirroring the
unification found for the gauge couplings.
Use of the RG equations to continue the mass
parameters to the GUT scale shows that,
although qualitatively in agreement with the
GUT\ predictions coming from simple Higgs
structures, there is a small quantitative
discrepancy. We have shown that these
discrepancies may be eliminated by finite
radiative threshold corrections involving the
supersymmetric partners of the Standard-Model
states. The required magnitude of these
corrections is what is expected at large
$\tan \beta $, and the form needed
corresponds to a supersymmetric spectrum in
which the gluino mass is negative with the
opposite sign to the wino mass. We have also
performed a fit to the recent data to extract
the underlying Yukawa coupling matrices for
the quarks and leptons. This is done in the
basis in which the mass matrices are
hierarchical in structure with the
off-diagonal elements small relative to the
appropriate combinations of on-diagonal
matrix elements, the basis most likely to be
relevant if the fermion mass structure is due
to a spontaneously broken family symmetry. We
have explored the effect of SUSY\ threshold
corrections for a variety of SUSY\ spectra.
The resulting structure has significant
differences from previous fits, and we hope
will provide the \textquotedblleft
data\textquotedblright\ for developing models
of fermion masses such as those based on a
broken family symmetry.

Since this work was first published, its
conclusions have been confirmed by studies
of other research groups \cite{Antusch:2008tf}.
The updated fits to the Yukawa textures
 and viability of the Georgi-Jarlskog relations
have been used in numerous string theory and family symmetry models.

\chapter{Mass Determination Toolbox at Hadron Colliders}
\label{ChapterMassDeterminationReview}

\section*{Chapter Overview}
In the previous chapter, we presented arguments that predict the sign of the gluino mass relative to the wino mass and relationships between these masses needed to satisfy classic Georgi-Jarlskog mass relationships at the GUT scale.

This chapter and the remaining chapters discuss mostly model-independent experimental techniques to determine the mass of a pair-produced dark-matter particle.
As a test case, we take the dark-matter to be the lightest supersymmetric particle (LSP) which we assume is the
neutralino $\tilde{\chi}^o_1$\footnote{The four neutralinos
are superpositions of the two Higgsinos, the bino, and the neutral wino ($\tilde{h}^o_u,\tilde{h}^o_d,\tilde{B},\tilde{W}^3$).  They are numbered from
$1$ having the smallest mass to $4$ having the largest mass.}.
Determining the mass of the dark-matter particle is a necessary prerequisite to determining the entire mass spectrum of the new particle states and to determining the sign of the gluino's mass (given that supersymmetry is the correct model) predicted in the previous chapter.

This chapter reviews challenges of hadron-collider mass-determination
and mostly
model-independent techniques to address these challenges existent in the literature.
In the subsequent chapters of this thesis, we will improve on these techniques and develop new techniques that can perform precision measurements of the dark-matter's mass with only a few assumptions about the underlying model.

\section{Mass Determination Challenges at Hadron Colliders}

Measuring all the masses and phases associated with the predictions in
 Chapter \ref{ChapterUnificationAndFermionMassStructure} or the predictions
of any of the many competing models depends on the
successful resolution of several challenges present at any hadron collider.
First to avoid selection bias and because of the large number of possible models, we would prefer kinematic model-independent
techniques to measure parameters like the mass instead of model-dependent techniques.
Kinematic techniques are made complicated because of the possibility of producing dark matter in the collider which would lead to new sources (beyond neutrinos) of missing transverse momentum.
Kinematic techniques are also made complex because the reference frame and center-of-mass energy of the parton collision is only known statistically \footnote{To the extent we understand the uncertainties in the measured parton distribution functions.}.
Last our particle detector has fundamental limitations on the shortest track that can be observed leading to combinatoric ambiguities identifying decay products from new particle states that are created and decay at the ``effective" vertex of the collision.

\subsection{Kinematic versus Model-Dependent Mass Determination}

Any new physics discovered
will come with unknown couplings, mixing angles, spins, interactions, and masses.
One approach is to assume a model and fit the model parameters to the observed data.
However, we have hundreds of distinct anticipated theories possessing tens of new particle states with differing spins and couplings each with tens of parameters.
Global fits to approximately $10^{9}$ events recorded per year are an enormous amount of work, and
assume that the `correct model' has been anticipated.

We would like to measure the particle properties with minimal model assumptions.
The term `model-independent' technique is misleading because we are always assuming some model.  When we say model-independent, we mean that we assume broadly applicable assumptions such as special relativity. We also try to ensure that our techniques are largely independent of \emph{a priori} unknown coupling coefficients and model parameters.  For this reason our desire for model-independence constrains our toolbox to kinematic properties of particle production and decay.

If we are able to determine the mass of the dark-matter particle, other properties follow more easily.
Knowing the dark-matter particle's mass, the remaining mass spectrum follows from kinematic edges.
If we know the dark-matter's mass and gluino's mass, then the sign of the gluino's mass predicted in
Chapter \ref{ChapterUnificationAndFermionMassStructure}
can be determined from the distribution of the invariant mass of jet pairs from the decay of gluino pair production \cite{Mrenna:1999ai}.
Another avenue to the sign of the gluino's mass requires measuring the masses of the gluino and the two stop squarks and the measuring the decay width of gluino to these states \cite[Appendix B]{BaerTata:2006}.
In addition, the measurement of spin correlations, a study that can contribute to spin determination,  has historically relied on knowing the masses to reconstruct the event kinematics \cite{Abbott:2000dt}.

In short, mass determination is a key part
to identifying the underlying theory which lies beyond the Standard Model. Newly discovered particles could be Kaluza-Klein (KK) states from extra-dimensions, supersymmetric partners of known states,
technicolor hadrons, or something else that we have not anticipated.  Models predict relationships between parameters. Supersymmetry relates the couplings of current fermions
to the couplings of new bosons,
and the supersymmetric particle masses reflect the origin of supersymmetry breaking. Masses of KK states tell us about the size of the extra dimensions.
If these parameters, such as the masses and spins,
can be measured without assuming the model, then these observations
exclude or confirm specific models.
In general, mass determination of new particle states is central to discerning what lies beyond the Standard Model.

\subsection{Dark matter: particle pairs carrying missing transverse momentum}

If dark matter is produced at a hadron collider, the likely signature will be missing transverse momentum.
The lightest supersymmetric particle (LSP) or lightest Kaluza-Klein particle (LKP) is
expected to be neutral, stable, and invisible to the particle detectors.
The astrophysical dark matter appears to be stable.
Whatever symmetry makes dark matter
stable and / or distinguishes superpartners from their Standard-Model cousins
will likely also require that dark matter particles be pair produced.
Therefore, events with supersymmetric new particle states are expected to end in two
LSP's leaving the detector unnoticed.  Events with Kaluza-Klien particles are expected to end with two LKP's leaving the detector unnoticed.
The presence of two invisible particles prevents complete reconstruction of the kinematics of any single event and leads to missing transverse momentum in the events.

\subsection{Reference frame unknown due to colliding hadrons}

At a hadron collider, because the partons colliding each carry an unknown
fraction of the hadron's momentum we do not know the reference frame of the collision.
Although the individual parton momentum is unknown, the statistical distribution of the parton's momentum can be measured from deep inelastic scattering \cite{Devenish:2004pb}.
The measured parton distribution functions (PDFs) $f^H_i(x,Q)$ give the distribution of parton $i$ within the hadron $H$ carrying $x$ fraction of the hadron's momentum when probed by a space-like momentum probe $q^\mu$ with $Q^2=-q^2$.  $Q$ is also the factorization scale at which one cuts off collinear divergences. The dominant parton types $i$ are $u$, $d$, $s$, their antiparticles $\bar u$, $\bar d$, $\bar s$, and gluons $g$.

The events produced in the collider follow from a convolution of the cross section over the parton distribution functions. If the two colliding protons have 4-momentum $p_1=\sqrt{s_{LHC}}(1,0,0,1)/2$ and $p_2=\sqrt{s_{LHC}}(1,0,0,-1)/2$,
then for example the $u$ and $\bar{u}$ quarks colliding would have 4-momentum $x_u p_1$ and $x_{\bar{u}} p_2$.
The spatial momentum of the collision along the beam axis is given by $(x_u - x_{\bar{u}})\sqrt{s_{LHC}}/2$, and the center-of-mass energy of the parton collision is $\sqrt{\,x_u\, x_{\bar{u}} s_{LHC}}$.
Because $x_u$ and $x_{\bar{u}}$ are only known statistically, any individual
collision has an unknown center-of-mass energy and an unknown momentum along the beam axis.

LHC processes are calculated by convolutions over the {PDF}s as in
\begin{equation}
 \sigma = \int \,dx_u\, dx_{\bar{u}}\,  f^p_u(x_u,Q) f^p_{\bar{u}}(x_{\bar{u}},Q) \sigma( x_u \, x_{\bar{u}} s_{LCH},Q)_{u \bar{u} \rightarrow final}
 \end{equation}
where $\sigma(s,Q)_{u \bar{u} \rightarrow final}$ is the total cross section for a given parton collision center-of-mass squared $s$, factorization scale $Q$, and $\sqrt{s_{LHC}}$ is the center-of-mass collision energy of the protons colliding at the LHC.  The region of integration of $x_u$ and $x_{\bar{u}}$ is based on the kinematically allowable regions.  The factorization scale $Q$ is chosen to minimize the size of the logarithms associated with regulating collinear divergences in the cross section.  See Ref~\cite{Maltoni:2007tc} for a recent review on choosing the factorization scale.

Calculating these input cross sections even at tree level  is an arduous process
that has been largely automated.  Performing these convolutions
over experimentally determined {PDF}s adds even further difficulties.
Monte-Carlo generators typically perform these calculations.
The calculations in subsequent chapters made use of {MadGraph} and {MadEvent} \cite{Alwall:2007st}, \herwig \cite{Corcella:2002jc}, {CompHEP} \cite{Boos:2004kh}.

Despite this automation, the author has found it useful to reproduce these tools so that we can deduce and control what aspects of the observed events are caused by what assumptions. For this purpose, parton distributions can be downloaded\footnote{One should also look for the Les Houches Accord PDF Interface (LHAPDFs) interfaces to interface PDFs to different applications (\tt{http://projects.hepforge.org/lhapdf/}).} from \cite{DurhamPDFFarm}.  Analytic cross sections for processes can be produced by {CompHEP}\cite{Boos:2004kh}.

These review studies have
led to an appreciation of uncertainties in the parton distribution functions,  uncertainties about the correct $Q$ to use\footnote{Ref~\cite{BaerTata:2006} suggests using $Q \approx 2 M_{SUSY}$ or whatever energy scale is relevant to the particles being produced.}, and uncertainties in the process of hadronization of the outgoing partons. All these uncertainties contribute to significant uncertainties in the background of which new physics must be discovered, and to uncertainties underling model-dependent, cross-section-dependent determinations of particle masses.

\subsection{Detector Limitations}

Particle detectors have finite capabilities which limit what we can learn.
We will discuss the effects of finite energy resolution later in this thesis.
Here I focus on the lack of the traditional tracks one sees in event representations.

The early history of particle physics shows beautiful
bubble chamber tracks of particles.
Studies of Pions and Kaons relied on tracks left in bubble chambers or modern digitized detectors.  In the case of the $K_S$ the lifetime is $\tau = 0.9 \times 10^{-10}$ seconds or $c \tau = 2.7$ cm, or equivalently the decay width is $\Gamma=7 \times 10^{-6}$ eV.
A particle with a width of $\Gamma=1$ eV has a lifetime of $\tau=6 \times 10^{-16}$ seconds with $c \tau = 0.2 \mu\rm{m}$.
The more massive known states of $W^\pm$, $Z^o$ bosons and the top quark have widths of $\Gamma_W =2.141 \GeV$, $\Gamma_Z=2.50 \GeV$, and $\Gamma_t \approx 1.5 \GeV$.
These decay widths give these states tracks with $c \tau \approx 0.1$ fm or about $1/10$ the size of an atomic nuclei.

Most supersymmetric states, depending on the model, have decay widths between around $1$ MeV and $2$ GeV \footnote{This is not a universal statement. Some viable models have long lived charged states ($c \tau >> 1$ cm)  \cite{Feng:1999fu,Krasnikov:1996fu,Jedamzik:2007cp}.}.
These observations, while trivial, make us realize that new physics discoveries will most likely need to deduce properties of the new states from their decay products only.
The fact that all events effectively occur at the origin creates combinatoric problems because we cannot know the order in which the visible particles came off of a decay chain or even to which decay chain they belong.

\section{Invariant Mass Edge Techniques}

\label{SecEdgeTechniques}

We now turn to existent tools to overcome the challenges inherent in
mass determination at hadron colliders in the presence of missing transverse momentum.

\subsection{Invariant Mass Edges}

\label{SubSecEdges}
\begin{figure}
 \centerline{\includegraphics[width=3.2in]{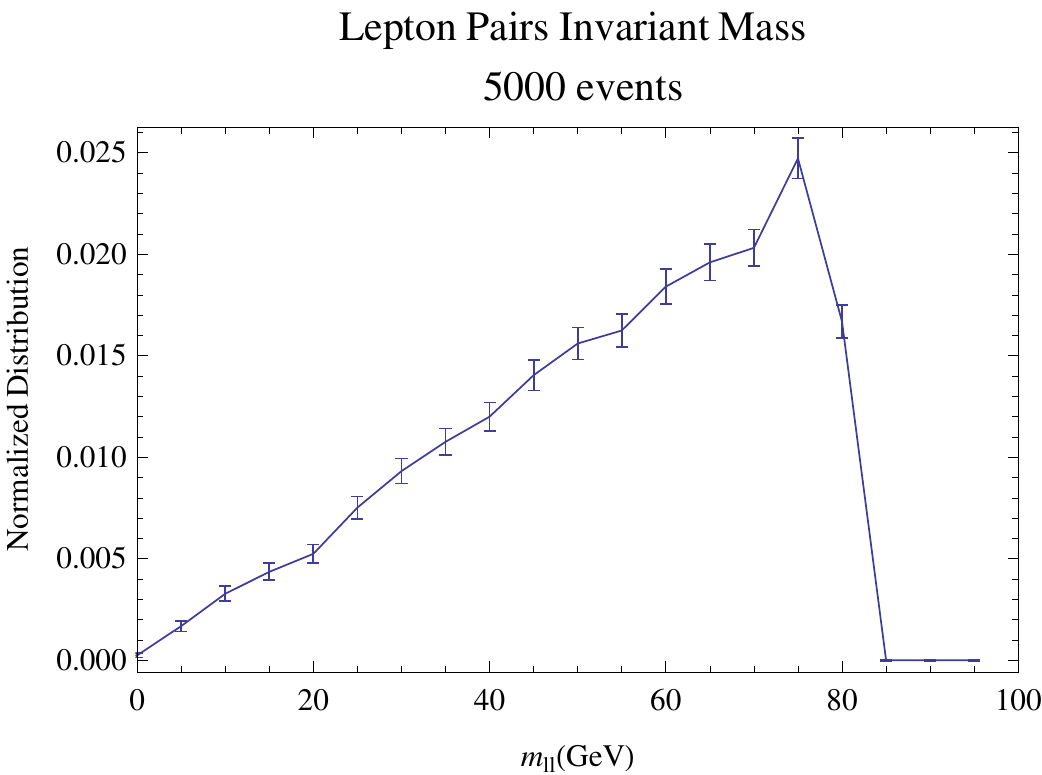} \includegraphics[width=3.5in]{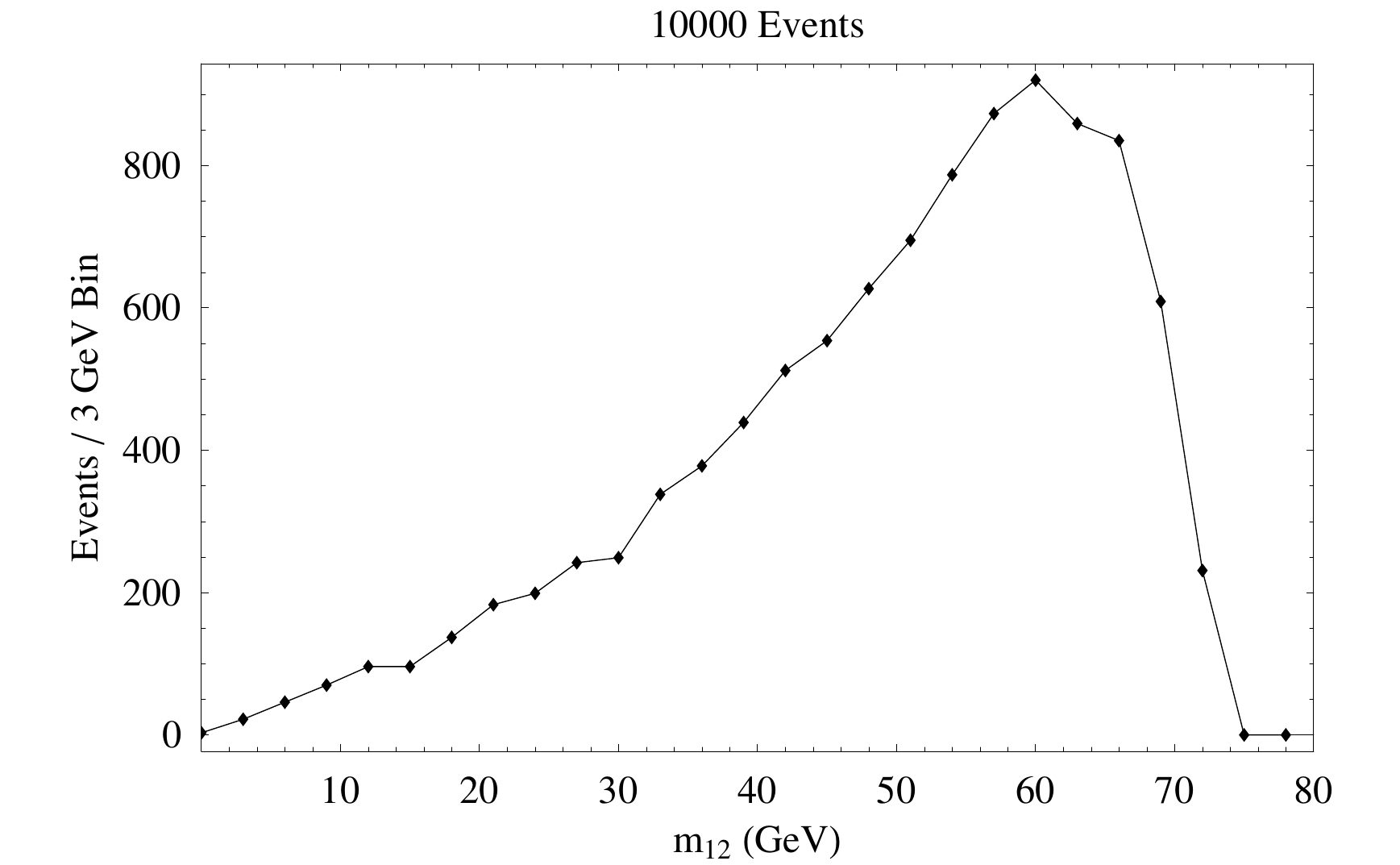}}
 \centerline{\includegraphics[width=2.3in]{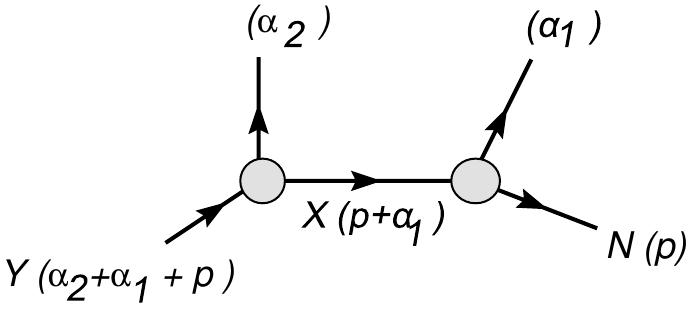} \ \ \ \ \ \ \includegraphics[width=2in]{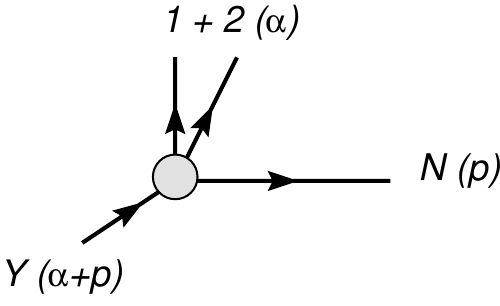}}
 \caption{\label{FigDecayTypes} ({\bf{Left:}}) Two body decay and it's associated $m_{ll}$ distribution.  ({\bf{Right:}}) Three body decay and it's associated $m_{ll}$ distribution. }
\end{figure}

Lorentz invariant distributions are optimal for making a distribution independent of the unknown frame of reference.
Fig~\ref{FigDecayTypes} shows a two-body decay and a three-body decay.
The decay products from $Y$ in both cases involve a dark-matter particle $N$ and two visible states with 4-momenta $\alpha_1$ and $\alpha_2$; we will abbreviate $\alpha=\alpha_1+\alpha_2$. The Lorentz invariant mass of these two visible states is defined as
 \begin{equation}
   m^2_{12} = \alpha^2= (\alpha_1 + \alpha_2)^2.
 \end{equation}
Assuming no spin correlation between $\alpha_1$ and $\alpha_2$, the distribution for the two-body decay looks more like a right-triangle whereas the three-body decay case looks softer.
Spin correlations between these states may change the shape and might be used to measure spin in some cases \cite{Athanasiou:2006ef}.
The distribution shape can also be affected by competing processes in the three-body decay.
For example, the degree of interference with the slepton can push the peak to a smaller value\cite{Phalen:2007te}.

The end-point of the distribution gives information about the masses of the new particle states.
In the two-body decay case the endpoint gives
\begin{equation}
 \max m^2_{12} = \frac{(M_Y^2 -M_X^2)(M_X^2-M_N^2)}{M_X^2},
 \label{EqTwoBodyDecayEdge}
\end{equation}
and in the three-body decay case
\begin{equation}
 \max m_{12} = M_Y -M_N.
 \label{EqThreeBodyDecayEdge}
\end{equation}
These edges in invariant mass combinations provide information about mass differences or mass squared differences but not about the mass-scale itself.
Measuring these edges requires some model of the events creating the distribution in order to simulate the effect of the detectors energy resolution.
For this reason, even end-point techniques,
although mostly model-independent, still require some
estimate of the distribution near the end point to get an accurate end-point measurement.
Radiative corrections also
play a role in shifting this endpoint slightly\cite{Drees:2006um}.

These techniques have been studied to select a model if one assumes a particular set of starting models \cite{Allanach:2000kt}.  They have also been used with Bayesian techniques to measure the mass difference between slepton states \cite{Allanach:2008ib}.

\subsection{Constraints from Cascade Decays}

\begin{figure}
 \centerline{\includegraphics[width=4in]{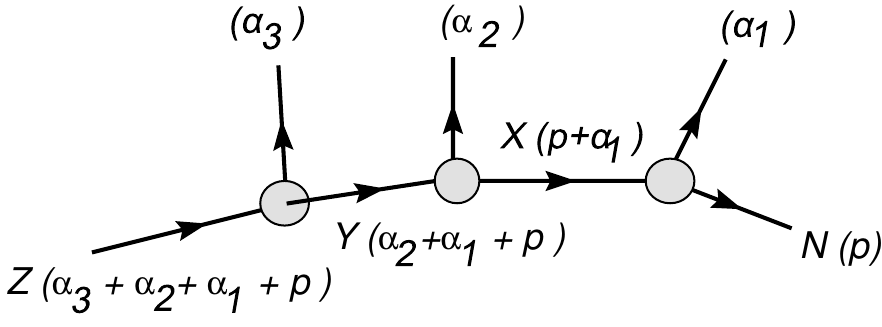}}
 \caption{\label{FigCascadeDecay} Cascade decay from $Z$ to $Y$ to $X$ ending in $N$ and visible particle momenta $\alpha_1$, $\alpha_2$, $\alpha_3$. }
\end{figure}

If there are many new particle states produced at the collider,
then we can use edges from cascade decays between these new states
to provide constraints between the masses.
Given some luck regarding the relative masses, these constraints may be inverted to obtain the mass scale.
Fig~\ref{FigCascadeDecay} shows a cascade decay from $Z$ to $Y$ to $X$ ending in $N$ and visible particle momenta $\alpha_1$, $\alpha_2$, $\alpha_3$.

There are four unknown masses and potentially four linearly-independent endpoints.
In addition to Eq(\ref{EqTwoBodyDecayEdge}), we also have \cite{phdthesis-lester,Allanach:2000kt}
 \begin{equation}
  \max m^2_{32} = (M^2_Z-M^2_Y)(M^2_Y-M^2_X) / M_Y^2,
 \end{equation}
and
\begin{eqnarray}
 \max  m^2_{123} = \begin{cases}
 (M^2_Z-M^2_Y)(M^2_Y-M^2_N) / M_Y^2 & {\rm{iff}} M_Y^2 < M_N M_Z \\
 (M^2_Z-M^2_X)(M^2_X-M^2_N) / M_X^2 & {\rm{iff}} M_N M_Z < M_X^2 \\
 (M^2_Z M_X^2-M_Y^2 M_N^2)(M_Y^2-M_X^2)/(M^2_Y M^2_X) & {\rm{iff}} M_X^2 M_Z < M_N M_Y^2 \\
 (M_Z-M_N)^2 &   {\rm{otherwise}} .     \end{cases}
 \label{EqM123Case}
\end{eqnarray}
The fourth endpoint is
 \begin{equation}
  \max m^2_{13} = (M^2_Z-M^2_Y)(M^2_X-M^2_N) / M_X^2.
 \end{equation}
Depending on whether we can distinguish
visible particles $1$, $2$ and $3$ from each other,
some of these endpoints may be obscured by combinatorics problems.
If the visible particles and the masses are such that these four can be disentangled from combinatoric problems, and if the masses are such that these four end-points provide independent constraints, and if there is sufficient statistics to measure the endpoints well-enough, then we can solve for the mass spectrum.

This technique has been studied in \cite{Bachacou:1999zb,phdthesis-lester,Allanach:2000kt}
and subsequently by many others.  In some cases, there is more than one solution to
the equations.  This degeneracy can be lifted by using the
shape of these distributions \cite{Gjelsten:2004ki,Lester:2006yw,Gjelsten:2006tg}.
Using the Supersymmetric benchmark point SPS 1a,
this technique has been shown to determine the LSP mass to $\pm 3.4 \GeV$ with about 500 thousand events from $300 \fb^{-1}$.
They are able to determine the mass difference to $\sigma_{M_{\N{2}}-M_{\N{1}}}=0.2 \GeV$ and $\sigma_{M_{\tilde{l}_R}-M_{\N{1}}}=0.3 \GeV$ \cite{Gjelsten:2004ki}.

\section{Mass Shell Techniques}

What if there are not as many new-states accessible at the collider, or what if the masses arrange themselves to not enable a solution to the four invariant mass edges listed above?
There is also a series of approaches called Mass Shell Techniques (MST)\footnote{The title MST is suggested in Ref.~\cite{Bisset:2008hm}.} where assumption are used about the topology and
on-shell conditions to solve for the unknown masses.

\begin{figure}
\centerline{\includegraphics[width=4in]{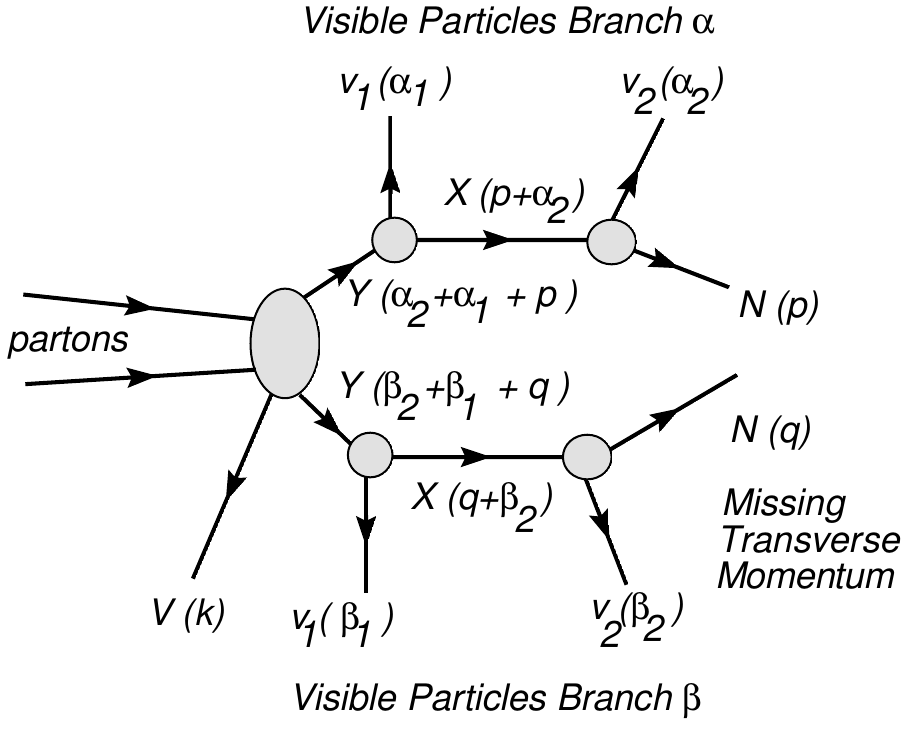}}
\caption{\label{FigEventTopologyTovey} Events with the new state $Y$ is pair produced and in which each $Y$ decays
through a two-body decay to a massive new state $X$ and a visible state $1$ and then where $X$ subsequently decays to a massive state $N$ invisible to the detector
and visible particles  $2$. All previous decay products are grouped into the upstream transverse momentum, $k$. }%
\end{figure}
An MST was used in the early 1990s by Goldstein, Dalitz (GD) \cite{Goldstein:1993mj}, and Kondo, Chikamatsu, Kim (KCK)\cite{Kondo:1993in} as a suggested method to measure the mass of the top quark.
The top-quark pair production and decay has the topology
shown in Fig~$\ref{FigEventTopologyTovey}$.
Here $Y=t$ or $\bar{t}$, $X=W^\pm$, $(1)=b$, $(2)=e$ and $N=\nu_e$.
In each event, one measures $\alpha_{1,2}$, $\beta_{1,2}$, and the
missing transverse momentum $\slashed{P}_T=(p+q)_T$.
The resulting mass shell and missing momentum equations are
\begin{eqnarray}
  M_N^2 & = & p^2=q^2, \\
  M_X^2 &  =& (p+\alpha_2)^2=(q+\beta_2)^2,\\
  M_Y^2 & = & (p+\alpha_1+\alpha_2)^2=(q+\beta_1+\beta_2)^2, \\
  \slashed{P}_T & = & (p+q)_T.
\end{eqnarray}
If we assume a mass for $M_Y$, then we have $8$ unknowns
(the four components of each $p$ and $q$) and
$8$ equations.  The equations can be reduced
to the intersection of two ellipses which have between zero
 and four distinct solutions.
In GD and KCK, they supplemented these solutions with some data from the
cross section and used Bayes' theorem to create a likelihood function
for the unknown mass of the top quark.  The approach has a systematic bias \cite{Raja:1996vz}\cite{Raja:1997qs} that must be removed by modeling \cite{Brandt:2006uc}\footnote{Details of this bias are discussed more in Section \ref{SecM3CComparisonToOtherTechniques}}.

A modern reinvention of this MST is found in Cheng, Gunion, Han, Marandella and McElrath (CHGMM)~\cite{Cheng:2007xv}
where they assume a
symmetric topology with an on-shell intermediate state
such as in Fig~$\ref{FigEventTopologyTovey}$. This reinvention assumes $Y=\N{2}$,
$X=\tilde{l}$ and $N=\N{1}$.
CHGMM do a test of each event's compatibility with all values of $(M_Y,M_X,M_N)$.
Using this approach, they find that with $300 \fb^{-1}$ of events that they can determine the LSP mass to $\pm 12$ GeV.  Unlike GD and KCK, CHGMM make no reference to Bayesian techniques.

Another MST assumes a longer symmetric decay chain and assumes the masses in two events must be equal \cite{Cheng:2008mg}.
This results in enough equations to equal the number of unknowns to solve for the masses directly.
For SPS 1a after $300 \fb^{-1}$, they reach $\pm 2.8 \GeV$ using $700$ events satisfying their cuts but have a $2.5 \GeV$ systematic bias that needs modeling to remove.

Finally there is a suggestion for hybrid MST where one combines on-shell conditions with the information from the many edges in cascade decays \cite{Nojiri:2007pq}. The $M_{2C}$ variable introduced in Chapters \ref{ChapterM2Cdirect} is a simple
example of such a hybrid technique that predated their suggestion.

\section{Transverse Mass Techniques}
 \label{SecWMassDetermination}

\subsection{$M_T$ and Measuring $M_W$}

The invariant mass distributions are Lorentz invariant.
In events with one invisible particle, like the neutrino,
leaving the detector unnoticed, we only know
the neutrino's transverse momentum inferred from the missing transverse momentum.
If a parent particle $X$ decays into a visible particle with observed 4-momentum $\alpha_1$ and an invisible particle $p$, then what progress
 can be made in mass determination?
There is a class of techniques which, although not Lorentz invariant,
is invariant with respect
to longitudinal boosts along the beam line.
The 4-momentum $p_\mu = (p_0, p_x,p_y,p_z)$
is cast in new coordinates
 \begin{eqnarray}
  \eta(p) &=& \frac{1}{2} \ln \frac{p_0+p_z}{p_0-p_z} \\
  E_T(p) &=& \sqrt{m_p^2 + p^2_x + p_y^2} \\
  p_\mu &=& (E_T \cosh \eta, p_x, p_y, E_T \sinh \eta)
 \end{eqnarray}
where $m_p^2=p^2$.
The mass of $X$ in the decay $X(\alpha_1+p) \rightarrow N(p) + 1(\alpha_1)$ in Fig.~\ref{FigDecayTypes} is
 \begin{eqnarray}
 M_X^2  =  (p+\alpha_1)^2 & = & M_N^2 + M_1^2 + 2 p \cdot \alpha_1 \\
                          & = & M_N^2 + M_1^2 + 2 E_T(p) E_T(\alpha_1) \cosh (\eta(p) - \eta(\alpha_1)) - 2 p_T \cdot (\alpha_1)_T.
 \end{eqnarray}
We observe the 4-momentum $\alpha_1$ in the detector.
From the missing transverse momentum, we can deduce the transverse components $p_x$ and $p_z$.  The components $p_o$ and $p_z$ remain unknown.
If we know the mass of $N$, we can bound the unknown mass of $X$ from below by minimizing with respect to the unknown rapidity $\eta(p)$.
The minimum
 \begin{eqnarray}
 M_X^2  \ge  M^2_T(p,\alpha_1) \equiv M_N^2 + M_1^2 + 2 E_T(p) E_T(\alpha_1) - 2 p_T \cdot (\alpha_1)_T
\label{EqMT}
 \end{eqnarray}
is at $\eta(p)=\eta(\alpha_1)$ and is guaranteed to be less than the true mass of $X$. This lower bound is called the transverse mass $M_T$.

This technique is used to measure the $W^\pm$ boson's mass with the identification $X=W^\pm$, $N=\nu_e$, and $(1)=e^\pm$.
The first observation of the $W^\pm$ was reported in 1983 at {UA1} Collaboration at Super Proton Synchrotron (SPS) collider at CERN\cite{Arnison:1983rp}.
They used $M_T$ to bound $M_W \ge 73 \GeV$ at the $90 \%$ confidence level.
Assuming the Standard Model and fitting the data to the model with $M_W$ as a free parameter gave the {UA1} collaboration the measurement $M_W = 81 \pm 5 \GeV$, which agreed with the prediction described by Llewellyn Smith and Wheater \cite{LlewellynSmith:1981yk}.
$M_T$ is still used even for the more recent 2002 {D0}-Collaboration model-independent measurement of $M_W= 80.483 \pm 0.084 \GeV$ \cite{Abazov:2002bu}.

\subsection{The Stransverse Mass Variable $M_{T2}$}

If a hadron collider pair produces dark-matter particles, then there are two sources of missing transverse momentum, each with the same unknown mass. The `stransverse mass' $m_{T2}$ introduced by Lester and Summers \cite{Lester:1999tx,Barr:2003rg} adapts the transverse mass $m_T$ to this task.
$M_{T2}$ techniques have become widely used with Refs.~\cite{Lester:1999tx,Barr:2003rg}
having more than 45 citations.

The stransverse mass is used to determine the mass {\em difference} between a parent particle and a dark-matter candidate particle given an assumed mass for the dark-matter candidate based on a topology similar to Figs~\ref{FigEventTopologyThreeBody} or \ref{FigEventTopologyTovey}.
The variable $m_{T2}$ accepts three inputs: $\chi_N$ (an assumed mass of the two particles carrying away missing transverse momenta),
$\alpha$ and $\beta$ (the visible momenta of each branch), and $\slashed{P}_T=(p+q)_T$
(the missing transverse momenta two-vector).  We can define $m_{T2}$ in terms of the transverse mass of each branch where we minimize the maximum of the two transverse masses over the unknown split between $p$ and $q$ of the overall missing transverse momenta:
 \begin{equation}
  M^2_{T2}(\chi_N,\alpha,\beta,\slashed{P}_T) = \min_{{p}_T+{q}_T=\slashed{{P}}_T} \left[ \max \left\{ M^2_T(\alpha,p), M^2_T(\beta,q) \right\} \right].
 \end{equation}
In this expression $\chi_N$ is the assumed mass of $N$, $\alpha$ and $\beta$ are the four momenta of the visible particles in the two branches, the transverse mass is given by $M^2_T(\alpha,p) = m^2_\alpha + \chi_N^2 + 2 (E_T(p) E_T(\alpha) - p_T \cdot \alpha_T)$ and the transverse energy $E_T(\alpha) = \sqrt{p_T^2 + \chi_N^2}$ is determined from the transverse momentum of $p$ and the assumed mass of $p$.
The $M_T$ here is identical to Eq(\ref{EqMT}) with $M_N$ replaced by the assumed mass $\chi_N$.
An analytic formula for the case with no transverse
upstream momentum $k_T$ can be found in the appendix of \cite{Lester:2007fq}.
For each event, the quantity $M_{T2}(\chi_N=m_N,\alpha_1+\alpha_2, \beta_1 + \beta_2, \slashed{P}_T)$ gives the smallest mass for the parent particle compatible with the event's kinematics.  Under ideal assumptions, the mass of the parent particle $Y$ is given by the end-point of the distribution of this $m_{T2}$ parameter over a large number of events like Figs.~\ref{FigEventTopologyThreeBody} or \ref{FigEventTopologyTovey}.  Because a priori we do not know $M_N$, we need some other mechanism to determine $M_N$.
We use $\chi$ to distinguish assumed values of the masses ($\chi_Y$, $\chi_X$, $\chi_N$) from the true values for the masses ($m_Y$, $m_X$, $m_N$).
Because of this dependence on the unknown mass, we should think of $\max\, m_{T2}$ as providing a relationship or constraint between the mass of $Y$ and the mass of $N$.  This forms a surface in the ($\chi_Y$, $\chi_X$, $\chi_N$) space on which the true mass will lie.
We express this relationship as $\chi_Y(\chi_N)$ \footnote{In principle this surface would be considered a function of $\chi_Y(\chi_X,\chi_N)$, but $m_{T2}$ makes no reference to the mass of $X$ and the resulting constraints are therefore independent of any assumed value of the mass of $X$.}.

In addition to invariance with respect to longitudinal boosts,  $M_{T2}$, in the limit where $k_T=0$, is also invariant with respect to back-to-back boosts of the parent particle pair in the transverse plane \cite{Cho:2007dh}.

$M_{T2}$ is an ideal tool for top-mass determination at the LHC \cite{Cho:2008cu}.
$M_{T2}$ would apply to events where both pair-produced top quarks decay to a $b$ quark and a $W^\pm$ and where in both branches the $W^\pm$ decays leptonically to $l$ and $\nu_l$.
To use $M_{T2}$ to find the mass of the parent particle, a value must be assumed for $\chi_N=M_N$.  For a $\nu$ we approximate $\chi_N=0$.
However, even in models of new physics where new invisible particles are nearly massless (like the gravitino studied in \cite{Hamaguchi:2008hy}), we would rather not just assume the mass of the lightest supersymmetric particle (LSP), which is needed as
an input to the traditional $m_{T2}$ analysis, without measuring it in some model independent way.

\subsection{Max $M_{T2}$ Kink Technique}

\begin{figure}[ptb]
\centerline{\includegraphics[width=3in]{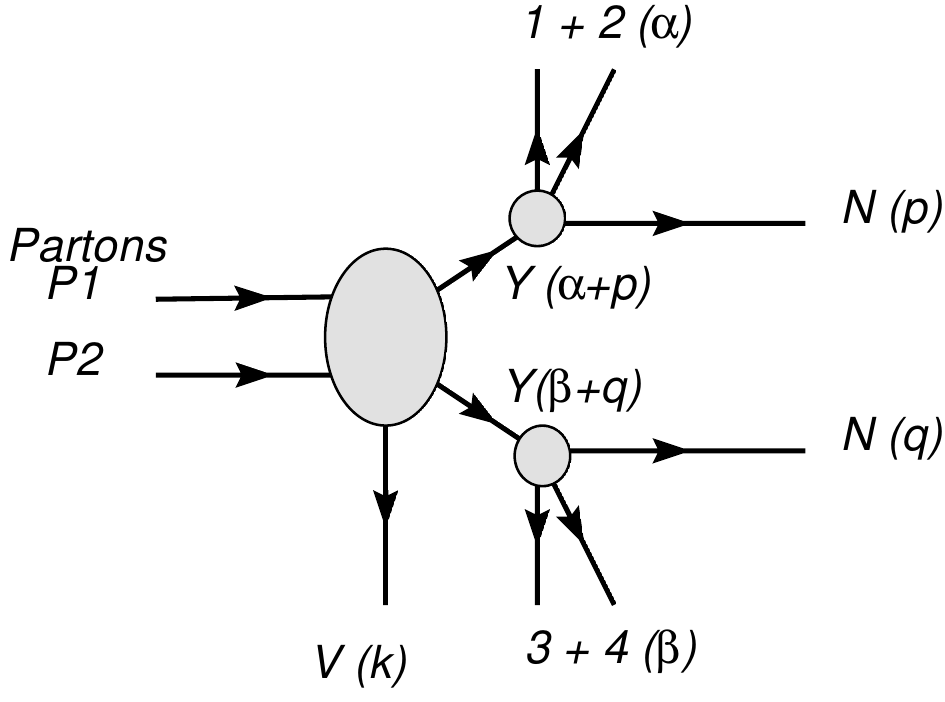}}
\caption{Events with the new state $Y$ is pair produced and in which each $Y$ decays
through a three-body decay to a massive state $N$ invisible to the detector
and visible particles $1$, $2$, $3$, and $4$. All previous decay products are grouped into the upstream transverse momentum, $k$. }%
\label{FigEventTopologyThreeBody}
\end{figure}

\begin{figure}
\centerline{\includegraphics[width=4in]{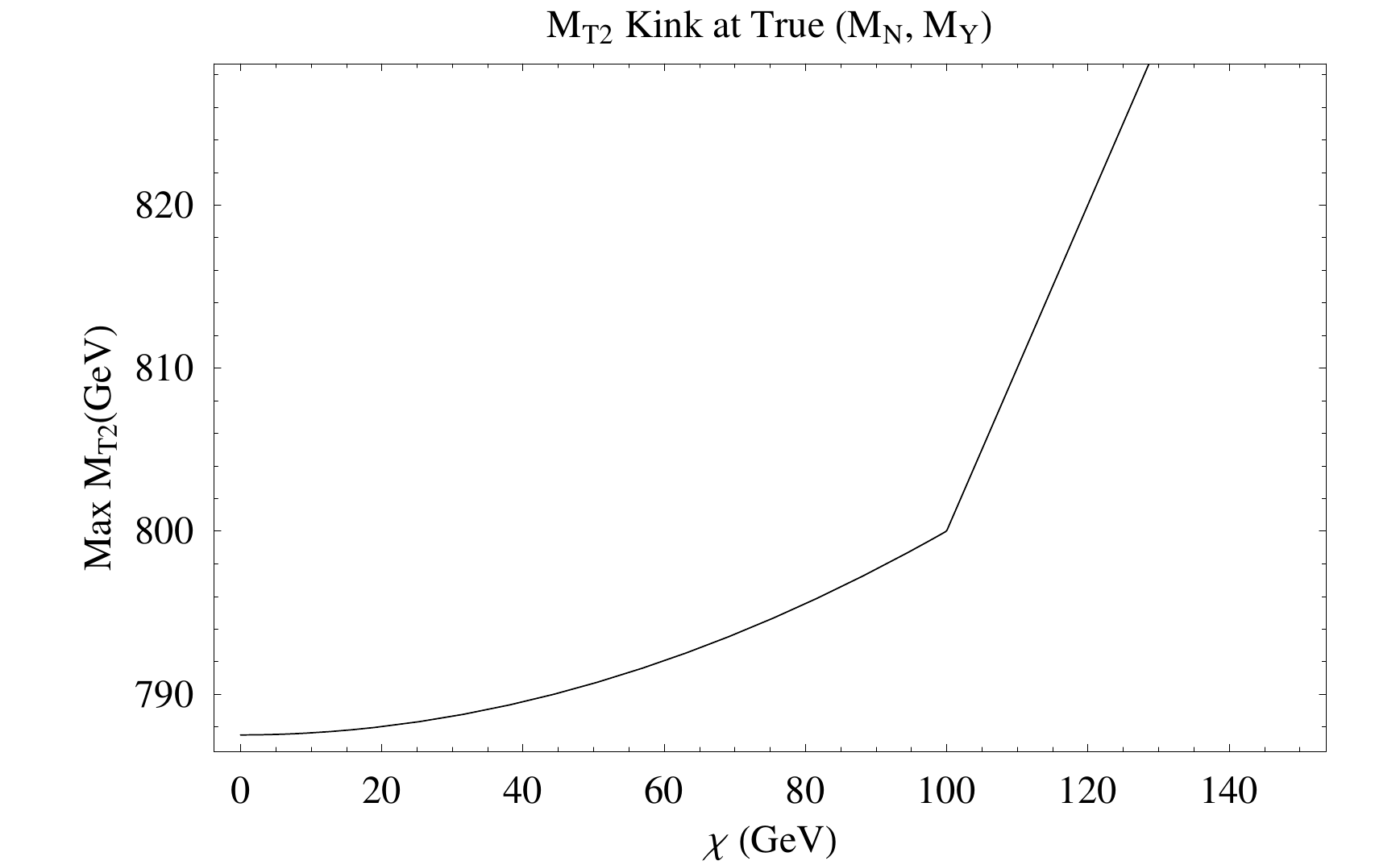}}
\caption{\label{FigKink}
The $\max M_{T2}(\chi)$ shows a kink at the true $M_N$ and $M_Y$.
For this simulation, $m_Y=200$ GeV, and $m_N=99$ GeV. }
\end{figure}

$M_{T2}$ assumes a mass $\chi_N$ for $N$.  Is there any way that one can determine $M_N$ from $M_{T2}$ alone?
At the same time as the author was completing work on  $M_{2C}$  described in Chapter \ref{ChapterM2Cdirect}, Cho, Choi Kim, Park (CCKP) \cite{Cho:2007qv,Cho:2007dh} published an observation on how the mass could be deduced from $M_{T2}$ alone.
If $Y$ is pair produced and each undergoes a three-body decay\footnote{The presence of a $\ge 3$-body decay is a sufficient but not necessary condition. Two-body decays can also display kinks
\cite{Gripaios:2007is,Barr:2007hy} provided the decaying particles have sufficiently large transverse boosts.} to $N$ as shown in Fig.~\ref{FigEventTopologyThreeBody},
then a `kink' in the $\Max m_{T2}$\cite{Lester:1999tx,Barr:2003rg}, will occur at a position which indicates the invisible particle mass.
This `kink' is corroborated in Refs.~\cite{Barr:2007hy,Nojiri:2008hy}.
Fig~\ref{FigKink} shows the
`kink' in $\Max m_{T2}$ in an idealized simulation where $M_Y=800$ GeV and $M_N=100$ GeV.
In the CCKP example, they study a mAMSB senario where a gluino
decays via a three-body decay to the LSP and quarks.
In this case, they determined the LSP's mass to $\pm 1.7$ GeV.
The difficulty in mass determination with this technique scales as $(M_+/M_-)^2 =(M_Y+M_N)^2/(M_Y-M_N)^2$.  For CCKP's example $M_{+}/M_{-}=1.3$; we will consider
examples where $M_{+}/M_{-} \approx 3$.  The $(M_{+}/M_{-})^2$ scaling behavior follows by propagating error in $M_{-}$ to position of the intersection of the two curves that form the kink.

This kink is quantified by the constrained mass variable $m_{2C}$  \cite{Ross:2007rm}\cite{Barr:2008ba} that is the subject of later chapters of this thesis.  We will be applying it to a case about $5$ or $6$ times more difficult because $M_{+}/M_{-} \approx 3$ for the neutralino studies on which we choose to focus.

\section*{Chapter Summary}

In this chapter, we have introduced the need for making mass measurements in a model independent manner of collider-produced dark matter.  We have outlined many approaches based on edges of invariant mass distributions, mass shell techniques, and transverse mass techniques.  Each technique has distinct domains of validity and requirements.  In the remaining chapters of this thesis, we introduce and test new model-independent techniques to determine mass of dark-matter particles.
Which approach turns out to be best is likely to depend on what scenario nature hands us, since the various techniques involve different assumptions. Having different approaches also offers the advantage of providing a system of redundant checks.

\chapter{Using $M_{T2}$ with Cascade decays}

\label{ChapterMT2onCascadeDecays}
\providecommand{\GeV}{\,{\rm{GeV}}}
\providecommand{\etal}{\textit{et.al.}}

\section*{Chapter Overview}
Recently, a paper by Tovey\cite{Tovey:2008ui} introduced a new variable, $M_{CT}$, with the powerful concept of extracting mass information from intermediate stages of a symmetric decay chain.  In this chapter, we compare $M_{T2}$  with $M_{CT}$.
The variable $M_{CT}$ is new but shares many similarities and differences with $M_{T2}$.
We briefly define  $M_{CT}$ and explain when it gives identical results to $M_{T2}$ and when it gives different results. We comment on benefits of each variable in its intended applications.  We find that for massless visible particles $M_{CT}$ equals  $M_{T2}$ in a particular limit, but $M_{T2}$ has better properties when there is initial state radiation (ISR) or upstream transverse momentum (UTM)\footnote{ISR and UTM both indicate $k_T\neq0$.  We take ISR to mean the case when $k_T$ is small compared to the energy scales involved in the collision.}.
We argue that $M_{T2}$ is a more powerful tool for extracting mass information from intermediate stages of a symmetric decay chain.  This chapter is based on work first published by the author in Ref.~\cite{Serna:2008zk}.

\section{$M_{T2}$ and $M_{CT}$ in this context}

Both $M_{T2}$ and $M_{CT}$  assume a pair-produced new-particle state followed by each branch decaying symmetrically to visible states and dark-matter candidates which escape detection and appear as missing transverse momentum. Fig \ref{FigEventTopologyTovey} is the simplest example on-which we can meaningfully compare the two kinematic quantities.  The figure shows two partons colliding and producing some observed initial state radiation (ISR) or upstream transverse momentum (UTM) with four momenta $k$ and an on-shell, pair-produced new state $Y$.  On each branch, $Y$ decays to on-shell states $X$ and $v_1$ with masses $M_{X}$ and $m_{v_1}$, and $X$ then decays to on-shell states $N$ and $v_2$ with masses $M_N$ and $m_{v_2}$.  The  four-momenta of $v_1$, $v_2$ and $N$ are respectively $\alpha_1$, $\alpha_2$ and $p$ on one branch and $\beta_1$, $\beta_2$ and $q$ in the other branch.  The missing transverse momenta $\slashed{P}_T$ is given by the transverse part of $p+q$.

Tovey \cite{Tovey:2008ui} recently defined a new variable $M_{CT}$ which has many similarities to $M_{T2}$.  The variable is defined as
 \begin{eqnarray}
   M_{CT}^2(\alpha_1,\beta_1) & = & (E_T(\alpha_1)+E_T(\beta_1))^2 - (\alpha_{1T}-{\beta}_{1T})^2.
 \end{eqnarray}
Tovey's goal is to identify another constraint between masses in the decay chain.  He observes that in the rest frame of $Y$ the momentum of the back-to-back decay products $X$ and $v_1$ is given by
 \begin{equation}
   \left(k_*(M_Y,M_X,m_{v_1}) \right)^2 = \frac{(M_Y^2 - (m_{v_1}+M_X)^2)(M_Y^2 - (m_{v_1}-M_X)^2)}{4 M_Y^2} \label{Eq2BEMP}
 \end{equation}
where $k_*$ is the two-body excess-momentum parameter (2BEMP) \footnote{Tovey refers to this as the 2-body mass parameter ${\mathcal{M}}_i$.  We feel calling this a mass is a bit misleading so we are suggesting 2BEMP.}.  In the absence of transverse ISR ($k_T=0$) and if the visible particles are effectively massless ($m_{v_1}=0$), Tovey observes that $\max M_{CT}(\alpha_1,\beta_1)$ is given by $2 k_*$; this provides an equation of constraint between $M_Y$ and $M_X$.  Tovey observes that if we could do this analysis at various stages along the symmetric decay chain, then all the masses in principle could be determined.

The big advantage of $M_{CT}$ is in its computational simplicity.  It is a simple one line formula to evaluate.  Also, $M_{CT}$ is intended to only be calculated once per event instead of at a variety of choices of the hypothetical LSP mass $\chi$.  In contrast, $M_{T2}$ is a more computationally intensive parameter to compute; but this is aided by the use of a common repository of community tested C++ libraries found at \cite{AtlasMT2Wiki}.

How are these two variables similar?  Both $M_{CT}$ and $M_{T2}$, in the limit of $k_T=0$, are invariant under back-to-back boosts of the parent particles momenta \cite{Cho:2007dh}.  The variable $M_{CT}$ equals $M_{T2}$ in the special case where $\chi=0$, and the visible particles are massless $(\alpha_1^2=\beta_1^2=0)$, and there is no transverse ISR or UTM ($k_T=0$)
 \begin{eqnarray}
   M_{CT}(\alpha_1, \beta_1) & = & M_{T2}(\chi=0,\alpha_1,\beta_1, \slashed{P}_T = (p+q+\alpha_2 +\beta_2)_T) \ \ \ {\rm{if}} \ \ \alpha_1^2=\beta_1^2=0. \label{EqMCTMT2Equality} \\
    & = & 2 ( {\alpha_1}_T \cdot {\beta_1}_T + |{\alpha_1}_T |\,| {\beta_1}_T |).  \label{EqMCTMT2Equality2}
 \end{eqnarray}
The $M_{CT}$ side of the equation is straight forward. The $M_{T2}$ side of the expression can be derived analytically using the formula for $M_{T2}$ given in \cite{Lester:2007fq}; we also show a short proof in  Appendix \ref{AppendixMT2Eq54}. Eq(\ref{EqMCTMT2Equality}) uses a $M_{T2}$ in a unconventional way;  we group the observed momenta of the second decay products into the missing transverse momenta.    In this limit, both share an endpoint of $2 k_* = (M_Y^2 - M_X^2)/M_Y$. To the best of our knowledge, this endpoint was first pointed out by CCKP \cite{Cho:2007qv} \footnote{The endpoint given by CCKP is violated for non-zero ISR at $\chi_N < M_N$ and $\chi_N > M_N$.}.  We find it surprising that a physical relationship between the masses follows from $M_{T2}$ evaluated at a non physical $\chi$.  In the presence of ISR or UTM, Eq(\ref{EqMCTMT2Equality}) is no longer an equality.  Furthermore in the presence of ISR or UTM, the end point of the distribution given by either side of Eq(\ref{EqMCTMT2Equality}) exceeds $2 k_*$.  In both cases, we will need to solve a combinatoric problem of matching visible particles to their decay order and branch of the event which we leave for future research.


In the case where the visible particle $v_1$ is massive, the two parameters give different end-points

\noindent
\small
 \begin{eqnarray}
   \max M_{CT} (\alpha_1,\beta_1) & = &  \frac{M_Y^2 - M_X^2}{M_Y} + \frac{m_{v_1}^2}{M_{Y}} \label{EqMassiveMCTLimit},\ {\rm{and}} \\
   \max M_{T2}(\chi=0,\alpha_1,\beta_1, \slashed{P}_T = (p+q+\alpha_2 +\beta_2)_T) & = &   \sqrt{ m_{v_1}^2 + 2(k_*^2 + k_* \sqrt{k_*^2+m^2_{v_1}})} \label{EqMassiveMT2Limit}
 \end{eqnarray}
\normalsize

\baselineskip=21pt plus1pt
\noindent
where $k_*$ is given by Eq(\ref{Eq2BEMP}).  Unfortunately, there is no new information about the masses in these two endpoints.  If we solve Eq(\ref{EqMassiveMCTLimit}) for $M_X$ and substitute this into Eq(\ref{EqMassiveMT2Limit}) and (\ref{Eq2BEMP}), all dependence on $M_Y$ is eliminated.

\section{Application to symmetric decay chains}

Tovey's idea of analyzing the different steps in a symmetric decay chain to extract the masses is powerful.  Up to now, we have been analyzing both variables in terms of the first decay products of $Y$.  This restriction is because $M_{CT}$ requires no transverse ISR to give a meaningful endpoint.  If we were to try and use $\alpha_2$ and $\beta_2$ to find a relationship between $M_X$ and $M_N$, then we would need to consider the transverse UTM to be $(k+ \alpha_1+\beta_1)_T$, which is unlikely to be zero.

\begin{figure}
\centerline{\includegraphics[width=3in]{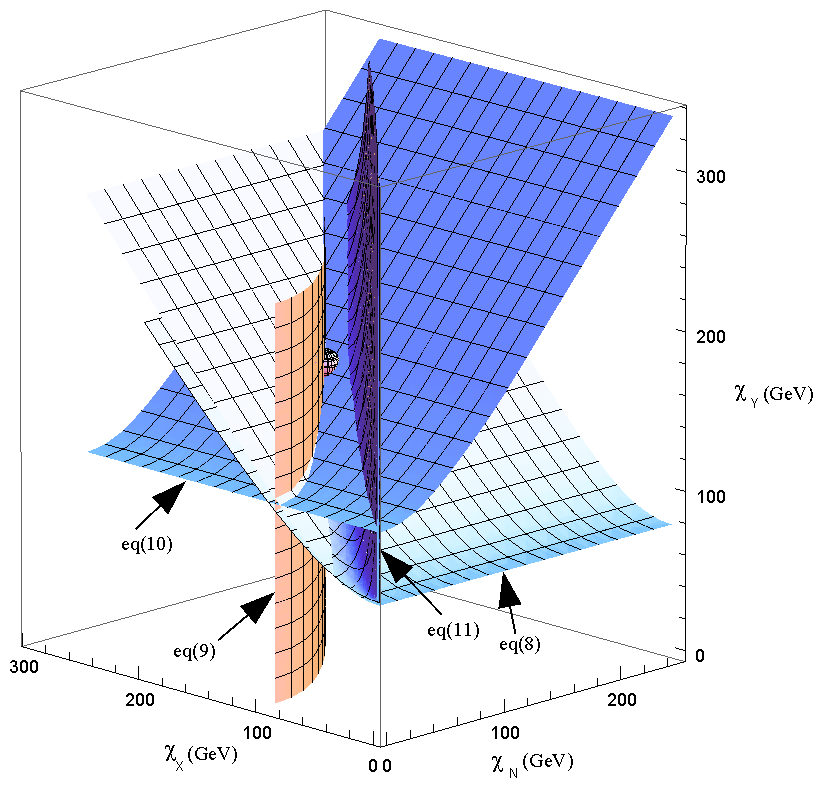}}
\caption{\label{FigCombinedConstraints} Shows constraints from $\max\,M_{T2}$ used with different combinations as described in eqs(\ref{EqYGivenX},\ref{EqXGivenN},\ref{EqYGivenN}) and the $\max m_{12}$ described in Eq(\ref{EqmllConstraint}).  Intersection is at the true mass $(97 \GeV,144 \GeV,181 \GeV)$ shown by sphere. Events include ISR but otherwise ideal conditions: no background, resolution, or combinatoric error.}
\end{figure}
We suggest $M_{T2}$ is a better variable with which to implement Tovey's idea of analyzing the different steps in a symmetric decay chain because of its ISR properties.
With and without ISR, $M_{T2}$'s endpoint gives the correct mass of the parent particle when we assume the correct value of the missing-energy-particle's mass \footnote{In principle we could plot the $\max M_{T2}(\chi_X,\alpha_1,\beta_1,\slashed{P}_T=(\alpha_2+\beta_2+p+q)_T)$ vs $\chi_X$ as a function of transverse ISR and the value of $\chi_X$ at which the end point is constant would give the correct value of $M_X$; at which point the distributions end point would give the correct $M_Y$.  In practice we probably will not have enough statistics of ISR events.}.
For this reason, $\max M_{T2}$ gives a meaningful relationship between masses $(M_Y,M_X,M_N)$ for all three symmetric pairings of the visible particles across the two branches.  A relationship between $M_Y$ and $M_X$ is given by
 \begin{equation}
 \chi_Y(\chi_X) =  \max M_{T2}(\chi_X,\alpha_1,\beta_1, \slashed{P}_T = (p+q+\alpha_2 +\beta_2)_T). \label{EqYGivenX}
 \end{equation}
A relationship between $M_X$ and $M_N$  can be found by computing
 \begin{equation}
 \chi_X(\chi_N) =  \max M_{T2}(\chi_N,\alpha_2,\beta_2, \slashed{P}_T = (p+q)_T)  \label{EqXGivenN}
 \end{equation}
where we have grouped $\alpha_1+\beta_1$ with the $k$ as a part of the ISR.
A relationship between $M_Y$ and $M_N$ can be found by using $M_{T2}$ in the traditional manner giving
 \begin{equation}
 \chi_Y(\chi_N) =  \max M_{T2}(\chi_N,\alpha_1+\alpha_2,\beta_1+\beta_2, \slashed{P}_T = (p+q)_T).  \label{EqYGivenN}
 \end{equation}
Lastly, we can form a distribution from the invariant mass of the visible particles on each branch $m^2_{12}=(\alpha_1+\alpha_2)^2$ or  $m^2_{12}=(\beta_1+\beta_2)^2$.  The endpoint of this distribution gives a relationship between $M_Y$, $M_X$, and $M_N$ given by
 \begin{equation}
   \max m^2_{12} = \frac{(M_Y^2-M_X^2)(M_X^2-M_N^2)}{M_X^2}.  \label{Eqmll}
 \end{equation}
Solving this expression for $M_Y$ gives the relationship
 \begin{equation}
  \chi^2_Y (\chi_N,\chi_X) = \frac{\chi_X^2 ( (\max m^2_{12}) +\chi_X^2 - \chi_N^2)}{\chi_X^2 - \chi_N^2} \label{EqmllConstraint}.
 \end{equation}
Fig \ref{FigCombinedConstraints} shows the constraints from Eqs(\ref{EqYGivenX}, \ref{EqXGivenN}, \ref{EqYGivenN}, \ref{EqmllConstraint}) in an ideal simulation using $(M_Y=181 \, {\rm{GeV}}$, $M_X=144\, {\rm{GeV}}$, $M_N=97\, {\rm{GeV}})$, 1000 events, and massless visible particles, and ISR added with an exponential distribution with a mean of $50$ GeV.  These four surfaces in principle intersect at a single point $(M_Y, M_X, M_N)$ given by the sphere in the figure \ref{FigCombinedConstraints}.  Unfortunately, all these surfaces intersect the correct masses at a shallow angles so we have a sizable uncertainty along the direction of the sum of the masses and
tight constraints in the perpendicular directions.
In other words, the mass differences are well-determined but not the mass scale.
From here one could use a shape fitting technique like that described in Chapters
\ref{ChapterM2Cdirect} - \ref{ChapterM3C}.
to find a constraint on the sum of the masses.
Tovey's suggestion for extracting information from these intermediate stages of a symmetric cascade chain clearly provides more constraints to isolate the true mass than one would find from only using the one constraint of Eq(\ref{EqYGivenN}) as described in \cite{Cho:2007qv}.  However, Tovey's suggestion
is more feasible using the $M_{T2}$ rather than $M_{CT}$ because the constraint surfaces derived from $M_{T2}$ intersect the true masses even with UTM.

\section*{Chapter summary}

In summary, we have compared and contrasted $M_{CT}$ with $M_{T2}$.  The variable $M_{CT}$ is a special case of $M_{T2}$ given by Eq(\ref{EqMCTMT2Equality}) when ISR can be neglected and when the visible particles are massless.  In this case, the end-point of this distribution gives $2 k_*$, twice the two-body excess momentum parameter (2BEMP). If $m_{v_1} \neq 0$, the two distributions have different endpoints but no new information about the masses.   In the presence of ISR the two functions are not equal; both have endpoints that exceed $2 k_*$.  Because of it's better properties in the presence of UTM or ISR, $M_{T2}$ is a better variable for the task of extracting information for each step in the decay chain.  Extracting this information requires solving combinatoric problems which are beyond the scope of this chapter.
%

\providecommand{\Mvariable}[1]{}
\providecommand{\etal}{\textit{et.al.}}
\providecommand{\MYMin}{M_{2C}}
\providecommand{\MY}{M_{Y}}
\providecommand{\MN}{M_{N}}
\providecommand{\Mm}{M_{-}}
\providecommand{\Mp}{M_{+}}
\providecommand{\chiTwo}{\tilde{\chi}^o_2}
\providecommand{\chiOne}{\tilde{\chi}^o_1}
\providecommand{\MeV}{{\rm{MeV}\ }}
\providecommand{\GeV}{{\rm{GeV}\ }}
\providecommand{\eV}{{\rm{eV}\ }}
\providecommand{\imag}[1]{\,i\,}

\chapter{The Variable $M_{2C}$: Direct Pair Production}
 \label{ChapterM2Cdirect}

\section*{Chapter Overview}

In this chapter, we propose an improved method for hadron-collider mass determination of new
states that decay to a massive, long-lived state like the LSP in the MSSM. We
focus on pair-produced new states which undergo three-body decay to a pair of
visible particles and the new invisible long-lived state. Our approach is to
construct a kinematic quantity which enforces all known physical constraints
on the system. The distribution of this quantity calculated for the observed
events has an endpoint that determines the mass of the new states. However we
find it much more efficient to determine the masses by fitting to the entire
distribution and not just the end point. We consider the application of the
method at the LHC for various models under the ideal assumption of effectively
direct production with minimal
ISR and demonstrate that the method can
determine the masses within about $6$ GeV using only $250$ events. This
implies the method is viable even for relatively rare processes at the LHC
such as neutralino pair production.

This chapter, which  is based on
work first published by the author and his supervisor in Ref.~\cite{Ross:2007rm},
 concentrates on mass determination involving the
production of only two new states. Our particular concern is to use the
available information as effectively as possible to reduce the number of
events needed to make an accurate determination of $M_{Y}$ and $M_{N}$. The
main new ingredient of the method proposed is that it does not rely solely on
the events close to the kinematic boundary but makes use of all the events.
Our method constrains the unobserved energy and momentum such that all the
kinematical constraints of the process are satisfied including the mass
difference, Eq(\ref{EqThreeBodyDecayEdge}), which can be accurately measured from the
$ll$ spectrum discussed in Sec \ref{SubSecEdges}. This increases the information that events far from the kinematic
boundary can provide about $M_{Y}$ and significantly reduces the number of
events needed to obtain a good measurement of the overall mass scale.
We develop the method for the case where $Y$ is directly pair produced in the parton collision with minimal ISR and where each $Y$ decays via a three-body decay to
a on-shell final states $N + l^{+} + l^{-}$. Its
generalization to other processes is straightforward and considered later in this thesis\footnote{We note that the
on-shell intermediate case studied by CGHMM is also improved by including the
relationship measured by the edge in the $ll$ distribution on each event's
analysis. The $Y$ decay channel with an on-shell intermediate state $X$ has an
edge in the $ll$ invariant mass distribution which provides a good
determination of the relationship $\max m_{ll}^{2}=(M_{Y}^{2}-M_{X}^{2}%
)(M_{X}^{2}-M_{N}^{2})/M_{X}^{2}$. This relationship forms a surface in
$M_{N}$,$M_{X}$,$M_{Y}$ space that only intersects the allowed points of
CGHMM's Fig 3 near the actual masses.  We will investigate this case in Chapter \ref{ChapterM3C}.}.

The chapter is structured as follows: In Section \ref{SecImprovedDistribution}, we introduce the $M_{2C}$
distribution whose endpoint gives $M_{Y}$, and whose distribution can be
fitted away from the endpoint to determine $M_{Y}$ and $M_{N}$ before we have
enough events to saturate the endpoint.
Section \ref{SecAppendixRelateToMT2}
discusses the relationship between our distribution and the kink in
$M_{T2}(\chi)$ of CCKP and how this relationship can be used to calculate $M_{2C}$
in a computationally efficient manner.
We then in \ref{SecSymmetryUnderSqrtS} discuss
symmetries and dependencies of the $M_{2C}$ distribution.
Section \ref{SecEstimatedPerformance}
estimates the performance for a few SUSY models where we include approximate
detector resolution effects and where we expect backgrounds to be minimal but where
we assume ($k_T \approx 0$).
Finally we summarize the chapter's findings.

\section{An improved distribution from which to determine $M_{Y}$}

\label{SecImprovedDistribution}

We consider the event topology shown in Fig \ref{FigEventTopologyThreeBody}. The new
state $Y$ is pair produced. Each branch undergoes a three-body decay to the
state $N$ with 4-momentum $p$ ($q$) and two visible particles $1+2$ ($3+4$)
with 4-momentum $\alpha$ ($\beta$). The invariant mass $m_{12}$ ($m_{34}$) of
the particles $1+2$ ($3+4$) will have an upper edge from which we can
well-determine $M_{-}$. Other visible particles not involved can be grouped
into $V$ with 4-momentum $k$.  In general we refer to any process that creates non-zero $k_T$ as upstream transverse momentum (UTM).  One type of UTM is initial state radiation (ISR), which tends to be small compared to the mass scales involved in SUSY processes. Another type of UTM would be decays of heavier particles earlier in the decay chain.
In the analysis presented in this chapter, we tested the concepts against both $k=0$
and $k \lesssim 20$ GeV commiserate with what we might expect for ISR.

We adapt the concept from $M_{T2}$ of minimizing the transverse mass over the
unknown momenta to allow for the incorporation of all the available
information about the masses. To do this we form a new variable $M_{2C}$ which
we define as the minimum mass of the second to lightest new state in the event
$M_{Y}$ constrained to be compatible with the observed 4-momenta of $Y$'s
visible decay products with the observed missing transverse energy, with the
four-momenta of $Y$ and $N$ being on shell, and with the constraint that
$M_{-}=M_{Y}-M_{N}$ is given by the value determined by the end point of the
$m_{12}$ distribution.  The minimization is performed over the eight relevant unknown parameters which may be taken as the 4-momenta $p$ and $q$ of the particle $N$. We neglect any contributions from unobserved initial state radiation
(ISR). Thus we have%
\begin{eqnarray}
M_{2C}^{2}  & = & \min_{p,q}(p+\alpha)^{2}\label{mymin}\\
& & \mathrm{subject}\ \mathrm{to}\ \mathrm{the}\ 5\ \mathrm{constraints}%
\nonumber\\
& &(p+\alpha)^{2}  = (q+\beta)^{2}, \label{eqC1} \\
& & p^{2}=q^{2}, \label{eqC2} \\
& & \slashed{P}_T  = (p+q)_T \label{eqC3} \\
& & \sqrt{(p+\alpha)^{2}}-\sqrt{(p)^{2}} = M_{-} \label{eqC4} \label{EqDeltaMConstraint}%
\end{eqnarray}
where $\slashed{P}_T$ is the missing transverse momentum and $(p+q)_T$ are the transverse components of $p+q$.
Although we can implement the minimization numerically or by using Lagrange
multipliers, we find the most computationally efficient approach is to modify
the $M_{T2}$ analytic solution from Lester and Barr \cite{Lester:2007fq}.
Details regarding implementing $M_{2C}$ and the relation of $M_{2C}$ to
$M_{T2}$ and the approach of CCKP are in Sec.~\ref{SecAppendixRelateToMT2}.

Errors in the determined masses propagated from the error in the mass
difference in the limit of $k=0$ are given by
\begin{equation}
\delta M_{Y} = \frac{\delta M_{-}}{2} \left(  1- \frac{M_{+}^{2}}{M_{-}^{2}}
\right)  \ \ \ \delta M_{N} = - \frac{\delta M_{-}}{2} \left(  1+ \frac
{M_{+}^{2}}{M_{-}^{2}} \right) \label{EqDeltaMmErrorEffectsDirect}%
\end{equation}
where $\delta M_{-}$ is the error in the determination of the mass difference
$M_{-}$. To isolate this source of error from those introduced by low
statistics, we assume we know the correct $M_{-}$, and we should consider the
error described in Eq(\ref{EqDeltaMmErrorEffectsDirect}) as a separate uncertainty
from that reported in our initial performance estimates in the Section \ref{SecEstimatedPerformance}.

Because the true $p$, $q$ are in the domain over which we
are minimizing, $M_{2C}$ will always satisfy $M_{2C}\leq M_{Y}$. The equality
is reached for events with either $m_{12}$ or $m_{34}$ smaller than $M_{-},$
with $p_{z}/p_{o}=\alpha_{z}/\alpha_{o}$, and $q_{z}/q_{o}=\beta_{z}/\beta
_{o}$, and with the transverse components of $\alpha$ parallel to the
transverse components of $\beta$.  When $m_{12}=m_{34}=M_{-}$ the event only gives information about about the mass difference.

The events that approximately saturate the bound have the added benefit that
they are approximately reconstructed ($p$ and $q$ are known).  We present a proof of this in Appendix \ref{AppendixUniquenessOfReconstruction}. If $Y$ is
produced near the end of a longer cascade decay, then this reconstruction
would allow us to determine the masses of all the parent states in the event. The
reconstruction of several such events may also aid in spin-correlation studies \cite{Abbott:2000dt}.

\begin{figure}[ptb]
\centerline{\includegraphics[width=6in]{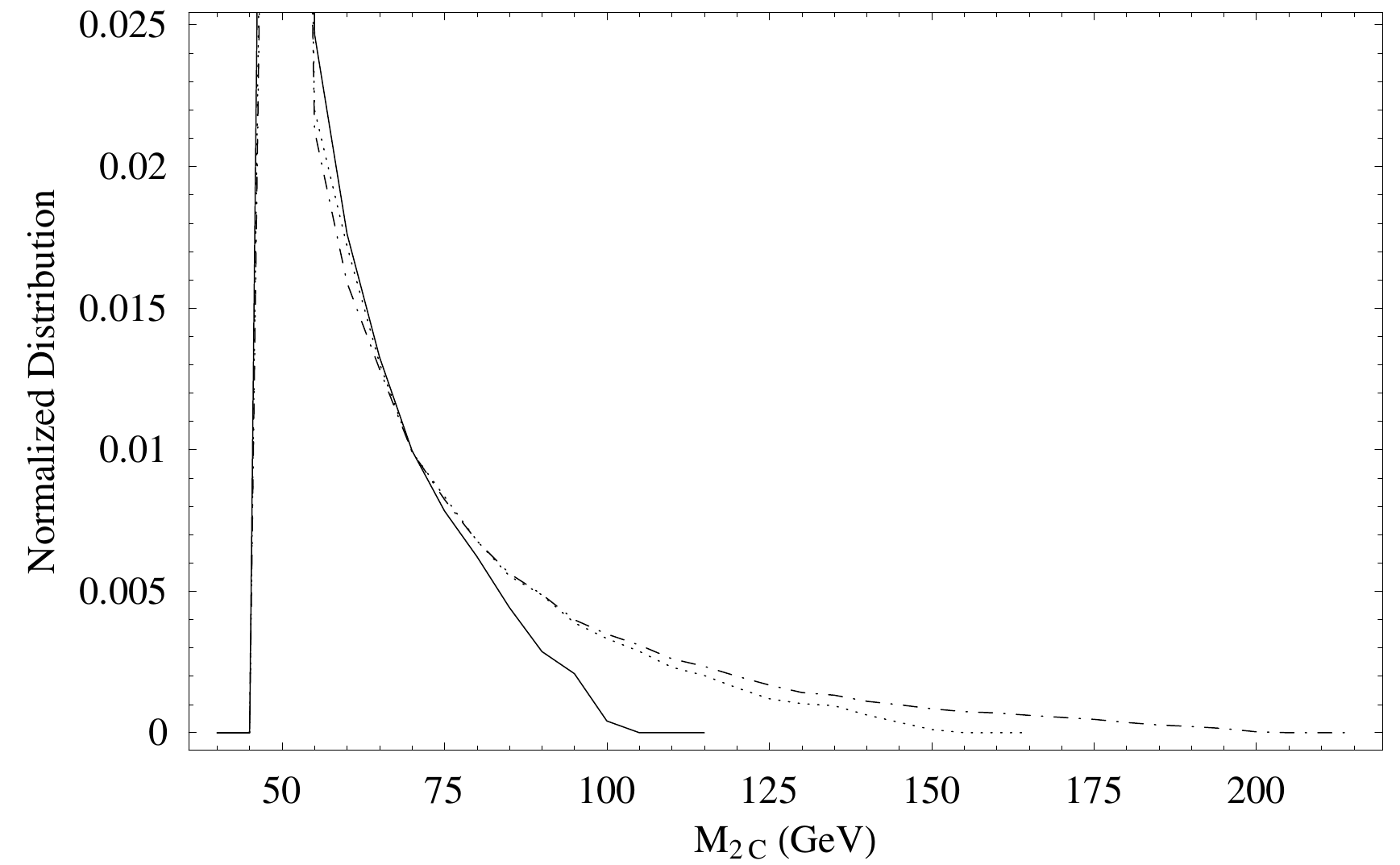}}\caption{The
distribution of 30000 events in 5 GeV bins with perfect resolution and no
background. The three curves represent $M_{Y}=200$ GeV (dot-dashed),
$M_{Y}=150$ GeV (dotted) and $M_{Y}=100$ GeV (solid) each with $M_{-}=50$ GeV.
Each distribution cuts off at the correct $M_{Y}$.}%
\label{FigMYMinIdealExample}%
\end{figure}

In order to determine the distribution of $M_{2C}$ for the process shown in
Fig \ref{FigEventTopologyThreeBody}, we computed it for a set of events generated using
the theoretical cross section and assuming perfect detector resolution and no
background.
Figure \ref{FigMYMinIdealExample} shows the resulting distribution for three cases:
$M_{Y}=200$ GeV, $M_{Y}=150$ GeV and $M_{Y}=100$ GeV each with $M_{-}=50$ GeV.
Each distribution was built from 30000 events. Note that the minimum $M_{Y}$
for an event is $M_{-}$.
The three examples each have endpoints that give the mass scale, and we
are able to distinguish between different $M_{Y}$ for a given $M_{-}$.
The end-point for $M_Y=100$ GeV is clear, and the endpoint for $M_Y=150$ GeV and $M_Y=200$ GeV becomes more difficult to observe.
The shape of the
distribution exhibits a surprising symmetry discussed in Sec~\ref{SecSymmetryUnderSqrtS}.

We can also see that as $M_{+}/M_{-}$ becomes large, the $M_{Y}$
determination will be hindered by the small statistics available near the
endpoint or backgrounds. To alleviate this, we should instead fit to the
entire distribution.
It is clear that events away from the endpoint
also contain information about the masses.
For this reason we propose to fit
the entire distribution of $M_{2C}$ and compare it to the `ideal' distribution
that corresponds to a given value of the masses. As we shall discuss this
allows the determination of $M_{Y}$ with a significant reduction in the number
of events needed. This is the most important new aspect of the method proposed here.

\section{Using $M_{T2}$ to Find $M_{2C}$ and the $\max M_{T2}$ Kink}

\label{SecAppendixRelateToMT2}
The calculation of $M_{2C}$ is greatly facilitated by understanding
its relation to $M_{T2}$.
The variable $M_{T2}$, which was introduced in
by Lester and Summers \cite{Lester:1999tx}, is equivalent to
\begin{align}
M_{T2}^{2}(\chi)  & =\min_{p,q} (p+\alpha)^{2}\label{mt2}\\
& \mathrm{subject}\ \mathrm{to}\ \mathrm{the}\ 5\ \mathrm{constraints}%
\nonumber\\
& (p+\alpha)^{2}=(q+\beta)^{2},\\
& p^{2}=q^{2}\\
& \slashed{P}_T=(p+q)_T\\
& p^{2}=\chi^{2}.\label{EqChiConstraint}%
\end{align}
As is suggested in the simplified example of \cite{Gripaios:2007is}, the
minimization over the longitudinal frame of reference and center-of-mass energy is equivalent to assuming $p$ and
$\alpha$ have equal rapidity and $q$ and $\beta$ have equal rapidity.
Implementing this Eq(\ref{mt2}) reduces to the traditional definition of the
Cambridge transverse mass \footnote{We have only tested and verified this equivalence numerically for events satisfying $M_{T2}(\chi=0)>M_{-}$.}.

\begin{figure}[ptb]
\centerline{\includegraphics[width=4in]{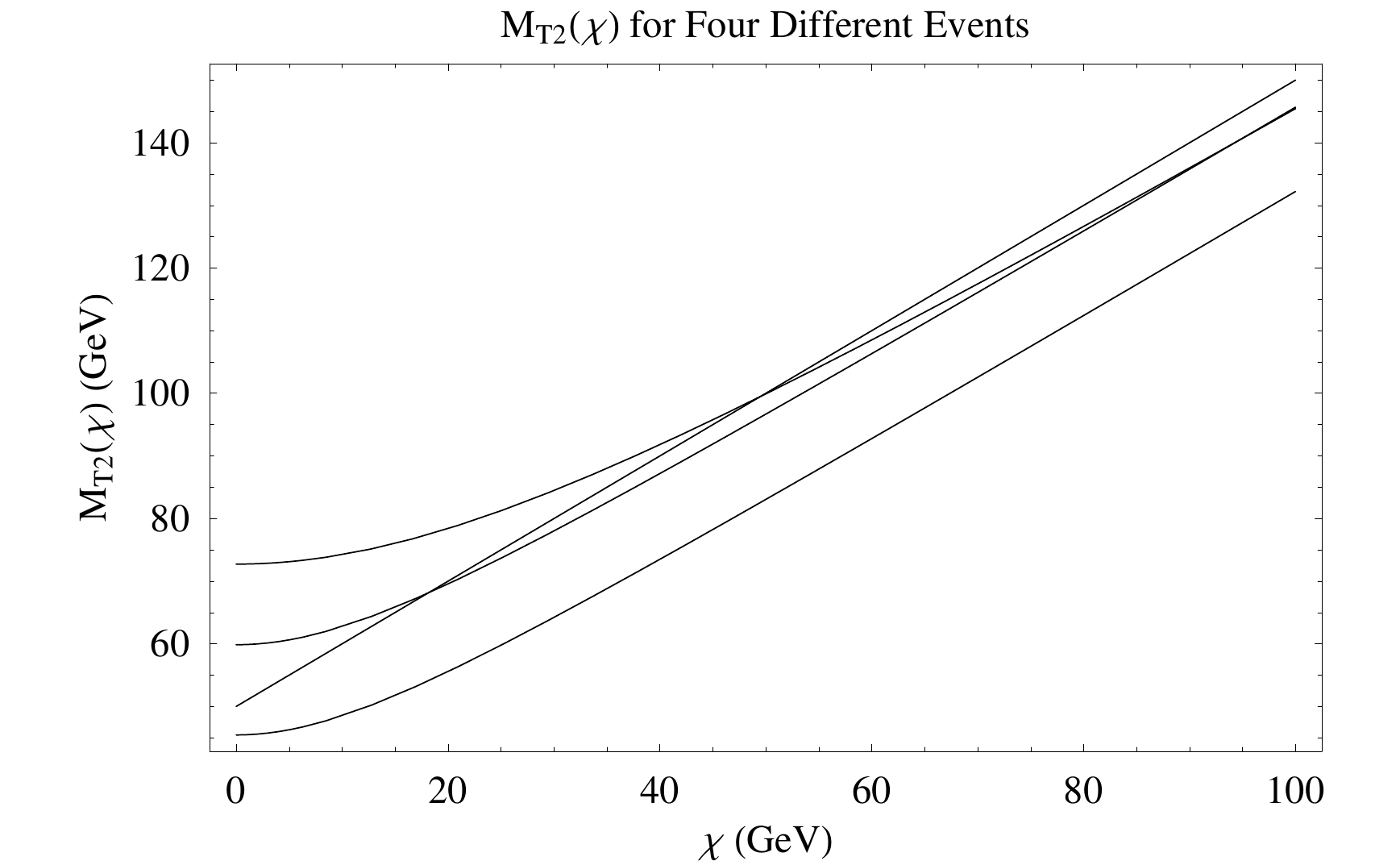}}\caption{ The
$M_{T2}(\chi)$ curves for four events with $M_{N}=50$ GeV and $M_{Y}=100$ GeV.
Only the events whose curves starts off at $M_{T2}(0) > M_{-}$ intersect the
straight line given by $M_{T2}(\chi) - \chi= M_{-}$. The $M_{T2}$ at the
intersection is $M_{2C}$ for that event. }%
\label{FigMT2ComparisonPlot}%
\end{figure}

By comparing $M_{T2}(\chi)$ as defined above to $M_{2C}$ defined in
Eq(\ref{mymin}), we can see that they are very similar with the exception
that the constraint Eq(\ref{EqDeltaMConstraint}) is replaced by the constraint
Eq(\ref{EqChiConstraint}). $M_{2C}$ can be found by scanning $M_{T2}(\chi)$
for the $\chi$ value that such that the constraint in
Eq(\ref{EqDeltaMConstraint}) is also satisfied.  At the value of $\chi$ that satisfies $M_{T2}(\chi)-\chi=M_{-}$, both Eq(\ref{EqDeltaMConstraint}) and Eq(\ref{EqChiConstraint}) are satisfied.

We can see the $M_{2C}$ and $M_{T2}$ relationship visually. Each event
provides a curve $M_{T2}(\chi)$; Fig \ref{FigMT2ComparisonPlot} shows curves
for four events with $M_{N}=50$ GeV and $M_{Y}=100$ GeV. 
For all events
$M_{T2}(\chi)$ is a continuous function of $\chi$.
CCKP point that
out that at $\chi>M_N$ and at $k=0$ the \textit{maximum }$M_{T2}(\chi)$ approaches
$\chi+M_{-}$. 
At $\chi<M_N$ and at $k=0$  the \textit{maximum} $M_{T2}(\chi)$ approaches 
\begin{equation}
 \max M_{T2}(\chi) = k^* + \sqrt{(k^*)^2+\chi^2} \label{EqMaxMT2}
 \end{equation}
where $k^*=(M_Y^2-M_N^2)/2 M_Y$ is the 2BEMP.  This maximum occurs for events with $\alpha^2=\beta^2=0$.
Putting this together, if $M_{T2}( \chi=0)>M_{-}$,
as is true for two of the four events depicted in
Fig.~\ref{FigMT2ComparisonPlot}, then because the event is bounded above by Eq(\ref{EqMaxMT2}) and is continuous, then it must cross $\chi+M_{-}$ at $\chi \leq M_N$.
At this intersection there is a solution to $M_{T2}(\chi
)=\chi+M_{-}$ where $M_{T2}(\chi)=\min M_{Y} |_{\mathrm{Constraints}}
\equiv M_{2C}$. Equivalently
\begin{align}
M_{2C}  & =M_{T2}\ \ \mathrm{at}\ \chi\ \mathrm{where}\ \ M_{T2}(\chi
)=\chi+M_{-}\ \ \ \mathrm{if}\ \ M_{T2}(\chi=0)>M_{-}\\
& =M_{-}\ \ \ \ \mathrm{otherwise}.
\end{align}
Assuming $k=0$, the maximum $\chi$ of such an intersection occurs for $\chi=M_{N}$ which is why the endpoint of $M_{2C}$ occurs at the correct $M_{Y}$ and why this
corresponds to the kink of CCKP. Because Barr and Lester have an analytic
solution to $M_{T2}$ in Ref.~\cite{Lester:2007fq} in the case $k=0$, this is
computationally very efficient as a definition.  We will study the intersection of these two curves when $k_T \neq 0$ in Chapter \ref{ChapterM2CwUTM}.

\section{Symmetries and Dependencies of the $M_{2C}$ Distribution}
\label{SecSymmetryUnderSqrtS}
\label{SecAppendixSimulationDetails}
\label{SecSpinCorrelationsM2C}

All transverse mass variables $M_T$ and $M_{T2}$ are invariant under longitudinal boosts.  $M_{T2}$ has an additional invariance.
CCKP \cite{Cho:2007dh} prove that if $k_T=0$ then $M_{T2}$ is invariant under back to back boosts of the parent particle pair in the plane perpendicular to the beam direction.  This means that for any event of the topology in Fig.~\ref{FigEventTopologyThreeBody} with $k_T=0$ and barring spin correlation effects,
the $M_{T2}$ distribution will be the same for a fixed center-of-mass energy as it is for a mixed set of collision energies.  We now verify this argument numerically.

In order to numerically
determine the distribution of $M_{2C}$ for the processes shown in
Fig \ref{FigEventTopologyThreeBody}, it is necessary to generate a large sample of
\textquotedblleft ideal\textquotedblright\ events corresponding to the
physical process shown in the figure. We assume $k_T=0$
where we expect the back-to-back boost invariance to lead to a symmetry of the
distribution under changes in the $\sqrt{s}$ of the collision.
We assume each branch decays via an
off-shell $Z^o$-boson as this is what could be calculated quickly and captures the
essential elements to provide an initial estimate of our approach's utility.

Even under these assumptions without knowing about the
invariance of $M_{T2}$ under back-to-back boosts,
we might expect that the shape of the distribution
depends sensitively on the parton distribution and many aspects of the
differential cross section and differential decay rates.
This is
not the case; in the case of direct pair production the shape of the distribution depends sensitively only on two properties:

(i) the shape of the $m_{12}$ (or equivalently $m_{34}$) distributions. In the
examples studied here for illustration, we calculate the $m_{12}$ distribution
assuming it is generated by a particular supersymmetric extension of the Standard Model, but in practice we should use the measured distribution which is accessible to accurate determination.  The particular shape of $m_{12}$ does not greatly affect the ability to determine the mass of $N$ and $Y$ so long as we can still find the endpoint to determine $M_Y-M_N$ and use the observed $m_{ll}$ distribution to model the shape of the $M_{2C}$ distribution.

(ii) the angular dependence of the $N$'s momenta in the rest frame of $Y$. In
the preliminary analysis presented here we assume that in the rest frame of
$\tilde{\chi}_{2}^{o}$, $\tilde{\chi}_{1}^{o}$'s momentum is distributed
uniformly over the $4\pi$ steradian directions. While this assumption is not
universally true it applies in many cases and hence is a good starting point
for analyzing the efficacy of the method.

Under what conditions is the uniform distribution true? Note that the
$\tilde{\chi}_{2}^{o}$'s spin is the only property of $\tilde{\chi}_{2}^{o}$
that can break the rotational symmetry of the decay products. For $\tilde
{\chi}_{2}^{o}$'s spin to affect the angular distribution there must be a
correlation of the spin with the momentum which requires a parity violating
coupling. Consider first the $Z^o$ contribution. Since we are integrating over the
lepton momenta difference\footnote{$M_{2C}$ only depends on the sum of the two OSSF lepton momenta that follow from a decay of $Y$.}, the parity violating term in the cross section coming from the
lepton-$Z^o$ vertex vanishes and a non-zero correlation requires that the parity
violating coupling be associated with the neutralino vertex. The $Z^o$-boson
neutralino vertex vanishes as the $Z^o$ interaction is proportional to
$\overline{\tilde{\chi}_{2}^{o}}\gamma^{5}\gamma^{\mu}\tilde{\chi}_{1}%
^{o}Z^o_{\mu}$ or $\overline{\tilde{\chi}_{2}^{o}}\gamma^{\mu}\tilde{\chi}%
_{1}^{o}Z^o_{\mu}$ depending on the relative sign of $M_{\tilde{\chi}_{2}^{o}}$
and $M_{\tilde{\chi}_{1}^{o}}$ eigenvalues. However if the decay has a
significant contribution from an intermediate slepton there are parity
violating couplings and there will be spin correlations. In this case there
will be angular correlations but it is straightforward to modify the method to
take account of correlations\footnote{Studying and exploiting the neutralino spin correlations
is discussed further in Refs
\cite{MoortgatPick:1999di,MoortgatPick:2000db,Choi:2005gt}.}.

\begin{figure}
\centerline{\includegraphics[width=3.2in]{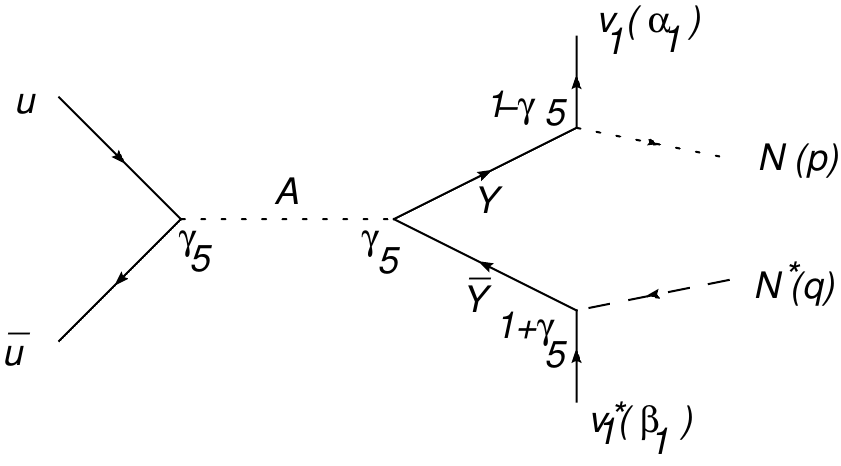}
\includegraphics[width=3.2in]{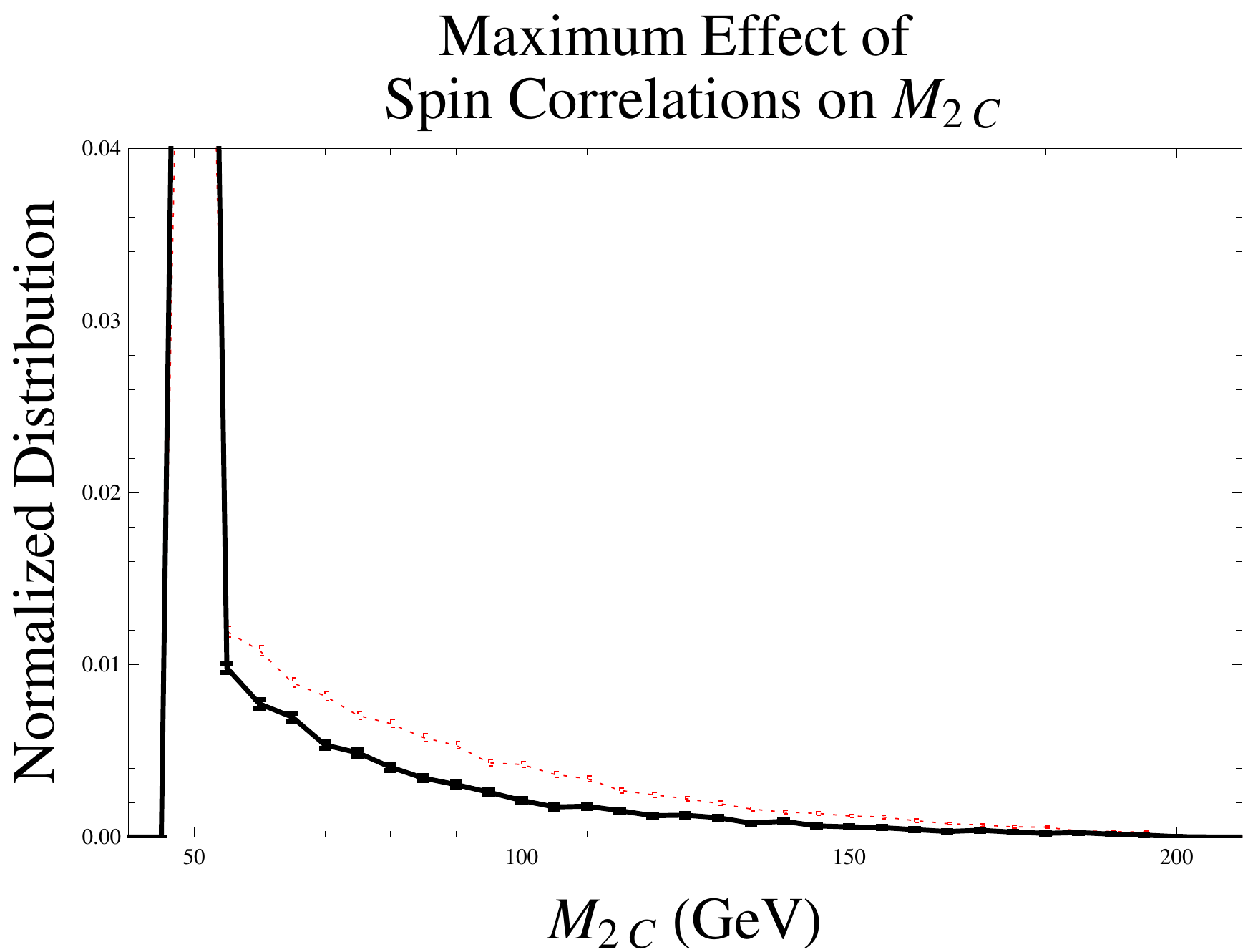}}
\caption{\label{FigM2CSpinCorrelations}Effect of this maximally spin correlated process on the $M_{2C}$ distribution. Modeled masses are $M_Y=200$ GeV and $M_N=150$ GeV.  The solid black distribution is the uncorrelated case and red dotted  distribution is maximally spin correlated.}
\end{figure}

How big an effect could spin correlations
have on the shape of the $M_{2C}$ distribution?
To demonstrate, we modeled a maximally spin-correlated direct production process.
Fig \ref{FigM2CSpinCorrelations} show the spin-correlated process that we consider, and the $M_{2C}$ distributions from this process compared to the $M_{2C}$ distribution from the same topology and masses but without spin correlations.  The modeled case has perfect energy resolution, $m_{v_1}=0$ GeV, $M_Y=200$ GeV, and $M_N=150$ GeV.  Our maximally spin-correlated process involves pair production of $Y$ through a pseudoscalar $A$.  The fermion $Y$ in both branches decays to a complex scalar $N$ and visible fermion $v_1$ through a purely chiral interaction.
The production of the pseudoscalar ensures that the $Y$ and $\bar{Y}$ are in a configuration $\sqrt{2}^{-1}(|\uparrow\downarrow\rangle + |\downarrow\uparrow\rangle)$.
The $Y$ then decays with $N$ preferentially aligned with the spin.
The $\bar{Y}$ decays with $N^*$ preferentially aligned against the spin.
This causes the two sources of missing transverse momentum to be preferentially parallel
and pushes $M_{2C}$ closer to the endpoint.
For this reason the spin-correlated distribution (red dotted distribution) is above the uncorrelated distribution (solid thick distribution) in Fig.~\ref{FigM2CSpinCorrelations}.
For the remainder of the chapter we assume no such spin correlations are present as is true for neutralino pair production and decay to leptons through a $Z^o$ boson.

How likely will we have no spin correlations in supersymmetric LHC events?
We showed earlier that $Z^o$-boson dominated three-body decays lack spin correlations.
Even in the case that the slepton contribution is significant the correlations
may still be largely absent.
Because we are worried about a distribution, the spin
correlation is only of concern to our assumption if a mechanism aligns the
spin's of the $\tilde{\chi}_{2}^{o}$s in the two branches.
Table \ref{TableEventCounts} shows that
most of the $\tilde{\chi}_{2}^{o}$s that we
expect follow from decay chains involving a squark, which being a scalar
should make uncorrelated the spin of the $\tilde{\chi}_{2}^{o} $ in the two
branches.
We would then average over the spin states
of $\tilde{\chi}_{2}^{o}$ and recover the uniform angular distribution of $\tilde{\chi}_{1}^{o}$'s
momentum in $\tilde{\chi}_{2}^{o}$'s rest frame.

\begin{figure}[ptb]
\centerline{
\includegraphics[width=3.2in]{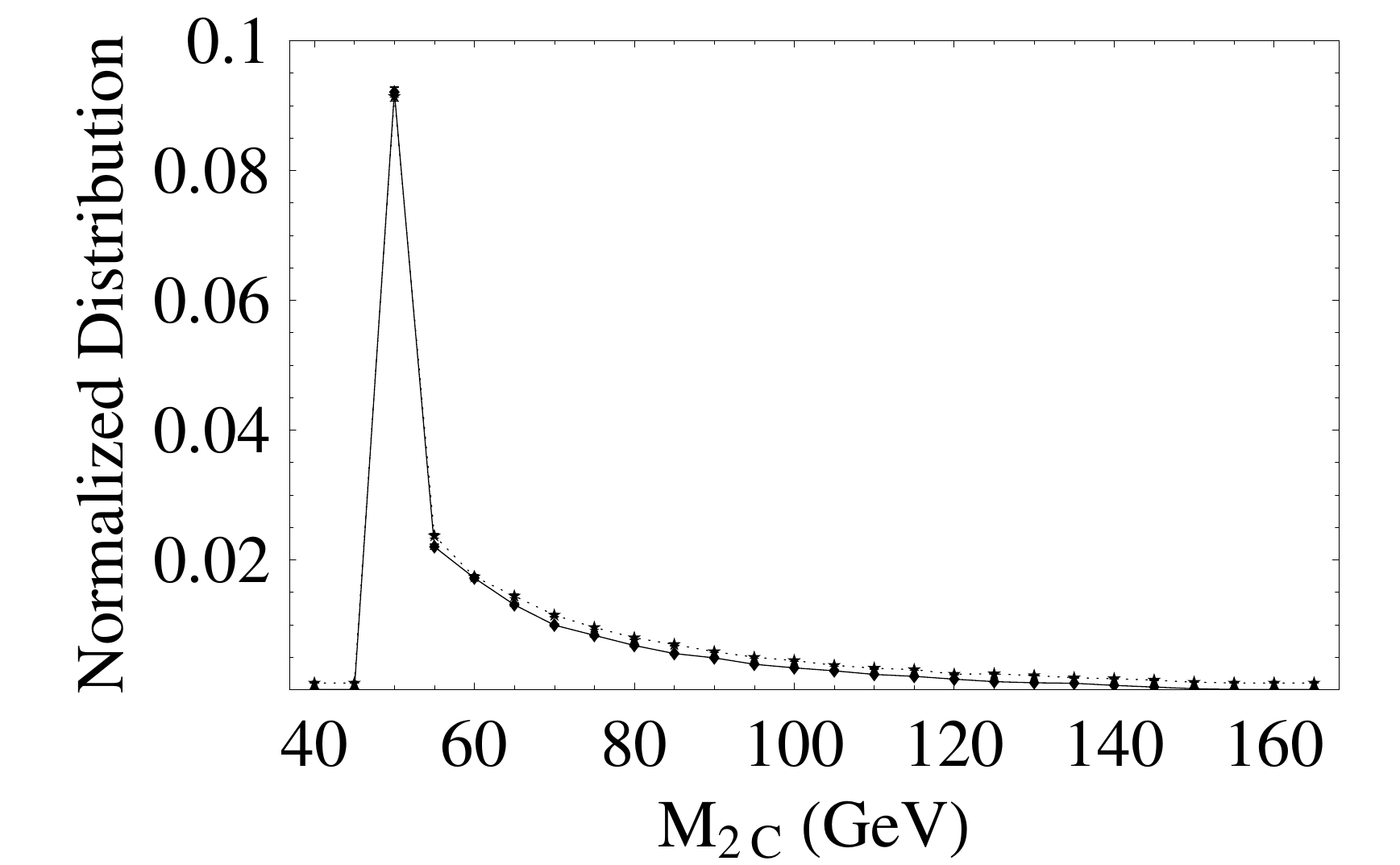}
\includegraphics[width=3.2in]{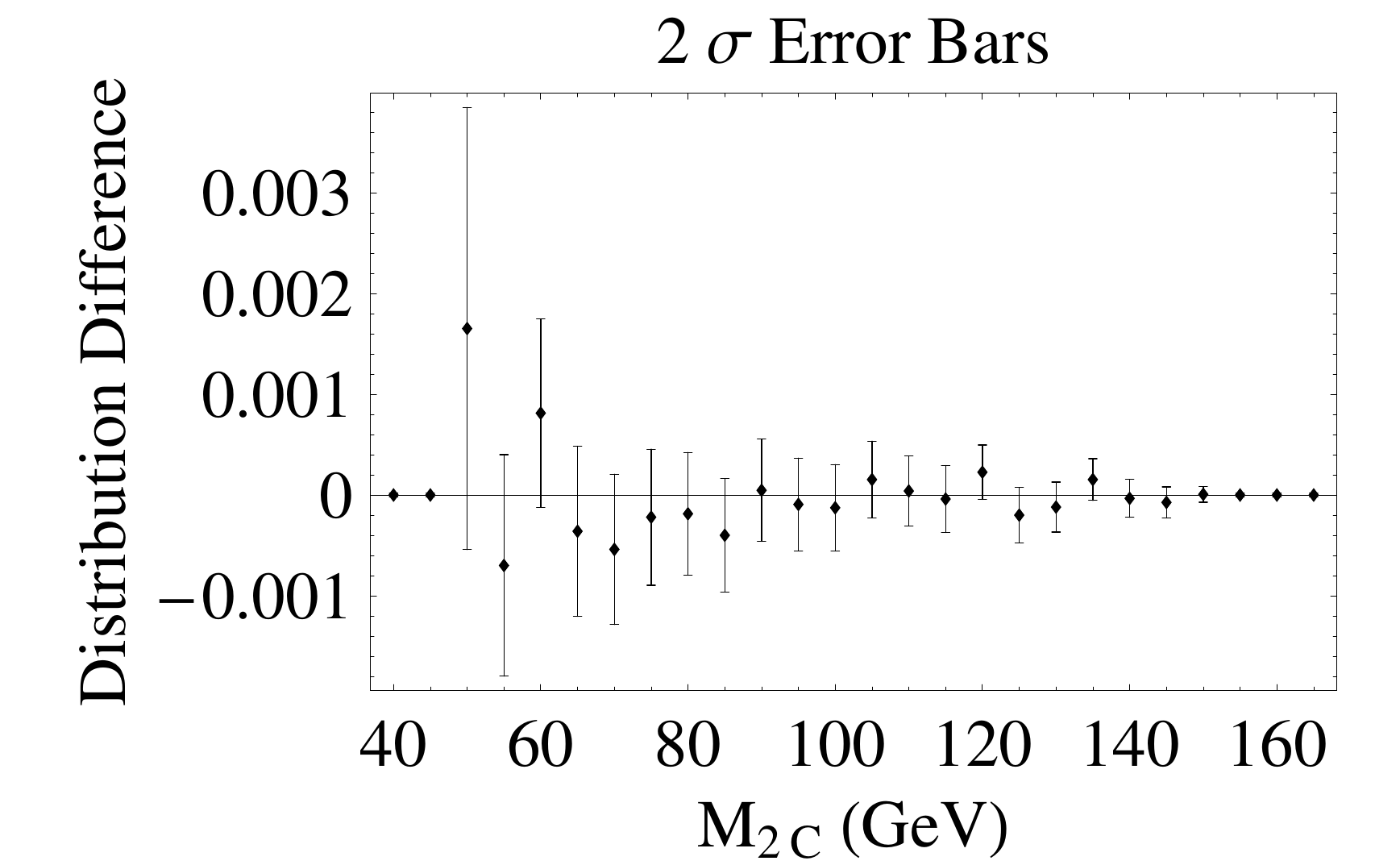}}\caption{Demonstrates
the distribution is independent of the COM energy, angle with which the pair
is produced with respect to the beam axis, and the frame of reference.}%
\label{FigShapeIndenpendence}%
\end{figure}

Once we have fixed the dependencies (i) and (ii) above, the shape of the
distribution is essentially independent of the remaining parameters.
We calculate the \textquotedblleft ideal\textquotedblright\ distributions for $M_{2C}$ assuming
that $k=0$ and that in the rest frame of $Y$ there is an equal likelihood of
$N$ going in any of the $4\pi$ steradian directions. The observable invariant
$\alpha^{2}$ is determined according to the differential decay probability of
$\chi_{2}^{o}$ to $e^{+}$ $e^{-}$ and $\chi_{1}^{o}$ through a $Z^o$-boson
mediated three-body decay. Analytic expressions for cross sections were
obtained from the Mathematica output options in {CompHEP} \cite{Boos:2004kh}
To illustrate the symmetry with respect to changes in $\sqrt{s}$ and
 the angle of $Y$'s production, we show in Fig \ref{FigShapeIndenpendence} two cases:

(1) The case that the collision energy and frame of reference and angle of the
produced $Y$ with respect to the beam axis are distributed according to the
calculated cross section for the process considered in Section
\ref{SecEstimatedPerformance} in which $\tilde{\chi}_{2}^{o}$ decays via $Z^o$
exchange to the three-body state $l^{+}+l^{-}+\tilde{\chi}_{1}^{o},$
convoluted with realistic parton distribution functions.

(2) The case that the angle of the produced $Y$ with respect to the beam axis
is arbitrarily fixed at $\theta=0.2$ radians, the azimuthal angle $\phi$ fixed
at $0$ radians, and the center-of-mass energy set to $\sqrt{s}=500$ GeV.

The left plot of Fig \ref{FigShapeIndenpendence} shows the two distributions
intentionally shifted by 0.001 to allow us to barely distinguish the two
curves. On the right side of Fig \ref{FigShapeIndenpendence} we show the
difference of the two distributions with the 2 $\sigma$ error bars within
which we expect 95\% of the bins to overlap $0$ if the distributions are
identical.

Analytically, this symmetry is because both the $M_{T2}$ and $M_{2C}$ are invariant under back to back boosts when $k_T=0$.
This symmetry implies that the distributions (1) and (2) above should be the same
which has been verified numerically in Fig~\ref{FigShapeIndenpendence}.
In this way we have identified both the mathematical origin of the symmetry and numerically verified the symmetry.
Thus the $M_{2C}$ distribution has a symmetry with respect to changes in $\sqrt{s}$ when $k_T=0$.

In addition to tests with $k_T=0$, we also tested that $k \lesssim 20$ GeV does not change the shape of the distribution to within our numerical uncertainties.
We constructed events with $M_Y=150$ GeV, $M_N=100$ GeV,
 $\sqrt{k^2}$ uniformly distributed between $2$ and $20$ GeV,  $\vec{k}/k_o = 0.98$, and with uniform angular distribution.
We found the $M_{2C}$ distribution agreed with the distribution shown in
Fig.~\ref{FigShapeIndenpendence} within the expected error bars after 10000 events.
Scaling this down to the masses studied in model $P1$ we trust these results remain unaffected for $k_T \lesssim 20$ GeV.
Introduction of cuts on missing traverse energy and and very large UTM ($k_T \gtrsim M_{-}$) changes the shape of the distribution.  These effects will be studied in Chapter \ref{ChapterM2CwUTM}.

Inclusion of backgrounds also changes the shape.
Backgrounds that we can anticipate or measure, like di-$\tau$s or leptons from other neutralino decays observed with different edges can be modeled and included in the ideal shapes used to perform the mass parameter estimation.
A more complete study of the background effects also follows in Chapter \ref{ChapterM2CwUTM}.

\section{Application of the method: SUSY model examples}

\label{SecEstimatedPerformance} To illustrate the power of fitting the full
$M_{2C}$ distribution, we now turn to an initial estimate of our ability to
measure $M_{Y}$ in a few specific supersymmetry scenarios. Our purpose here is
to show that fitting the $M_{2C}$ distribution can determine $M_{Y}$ and
$M_{N}$ with very few events.
We include detector resolution effects and assume $k=0$ (equivalent to assuming direct production), but
neglect backgrounds until Chapter \ref{ChapterM2CwUTM}.
We calculate $M_{2C}$ for the case where the analytic
$M_{T2}$ solution of Barr and Lester can be used to speed up the calculations
as described in Sec~\ref{SecAppendixRelateToMT2}.
We make the same modeling assumptions described in Sec.~\ref{SecSymmetryUnderSqrtS}.

Although fitting the $M_{2C}$ distribution could equally well be applied to
the gluino mass studied in CCKP, we explore its applications to pair-produced
$\tilde{\chi}^{o}_{2}$. We select SUSY models where $\tilde{\chi}^{o}_{2}$
decays via a three-body decay to $l^{+} + l^{-} + \tilde{\chi}^{o}_{1}$. The
four momenta $\alpha=p_{l^{+}} + p_{l^{-}}$ for the leptons in the top branch,
and the four momenta $\beta=p_{l^{+}} + p_{l^{-}}$ for the leptons in the
bottom branch.


The production and decay cross section estimates in this section are
calculated using {MadGraph/MadEvent} \cite{Alwall:2007st}
and using SUSY mass spectra inputs from {SuSpect} \cite{Djouadi:2002ze}.
We simulate the typical LHC detector lepton energy resolution
\cite{AtlasTDR,CMSTDR} by scaling the $\alpha$
and $\beta$ four vectors by a scalar normally distributed about $1$ with the
width of
\begin{equation}
\frac{\delta\alpha_{0}}{\alpha_{0}} = \frac{0.1}{\sqrt{\alpha_{o}
(\mathrm{GeV})}} + \frac{0.003}{ \alpha_{o} (\mathrm{GeV})} + 0.007.
\end{equation}
The missing transverse momentum is assumed to be whatever is missing to conserve the transverse momentum after the smearing of the leptons momenta.  We do not account for the greater uncertainty in missing momentum from hadrons or from muons which do not deposit all their energy in the calorimeter and whose energy resolution is therefore correlated to the missing momentum.  Including such effects is considered in Chapter \ref{ChapterM2CwUTM}.
These finite resolution effects are simulated in both the
determination of the ideal distribution and in the small sample of events that
is fit to the ideal distribution to determine $M_{Y}$ and $M_{N}$.  We do not expect expanded energy resolutions to greatly affect the results because the resolution effects are included in both the simulated events and in the creation of the ideal curves which are then fit to the low statistics events to estimate the mass.

We consider models where the three-body decay channel for $\tilde{\chi}%
_{2}^{o}$ will dominate. These models must satisfy $M_{\tilde{\chi}_{2}^{o}%
}-M_{\tilde{\chi}_{1}^{o}}<M_{Z}$ and must have all slepton masses greater
than the $M_{\tilde{\chi}_{2}^{o}}$. The models considered are shown in Table
\ref{TableModels}. The Min-Content model assumes that there are no other SUSY
particles accessible at the LHC other than $\tilde{\chi}_{2}^{o}$ and
$\tilde{\chi}_{1}^{o}$ and we place $M_{\tilde{\chi}_{1}^{o}}$ and
$M_{\tilde{\chi}_{2}^{o}}$ at the boundary of the PDG Live exclusion limit
\cite{PDBook2006}. SPS 6, P1, and $\gamma$ are models taken from references
\cite{Allanach:2002nj}, \cite{VandelliTesiPhD}, and \cite{DeRoeck:2005bw}
respectively. Each has the $\tilde{\chi}^o_{2}$ decay channel to leptons via a
three-body decay kinematically accessible. We will only show simulation results for the masses in model P1 and SPS 6 because they have the extreme values of $M_{+}/M_{-}$ with which the performance scales.   The Min-Content model and the $\gamma$ model are included to demonstrate the range of the masses and production cross sections that one might expect.

Bisset, Kersting, Li, Moortgat, Moretti, and Xie (BKLMMX) \cite{Bisset:2005rn}
have studied the four lepton with missing transverse momentum Standard-Model background for the
LHC. They included contributions from jets misidentified as leptons and
estimated about $190$ background events at a ${\mathcal{L}}=300\ \mathrm{fb}%
^{-1}$ which is equivalent to $0.6$ fb. Their background study made no
reference to the invariant mass squared of the four leptons, so we only
expects a fraction of these to have both lepton pairs to have invariant masses
less than $M_{-}$. Their analysis shows the largest source of backgrounds will
most likely be other supersymmetric states decaying to four leptons. Again,
we expect only a fraction of these to have both lepton pairs with invariant
masses within the range of interest. The background study of BKLMMX is
consistent with a study geared towards a $500$ GeV $e^{+}$ $e^{-}$ linear
collider in Ref.~\cite{Ghosh:1999ix} which predicts $0.4$ fb for the standard
model contribution to 4 leptons and missing transverse momentum.   The neutralino decay to $\tau$ leptons also provide a background because the $\tau$ decay to a light leptons $l=e,\mu$  ($\Gamma_{\tau \rightarrow l \bar{\nu}_l} / \Gamma \approx 0.34$) cannot be distinguished from prompt leptons. The neutrinos associated with these light leptons will be new sources of missing transverse momentum and will therefore be a background to our analysis.  The di-$\tau$ events will only form a background when both opposite sign same flavor $\tau$s decay to the same flavor of light lepton which one expects about 6\% of the time.

\begin{table}[ptb]%
\footnotesize
\scriptsize
\begin{tabular}
[c]{|c|c|c|c|c|}\hline
Model & Min Content (ref \cite{PDBook2006}) & SPS 6 (ref
\cite{Allanach:2002nj}) & P1 (ref \cite{VandelliTesiPhD}) & $\gamma$ ( Ref.~\cite{DeRoeck:2005bw})\\\hline
Definition &
\begin{tabular}
[c]{l}%
$\tilde{\chi}^{o}_{1}$ and $\tilde{\chi}^{o}_{2}$\\
are the only\\
LHC accessible\\
SUSY States\\
with smallest\\
allowed masses.
\end{tabular}
&
\begin{tabular}
[c]{l}%
Non Universal\\
Gaugino Masses\\
$m_{o}=150$ GeV\\
$m_{1/2} = 300$ GeV\\
$\tan\beta= 10$\\
$\mathrm{sign}(\mu) = +$\\
$A_{o}=0$\\
$M_{1}=480$ GeV\\
$M_{2}=M_{3}=300$ GeV
\end{tabular}
&
\begin{tabular}
[c]{l}%
mSUGRA\\
$m_{o}=350$ GeV\\
$m_{1/2} = 180$ GeV\\
$\tan\beta= 20$\\
$\mathrm{sign}(\mu) = +$\\
$A_{o}=0$%
\end{tabular}
&
\begin{tabular}
[c]{l}%
Non-Universal\\
Higgs Model\\
$m_{o} = 330$ GeV\\
$m_{1/2}=240$ GeV\\
$\tan\beta= 20$\\
$\mathrm{sign}(\mu) = +$\\
$A_{o}=0$\\
$H_{u}^{2} = -(242\,\mathrm{GeV})^{2}$\\
$H_{d}^{2} = +(373\,\mathrm{GeV})^{2}$\\
\end{tabular}
\\\hline
$M_{\tilde{\chi}^{o}_{1}}$ & $46$ GeV & $189$ GeV & $69$ GeV & $95$
GeV\\\hline
$M_{\tilde{\chi}^{o}_{2}}$ & $62.4$ GeV & $219$ GeV & $133$ GeV & $178$
GeV\\\hline
$M_{+}/M_{-}$ & $6.6$ & $13.6$ & $3.2$ & $3.3$\\\hline
\end{tabular}
\normalsize
\caption{Models with $\tilde{\chi}^{o}_{2}$ decaying via a three-body decay to
leptons.  We only show simulation results for the masses in model P1 and SPS 6 because they have the extreme values of $M_{+}/M_{-}$ with which the performance scales. }%
\label{TableModels}%
\end{table}

\begin{table}[ptb]%
\scriptsize
\begin{tabular}
[c]{|c|c|c|c|}\hline
Model &
\begin{tabular}
[c]{l}%
$\sigma_{\tilde{\chi}^{o}_{2}\,\tilde{\chi}^{o}_{2}}$ Direct\\
$\sigma_{\tilde{\chi}^{o}_{2}\,\tilde{\chi}^{o}_{2}}$ Via $\tilde{g}$ or
$\tilde q$\\
\end{tabular}
&
\begin{tabular}
[c]{l}%
$\mathrm{BR}_{\tilde{\chi}^{o}_{2} \rightarrow l + \bar{l} + \tilde{\chi}%
^{o}_{1}}$\\
$\mathrm{BR}_{\tilde{\chi}^{o}_{2} \rightarrow q + \bar{q} + \tilde{\chi}%
^{o}_{1}}$%
\end{tabular}
&
\begin{tabular}
[c]{l}%
Events with\\
$4\,$ leptons $+ E_{T}$ missing\\
+ possible extra jets\\
${\mathcal{L}}=300\ \mathrm{fb}^{-1}$%
\end{tabular}
\\\hline
Min Content &
\begin{tabular}
[c]{l}%
$2130$ fb\\
N/A
\end{tabular}
&
\begin{tabular}
[c]{l}%
0.067\\
0.69
\end{tabular}
& 2893\\\hline
SPS 6 &
\begin{tabular}
[c]{l}%
$9.3$ fb\\
$626$ fb
\end{tabular}
&
\begin{tabular}
[c]{l}%
0.18\\
0.05
\end{tabular}
& 6366\\\hline
P1 &
\begin{tabular}
[c]{l}%
$35$ fb\\
$12343$ fb
\end{tabular}
&
\begin{tabular}
[c]{l}%
0.025\\
0.66
\end{tabular}
& 2310\\\hline
$\gamma$ &
\begin{tabular}
[c]{l}%
$17$ fb\\
$4141$ fb
\end{tabular}
&
\begin{tabular}
[c]{l}%
0.043\\
0.64
\end{tabular}
& 2347\\\hline
\end{tabular}
\normalsize
\caption{The approximate breakdown of signal events. }%
\label{TableEventCounts}%
\end{table}

\begin{figure}[ptb]
\centerline{\includegraphics[width=4in]{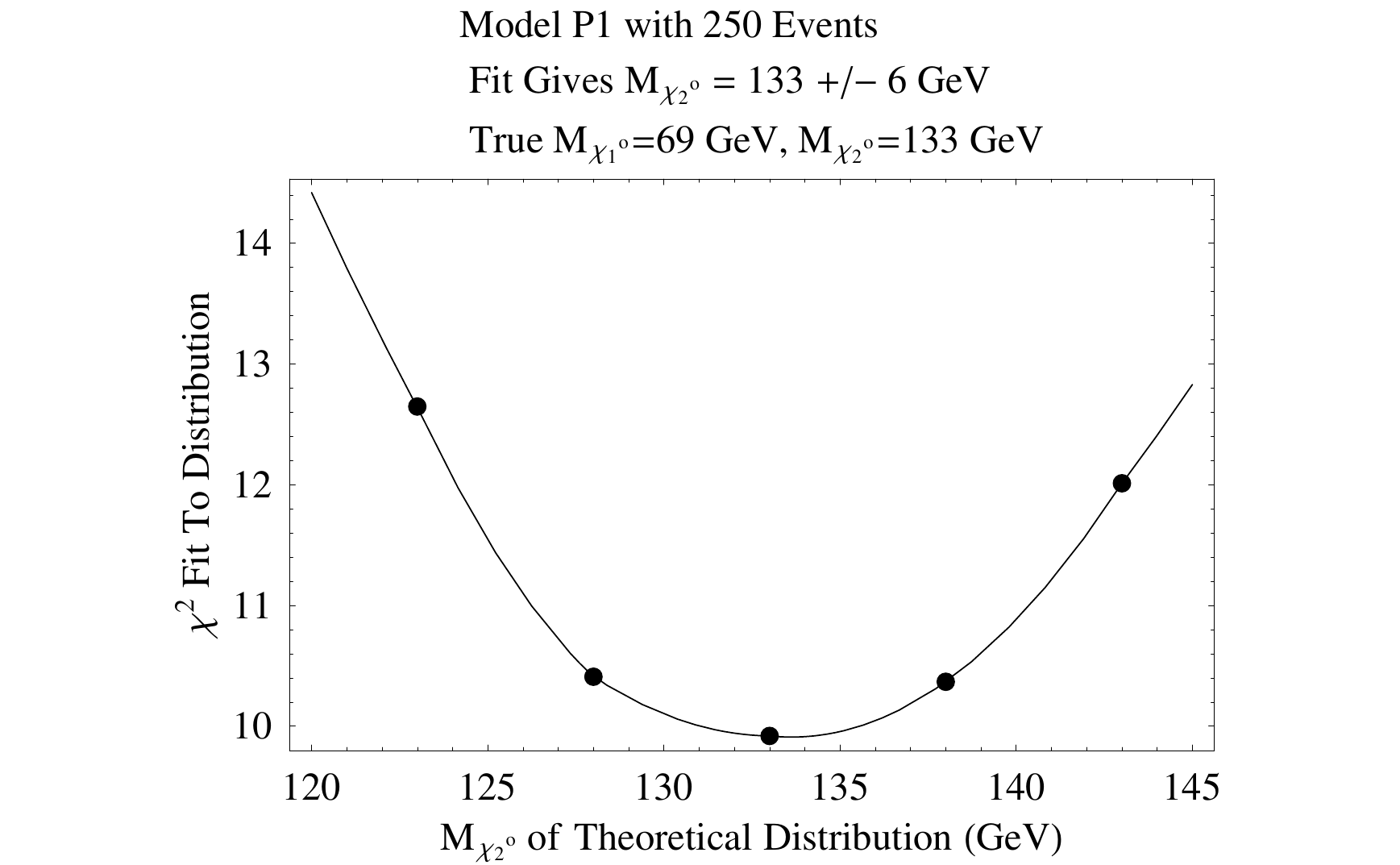}}\caption{$\chi^{2}$
fit of 250 events from model P1 of Ref \cite{VandelliTesiPhD} to the
theoretical distributions calculated for different $M_{\chi_{2}^{o}}$ values
but fixed $M_{\chi_{2}^{o}}-M_{\chi_{1}^{o}}$. \ The fit gives $M_{\chi
_{2}^{o}}=133\pm6$ GeV. }%
\label{FigP1ChiSqFitExample}%
\end{figure}

\begin{figure}[ptb]
\centerline{\includegraphics[width=4in]{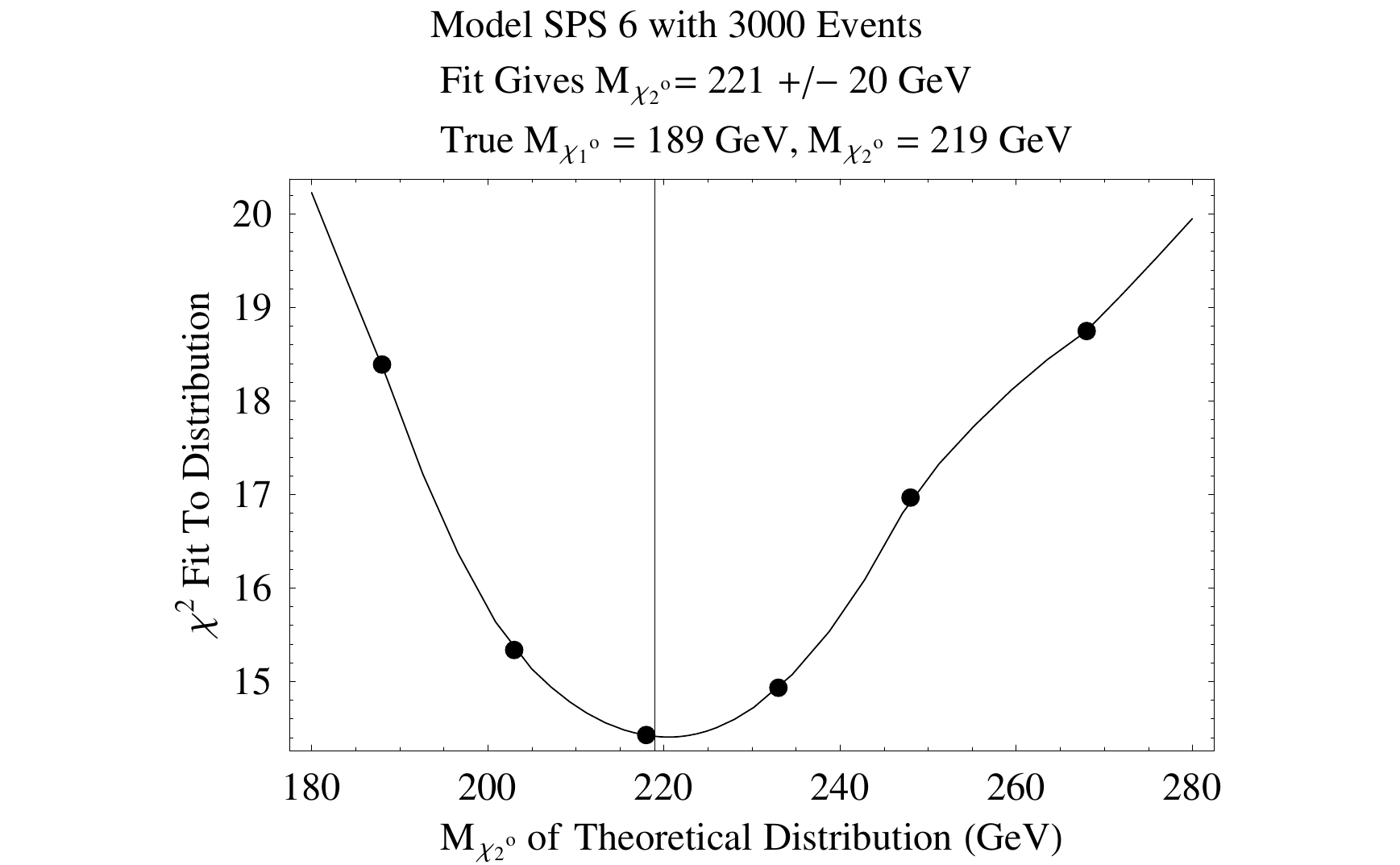}}\caption{$\chi^{2}$ fit
of 3000 events from model SPS 6 of Ref \cite{Allanach:2002nj} to the
theoretical distributions calculated for different $M_{\chi_{2}^{o}}$ values
but fixed $M_{\chi_{2}^{o}}-M_{\chi_{1}^{o}}$. The fit gives $M_{\chi_{2}^{o}%
}=221\pm20$ GeV. }%
\label{FigSPS6ChiSqFitExample}%
\end{figure}

Table \ref{TableEventCounts} breaks down the LHC production cross section for
pair producing two $\tilde{\chi}^{o}_{2}$ in each of these models. In the
branching ratio to leptons, we only consider $e$ and $\mu$ states as the
$\tau$ will decay into a jet and a neutrino introducing more missing transverse momentum.
Direct pair production of $\tilde{\chi}^{o}_{2}$ has a rather modest cross
section, but production via a gluino or squark has a considerably larger cross
section but will be accompanied by additional QCD jets. We do expect to be
able to distinguish QCD jets from $\tau$ jets \cite{2005NuPhS.144..341T}.
The events with gluinos and jets will lead to considerable $k_T$.  In this chapter we assume $k_T \lesssim 20$ GeV, but we take up the case of large $k_T$ in the following
chapter.

We now estimate how well we may be able to measure $M_{\tilde{\chi}_{1}^{o}}
$ and $M_{\tilde{\chi}_{2}^{o}}$ in these models under these simplifying ideal
assumptions. Figures
\ref{FigP1ChiSqFitExample} and \ref{FigSPS6ChiSqFitExample} show a $\chi^{2}$
fit\footnote{See Appendix \ref{AppendixChiSqFitting} for details of how $\chi^{2}$ is calculated.} of
the $M_{2C}$ distribution from the observed small set of events to `ideal'
theoretical $M_{2C}$ distributions parameterized by $M_{\tilde{\chi}_{2}^{o}}%
$. The `ideal' theoretical distributions are calculated for the observed value
of $M_{-}$ using different choices for $M_{\tilde{\chi}_{2}^{o}}$. A
second-order interpolation is then fit to these points to estimate the value
for $M_{\tilde{\chi}_{2}^{o}}$. The $1\,\sigma$ uncertainty for $M_{\tilde
{\chi}_{2}^{o}}$ is taken to be the points where the $\chi^{2}$ increases from
its minimum by one.

The difficulty of the mass determination from the distribution grows with the
ratio $M_{+}/M_{-}.$ Figures \ref{FigP1ChiSqFitExample} and
\ref{FigSPS6ChiSqFitExample} show the two extremes among the cases we
consider. For the model P1 $M_{+}/M_{-}=3.2$, and for model $\gamma$
$M_{+}/M_{-}=3.3$. Therefore these two models can have the mass of
$M_{\tilde{\chi}_{2}^{o}}$ and $M_{\tilde{\chi}_{1}^{o}}$ determined with
approximately equal accuracy with equal number of signal events. Figure
\ref{FigP1ChiSqFitExample} shows that we may be able to achieve $\pm6$ GeV
resolution after about $30\ \mathrm{fb}^{-1}$. Model SPS 6 shown in Fig \ref{FigSPS6ChiSqFitExample} represents a much harder case because
$M_{+}/M_{-}=13.6$. In this scenario we can only achieve $\pm20$ GeV
resolution with $3000$ events corresponding to approximately $150\,\mathrm{fb}%
^{-1}$.
In addition to these uncertainties, we need to also consider the
error propagated from $\delta M_{-}$ in Eq(\ref{EqDeltaMmErrorEffectsDirect}).


\section*{Chapter Summary}

\label{SecConclusions}

We have proposed a method to extract the masses of new pair-produced states
based on a kinematic variable, $M_{2C}$, which incorporates all the known
kinematic constraints on the observed process and whose endpoint determines
the new particle masses.
However the method does not rely solely on the
endpoint but uses the full data set, comparing the observed distribution for
$M_{2C}$ with the ideal distribution that corresponds to a given mass. As a
result the number of events needed to determine the masses is very
significantly reduced so that the method may be employed at the LHC event for
processes with electroweak production cross sections.

This chapter is an initial feasibility study of the method for several
supersymmetric models.
We have made many idealized assumptions which amount to conditions present if the new particle states are directly produced.
We have included  the effect of detector resolution but not
backgrounds, or cuts. Our modeling assumed that $k=0$.
We demonstrated that for
some of the models studied we are able to determine the masses to within 6 GeV
from only 250 events.
This efficiency is encouraging,
a study
including more of the real-world complications follows in Chapter \ref{ChapterM2CwUTM}.

The constrained mass variables we advocate here can be readily extended to other processes.
By incorporating all the known kinematical constraints, the information away from
kinematical end-points can, with some mild process-dependent information, be
used to reduce the number of events needed to get mass measurements. We shall
illustrate an extension to three on-shell states in Chapter \ref{ChapterM3C}.

\chapter{The Variable $M_{2C}$: Significant Upstream Transverse Momentum}
\label{ChapterM2CwUTM}

\section*{Chapter Overview}



In the previous chapter, we introduced the $M_{2C}$ kinematic variable which gives an event-by-event lower
bound on the dark-matter particle's absolute mass given the mass difference between the dark matter candidate and its parent.  The previous chapter focused on direct pair production with minimal initial state radiation (ISR).
In this chapter, we introduce a complementary variable $M_{2C,UB}$ which gives an event-by-event \emph{upper} bound on the same absolute mass.
The complementary variable is only relevant in the presence of large upstream transverse momentum (UTM).
Our study shows that the technique presented is as good if not better than other model-independent invisible-particle mass determination techniques in precision and accuracy.

In this chapter, which is based on work first published by Barr, Ross, and the author in Ref~\cite{Barr:2008ba}, we demonstrate the use of the variable $M_{2C}$ and $M_{2C,UB}$ in LHC conditions.
The variables $M_{2C}(\Deltam)$ and $M_{2C,UB}(\Deltam)$ give an event-by-event lower-bound and upper-bound respectively on the mass of $Y$ assuming the topology in Fig.~\ref{FigEventTopologyThreeBody} and the mass difference $\Deltam=M_Y-M_N$.  To get the mass difference, we use events where $Y$ decays into $N$ and two visible states
via a three-body decay
in which we can easily determine the mass difference from the end point of the visible-states
invariant-mass distribution, $m^2_{12}$.
One might also conceive of a situation with $M_{2C}$ supplementing an alternative technique that gives a tight constraint on the mass difference but may have multiple solutions or a weaker constraint on the mass scale \cite{Gjelsten:2004ki}\cite{Serna:2008zk}.
Given this mass difference and enough statistics, $M_{2C}$'s endpoint gives the mass of $Y$.
However the main advantage of the $M_{2C}$ method is that it does not rely simply on the position at the endpoint but it uses the additional information contained in events which lie far from the endpoint.  As a result it gives a mass determination using significantly fewer events and is less sensitive to energy resolution and other errors.

To illustrate the method, in this chapter we study in detail the performance of the $M_{2C}$ constrained mass variable in a specific supersymmetric model.
We study events where each of the two branches have decay chains that end with a $\N{2}$ decaying to a $\N{1}$ and a pair of opposite-sign same-flavor (OSSF) leptons.
Thus the final states of interest contain four isolated leptons
(made up of two OSSF pairs) and missing transverse momentum.
Fig~\ref{FigEventTopologyThreeBody} defines the four momentum of the particle states with $Y=\N{2}$, $N=\N{1}$, and the OSSF pairs forming the visible particles $1-4$.
Any decay products early in the decay chains of either branch are grouped into $k$ which we generically refer to as upstream transverse momentum (UTM).
Nonzero $k$ could be the result of initial state radiation (ISR) or decays of heavier particles further up the decay chain.
Events with four leptons and missing transverse momentum have a very small Standard-Model background.
To give a detailed illustration of the $M_{2C}$ methods, we have chosen to analyze
the benchmark point P1 from \cite{VandelliTesiPhD} which corresponds to mSUGRA with $m_o=350$ GeV, $m_{1/2}=180$ GeV,
$\tan \beta=20$, ${\rm{sign}}(\mu)=+$, $A_o=0$. Our SUSY particle spectrum was calculated with ISAJET \cite{Paige:2003mg} version~7.63.
We stress that the analysis technique employed applies generically to models involving decays to a massive particle state that leaves the detector unnoticed.

A powerful feature of the $M_{2C}$ distribution is that, with some mild assumptions, the shape away from the endpoint can be entirely determined from the
unknown mass scale and quantities that are measured.   The ideal shape fit against early data therefore provides an early mass estimate for the invisible particle.
This study is meant to be a guide on how to overcome difficulties in establishing and fitting the shape: difficulties from combinatoric issues, from differing energy resolutions for the leptons, hadrons,
and missing transverse momentum, from backgrounds, and from large upstream transverse momentum (UTM) \footnote{Our references to UTM correspond to the Significant Transverse Momentum (SPT), pair production category in \cite{Barr:2007hy} where SPT indicates that the relevant pair of parent particles can be seen as recoiling against a significant transverse momentum.}.  As we shall discuss, UTM actually provides surprising benefits.

The chapter is structured as follows:  In Section \ref{SecUB}, we review $M_{2C}$ and introduce the new observation that, in addition to an event-by-event lower bound on $M_Y$,  large recoil against UTM enables one also to obtain an event-by-event \emph{upper} bound on $M_Y$.  We call this quantity $M_{2C,UB}$.   Section
\ref{SecModeling} describes the modeling and simulation employed.   Section \ref{SecShapeFactors} discusses the implications of several effects on the shape of the distribution including the $m_{12}$ (in our case $m_{ll}$) distribution, the UTM distribution, the backgrounds, combinatorics, energy resolution, and missing transverse
momentum cuts.  In Section \ref{SecPerformance}, we put these factors together and estimate the performance.  We conclude in Section \ref{M2CUTMSecConclusions} with a discussion about the performance in comparison to previous work.


%



\section{Upper Bounds on $M_Y$ from Recoil against UTM}
\label{SecUB}

We will now review the definition of $M_{2C}$ as providing an event-by-event lower bound on $M_Y$.
In generalizing this framework, we find a new result that one can
also obtain an upper bound on the mass $M_Y$ when the
two parent particles $Y$ recoil against some large upstream transverse momentum $k_T$.

\subsection{Review of the Lower Bound on $M_Y$}

Fig \ref{FigEventTopologyThreeBody} gives the relevant topology and the momentum assignments.  The visible particles $1$ and $2$ and invisible particle $N$ are labeled with momentum $\alpha_1$ and $\alpha_2$ (which we group into $\alpha=\alpha_1+\alpha_2$) and $p$, respectively $\beta=\beta_1+\beta_2$ and $q$ in the other branch.  We assume that the parent particle $Y$ is the same in both branches so $(p+\alpha)^2=(q+\beta)^2$.  Any earlier decay products of either branch are grouped into the upstream transverse momentum (UTM) 4-vector momentum, $k$.

In the previous chapter  we  showed how to find an event-by-event lower bound on the true mass of $M_N$ and $M_Y$.    We assume that the mass difference $\Deltam = M_Y - M_N$ can be accurately measured from the invariant mass edges $\Max m_{12}$ or $\Max m_{34}$.
For each event, the variable $M_{2C}$ is the minimum value of the mass of $Y$ (the second lightest state) after minimizing over the unknown division of the
missing transverse momentum $\slashed{P}_T$ between the two dark-matter particles $N$ as described in Eq(\ref{mymin}-\ref{eqC4}).

One way of calculating $M_{2C}$ for an event is to use $M_{T2}(\chi_N)$ \cite{Lester:1999tx,Barr:2003rg,Ross:2007rm}, which provides a lower bound on the mass of $Y$ for an assumed mass $\chi_N$ of $N$.  The true mass of $Y$ lies along the line $\chi_Y( \chi_N)=\Deltam + \chi_N$ where we use $\chi_Y$ to denote the possible masses of $Y$ and to distinguish it from the true mass of $Y$ denoted with $M_Y$.
In other words $M_{T2}$ provides the constraint $\chi_Y( \chi_N) \ge M_{T2}(\chi_N)$.
Thus we can see that for $\chi_N$ to be compatible with an event, we must have $M_{T2}(\chi_N) \leq \chi_Y( \chi_N) = \chi_N + \Deltam$.

For a given event, if one assumes a mass $\chi_N$ for $N$, and if the inequality $M_{T2}(\chi_N) \le \chi_N + \Deltam$ is satisfied, then there is no contradiction, and the event is compatible with this value of $\chi_N$.
If however, $M_{T2}(\chi_N) > \Deltam+\chi_N$, then we have a contradiction, and the event excludes this value $\chi_N$ as a viable mass of $N$.   Using this observation, $M_{2C}$ can be found for each event by seeking the intersection between $M_{T2}(\chi_N)$ and $\chi_N+\Deltam$ \cite{Ross:2007rm}.
Equivalently, the lower bound on $M_Y$ is given by $M_Y \ge \Deltam +\chi_N^o$ where $\chi_N^o$ is the zero of
  \begin{eqnarray}
    g(\chi_N)=M_{T2}(\chi_N) - \chi_N - \Deltam \label{EqgChi} \nonumber \\
    {\rm{with}}\ \ \ g'(\chi_N^o) < 0. \label{EqGprime}
     \end{eqnarray}
In the case $k=0$, the extreme events analyzed in CCKP \cite{Cho:2007qv} demonstrate that $g(\chi)$ will only have one positive zero or no positive zeros, and the slope at a zero will always be negative.
For no positive zeros, the lower bound is the trivial lower bound given by $\Deltam$.
 Note that a lower bound on the value of $M_Y$ corresponds to a lower bound on the value of $M_N$.
The Appendix in Ref.~\cite{Ross:2007rm} shows that at the
zeros of $g(\chi_N)$ which satisfy Eq(\ref{EqGprime}), the momenta satisfy Eqns(\ref{eqC1}-\ref{eqC4}).

\subsection{A New Upper Bound on $M_Y$}

If there is large upstream transverse momentum (UTM) ($k_T \gtrapprox \Deltam $)
against which the system recoils, then we find a new result.
Using the $M_{T2}$ method to calculate $M_{2C}$ gives one the immediate ability to see that $M_Y$ can also have an upper bound when requiring Eqns(\ref{eqC1}-\ref{eqC4}).
This follows because for large UTM the function $g(\chi_N)$ may have two zeros\footnote{There may be regions in parameter space where function $g(\chi)$ has more than two zeros, but we have not encountered such cases in our simulations.}
which provides both an upper and a lower bound for $M_Y$ from a single event. We have also found regions of parameter space where $g(\chi)$ has a single zero but $g'(\chi_N^o) > 0$ corresponding to an upper bound on the true mass of $M_N$ ( and $M_Y$) and
only the trivial lower bound of $M_N \ge 0$.

We can obtain some insight into the cases in which events with large UTM provides upper bounds on the mass by studying a class of extreme event with two hard jets, $j_\alpha$ and $j_\beta$ against which $Y$ recoils ($k=j_\alpha+j_\beta$).  We will describe this extreme event and solve for the regions of parameter space for which one can analytically see the intersection points giving a lower bound and/or an upper bound.
The event is extremal in that $M_{T2}(\chi_N)$, which gives a lower bound on $M_Y$, actually gives the true value of $M_Y$ when one selects $\chi_N$ equal to the true mass $M_N$.

The ideal event we consider is where a heavier state $G$ is pair-produced on shell at threshold.  For simplicity we assume the lab-frame is the collision frame.
Assume that the $G$s, initially at rest, decay into
visible massless jets $j_\alpha$, $j_\beta$ and the two $Y$ states with the decay product momenta $\alpha+p$ and $\beta+q$.
Both jets have their momenta in the same transverse plane along the negative $\hat{x}$-axis,
and both $Y$'s momentum are directed along the $\hat{x}$-axis.  Finally, in the rest frame of the two $Y$s, both decay such that the decay products visible states have their momentum $\alpha$ and $\beta$ along the $\hat{x}$-axis and both invisible massive states $N$ have their two momenta along the negative $\hat{x}$-axis. In the lab frame, the four-vectors are given by
 \begin{eqnarray}
   j_\alpha = j_\beta & = &  \frac{M_G}{2}\left(1 - \frac{M_Y^2}{M_G^2} \right) \left\{ 1 , -1 , 0, 0 \right\} \label{eqEvent1A} \\
   \alpha = \beta & = & \frac{M_G}{2}\left(1 - \frac{M_N^2}{M_Y^2} \right) \left\{ 1 , 1 ,0 ,0  \right\} \label{eqEvent1B} \\
   p=q & = & \frac{M_G}{2} \left\{
    \left( \frac{M_N^2}{M_Y^2} + \frac{M_Y^2}{M_G^2} \right),
   \left( \frac{M_N^2}{M_Y^2} - \frac{M_Y^2}{M_G^2} \right)  ,0 ,0\right\} \label{eqEvent1C}.
 \end{eqnarray}

For the event given by Eqns(\ref{eqEvent1A}-\ref{eqEvent1C}), we can exactly calculate $M_{T2}(\chi_N)$:
 \begin{eqnarray}
    M^2_{T2}(\chi_N) & = &  \chi_N ^2+ \frac{\left(M_N^2-M_Y^2\right)\left(M_N^2 M_G^2-M_Y^4\right)}{2\,  M_Y^4} \nonumber \\
   & & +\frac{\left(M_Y^2-M_N^2\right)\sqrt{4\,
   M_G^2 \,\chi_N ^2 M_Y^4+\left(M_Y^4-M_N^2
   M_G^2\right)^2} }{2\,  M_Y^4}.
 \end{eqnarray}
This is found by calculating the transverse mass for each branch while assuming $\chi_N$ to be the mass of $N$.
The value of $p_x$ is chosen so that the transverse masses of the two branches are equal.  Substituting this value back into the transverse mass of either branch gives $M_{T2}(\chi_N)$.

\begin{figure}
\centerline{\includegraphics[width=4in]{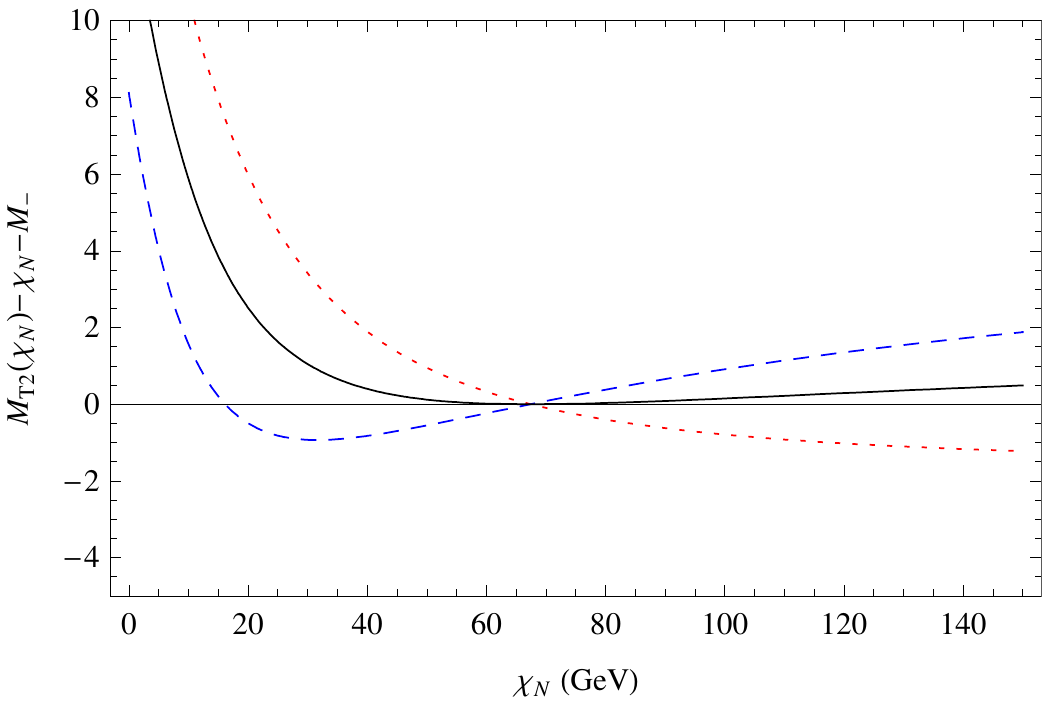}}
\caption{\label{FigExtremeMT2} Shows $g(\chi_N)$ for the extreme event in Eq(\ref{eqEvent1A}-\ref{eqEvent1C}) with $\Deltam=53 \GeV$ and $M_N=67.4 \GeV$. The red dotted line has $M_G=150 \GeV$ and shows an event providing a lower bound on $M_Y$. The blue dashed line $M_G=170 \GeV$ and shows an event with both a lower bound and an upper-bound on $M_Y$. The black solid line shows $M_G=\sqrt{M_Y^3/M_N}$ where the lower bound equals the upper bound.}
\end{figure}

Fig \ref{FigExtremeMT2} shows $g(\chi_N)$, given in Eq(\ref{EqgChi}), for several choices of $M_G$ for the process described by Eqs(\ref{eqEvent1A}-\ref{eqEvent1C}) with $\Deltam=53 \GeV$ and $M_N=67.4 \GeV$.
Because $G$ is the parent of $Y$, we must have $M_G > M_Y$.  If $M_Y < M_G < 2 M_Y^2 /(M_N+M_Y)$, then $M_{T2}(\chi_N < M_N)$ is larger than $\chi_N + \Deltam$ up until their point of intersection at $\chi_N=M_N$. In this case their point of intersection provides a lower bound as illustrated by the dotted line in Fig.~\ref{FigExtremeMT2} for the case with $M_G=150 \GeV$.
For $ 2 M_Y^2 /(M_N+M_Y) < M_G < \sqrt{M_Y^3/M_N}$ there are two solutions
 \begin{eqnarray}
 \chi_{N,{\rm{Min}}} & = & M_N \\
 \chi_{N,{\rm{Max}}} & = & \frac{(M_N-M_Y) \left(-2 M_Y^4+M_N M_G^2 M_Y+M_N^2 M_G^2\right)}{(M_N M_G+(M_G-2 M_Y) M_Y) (M_N M_G+M_Y (M_G+2 M_Y))}.
 \end{eqnarray}
When $M_G=\sqrt{M_Y^3/M_N}$, the function $g(\chi_N)$ has only one zero with the lower bound equalling the upper bound at $M_N$. The solid line in Fig.~\ref{FigExtremeMT2} shows this case.
Between $\sqrt{M_Y^3/M_N} < M_G < \sqrt{6} M_Y^2 / \sqrt{(M_N+M_Y)(2 M_N + M_Y)}$ we again have two solutions but this time with
\begin{eqnarray}
 \chi_{N,{\rm{Min}}} & = & \frac{(M_N-M_Y) \left(-2 M_Y^4+M_N M_G^2 M_Y+M_N^2 M_G^2\right)}{(M_N M_G+(M_G-2 M_Y) M_Y) (M_N M_G+M_Y (M_G+2 M_Y))} \\
 \chi_{N,{\rm{Max}}} & = & M_N.
 \end{eqnarray}
The dashed line in Fig.~\ref{FigExtremeMT2} shows this case with  $M_G=170 \GeV$.
For $M_G$ greater than this, we have $\chi_{N,{\rm{Max}}}=M_N$ and  $\chi_{N,{\rm{Min}}}=0$.

This example illustrates how $M_{2C}$ can provide both a
lower-bound and an upper-bound on the true mass
for those events with large UTM.
The upper-bound distribution provides extra information that can also be used to improve early mass determination,
and in what follows we will refer to the upper bound as $M_{2C,UB}$.  We now move on to discuss  modeling and simulation of this new observation.

\section{Modeling and Simulation}
\label{SecModeling}
\label{M2CSecModeling}

As a specific example of the application of the $M_{2C}$ method, we have chosen a supersymmetry model  mSUGRA, $m_o=350$ GeV, $m_{1/2}=180$ GeV, $\tan \beta=20$, ${\rm{sign}}(\mu)=+$, $A_o=0$ \footnote{ This was model $P1$ from \cite{VandelliTesiPhD} which we also used in \cite{Ross:2007rm}.}.  The spectrum used in the simulation has $M_{\N{1}}=67.4 \GeV$ and 
$M_{\N{2}}=120.0 \GeV$.
We have employed two simulation packages.  One is a Mathematica code that creates the `ideal' distributions based only on very simple assumptions and input data.  The second is \herwig\ \cite{Corcella:2002jc,Moretti:2002eu,Marchesini:1991ch} which simulates events based on a SUSY spectrum, MSSM cross sections, decay chains, and appropriate parton distribution functions.
If the simple Mathematica simulator predicts `ideal' shapes that agree with \herwig\ generator, then we have reason to believe that
all the relevant factors relating to the shape are identified in the simple Mathematica simulation.
This is an important check in validating the benefits of fitting the $M_{2C}$ and $M_{2C,UB}$ distribution shape as a method to measure the mass of new invisible particles produced at hadron colliders.

\subsection{Generation of ``Ideal'' Distributions}

Our `ideal' distributions are produced from a home-grown Monte Carlo event generator written in Mathematica.  This generater serves to ensure that we understand the origin of the distribution shape. It also ensures that we have control over measuring the parameters needed to determine the mass without knowing the full model, coupling coefficients, or parton distribution functions. We also use this simulation to determine on what properties the ideal distributions depends.

The simulator is used to create events satisfying the topology shown in Fig~\ref{FigEventTopologyThreeBody} for a set of specified masses.
We take as given the previously measured  mass difference $M_{\N{2}}-M_{\N{1}} = 52.6 \GeV$, which we use in this chapter's simulations.
We neglect finite widths of the particle states as most are in the sub GeV range for the  model we are considering. We neglect spin correlations between the two branches.  We perform the simulations in the center-of-mass frame because $M_{2C}$ and $M_{2C,UB}$ are transverse observables and are invariant under longitudinal boosts.  The collision energy $\sqrt{s}$ is distributed according to normalized distribution
 \begin{equation}
 \rho(\sqrt{s}) = 12 \,M_{\N{2}}^2 \frac{\sqrt{s- 4\, M_{\N{2}}^2}}{s^2} \label{EqSdep}
 \end{equation}
unless otherwise specified.  The $\N{2}$ is produced with a uniform angular distribution, and all subsequent decays have uniform angular distribution in the rest frame of the parent.  The UTM is simulated by making $k_T$ equal to the UTM with $k^2=(100 \GeV)^2$ (unless otherwise specified), and boosting the other four-vectors of the event such that the total transverse momentum is zero.
As we will show, these simple assumptions capture the important elements of the process.
Being relatively model independent, they provide a means of
determining the mass for various production mechanisms.
If one were to assume detailed knowledge of the production process, it would be
possible to obtain a better mass determination by using a more complete simulation
like \herwig\
to provide the `ideal' distributions against which one compares with the data.
Here we concentrate on the more model independent simulation to demonstrate that it predicts
the $M_{2C}$ and $M_{2C,UB}$ distributions well-enough to perform the mass determination that we demonstrate in this case-study.

\subsection{\herwig\  ``Data''}

\begin{figure}
\centerline{\includegraphics[width=3.1in]{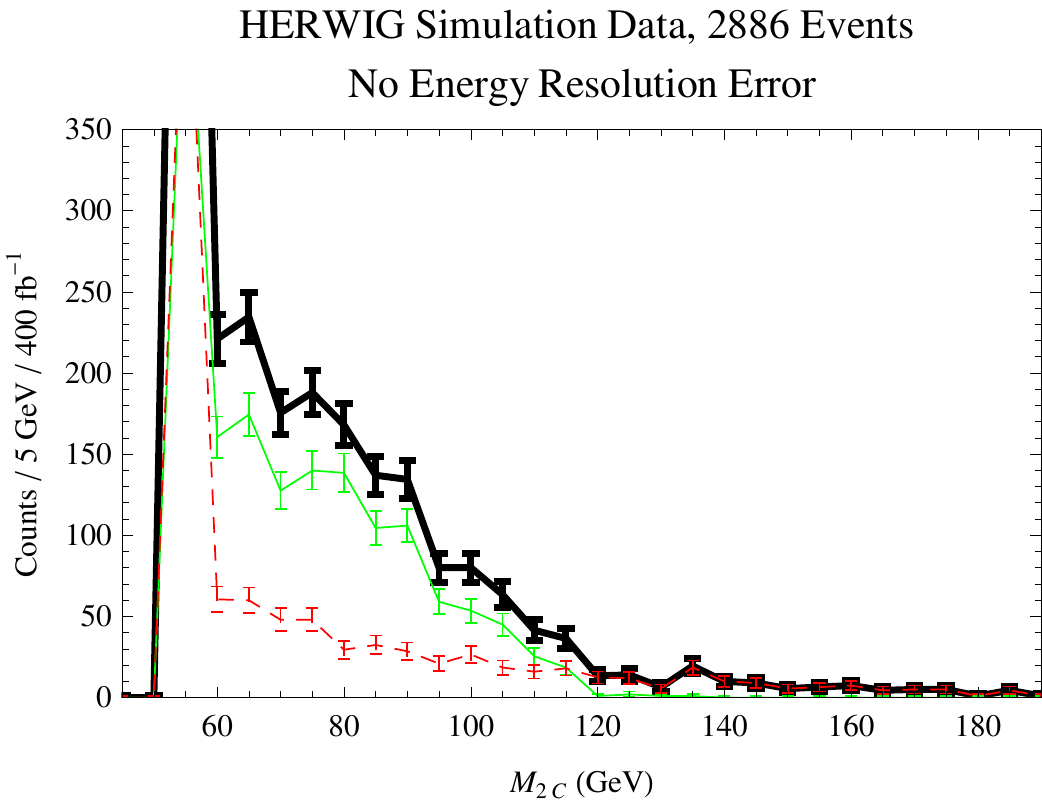}\ \includegraphics[width=3.1in]{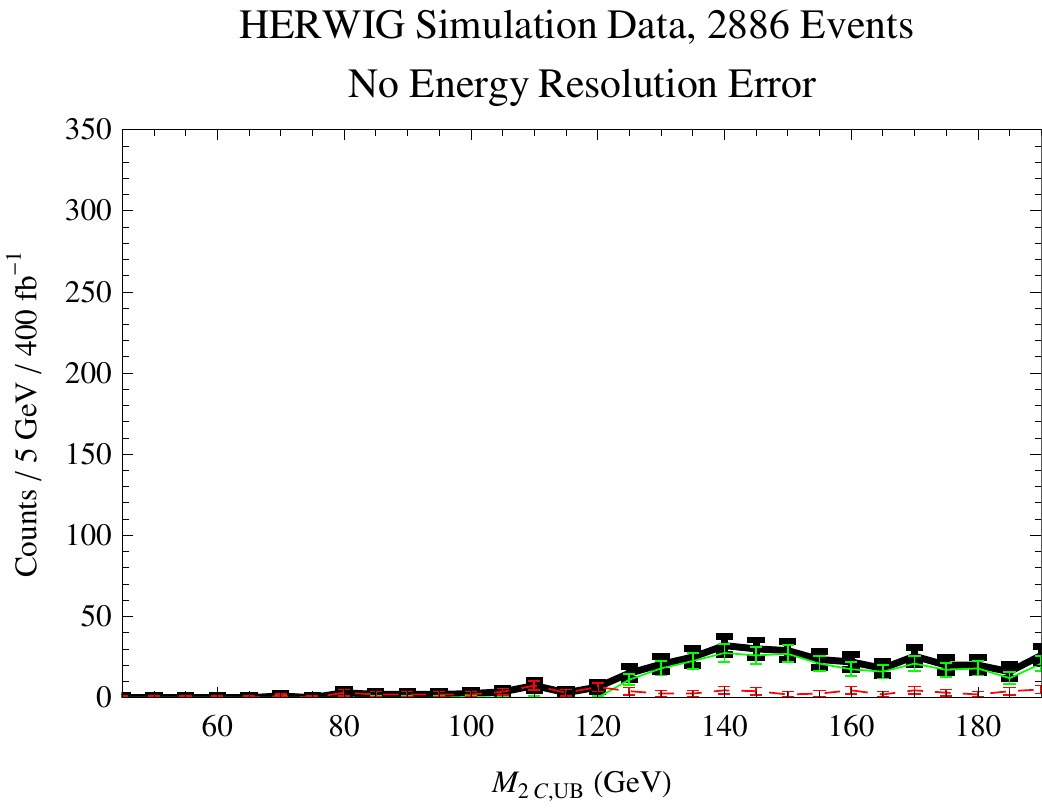}}
\caption{\label{FigHerwigResults}The $M_{2C}$ and $M_{2C,UB}$ distributions of \herwig\ events before smearing (to simulate detector resolution) is applied. The distributions' end-points show $M_{\N{2}} \approx 120$ GeV. The top thick curve shows the net distribution, the next curve down shows the contribution of only the signal events, and the bottom dashed curve shows the contribution of only background events.}
\end{figure}

In order to obtain a more realistic estimate
of the problems associated with collision data,
we generate samples of unweighted inclusive supersymmetric particle pair production,
using the \herwig\ Monte Carlo program with LHC beam conditions.
These samples produce a more realistic simulation of the event
structure that would be obtained for the supersymmetric model studied here,
including the (leading order) cross sections and parton distributions.
It includes all supersymmetric processes and so contains the relevant background processes
as well as the particular decay chain that we wish to study.
Figure \ref{FigHerwigResults} shows the $M_{2C}$ and $M_{2C,UB}$ distributions of a sample of \herwig\ generated signal and background events.

Charged leptons ($e^\pm$ and $\mu^\pm$)
produced in the decay of heavy objects (SUSY particles and $W$ and $Z$ bosons)
were selected for futher study
provided they satisfied basic selection criteria on transverse momentum
($p_T>10$~GeV) and pseudorapdity ($|\eta|<2.5$).
Leptons coming from hadron decays are usually contained within hadronic jets and
so can be experimentally rejected with high efficiency using energy or track isolation criteria.
This latter category of leptons was therefore not used in this study.
The acceptance criterion used for the hadronic final state was $|\eta|<5$.
The detector energy resolution functions used are described in Section~\ref{sec:detector}.


\section{Factors for Successful Shape Fitting}
\label{SecShapeFactors}

There are several factors that control or
affect the shape of the $M_{2C}$ and $M_{2C,UB}$ distributions.  We divide the factors into those that
affect the in-principle distribution and the factors that affect
the observation of the distribution by the detector like
energy resolution and selection cuts.

The in-principle distribution of these events is influenced by
the presence or absence of spin-correlations between the branches, the
$m_{ll}$ distribution of the visible particles, any
significant upstream transverse momentum (UTM) against  which the system is recoiling (\emph{e.g.}  gluinos or squarks decaying further up the decay chain), and background coming from other new-physics processes or the Standard Model.  As all these processes effectively occur at the interaction vertex, there are some combinatoric ambiguities.
These are the factors that influence the in-principle distribution of events that impinges on the particle detector.

The actual distribution recorded by the detector will depend on further factors.  Some factors we are able to regulate -- for example cuts on the missing transverse momentum.  Other factors depend on how well we understand the detector's operation -- such as the energy resolution and particle identification.

Where the effect of such factors is significant,
for example for the $m_{12}$, $k_T$, and background distributions,
our approach has been to model their effect on the ideal distributions
by using appropriate information from the `data', much as one would do in a real LHC experiment.
For the present our `data' are provided by \herwig, rather than LHC events, but the principle is the same.

\subsection{Factors Affecting the In-principle Distribution}

\begin{itemize}
\item \textbf{Mass Difference and Mass Scale}
\end{itemize}

\begin{figure}
\centerline{\includegraphics[width=4in]{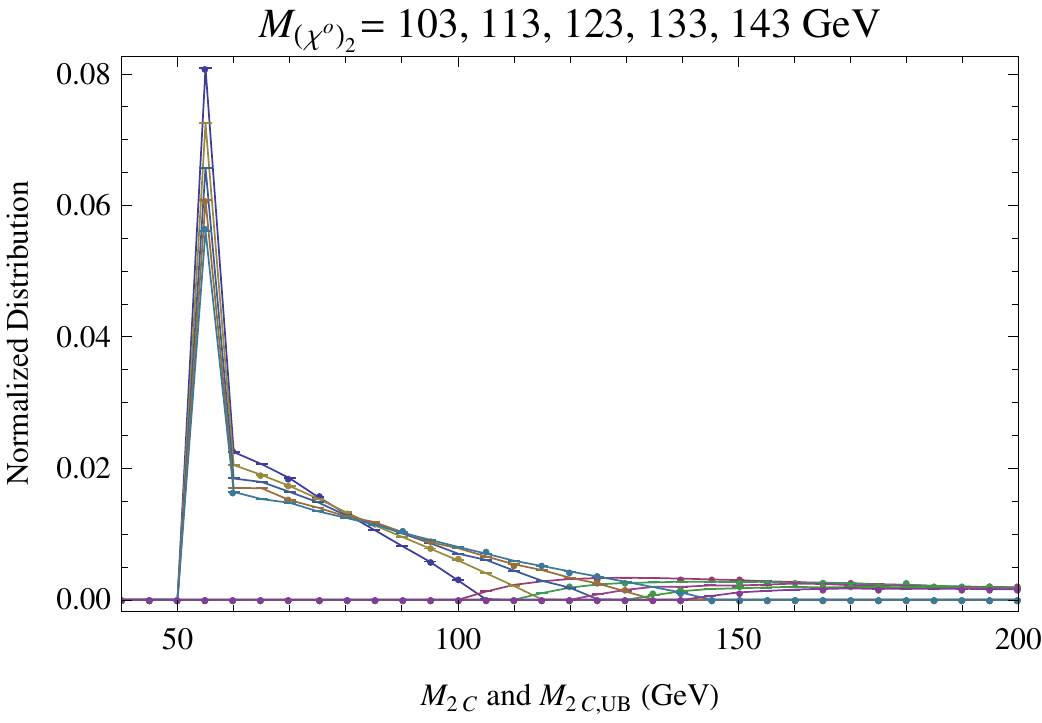}}
\caption{\label{FigLotsOfIdealCurves} We show the $M_{2C}$ and $M_{2C,UB}$ ideal
distributions for five choices of $M_{\N{2}}$
assuming the \herwig\ generated $m_{ll}$ and UTM distributions.}
\end{figure}

The end-point of $M_{2C}$ and $M_{2C,UB}$ distributions give the mass of $\N{2}$.  Therefore the mass scale, $M_{\N{2}}$, is a dominant factor
in the shape of the `ideal' distribution. This is the
reason we can use these distributions to determine the mass scale.
Fig \ref{FigLotsOfIdealCurves} shows the $M_{2C}$ and $M_{2C,UB}$ distributions
for five choices of $M_{\N{2}}$ assuming the \herwig\ generated
$m_{ll}$ and UTM distributions.

How does the shape change with mass scale?  The shape is typically sharply peaked at $M_{2C}=\Deltam$ followed by a tail that ends at the mass of $M_{\N{2}}$.
The peak at $\Deltam$ is due to events that are compatible with $M_{\N{1}}=0$.  We say these events give the trivial constraint. Because we bin the data, the height of the first bin depends on the bin size.  As $M_{+}/M_{-}=(M_{\N{2}} + M_{\N{1}})/(M_{\N{2}} - M_{\N{1}})$ becomes larger, then the non-trivial events are distributed over a wider range and the endpoint becomes less clear.  In general if all other things are equal, the larger the mass, the more events in the first bin and the longer and flatter the tail.

The distribution also depends on the mass difference $M_{-}$ which
we assume has been determined.
We expect that experimentally one should be able to read off the mass difference from the $m_{ll}$ kinematic end-point with very high precision.
Gjelsten, Miller, and Osland estimate this edge can be measured to better than $0.08 \GeV$  \cite{Gjelsten:2004ki,Gjelsten:2006tg}
using many different channels that
lead to the same edge, and after modeling energy resolution and background.

Errors in the mass determination
propagated from the error in the mass
difference in the limit of $k_T=0$ are given approximately by
\begin{equation}
\delta M_{\N{2}} = \frac{\delta M_{-}}{2} \left(  1- \frac{M_{+}^{2}}{M_{-}^{2}}
\right)  \ \ \ \delta M_{\N{1}} = - \frac{\delta M_{-}}{2} \left(  1+ \frac
{M_{+}^{2}}{M_{-}^{2}} \right) \label{EqDeltaMmErrorEffects}%
\end{equation}
where $\delta M_{-}$ is the error in the determination of the mass difference
$M_{-}$.  An error in  $M_{-}$ will lead to an $M_{2C}$ distribution with a shape and endpoint above or below the true mass in the direction indicated by Eq(\ref{EqDeltaMmErrorEffects}).

To isolate this source of error from the uncertainty in the fit, we assume that the mass difference is known exactly in our stated results. In our case an uncertainty of $\delta M_{-} = 0.08 \GeV$ would lead to an additional $\delta M_{\N{1}} = \pm 0.5 \GeV$ to be added in quadrature to the error from fitting.

\begin{itemize}
\item \textbf{Spin Correlations Between Branches}
\end{itemize}

The potential impact of spin correlation between branches on $M_{2C}$ were studied in Sec \ref{SecSpinCorrelationsM2C}.
There are also no spin
correlations if the $\N{2}$ parents are part of a longer decay chain which involves a scalar at some stage as is the case in most of the events in the model we study here.
In the simple Mathematica simulations, we have assumed no spin dependence in the production of the hypothetical ideal distribution.

\begin{itemize}
\item \textbf{Input $m_{12}$ Distributions}
\end{itemize}

\begin{figure}
\centerline{\includegraphics[width=3in]{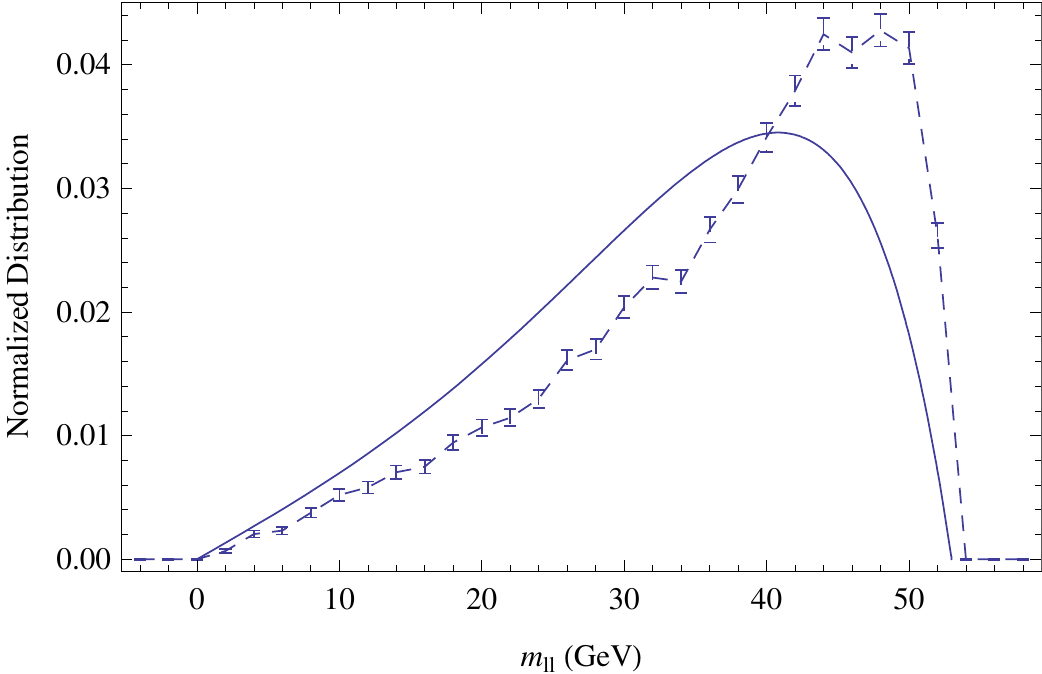} \includegraphics[width=3in]{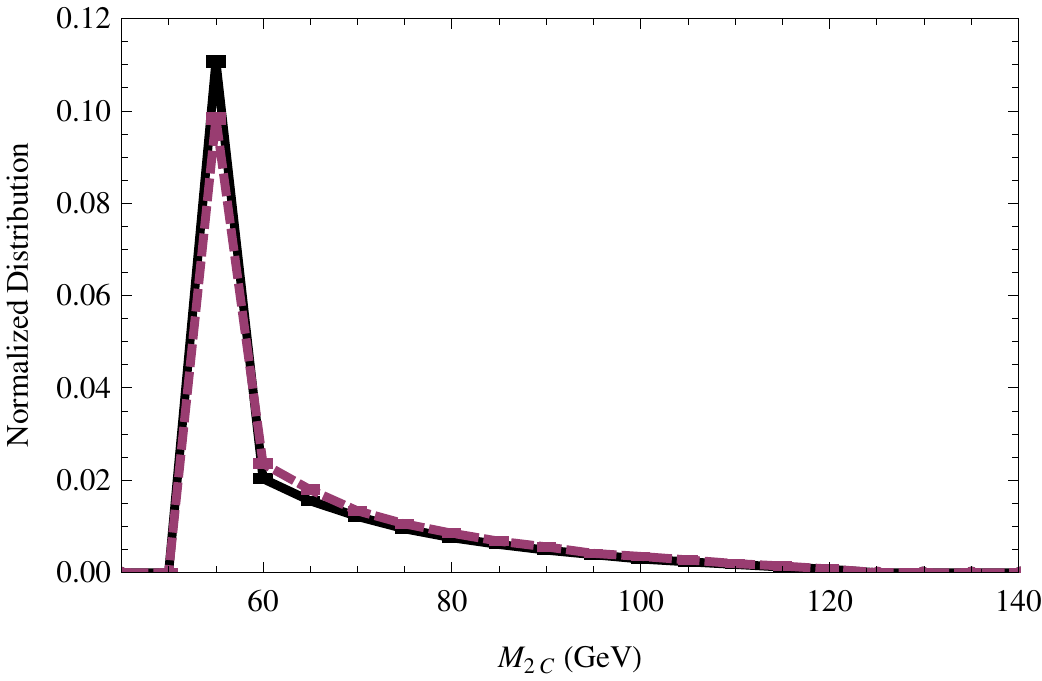}}
\caption{\label{FigMllDep} Dependence of $M_{2C}$ distribution on the $m_{ll}$ distribution.
{\textbf{Left:}} The $m_{ll}$ distributions. {\textbf{Right:}}  The corresponding $M_{2C}$ distributions. The solid curves show the case where the $m_{ll}$ distribution when the three-body decay is dominated by the $Z$ boson channel, and the dashed curves show the case where the $m_{ll}$ distribution is taken directly from  the \herwig\ simulation.}
\end{figure}

The $m_{ll}$ distribution affects the $M_{2C}$ distribution.
Fig \ref{FigMllDep} shows two $m_{ll}$ distributions and the corresponding $M_{2C}$ distributions with $k_T=0$ (no UTM).  The solid lines show the case where the three-body decay from $\N{2}$ to $\N{1}$ is completely dominated by a $Z$ boson.  The dashed line shows the case where the $m_{ll}$ distribution is extracted from the `realistic' \herwig\ simulation.
We can see that the $M_{2C}$ distribution
is affected most strongly in the first several non-zero bins.
If we were to determine the mass only from the shape of these first several bins
using only the $Z$ contribution for the $m_{ll}$ difference,
we would estimate of the mass to be about $4$ GeV below the true mass.
This can be understood because the shape change of the $m_{ll}$ distribution effectively took events out of the first bin and spread them over the larger bins simulating the effect of a smaller mass.

\begin{itemize}
 \item \textbf{Input Upstream Transverse Momentum Distribution}
\end{itemize}

\begin{figure}
\centerline{\includegraphics[width=3.1in]{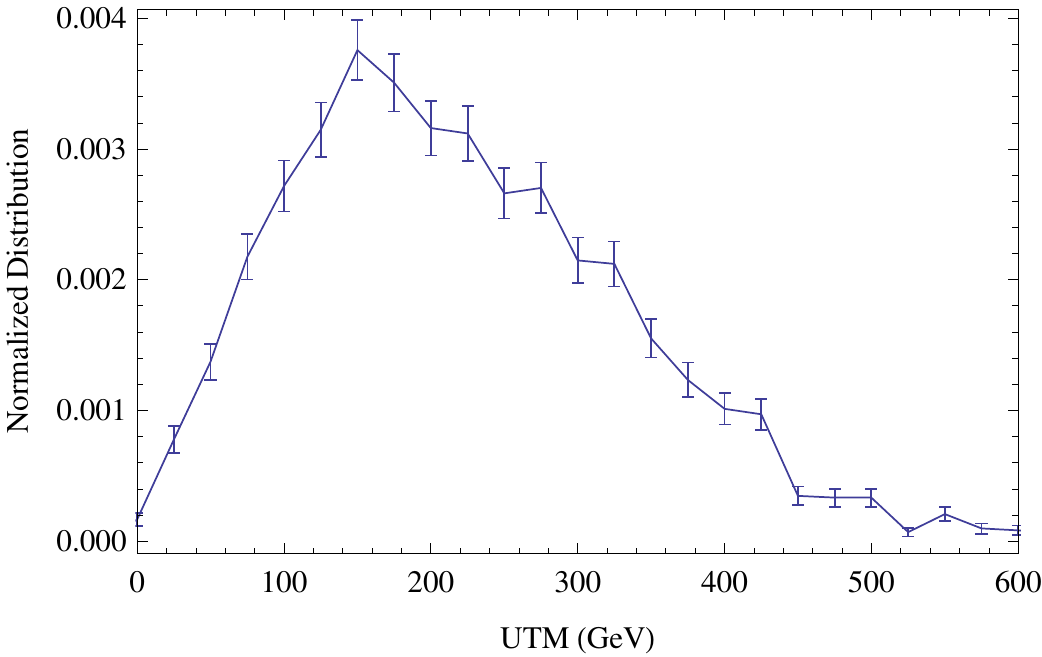}\ \includegraphics[width=3.1in]{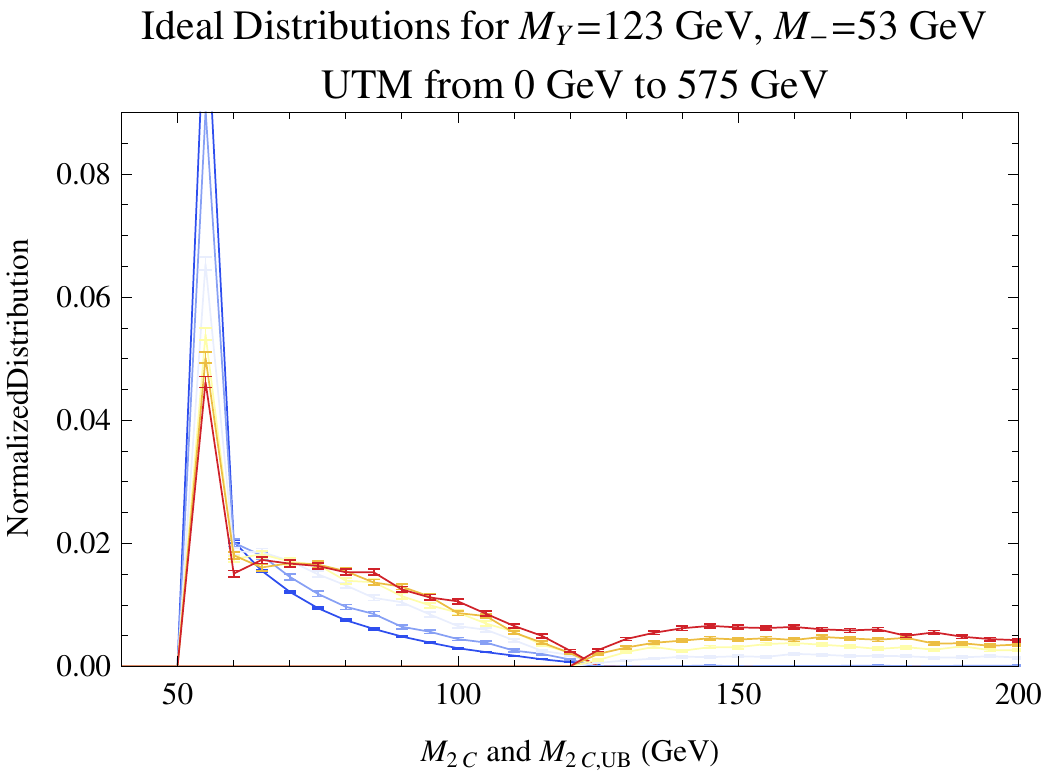} }
\caption{\label{FigM2CISRIdeal}
{\textbf{Left:}} The UTM distribution observed in the \herwig\ simulation.
{\textbf{Right:}} Ideal $M_{2C}$ upper bound and lower bound distribution for a range of upstream transverse momentum (UTM) values ($k_T = 0, 75, 175, 275, 375, 575 \GeV$) where $M_N=70$ GeV and $M_Y=123$ GeV.
}
\end{figure}

As we discussed in Section~\ref{SecUB}, if there is a large upstream transverse momentum (UTM) against which the two $\N{2}$'s recoil, then we have both an upper and lower bound on the mass scale.
The left frame of Fig.~\ref{FigM2CISRIdeal} shows the UTM distribution observed in the `realistic' \herwig\ data.
The right frame of Fig.~\ref{FigM2CISRIdeal} shows the $M_{2C}$ and $M_{2C,UB}$ distributions for fixed UTM ($k_T$) of $0$, $75$, $175$, $275$, $375$, and $575\ \GeV$ all with $k^2=(100 \GeV)^2$. As we discuss under the next bullet, we also find the distribution is not sensitive to the value of $k^2$. For $k_T > 275 \GeV$, these curves begin to approach a common shape. These are ideal $M_{2C}$ upper and lower bound distributions where $M_{N}=70 \GeV$ and $M_{Y}=123 \GeV$.
Notice that there is no upper-bound curve for the case with zero $k_T$ UTM.
The UTM  makes the distribution have a sharper endpoint and thereby make the mass easier to determine.  This is equivalent to having a sharper kink in $\Max M_{T2}$ in the presence of large UTM \cite{Barr:2007hy}.

How do we determine $k_T$ from the data?  Because we demand exactly four leptons (two OSSF pairs), we assume all other activity, basically the hadronic activity, in the detector is UTM. 
The shape used in the `ideal' distribution is a superposition of the different fixed UTM distributions, shown on right frame of Fig.~\ref{FigM2CISRIdeal}, weighted by the observed UTM distribution, shown on the left frame of Fig.~\ref{FigM2CISRIdeal}.  Equivalently, we obtain the ideal distribution by selecting $k_T$ in the Mathematica Monte Carlo according to the observed UTM distribution.

\begin{itemize}
\item \textbf{Shape Largely Independent of Parton Distributions and Collision Energy}
\end{itemize}

In the limit where there is no UTM, then $M_{2C}$ is invariant under back-to-back boosts of the parent particles; therefore, $M_{2C}$ is also invariant to changes in the parton distribution functions.

\begin{figure}
\centerline{\includegraphics[width=3.1in]{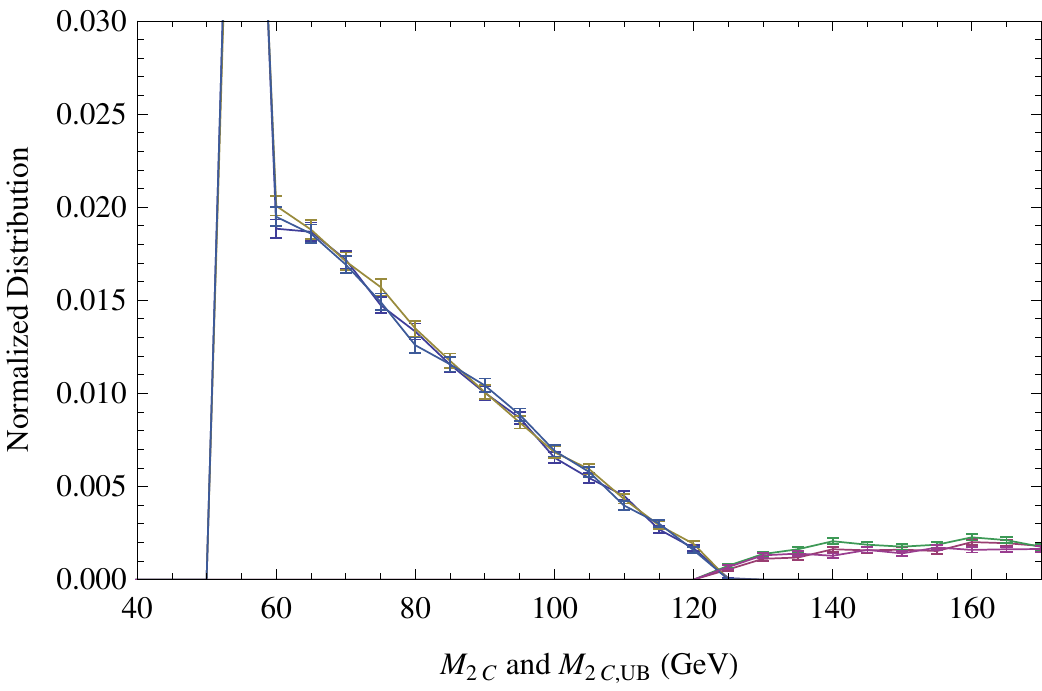}\ \includegraphics[width=3.1in]{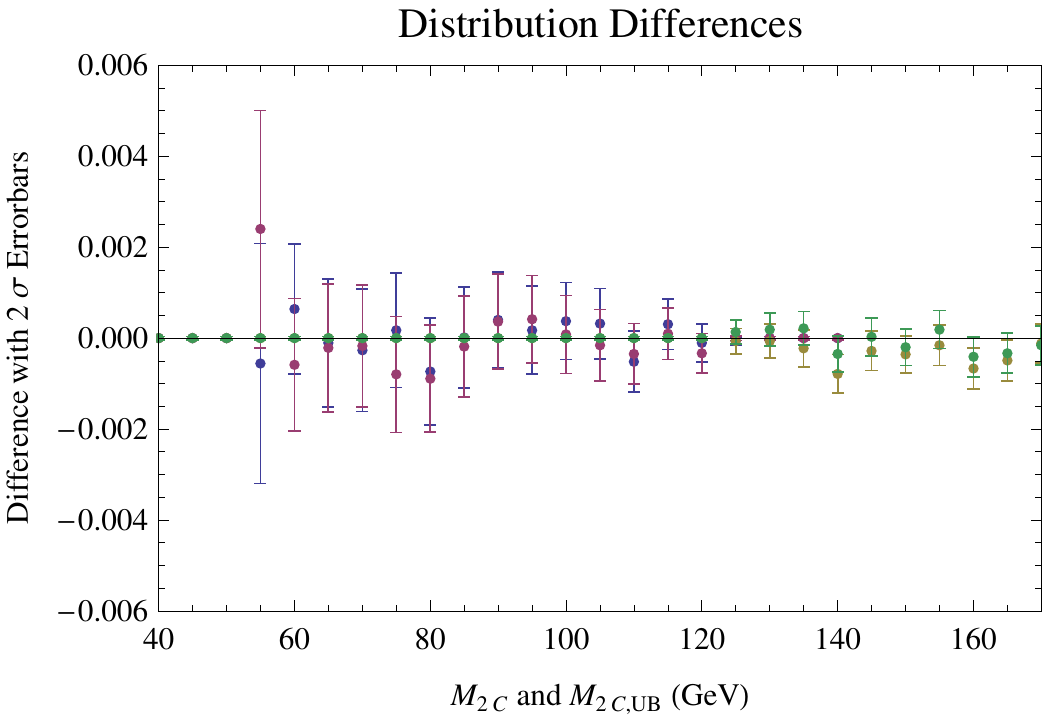}}
\caption{\label{FigDiff} Figure shows that even with large UTM, the distribution is independent of $k^2$ and the parton collision energy to the numerical
accuracies as calculated from 15000 events. Shown are three distributions and their difference. (1) $k_T=175 \GeV$, $k^2=(100 \GeV)^2$, $\sqrt{s}$ distributed via Eq(\ref{EqSdep}). (2)  $k_T=175 \GeV$, $k^2=(2000 \GeV)^2$, $\sqrt{s}$ distributed via Eq(\ref{EqSdep}). (3) $k_T=175 \GeV$, $k^2=(100 \GeV)^2$, $\sqrt{s}=549 \GeV$.}
\end{figure}

How much of this invariance survives in the presence of large UTM?  The answer is that it remains largely independent of the parton collision energy and largely independent of the mass $k^2$ as shown in Fig \ref{FigDiff} numerically.  On the left frame, we show three distributions and in the right frame their difference with $2 \ \sigma$ error bars calculated from $15000$ events. The first distribution assumes $k_T=175 \GeV$, $k^2=(100 \GeV)^2$, $\sqrt{s}$ distributed via Eq($\ref{EqSdep}$). The second distribution assumes  $k_T=175 \GeV$, $k^2=(2000 \GeV)^2$, $\sqrt{s}$ distributed via Eq(\ref{EqSdep}). The third distribution assumes $k_T=175 \GeV$, $k^2=(100 \GeV)^2$, and a fixed collision energy of $\sqrt{s}=549 \GeV$.

\begin{itemize}
\item \textbf{Backgrounds}
\end{itemize}

\begin{figure}
\centerline{\includegraphics[width=5.0in]{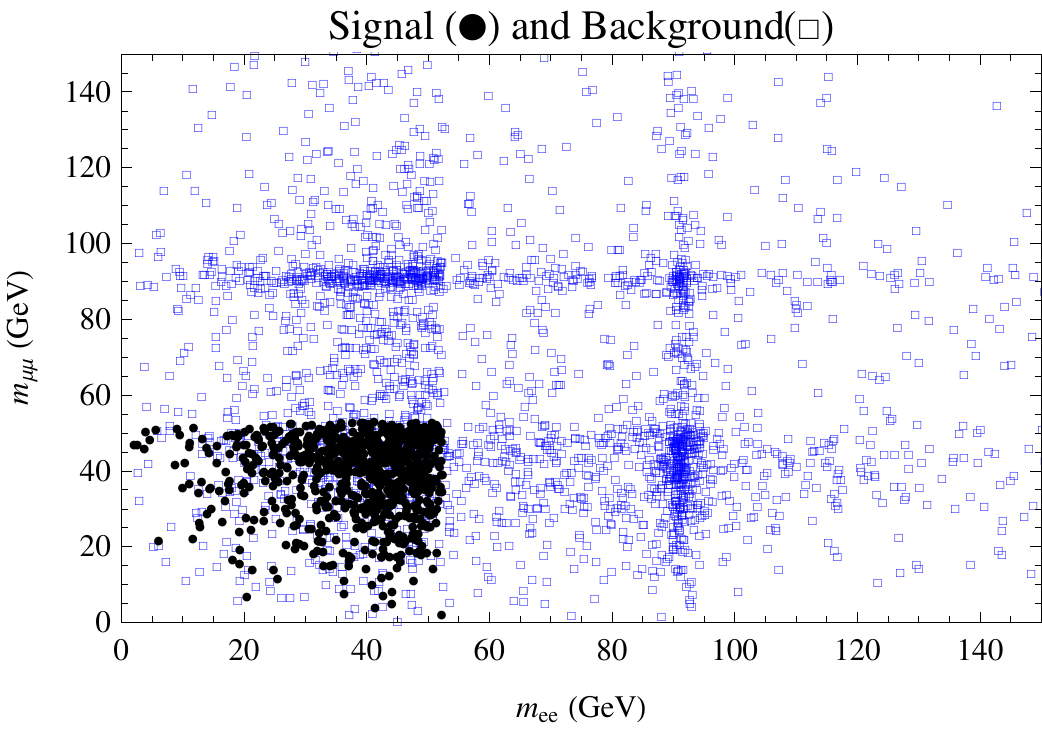}}
\caption{\label{FigMllDalitzLike} The invariant mass of the OSSF leptons from both branches of forming a Dalitz-like wedgebox analysis.
The events outside the $m_{ll} \le 53 \GeV$ signal rectangle provide
control samples from which we estimate the background shape and magnitude. The dark events are signal, the lighter events are background.}
\end{figure}

Backgrounds affect the shape and, if not corrected for,
could provide a systematic error in the estimated mass.  In Section \ref{SecPerformance} we will see that the position of the minimum $\chi^2$ in a fit to $M_{\N{1}}$ is barely affected by the background.  The main effect of the background is to shift the parabola up, giving a worse fit.  To improve the fit, we may be able to estimate the $M_{2C}$ and $M_{2C,UB}$ distribution and magnitude of the
background from the data itself.
We first discuss the sources of background, and then we describe a generic technique using a Dalitz-like wedgebox analysis to estimate a background model which gives approximately the correct shape and magnitude of the background.

One reason we study the four-lepton with missing transverse momentum channel is because of the very-low Standard-Model background \cite{Ghosh:1999ix,Bisset:2005rn}.  A previous study \cite{Bisset:2005rn} estimates about $120$ Standard-Model four-lepton events (two OSSF pairs) for $100\,\fb^{-1}$ with a $\slashed{P}_T > 20$ GeV cut. They suggest that we can further reduce the Standard-Model background by requiring several hadronic jets.  Because we expect very-little direct $\N{2}$ pair production, this would have very little effect on the number of signal events.   Also, because $Z^o$s are a part of the intermediate states of these background processes, very few of these events will have $m_{ll}$ significantly different from $M_Z$.

What is the source of these Standard-Model backgrounds? About $60 \%$ is from $Z$-pair production events with no invisible decay products, in which the missing transverse momentum can only arise from experimental particle identification and resolution errors.  This implies that a slightly stronger $\slashed{P}_T$ cut could further eliminate this background.   Another $40\%$ are due to $t,\bar{t}, Z$ production.
Not explicitly discussed in their study but representing another possible source of backgrounds are events containing heavy baryons which decay leptonically.
If we assume $b$-quark hadrons decay to isolated leptons with a branching ratio of $0.01$, then LHC $t,\bar{t}$ production will lead to about $10$ events passing these cuts for $100\,\fb^{-1}$ where both OSSF leptons pairs have $m_{ll} < M_{Z}$.

Tau decays also provide a background for our specific process of interest. The process $\N{2} \rightarrow \tau^+ \tau^- \N{1}$ will be misidentified as $e^{-}e^{+}$ or $\mu^{+}\mu^{-}$ about $3\%$ of the time ($6 \%$ total). Because the $\tau$ decays introduce new sources of missing transverse momentum ($\nu_\tau$), these events will distort the $M_{2C}$ calculation.  This suggests that the dominant background to the $\N{2},\N{2} \rightarrow 4 l + \slashed{P}_T + $ hadrons will be from other SUSY processes.

We now create a crude background model from which we estimate the magnitude and distribution of the background using the `true' \herwig\ data as a guide.
We follow the suggestion of Ref.~\cite{Bisset:2008hm,Bisset:2005rn} and use a wedgebox analysis plotting the invariant mass $m_{ee}$ against $m_{\mu\mu}$ to supplement our knowledge of the
background events mixed in with our signal events.  This wedgebox  analysis, seen in
Fig.~\ref{FigMllDalitzLike} for our \herwig\ simulation, shows patterns that tell about other SUSY states present. The presence of the strips along $91$ GeV indicate that particle states are being created that decay to two leptons via an on-shell $Z$.  The observation that the intensity changes above and below $m_{\mu\mu}=53$ GeV shows that many of the states produced have one branch that decays via a $\N{2}$ and the other branch decays via an on-shell $Z$.  The lack of events immediately above and to the right of the $(53 \GeV, 53 \GeV)$ coordinate but below and to the left of $(91 \GeV,91 \GeV)$ coordinate suggest that symmetric process are not responsible for this background.

We also see the density of events in the block above $53 \GeV$ and to the right of $53 \GeV$ suggest a cascade decay with an endpoint near enough to $91 \GeV$ that it is not distinguishable from $M_Z$.   Following this line of thinking, we model the background with a guess of an asymmetric set of events where one branch has new states $G$, $X$ and $N$ with masses such that the $m_{ll}$ endpoint is
\begin{equation}
  \Max m^2_{ll} ({\rm{odd}\ \rm{branch}}) = \frac{(M_G^2-M_X^2)(M_X^2-M_N^2)}{M_X^2} = (85 \GeV)^2
 \end{equation}
and the other branch is our $\N{2}$ decay.  The masses one chooses to satisfy this edge did not prove important so long as the mass differences were reasonably sized; we tried several different mass triplets ending with the LSP, and all gave similar answers.

We now describe the background model used in our fits.  One branch starts with a massive state with $M_G=160 \GeV$ which decays to a lepton and a new state $M_X=120 \GeV$ which in-turn decays to a lepton and the LSP. The second branch has our signal decay with the $\N{2}$ decaying to $\N{1}$ and two leptons via a three-body decay. We added UTM consistent with that observed in the events.

By matching the number of events seen outside the $m_{ll} < 53 \GeV$ region, we estimate the number of the events within the signal cuts that are due to backgrounds.  We estimate $0.33$ of the events with both OSSF pairs satisfying $m_{ll} < 53 \GeV$ are background events.
The model also gives a reasonable distribution for these events.  Inspecting the actual \herwig\ results showed that actual fraction of background events was $0.4$.  If we let the fraction be free and minimize the $\chi^2$ with respect to the background fraction, we found a minimum at $0.3$.

Our background model is simplistic and does not represent the actual processes, but it does a good job of accounting for the magnitude and the shape of the background mixed into our signal distribution. Most of the \herwig\ background events came from $W$ and charginos which introduce extra sources of missing transverse momentum.  Never the less, the shape fits very accurately and the performance is discussed in Section \ref{SecPerformance}. It is encouraging that our estimate of the background shape and magnitude is relatively insensitive to details of the full spectrum.  Even ignoring the background, as we will see in Section \ref{SecPerformance}, still leads to a minimum $\chi^2$ at the correct mass.

\begin{itemize}
\item \textbf{Combinatoric Ambiguities}
\end{itemize}
\label{SecM2CLBUBCombinatorics}

If we assume that the full cascade effectively occurs at the primary vertex (no displaced vertices), then the combinatoric question is a property of the ideal distribution produced in the collisions.
There are no combinatoric issues if the two opposite-sign same-flavor lepton pairs are each different flavors (\emph{e.g.} the four leptons are $e^+$, $e^-$, $\mu^+$ and $\mu^-$).  However if all four leptons are the same flavor, we have found that we can still identify unique branch assignments 90\% of the time.  The unique identification comes from the observation that both pairs must have an invariant mass $m_{ll}$ less than the value of the $ \Max m_{ll}$ edge. In $90 \%$ of the events, there is only one combination that satisfies this requirement.  This allows one to use $95 \%$ of the four lepton events without ambiguity.  The first $50 \%$ are identified from the two OSSF pairs being of different flavors and $90 \%$ of the remaining can be identified by requiring that both branches of OSSF lepton for an event's combinatoric pairing satisfy $m_{ll, \mathrm{Event}} < \Max_{\mathrm{All}\ \mathrm{Events}} \ m_{ll}$.  The events which remain ambiguous have two possible assignments, both of which are included with a weight of $0.5$ in the distribution.

\subsection{Factors Affecting Distribution Recorded by the Detector}
\label{sec:detector}

As just described, the `ideal' in-principle distribution is created from the observed $m_{ll}$ distribution and the observed UTM distribution.  We include combinatoric effects from events with four leptons of like flavors.  Last, we can estimate the magnitude of background events and their $M_{2C}$ and $M_{2C,UB}$ shape.
We now modify the in-principle distribution to simulate the
effects of the particle detector to form our final `ideal' distribution
that includes all anticipated effects.  The two main effects on the $M_{2C}$ and $M_{2C,UB}$ distributions are the energy resolution and the $\slashed{P}_T$ cuts.

\begin{itemize}
\item \textbf{Shape Dependence on Energy Resolution}
\end{itemize}

Energy resolution causes the $M_{2C}$ and $M_{2C,UB}$ distributions to be smeared.   Here we assume the angular resolution is negligible.
For both the Mathematica Monte Carlo model and the \herwig\ events
we simulate the detector's energy resolution by scaling the four vectors for electrons, muons, and hadrons by
\begin{eqnarray}
 \frac{\delta E_e}{E_e} & = & \frac{0.1}{\sqrt{E_e}} + \frac{0.003}{E_e} + 0.007 \label{EqDetectorEnergyResolutionElectron} \\
 \frac{\delta E_\mu}{E_\mu} & = & 0.03  \label{EqDetectorEnergyResolutionMuon}\\
 \frac{\delta E_H}{E_H} & = & \frac{0.58}{\sqrt{E_H}} + \frac{0.018}{E_H} + 0.025
 \label{EqDetectorEnergyResolutionHadron}
\end{eqnarray}
respectively
\cite{Akhmadalev:2001ar}\cite{AtlasTDR}.  The muon energy resolution is different because they are typically not contained by the calorimeter.
A more detailed detector simulation is of course possible,
but since we do not know the true behavior of any LHC detector until the device begins taking data,
a more sophisticated treatment would be of limited value here.
In practice the dependence of the ideal distribution shapes
on the missing transverse momentum resolution should reflect
the actual estimated uncertainty of the missing transverse momentum of the
observed events.

Smearing of the distributions decreases the area difference between two normalized distributions, thereby decreasing the precision with which one can determine the mass from a given number of signal events.  This expanded uncertainty can be seen in Section~\ref{SecPerformance}.

The $M_{2C}$ calculations depend on the mass difference, the four-momenta of the four leptons, and the missing transverse momentum.  As the lepton energy resolution is very tight, the missing transverse momentum's energy resolution is dominated by the hadronic energy resolution.  We model the energy resolution of the UTM as a hadronic jet.  This significantly increases the uncertainty in the missing transverse momentum because hadrons have about five times the energy resolution error.

In our Mathematica model, we represent the UTM as a single four-vector $k$, but in reality it will be the sum of many four-vectors.  Because we apply the energy resolution smearing to $k$, if $k$ is small the simple Mathematica model will have a smaller missing transverse momentum resolution error.
However, an events with almost $0$ UTM could have a large missing momentum energy resolution if it has a lot of hadronic jets whose transverse momentum mostly cancels.
Fig \ref{FigM2CISRIdeal} shows that most of the time we have considerable hadronic UTM, so this effect is a minor correction on our results.


\begin{itemize}
\item \textbf{Shape Dependence on Missing Transverse Momentum Cuts}
\end{itemize}

\begin{figure}
\centerline{\includegraphics[width=5.5in]{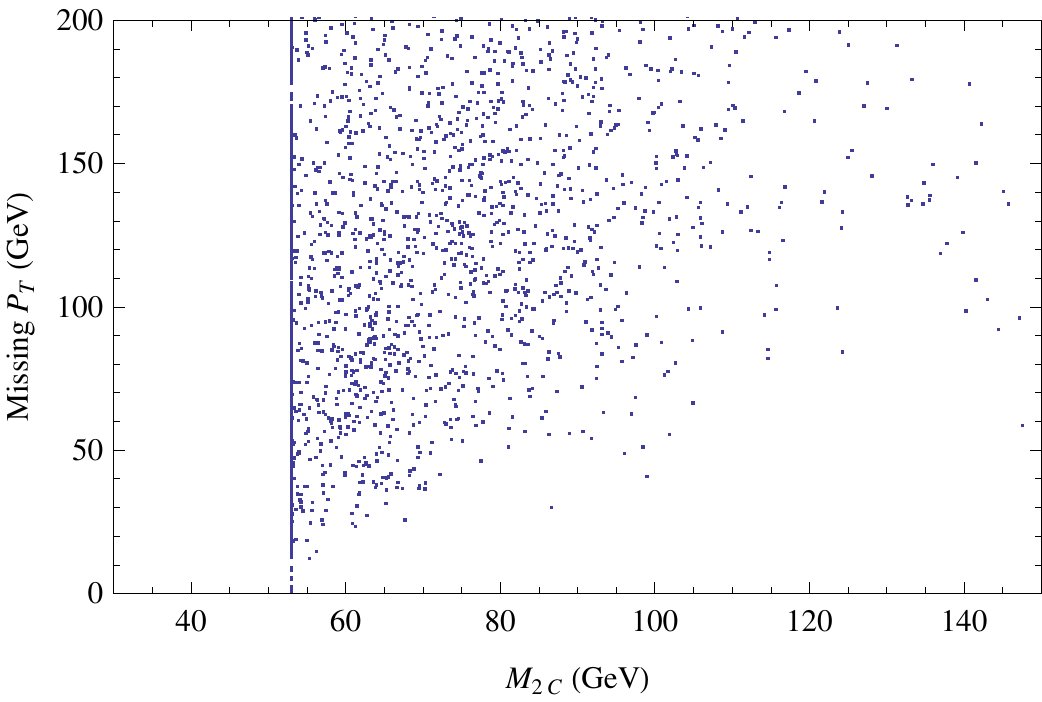}}
\caption{\label{FigPTCutDependence} The missing transverse momentum vs $M_{2C}$ values for \herwig\ data.  This shows that a $\slashed{P}_T>20 \GeV$ cut would not affect the distribution for $M_{2C} > 65 \GeV$.}
\end{figure}

A key distinguishing feature of these events is missing transverse momentum. To eliminate the large number of Standard Model events with four-lepton and with no $\slashed{P}_T$, we will need to cut on this parameter. Fig.~\ref{FigPTCutDependence} shows the \herwig\ simulation's missing transverse momentum versus the $M_{2C}$.  A non-trivial $M_{2C}$ requires substantial $\slashed{P}_T$.  Small $\slashed{P}_T$ of less than about $20 \GeV$ only affects the $M_{2C}$ shape below about $65 \GeV$.  The shape of the $M_{2C} < 65 \GeV$ therefore will require a higher fidelity model from which to train the shapes. Instead, we just choose to not fit bins with $M_{2C}< 65 \GeV$.

All events near the end of $M_{2C,UB}$ distribution require significant $\slashed{P}_T$, therefore $\slashed{P}_T$ cuts will not affect the part of this distribution which we fit.  The number of events with no non-trivial upper-bounds will also be affected by $\slashed{P}_T$ cuts.  We only fit the $M_{2C,UB}$ distribution up to about $233 \GeV$.

\section{Estimated Performance}
\label{SecPerformance}

\begin{figure}
\centerline{\includegraphics[width=3.1in]{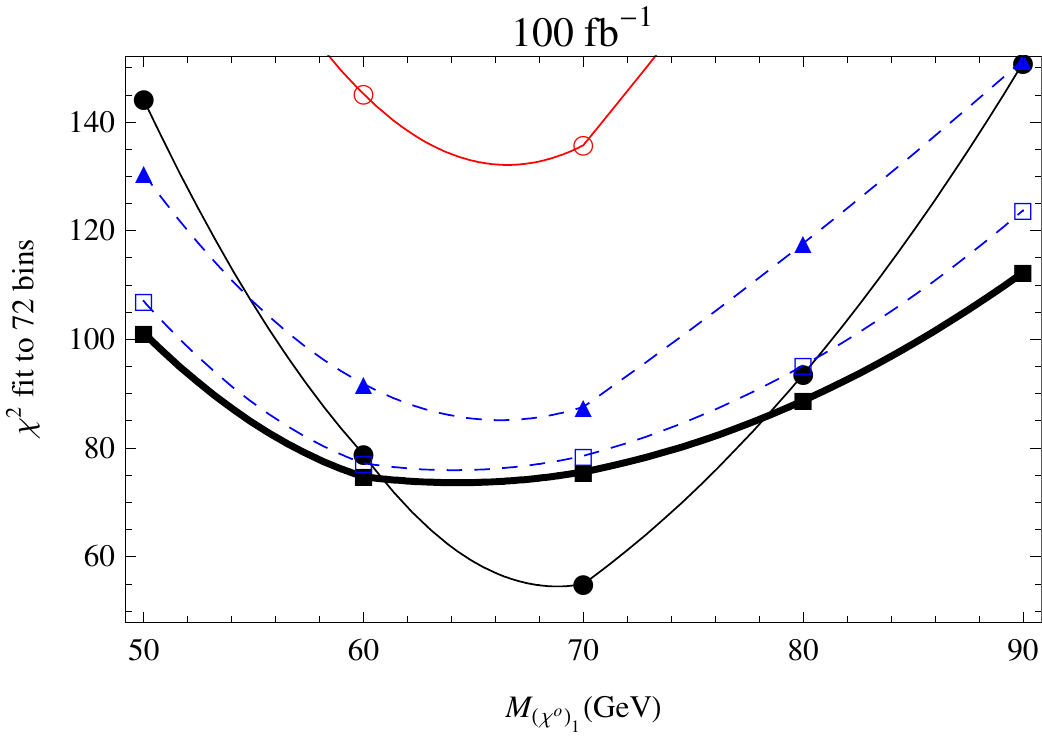}
\includegraphics[width=3.1in]{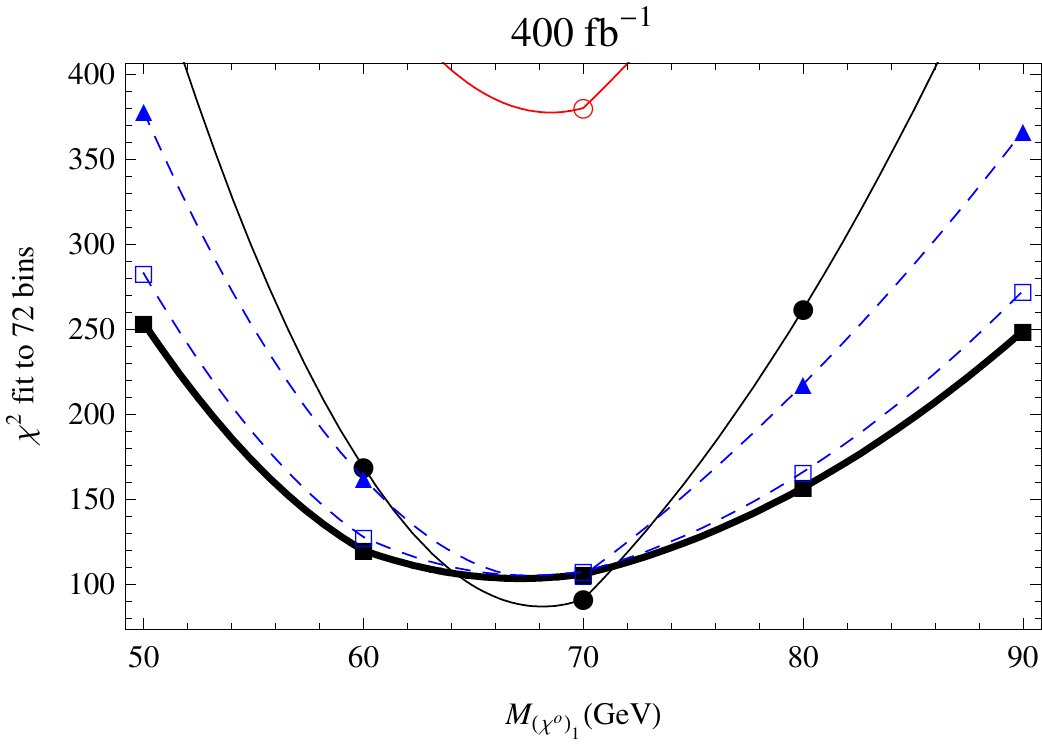}}
\caption{\label{FigPerformancePlots} The result of $\chi^2$ fits to the data with differing assumptions for $100 \fb^{-1}$ (left panel) and $400 \fb^{-1}$ (right panel). The thick line with filled squares shows the final result with all cuts, resolution error, combinatorics, and backgrounds included and estimated in the shape fitting.  This gives us $M_{\N{1}} = 63.2 \pm 4.1 \, \GeV$ with $700$ events (signal or background) representing $100 \fb^{-1}$.  After $400 \fb^{-1}$ this improves to $M_{\N{1}} = 66.0 \pm 1.8 \, \GeV$.  The error-free best case gives $M_{\N{1}} = 67.0 \pm 0.9 \, \GeV$. The correct value is $M_{\N{1}}=67.4 \GeV$.
}
\end{figure}

Determining the mass based on the shape of the distribution enables one to use all the events and not just those near the end point.
We fit both upper-bound and lower-bound shapes to the data as described in the Appendix \ref{AppendixChiSqFitting}. As one expects, fitting the lower-bound shape more tightly constrains the mass from below and fitting the upper-bound shape more tightly constrains the mass from above.  Combining the two gives approximately even uncertainty.  We calculate ideal distributions assuming $M_{\N{1}}$ at five values $50$, $60$, $70$, $80$, and $90$ GeV. We then fit a quadratic interpolation through the points.  Our uncertainties are based on the value where $\chi^2$ increases by $1$ from its minimum of this interpolation.  This uncertainty estimate agrees with about $2/3$ of the results falling within that range after repeated runs.
Our uncertainty estimates do not include the error propagated from the uncertainty in the mass difference (see Eq(\ref{EqDeltaMmErrorEffects})).

We present results for an early LHC run, about $100\ {\rm{fb}}^{-1}$, and for the longest likely LHC run before an upgrade, about $400\ {\rm{fb}}^{-1}$.  After about $100\  {\rm{fb}}^{-1}$, we have $700$ events ( about $400$ signal and $300$ background).  After $400\ {\rm{fb}}^{-1}$, we have about $2700$ events (about $1600$ signal and $1100$ background).  Only $4$ events out of $1600$ are
from direct pair production.  Most of our signal events follow at the end of different decay chains starting from gluinos or squarks.  The upstream decay products produce significant UTM against which the two $\N{2}$ parent particles recoil.

First for the ideal case.  After $400\,\fb^{-1}$, using only signal events and no energy resolution, the $\chi^2$ fits to the predicted shapes give
 $M_{\N{1}} = 67.0 \pm 0.9 \, \GeV$ (filled circles in Fig.~\ref{FigPerformancePlots}).  This mass determination can practically be read off from the endpoints seen in
 Fig.~\ref{FigHerwigResults}; the $M_{2C}$ endpoint is near $120 \GeV$ and subtracting the mass differences gives $M_{\N{1}} = 120 \GeV - \Deltam =67 \GeV$. We now explore how well we can do with
 fewer events and after incorporating the effects listed in Section \ref{SecShapeFactors}.

How does background affect the fit? If we ignore the existence of background in our sample, and we fit all the events to the signal-only shapes, then we find a poor fit shown as the empty circle curve in Fig.~\ref{FigPerformancePlots}. By poor fit, we mean the $\chi^2$ is substantially larger than the $72$ bins being compared
($36$ bins from each the upper-bound and lower-bound distributions).  Despite this worse fit, the shape fits still give a very accurate mass estimate: $M_{\N{1}}=65.4 \pm 1.8 \GeV$ after $100\ {\rm{fb}}^{-1}$ and $M_{\N{1}} = 67.4 \pm 0.9 \GeV$ after $400\ {\rm{fb}}^{-1}$.
At this stage, we still assume perfect energy resolution and no missing transverse momentum cut.

Next, if we create a background model as described in Section~\ref{SecShapeFactors}, we are able to improve the $\chi^2$ fit to nearly $1$ per bin; the mass estimate remains about the same, but the uncertainty increases by about $20\%$.  We find a small systematic shift (smaller than the uncertainty) in our mass
prediction as we increase the fraction of the shape due to the background model vs the signal model.  As we increased our fraction of background, we found the mass estimate was shifted down from $66.5$ at $0\%$ background to $65.6$ when we were at $60\%$ background.  The best $\chi^2$ fit occurs with $30\%$ background; which is very close to the $33 \%$ we use from the estimate, but farther from the true background fraction of about $40 \%$. With $400 \fb^{-1}$ of data, the systematic errors are all but eliminated with the endpoint dominating the mass estimate.
These fits are shown as the triangles with dashed lines and give $M_{\N{1}} = 65.1 \pm 2.4 \GeV$ which after the full run becomes $M_{\N{1}} = 67.3 \pm 1.1 \GeV$.

Including energy resolution as described in Section \ref{SecShapeFactors} shows a large increase in the uncertainty.  The dashed-line with empty square markers shows the $\chi^2$ fit when we include both a background model and the effect of including energy resolution.
These fits are shown as the empty squares with dashed lines and give $M_{\N{1}} = 63.0 \pm 3.6 \GeV$ which after the full run becomes $M_{\N{1}} = 66.5 \pm 1.6 \GeV$.

The final shape factor that we account for are the cuts associated with the missing transverse momentum. After we apply cuts requiring $ \slashed{P}_T > 20 \GeV$ and fit only $M_{2C} > 65 \GeV$ we have our final result shown by the thick lines with filled squares. This includes all cuts, resolution error, combinatorics and backgrounds.  We find $M_{\N{1}} = 63.2 \pm 4.1 \, \GeV$ with $700$ events (signal or background)
representing $100 \fb^{-1}$, and after $400 \fb^{-1}$ this improves to $M_{\N{1}} = 66.0 \pm 1.8 \, \GeV$.
The true mass on which the \herwig\ simulation is based is $M_{\N{1}}=67.4$, so all the estimates are within about $1\ \sigma$ of the true mass.

\begin{figure}
\centerline{\includegraphics{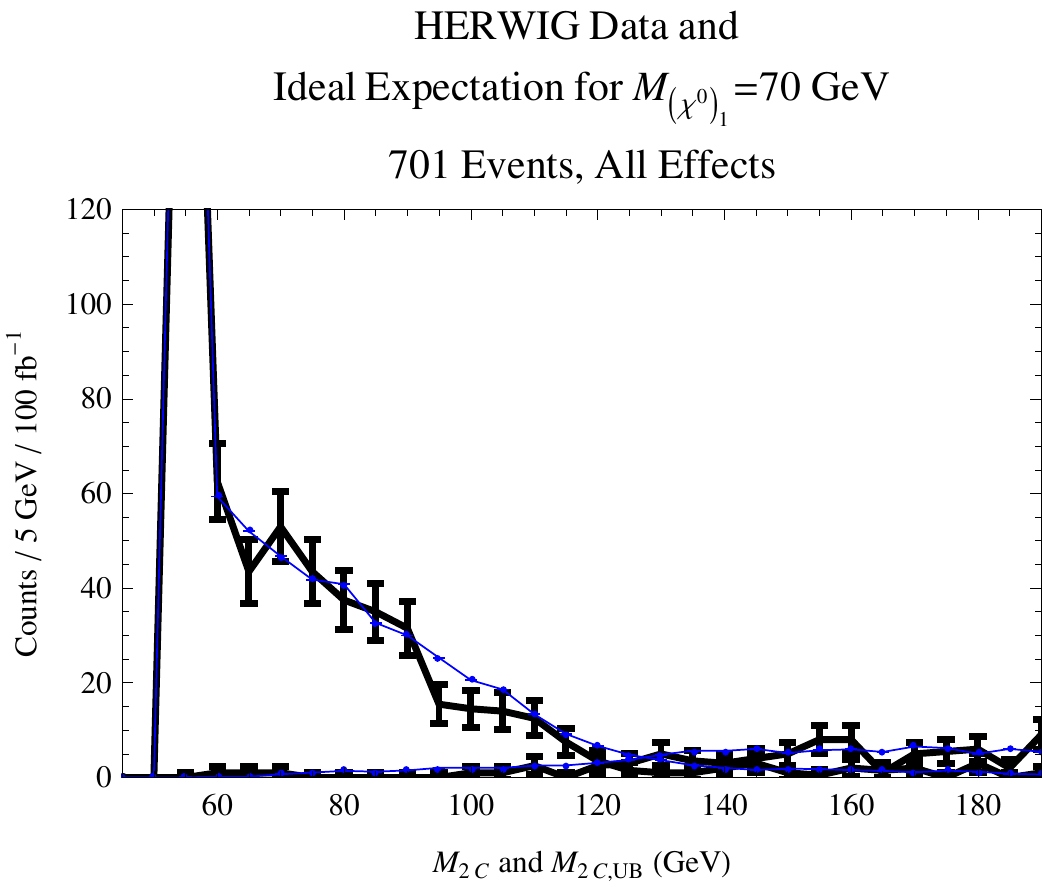}}
\caption{\label{FigHerwigWithExpectation} \herwig\ data for $100 \fb^{-1}$ (thick line) and the smooth ideal expectation assuming $M_{\N{1}}=70 \GeV$ generated by Mathematica with all resolution, background, and combinatoric effects included (thin line).
The $\chi^2$ of this curve to the \herwig\ gives the solid-black square on the left frame of Fig.~\ref{FigPerformancePlots}.}
\end{figure}
Fig \ref{FigHerwigWithExpectation} shows the ideal curve expected if $M_{\N{1}}=70 \GeV$ including all effects from energy resolution,
background, combinatoric, and $\slashed{P}_T$ cuts.
The $\chi^2$ corresponds to the solid square on the left panel of Fig.~\ref{FigPerformancePlots}.

The error in  mass determination obtained with limited statistics can be estimated using Poisson statistics.  In our studies we find that, as one would expect, increasing the number of events by a factor of four,
we bring down our error by about a factor of two.
This means that one could expect $\pm 8 \GeV$ after about $25 \fb^{-1}$ which represents $100$ signal events and $75$ background events.

\section*{Chapter Summary}
\label{M2CUTMSecConclusions}

Despite adding some of the complicating effects one would encounter with real data, we have discovered other factors which demonstrate that one could obtain an even better precision than estimated in Chapter \ref{ChapterM2Cdirect}.
There, we used only our simple Mathematica model and neglected most sources of realistic uncertainty.
We assumed all the events could be modeled as being direct production without spin-correlations.  With these simplifications, we argued the mass could be determined to $\pm 6 \GeV$ using $250$ signal events.


In this chapter, we performed a case study to show that that the relevant  $M_{2C}$ and $M_{2C,UB}$ shapes can be successfully determined from the mass difference $\Deltam$, the $m_{ll}$ distribution observation, the upstream transverse momentum (UTM) distribution observation.
We included and accounted for many realistic effects: we modeled the large energy-resolution error of hadronic jets.  We also included the effects of backgrounds, $\slashed{P}_T$ cuts, and combinatorics.  Our signal and backgrounds were generated with \herwig.
We discussed how a Dalitz-like plot can estimate the background fraction and shape. Observed inputs were used in a simple model to determine ideal distribution shapes that makes no reference to parton distribution functions, cross sections, or other model-dependent factors that we are not likely to know early in the process.

Despite these extra sources of uncertainty, we found a final mass determination of $\pm 4.1 \GeV$ with about $400$ signal events which is still better than the appropriately scaled result from RS \cite{Ross:2007rm}.
The sources of the mass determination improvement are twofold: (1) the prediction and fitting of upper-bound distribution, and (2) the sharper end-point in the presence of large UTM.
Under equivalent circumstances, the sharper endpoint is enough to give a factor of $2$ improvement in the uncertainty over the direct production case assumed in \cite{Ross:2007rm}.  Fitting the upper bound tends to improve the determination by an additional factor of $\sqrt{2}$. This improvement is then used to fight the large hadronic-jet energy resolution and background uncertainty.

Mass determination using $M_{2C}$ and $M_{2C,UB}$ applies to many other processes.  We have focused on cases where the mass difference is given by the end-point of an $m_{ll}$ distribution involving a three-body decay.
If there is not a three-body decay, then the mass difference may be found by applying other mass determination techniques like the mass shell techniques (MST) \cite{Cheng:2007xv,Cheng:2008mg,Nojiri:2007pq} or edges in cascade decays \cite{Bachacou:1999zb,Lester:2006yw,Gjelsten:2006tg} or $M_{T2}$ at different stages in symmetric decay chains \cite{Serna:2008zk}.

How does our method's performance compare to previous mass determination methods?  Firstly, this technique is more robust than the  $\Max M_{T2}$ `kink' because in fitting to the shape of the distribution, it does not rely entirely on identification of the events near kinematic boundary.  One can view $M_{2C}$ and $M_{2C,UB}$ as variables that event-by-event quantify the `kink'.  Other than the `kink' technique, the previous techniques surveyed in Chapter \ref{ChapterMassDeterminationReview} apply to cases where there is no three-body decay from which to measure the mass difference directly. However, each of those techniques still constrains the mass difference with great accuracy.
The technique of \cite{Bachacou:1999zb,Gjelsten:2004ki,Lester:2006yw,Gjelsten:2006tg} which uses edges from cascade decays determines the LSP mass to $\pm 3.4 \GeV$ with about 500 thousand events from $300 \fb^{-1}$.  The approach of \cite{Cheng:2008mg} assumes a pair of symmetric decay chains and assumes two events have the same structure.  They reach $\pm 2.8 \GeV$ using $700$ signal events after $300 \fb^{-1}$, but have a $2.5 \GeV$ systematic bias that needs modeling to remove.  By comparison, adjusting to $700$ signal events we achieve $\pm 2.9 \GeV$ without a systematic bias after propagating an error of $0.08 \GeV$ in the mass difference and with all discussed effects.
Uncertainty calculations differ amongst groups, some use repeated trial with new sets of Monte Carlo data, and others use $\chi^2$.
Without a direct comparison under like circumstance, the optimal method is not clear; but it is clear that fitting the $M_{2C}$ and $M_{2C,UB}$ distributions can determine the mass of
invisible particles at least as well, if not better than the other known methods in both accuracy and precision.

In summary, we have developed a mass determination technique, based on the constrained mass variables, which is able to determine the mass of a dark-matter particle state produced at the LHC in events with large missing transverse momentum.
The $M_{2C}$ method, which bounds the mass from below, was supplemented by
a new distribution $M_{2C,UB}$ which bounds the mass from above in events with large upstream transverse momentum.
A particular advantage of the method is that it also obtains substantial
information from events away from the end point allowing for a
significant reduction in the error.
The shape of the distribution away from the end-point can be determined without detailed knowledge of the underlying model, and as such, can provide an early estimate of the mass.
Once the underlying process and model generating the event has been identified the structure away from the end-point can be improved using, for example, \herwig\ to produce the process dependent shape.
We performed a case-study simulation under LHC conditions to demonstrate that mass-determination by fitting the $M_{2C}$ and $M_{2C,UB}$ distributions survives anticipated complications.  With this fitting procedure it is possible to get an early measurement of the mass - with just 400 signal events in our case study we found we would determine $M_{\N{1}} = 63.2 \pm 4.1$.  The ultimate accuracy obtainable by this method is $M_{\N{1}} = 66.0 \pm 1.8 \GeV$.  We conclude that this technique's precision is as good as, if not better than, the best existing techniques.




\chapter{The Variable $M_{3C}$: On-shell Interlineate States}
 \label{ChapterM3C}

\section*{Chapter Overview}

The main concept of the constrained mass variable $M_{2C}$ \cite{Ross:2007rm} \cite{Barr:2008ba} is that after studying several kinematic quantities we may have well determined the mass difference between two particle states but not the mass itself. We then incorporate these additional constraints in the analysis of the events. We check each event to test the lower bounds and upper bounds on the mass scale that still satisfies the mass difference and the on-shell conditions for the assumed topology.  Because the domain over which we are minimizing contains the true value for the mass, the end-points of the lower bounds and upper bounds distributions'  give the true mass.

The subject of this chapter is extending the constrained mass variable to the case with three new on-shell states as depicted in Fig~\ref{FigEventTopologyTovey}. The constrained mass variable for this case will be called $M_{3C}$.
We structure the chapter around a case study of the benchmark point SPS 1a \cite{Allanach:2002nj}.
In this study, the three new states are identified as  $Y=\N{2}$, $X=\tilde{l}$ and $N=\N{1}$.  The visible particles leaving each branch are all opposite-sign same-flavor (OSSF) leptons ($\mu$ or $e$).  This allows us to identify  any hadronic activity as upstream transverse momentum.

The chapter is structured as follows:
Sec.~\ref{SecM3CIntro} introduces the definition of $M_{3C}$ and how to calculate it.  Sec.~\ref{SecM3CDependence} discuses the dependence of $M_{3C}$ on complications from combinatorics, large upstream transverse momentum (UTM), $\slashed{P}_T$ cuts, parton distributions, and energy resolution.   Sec \ref{SecM3CPerformance} applies $M_{3C}$ variables to \herwig\ data from the benchmark supersymmetry spectrum SPS 1a.  Finally we summarize the chapter's contributions.

\section{Introducing $M_{3C}$}
\label{SecM3CIntro}
We will now introduce the definition of $M_{3C}$, how to calculate it, its relationship to previous mass shell techniques.

\subsection{Definition of $M_{3C}$}

The upper bound and lower bound
on the mass of the third lightest new particle state in the symmetric decay chain is the constrained mass variable $M_{3C,LB}$ and $M_{3C,UB}$.
This variable applies to the symmetric, on-shell intermediate state, topology from Fig.~\ref{FigEventTopologyTovey} which depicts
two partons that collide and produce some observed upstream transverse momentum (UTM) with four momenta $k$ and an on-shell, pair-produced new state $Y$.
On each branch, $Y$ decays to on-shell intermediate particle state $X$ and a visible particle $v_1$ with masses $M_{X}$ and $m_{v_1}$. Then $X$ decays to the dark-matter particle $N$ and visible particle $v_2$ with masses $M_N$ and $m_{v_2}$.
The  four-momenta of $v_1$, $v_2$ and $N$ are respectively $\alpha_1$, $\alpha_2$ and $p$ on one branch and $\beta_1$, $\beta_2$ and $q$ in the other branch.
The missing transverse momenta $\slashed{P}_T$ is given by the transverse part of $p+q$.

We initially assume that we have measured the mass differences from other techniques.  For an on-shell intermediate state, there is no single end-point that gives the mass difference. The short decay chain gives a kinematic endpoint $\max m_{12}$ described in Eq(\ref{EqTwoBodyDecayEdge}) that constrains a combination of the squared mass differences.  Unless two of the states are nearly degenerate, the line with constant mass differences lies very close to the surface given by Eq(\ref{EqTwoBodyDecayEdge}).  The two mass differences are often tightly constrained in other methods.
The mass differences are constrained to within $0.3$ GeV from studying long cascade decay chains where one combines constraints from several endpoints of different invariant mass combinations \cite{Gjelsten:2004ki}.
The concepts from Chapter \ref{ChapterMT2onCascadeDecays} also provide another technique to determine the mass differences.  After initially assuming that we know the mass difference, we show that our technique can also find the mass differences.
The $M_{3C}$ distribution shape is a function of both the mass scale and mass differences.
We can constrain both the mass differences and the mass scale by fitting the $\max m_{12}$ edge constrains
and the ideal $M_{3C}(M_N,\Delta M_{YN},\Delta M_{XN})$ distribution shapes to the observed $M_{3C}(\Delta M_{YN},\Delta M_{XN})$.  To find all three parameters from this fit, we will take $M_{N}$, $\Delta M_{YN}$, and $\Delta M_{XN}$  as independent variables.

For this first phase of the analysis, let's assume the mass differences are given.
For each event, the variable $M_{3C,LB}$ is the minimum value of the mass of $Y$ (third lightest state) after minimizing over the unknown division of the missing transverse energy $\slashed{P}_T$ between the two dark matter particles $N$:
 \begin{eqnarray}
   m^2_{3C,LB}(\Delta M_{YN},\Delta M_{XN}) &= & \min_{p,q}\  (p+\alpha_1+\alpha_2)^2 \label{EqM3CTop} \\
{\rm{Constrained}\ \rm{to}}  & &  \nonumber \\
    (p+q)_T & = & \slashed{P}_T \label{EqM3Cc1} \\
    \sqrt{(\alpha_1+\alpha_2+p)^2} -     \sqrt{(p^2)}& = &\Delta M_{YN} \\
        \sqrt{(\alpha_2+p)^2} -     \sqrt{(p^2)} & = & \Delta M_{XN} \\
    (\alpha_1+\alpha_2+p)^2 & =&  (\beta_1+\beta_2+q)^2 \\
        (\alpha_2+p)^2 & = & (\beta_2+q)^2 \\
                p^2 & = & q^2 \label{EqM3Cc7}
 \end{eqnarray}
where $\Delta M_{YN} = M_Y - M_N$ and $\Delta M_{XN} = M_X - M_N$.
There are eight unknowns in the four momenta of $p$ and $q$ and seven equations of constraint. Likewise we define $M_{3C,UB}$ as the maximum value of $M_Y$ compatible with the same constraints.   We discuss how to numerically implement this minimization and maximization in Sec.~\ref{SecNumericallyCalculatingM3C}.  Because the true $p$ and $q$ are within the domain over which we are minimizing ( or maximizing), the minimum (maximum) is guaranteed to be less than (greater than) or equal to $M_Y$.

\begin{figure}
\centerline{\includegraphics[width=5in]{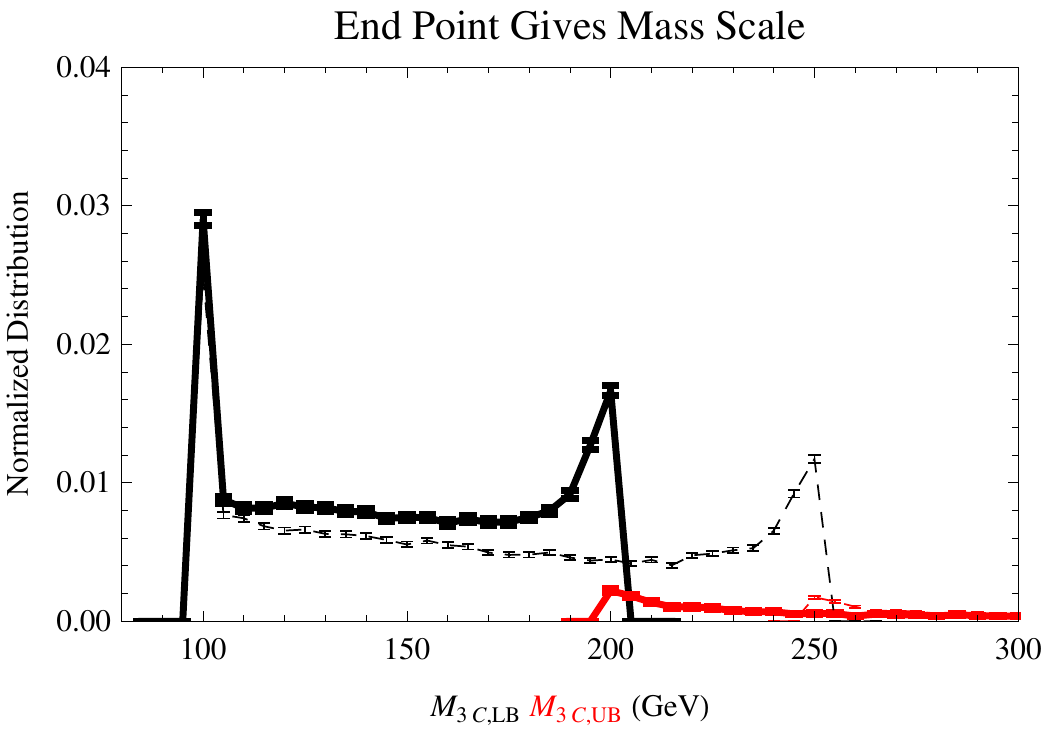}}
\caption{\label{FigM3CIdeal150100} Ideal $M_{3C,LB}$ and  $M_{3C,UB}$ distribution for 25000 events in two cases both sharing $\Delta M_{YN}=100$ GeV and $\Delta M_{XN}=50$ GeV.  The solid, thick line shows $M_Y=200$ GeV, and the dashed, thin line shows $M_Y=250$ GeV.}
\end{figure}

Figure \ref{FigM3CIdeal150100} shows an ideal $M_{3C,LB}$ and $M_{3C,UB}$ distributions for $25000$ events in two cases both sharing $\Delta M_{YN}=100$ GeV and $\Delta M_{XN}=50$ GeV. The dashed line represents the distributions from events with $M_Y=250$ GeV, and in the solid line represents the distributions from events with $M_{Y}=200$ GeV.  One can clearly see sharp end-points in both the upper bound and lower bound distributions that give the value $M_Y$. The upper-bound distribution is shown in red.

We expect an event to better constrain the mass scale
if we are given additional information about that event.
In comparison to $M_{2C}$ where $Y$ decays directly to $N$, the on-shell intermediate state has an additional state $X$.  The extra state $X$ and information about its mass difference $\Delta M_{XN}$ enables $M_{3C}$ to make an event-by-event bound on $M_Y$ stronger than in the case of $M_{2C}$.  We will see that this stronger bound is partially offset by greater sensitivity to errors in momentum measurements.

The variable $M_{3C}$, like other variables we have discussed $M_{2C}$, $M_{T2}$ $M_T$ and $M_{CT}$, is invariant under longitudinal boosts of its input parameters.
We can understand this because all the constraint equations are invariant under longitudinal boosts.  The unknown $p$ and $q$ are minimized over all possible values fitting the constraints so changing frame of reference will not change the extrema of the Lorentz invariant quantity $(p+\alpha)^2$.

\section{How to calculate $M_{3C}$}
\label{SecNumericallyCalculatingM3C}

To find the $M_{3C}$, we observe that if we assume masses of $Y$, $X$, and $N$ to be $(\chi_Y,\chi_X,\chi_N)$ \footnote{We again use $\chi$ to distinguish hypothetical masses $(\chi_Y,\chi_X,\chi_N)$ from the true masses $(M_Y,M_X,M_N)$.} with the given mass differences then there are eight constraints
 \begin{eqnarray}
    (p+q)_T & = & \slashed{P}_T \\
    (\alpha_1+\alpha_2+p)^2 =(\beta_1+\beta_2+q)^2 & = & \chi^2_Y   \\
        (\alpha_2+p)^2=(\beta_2+q)^2  & = & \chi^2_{X} = (\chi_Y - \Delta M_{YN}+\Delta M_{XN})^2 \\
              p^2 = q^2 & = & \chi_N^2 = (\chi_Y - \Delta M_{YN})^2
 \end{eqnarray}
and eight unknowns, $p_\mu$ and $q_\mu$.
The spatial momenta $\vec{p}$ and $\vec{q}$ can be found as linear functions of the  $0^{\rm{th}}$ component of $p$ and $q$ by solving the matrix equation

\footnotesize
 \begin{eqnarray}
 \left(\begin{matrix} 1 & 0 & 0 & 1 & 0 & 0 \cr
                      0 & 1 & 0 & 0 & 1 & 0 \cr
                      -2 \alpha_x & -2 \alpha_y & -2\alpha_z & 0 & 0 & 0 \cr
   0 & 0 & 0 & -2 \beta_x & -2\beta_y & -2 \beta_z  \cr
                        -2 (\alpha_2)_x & -2 (\alpha_2)_y & -2(\alpha_2)_z & 0 & 0 & 0 \cr
   0 & 0 & 0 & -2 (\beta_2)_x & -2(\beta_2)_y & -2 (\beta_2)_z \end{matrix} \right)
 \left( \begin{matrix} p_x \cr p_y \cr p_z \cr q_x \cr q_y \cr q_z \end{matrix} \right)
 = \left( \begin{matrix}-(k+\alpha+\beta)_x \cr -(k+\alpha+\beta)_y \cr
   -2 \alpha_o p_o + (\chi_Y^2 - \chi_N^2) - \alpha^2 \cr
   -2 \beta_o q_o + (\chi_Y^2 - \chi_N^2) -\beta^2 \cr
   -2 (\alpha_2)_o p_o + (\chi_X^2-\chi_N^2) -(\alpha_2)^2 \cr
   -2 (\beta_2)_o q_o +  (\chi_X^2-\chi_N^2) - (\beta_2)^2 \end{matrix} \right)
   \label{EqMatrixSolForpqVec}
 \end{eqnarray}
\normalsize
\baselineskip=21pt plus1pt

\noindent
where $\alpha=\alpha_1+\alpha_2$ and $\beta=\beta_1+\beta_2$.
We substitute $\vec{p}$ and $\vec{q}$ into the on-shell constraints
 \begin{eqnarray}
   p_o^2 - (\vec{p}(p_o,q_o))^2 = \chi_N^2 \label{Eqpo} \\
   q_o^2 - (\vec{q}(p_o,q_o))^2 = \chi_N^2 \label{Eqqo}
 \end{eqnarray}
giving two quadratic equations for $p_o$ and $q_o$.
These give four complex solutions for the pair $p_o$ and $q_o$.
We test each event for compatibility with a hypothetical triplet of masses $(\chi_Y,\chi_X,\chi_N)=(\chi_Y,\chi_Y-\Delta M_{YN}+\Delta M_{XN},\chi_Y-\Delta M_{YX})$.
If there are any purely real physical solutions where ($p_o>0$ and $q_o>0$), then we consider the mass triplet $(\chi_Y,\chi_X,\chi_N)$ viable.

As we scan $\chi_Y$, a solution begins to exist at a value less than or equal to $M_Y$ and then sometimes ceases to be a solution above $M_Y$. Sometimes there are multiple islands of solutions.
To find the $M_{3C}$, we can test each bin starting at $\chi_Y=\Delta M_{YN}$ along the path parameterized by $\chi_Y$ and the mass differences to find the first bin where at least one physical solution exists.
This is the lower bound value of $M_{3C}$ for the event.

Likewise for an upper bound.  We begin testing at the largest conceivable mass scale we expect for the $Y$ particle state.  If a solution exists, we declare this a trivial $M_{3C,UB}$.  If no solution exists, then we search downward in mass scale until a solution exists.

A faster algorithm involves a bisection search for a solution within the window that starts at $\Delta M_{YN}$ and ends at our highest conceivable mass. We then use a binary search algorithm to find at what $\chi_Y$ the solution first appears for $M_{3C,LB}$ or at what $\chi_Y$ the solution disappears giving $M_{3C,UB}$. There are rare events where there are multiple islands of solutions. This occurs in about $0.01\%$ of the events with $0$ UTM and in about $0.1\%$ for $k_T = 250$ GeV. In our algorithm we neglect windows of solutions more narrow than $15$ GeV.  We report the smallest edge of the lower-mass island as the lower bound and the upper edge of the larger-mass island as the upper bound. Because of the presence of islands, we are not guaranteed that solutions exist everywhere between $M_{3C,LB}$ and $M_{3C,UB}$.
With the inclusion of energy resolution errors and background events, we also find cases where there are no solutions anywhere along the path being parameterized. If there is no solutions anywhere in the domain we make $M_{3C,LB}$ to be the largest conceivable mass scale, and we set $M_{3C,UB}=\Delta M_{YN}$.

\subsection{Comparison to other Mass Shell Techniques}
\label{SecM3CComparisonToOtherTechniques}

The variable $M_{3C}$ is a hybrid mass shell technique\cite{Nojiri:2007pq}.
In Chapter \ref{ChapterMassDeterminationReview} we reviewed other mass shell technique that measure the mass in the case of three new states.
Cheng, Gunion, Han Marandella, McElrath (CGHMM) \cite{Cheng:2007xv} describe counting solutions at assumed values for the mass for $Y$, $X$, and $N$.
By incorporating a minimization or maximization, we enhance CGHMM's approach because we have a variable whose value changes slightly with slight changes of the inputs instead of the binary on-off that CGHMM has with the existence of a solution\footnote{I am grateful to Chris Lester for pointing out to me the importance of this feature.}.
We also incorporate knowledge of the added information from other measurements which accurately determine the mass differences.
Finally, the quantity $M_{3C}$ can form a distribution whose shape tells us information about the masses.
Because for most events there is only one `turn-on' point below $M_Y$, the distribution $M_{3C,LB}$ is very similar to the derivative of the figure 8 of CGHMM\cite{Cheng:2007xv} to the left of their peak and $M_{3C,UB}$ is similar to the negative of the derivative to the right of their peak.
They differ in that there may be multiple windows of solutions; also  CGHMM's Fig 8 is not exactly along the line of fixed mass differences; and the effect of backgrounds and energy resolution are dealt with differently.

We also hope to show that the use of the distribution's shape enables us
to exploit the essentially non-existent dependence of the distributions on the unknown collision energies and incorporate the dependency on UTM directly.  This diminishes the dependence of the measurement on the unknown model while still allowing us to exploit the majority of the distribution shape in the mass determination.

After studying previous MSTs, we were tempted to use Bayes theorem with  a parton distribution function as a likelihood function as was done in Goldstein and Dalitz \cite{Goldstein:1993mj} and Kondo, Chikamatsu, Kim \cite{Kondo:1993in} (GDKCK).
They used the parton distribution function to weight the different mass estimates of the top-quark mass ($M_Y$ in our topology).
We found that such a weighting leads to a prediction for $M_Y$ much smaller than the true value.  This can be understood because the parton distributions make collisions with smaller center-of-mass energies (small $x$) more likely, therefore the posterior will prefer smaller values of $M_Y$ which are only possible for smaller values of $x$.
Only if one includes the cross-section for production, i.e. the likelihood of the event existing at all, in the Bayes likelihood function will we have the appropriate factor that suppresses small values of $x$ and therefore small values of $M_Y$.  This balance therefore leads to the maximum likelihood (in the limit of infinite data) occurring at the correct $M_Y$.  Unfortunately, inclusion of the magnitude of the cross section introduces large model dependence.
In the case of the top-quarks mass determination, the GDKCK technique gives reasonable results. This is because they were not scanning the mass scale, but rather scanning $\chi_{Y}$ (the top quark mass) while assuming $\chi_N=M_N=0$ and $\chi_X=M_X=M_W$.
The likelihood of solutions as one scans $\chi_{Y}$ rapidly goes to zero below the true top-quark mass $M_{top}$.
The parton distribution suppresses the likelihood above the true $M_{top}$.
The net result gives the maximum likelihood near the true top-quark mass but suffers from a
systematic bias \cite{Raja:1996vz}\cite{Raja:1997qs} that must be removed by modeling \cite{Brandt:2006uc}.

\section{Factors for Successful Shape Fitting}
\label{SecM3CDependence}

One major advantage of using the $M_{3C}$ distribution (just as the $M_{2C}$ distribution) is that the bulk of the relevant events are used to determine the mass
and not just those near the endpoint.  To make the approach mostly model independent, we study on what factors the distributions shape depends.
We show that there is a strong dependence on upstream transverse momentum (UTM) which can be measured and used as an input and therefore does not increase the model dependency.  We show there is no numerically significant dependence on the collision energy which is distributed according to the parton distribution functions.
This makes the distribution shape independent of the production cross section and the details of what happens upstream from the part of the decay chain that we are studying.
We model these effects with a simple Mathematica Monte Carlo event generator
assuming $M_Y=200$ GeV, $M_X=150$ GeV, and $M_N=100$ GeV.

\begin{itemize}
\item {\bf{Effect of Combinatorics Ambiguities}}
\end{itemize}
Just as in the topology in Fig~\ref{FigEventTopologyThreeBody} studied earlier, where $\N{2}$ decays via a three body decay, the branch assignments can be determined by either distinct OSSF pairs or by studying which OSSF pairs have both $m_{12} \leq \max m_{12}$.
In $90 \%$ of the events, there is only one combination that satisfies this requirement.  This allows us to know the branch assignment of $95 \%$ of the four lepton events without ambiguity.

\begin{figure}
\centerline{\includegraphics[width=3.1in]{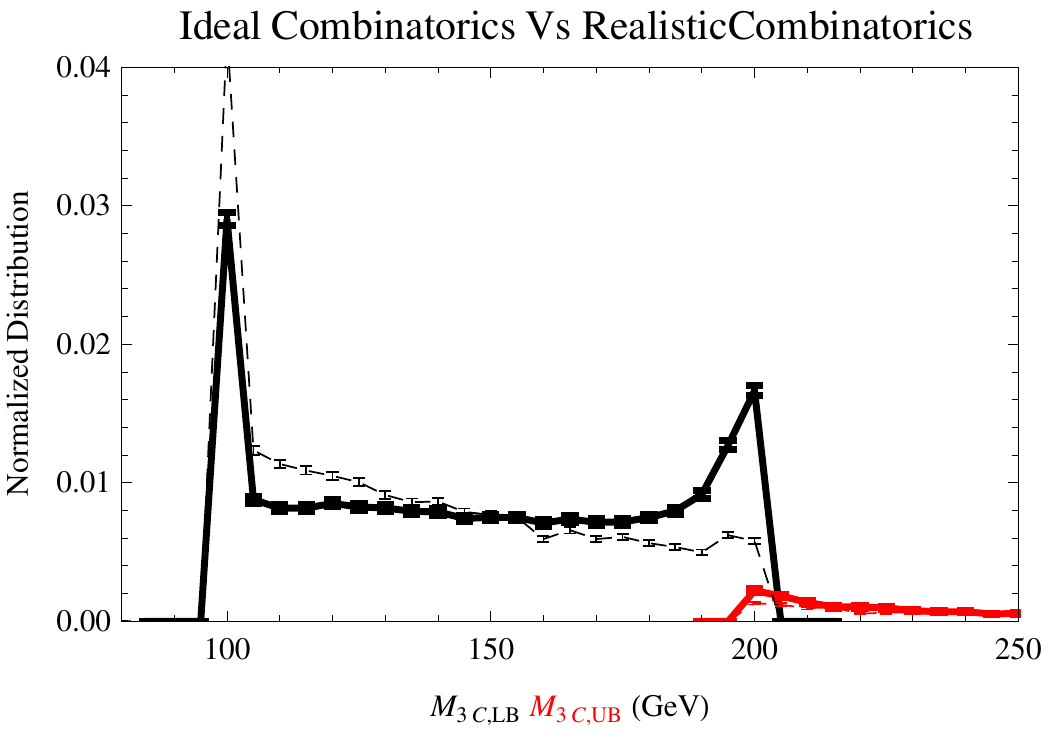}
\includegraphics[width=3.1in]{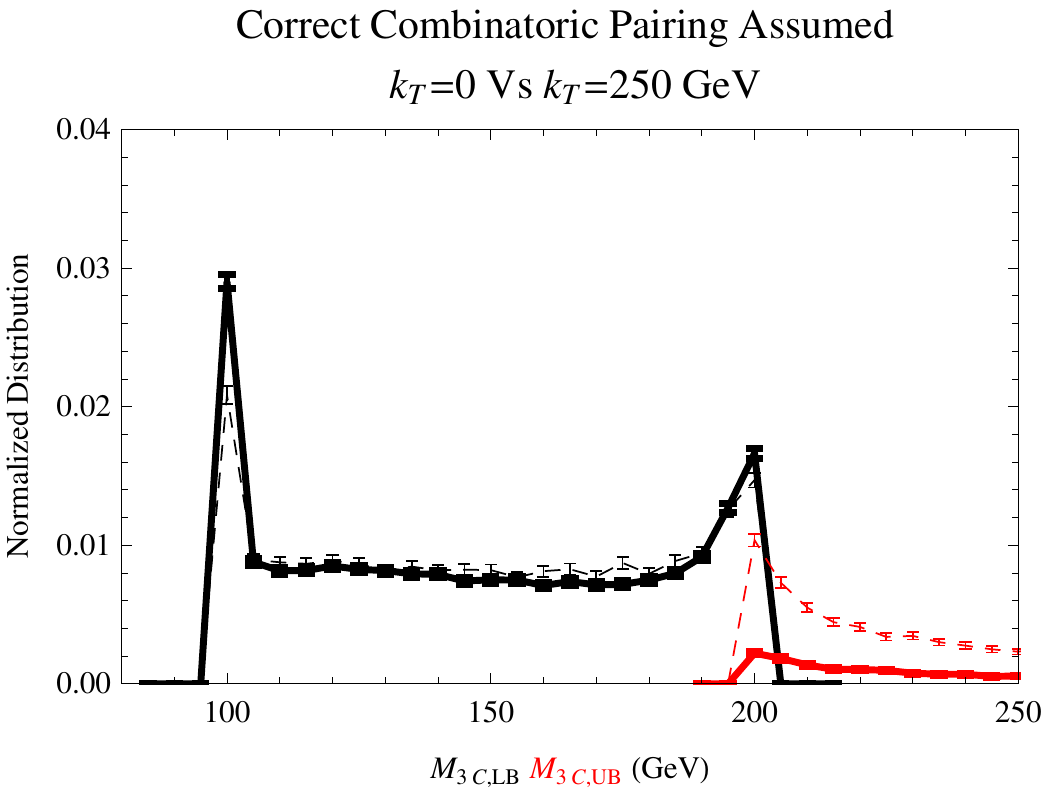}}
\caption{\label{FigM3CCombinatoricsDemo}\label{FigM3CUTMDistribution} (Left) The $M_{3C}$ distributions before (solid) and after (dashed) introducing the combinatoric ambiguity.  (Right) The $M_{3C}$ distributions with  and without UTM.
The no UTM case ($k_T=0$) is shown by the solid line; the large UTM case with $k_{T}=250$ GeV is shown by the dashed line.
}
\end{figure}

Unlike the three-body decay case, the order of the two leptons on each branch matters. The intermediate mass $M_X^2 = (\alpha_2 + p)^2$ depends on $\alpha_2$ and does not depend on $\alpha_1$.
To resolve this ambiguity we consider the four combinations that preserve the branch assignment but differ in their ordering.  The $M_{3C,LB}$ for the event is the minimum of these combinations.  Likewise the $M_{3C,UB}$ is the maximum of these combinations. As one expects, Fig.~\ref{FigM3CCombinatoricsDemo} (Left) shows how the combinatorics ambiguity degrades the sharpness of the cut-off at the true mass.  Not all applications share this ambiguity; for example in top-quark mass determination (pair produced with $Y=top$, $X=W^\pm$, $N=\nu$) the $b$-quark-jet marks $\alpha_1$ and the lepton marks $\alpha_2$.

\begin{itemize}
\item {\bf{Effect Large Upstream Transverse Momentum}}
\end{itemize}

In a similar behavior to $M_{2C}$, the distributions of the variable $M_{3C}$
show a strong dependence on large upstream transverse momentum (UTM).
In our case study this is identified as the combination of all the hadronic activity.
Fig~\ref{FigM3CUTMDistribution} (Right) shows the stronger upper-bound cut-off in the presence of large UTM.  Unlike $M_{2C}$, in $M_{3C}$ with $k_T=0$ we still have events with non-trivial upper bound values.

We also tested the distribution for different values of $k^2$.  In Fig~\ref{FigM3CUTMDistribution} (Right) we fixed $k^2=(100 \GeV)^2$.  We also performed simulations with $k^2=(500 \GeV)^2$ and found the difference of the two $M_{3C}$ distributions consistent with zero after 15000 events.  In other words, the distribution depends mostly on $k_x$ and $k_y$ and appears independent of $k_0$.

\begin{itemize}
\item {\bf{The Effects of Detector Energy Resolution}}
\end{itemize}

\begin{figure}
\centerline{
\includegraphics[width=3.1in]{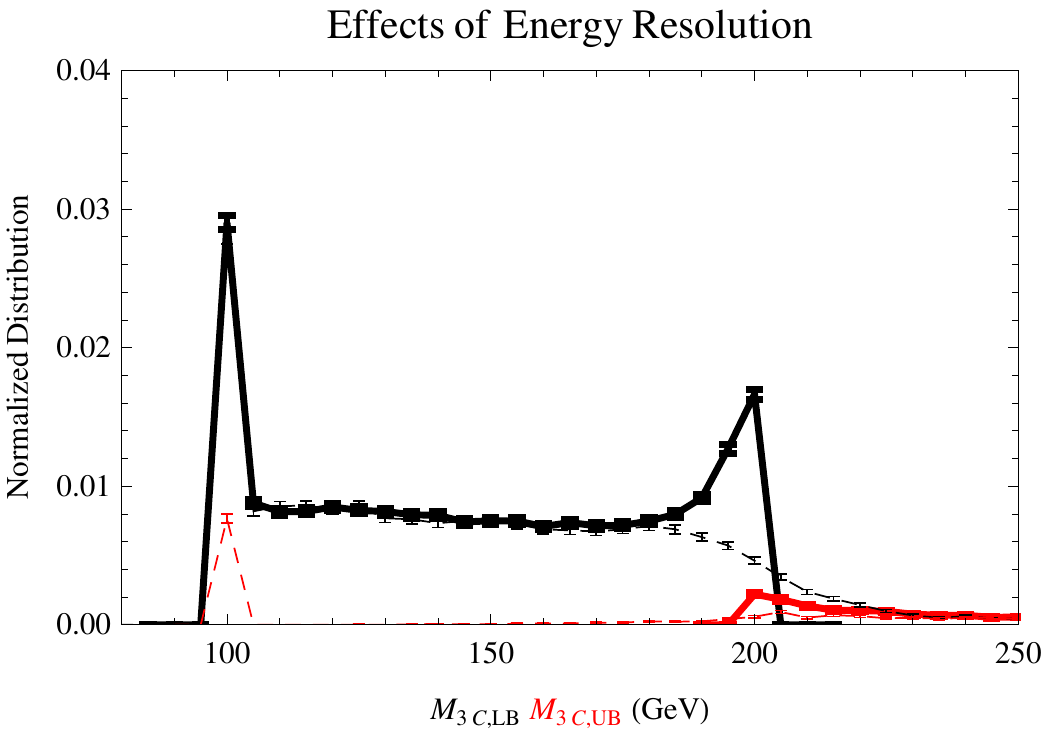}
\includegraphics[width=3.1in]{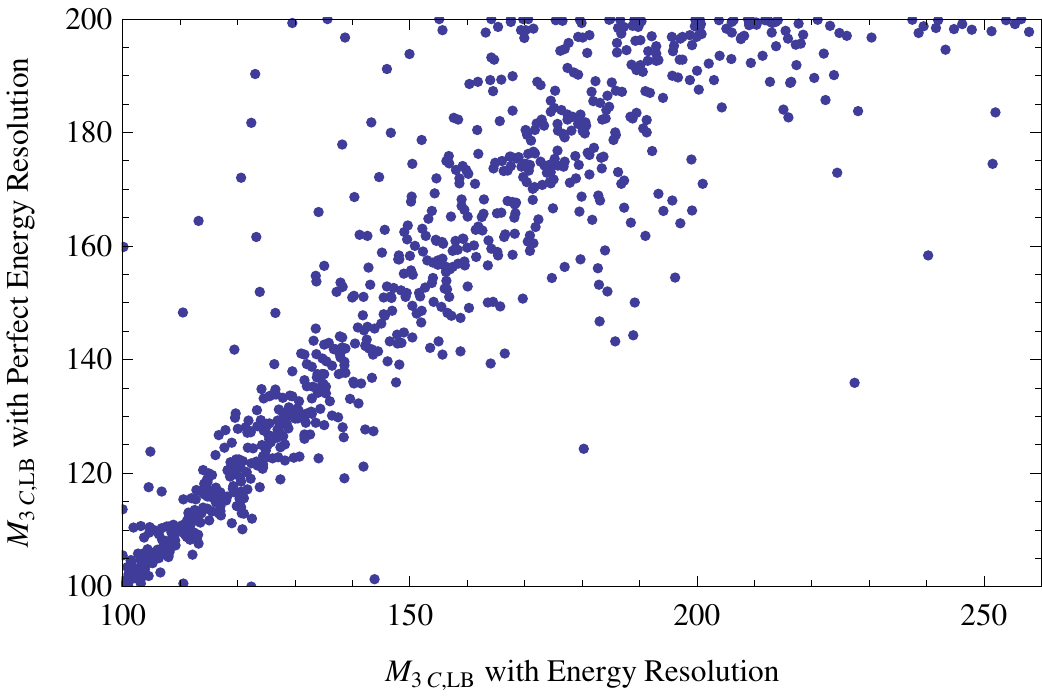}}
\caption{\label{FigM3CEnergyResolution}The effect of energy resolution on the $M_{3C}$ distribution.  (Left) The dotted line shows the energy resolution has washed out the sharp cut-off.
(Right) $M_{3C,LB}$ with perfect energy resolution plotted against the result with realistic energy resolutions.
}
\end{figure}

Compared to $M_{2C}$, the information about the extra states gives a stronger set of bounds.  Unfortunately, the solution is also more sensitive to momentum measurement error. We model the finite energy resolution using Eqs(\ref{EqDetectorEnergyResolutionElectron}-\ref{EqDetectorEnergyResolutionHadron}).
The hadronic energy resolution is larger than the leptonic energy resolution which will increase the uncertainty in the missing transverse momentum.

Fig.~\ref{FigM3CEnergyResolution} shows the effect on the $M_{3C}$ distribution of realistic leptonic energy
resolution while keeping $k_T=0$.
On the left we show the energy resolution (dashed line) compared to
 the perfect energy resolution (solid line).  The energy resolution washes out the sharp cut-off.
On the right we show $M_{3C,LB}$ with perfect energy resolution plotted against the result with realistic energy resolution.
This shows that the cut-off is strongly washed out because the events with $M_{3C}$ closer to the true value of $M_Y$ ($200$ GeV in this case) are more sensitive to energy resolution than the events with $M_{3C}$ closer to $\Delta M_{YN}$.  The peak in the upper-bound distribution at $M_{3C,UB}=100$ GeV comes from events that no longer have solutions after smearing the four momenta.

Because the energy resolution affects the distribution shape, its correct modeling is important.  In the actual LHC events the
 $\slashed{P}_T$ energy resolution will depend on the hadronic activity in the events being considered.
Two events with the same $k_T=0$ may have drastically different $\slashed{P}_T$ resolutions.
Modeling the actual detector's energy resolution for the events used is important to predict the set of ideal distribution shapes which are compared against the low-statistics observed data.

\begin{itemize}
\item {\bf{Parton Distribution Function Dependence}}
\end{itemize}

For a mostly model-independent mass determination technique, we would like to have a distribution that is independent of the specific production mechanism of the assumed event topology.
The parton distributions determine the center-of-mass energy $\sqrt{s}$ of the
hard collisions; but the cross section depends on model-dependent couplings and parameters.  The events we consider may come from production of different initial states (gluons or squarks) but end in the assumed decay topology.
The $M_{3C}$ distribution, like the $M_{2C}$ distribution, shows very little dependence on the underlying collision parameters or circumstances.

\begin{figure}
\centerline{
\includegraphics[width=3.1in]{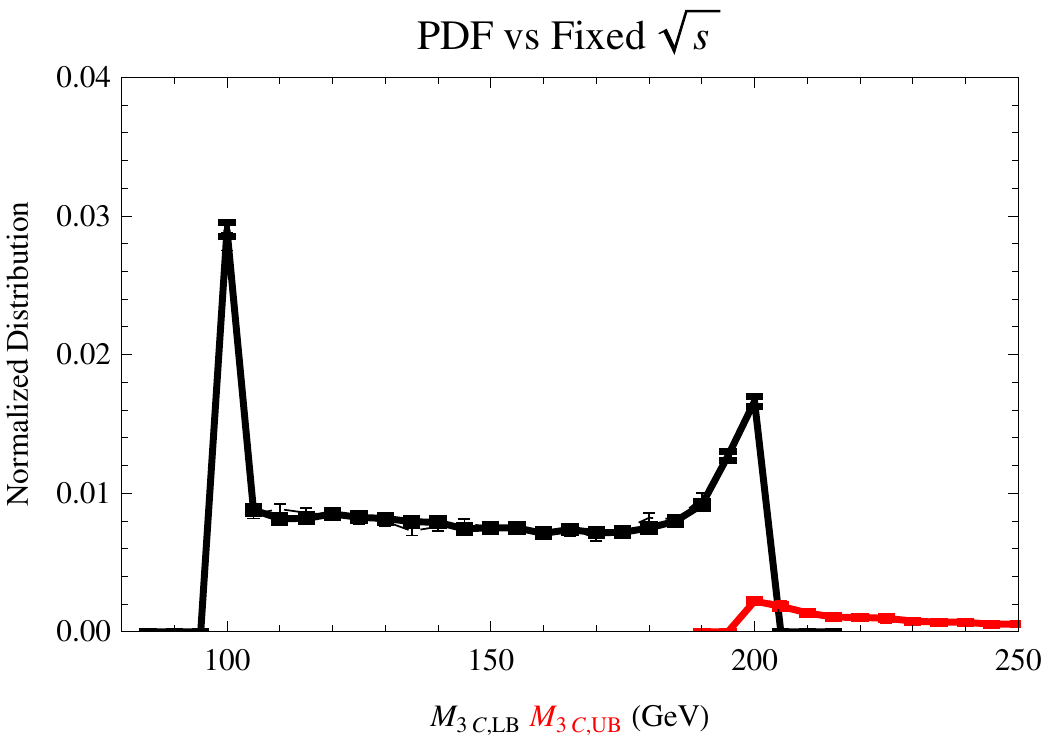}
\includegraphics[width=3.1in]{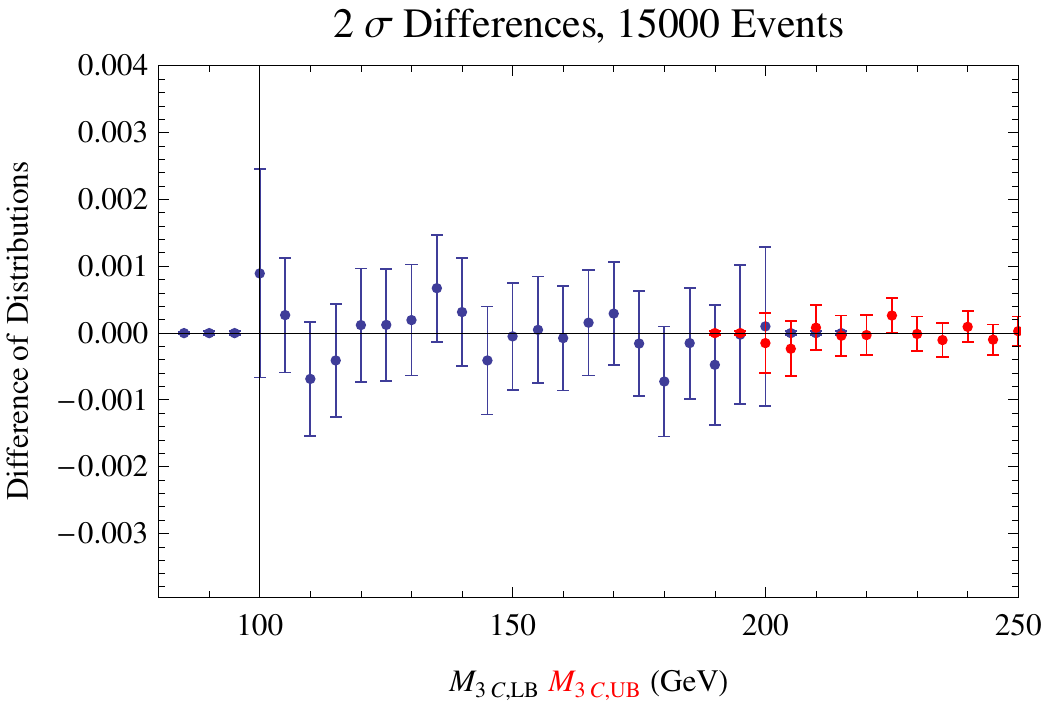}}
\caption{\label{FigM3CPdfDistributionDependence} The dependence of the $M_{3C}$ distributions on the parton collision energy.  The solid line shows the collision distributed according to Eq(\ref{EqSdep}), and the dashed line shows the collision energy fixed at $\sqrt{s}=600$ GeV. }
\end{figure}

Fig \ref{FigM3CPdfDistributionDependence} (Left) shows
the dependence of the $M_{3C}$ distributions on the parton collision energy.  The solid line shows the $M_{3C}$ distributions of events with collision energy $\sqrt{s}$ distributed according to Eq(\ref{EqSdep}), and the dashed line shows the $M_{3C}$ distributions of events with fixed $\sqrt{s}=600$ GeV.
Fig \ref{FigM3CPdfDistributionDependence} (Right) shows the difference of these two distributions with $2 \sigma$ error bars as calculated from $15000$ events.  The two distributions are equal to within this numerical precision.

\begin{itemize}
\item {\bf{Effects of $\slashed{P}_T$ Cuts}}
\end{itemize}

\begin{figure}
\centerline{
\includegraphics[width=3.1in]{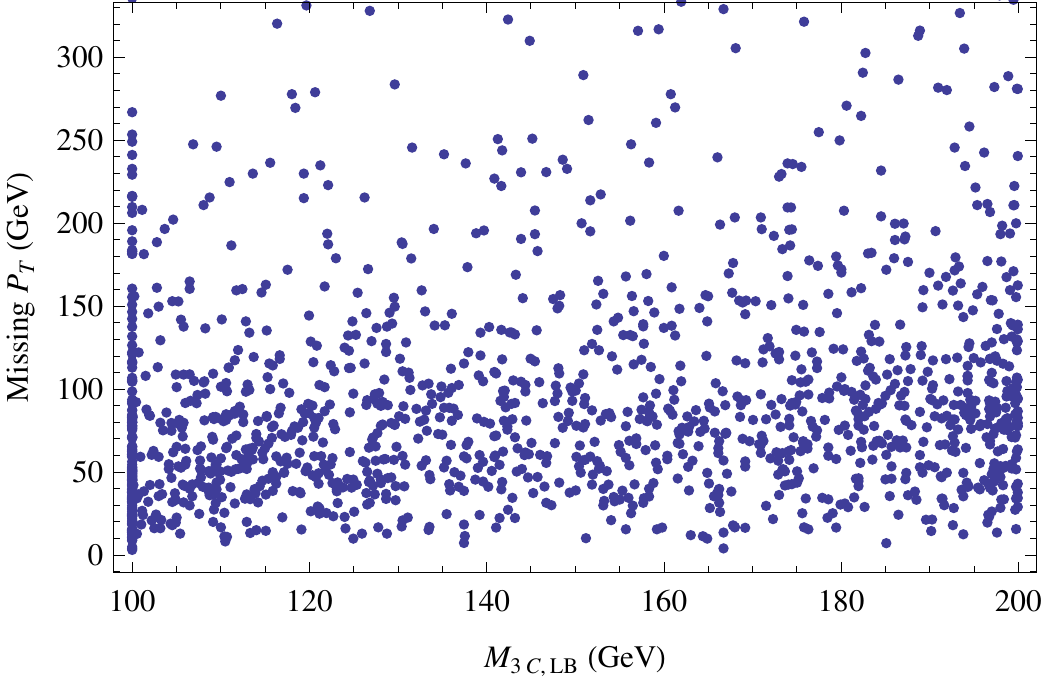}
\includegraphics[width=3.1in]{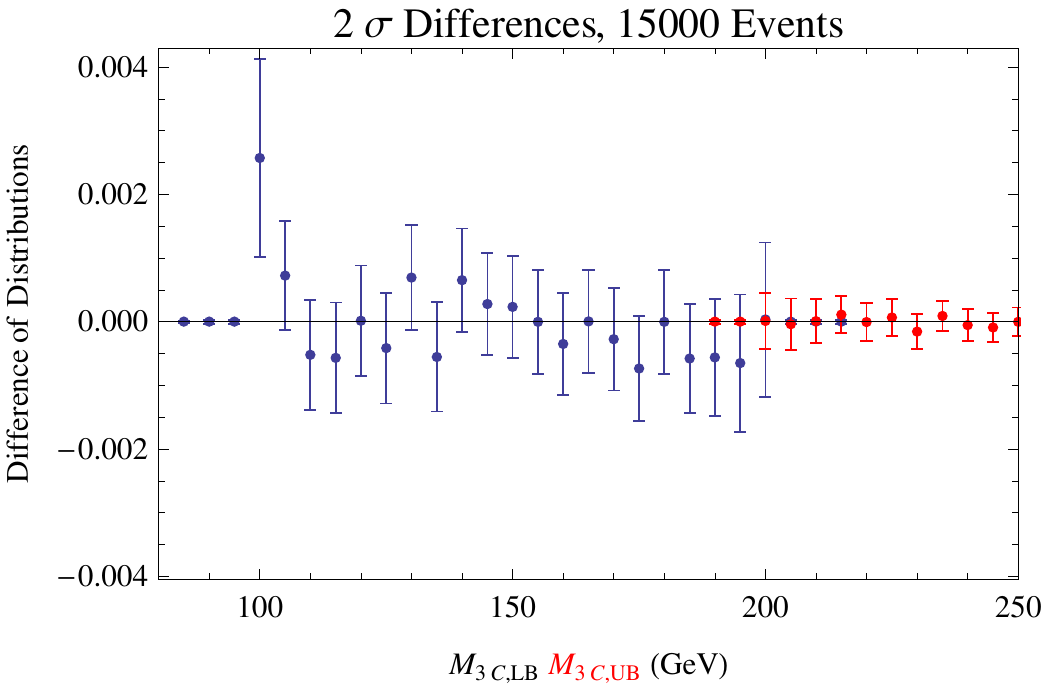}
}
\caption{\label{FigM3CPtCutDependenceDifference} The effect of missing transverse momentum
 cuts on the $M_{3C}$ distributions. (Left) The $M_{3C,LB}$ result versus the
 $\slashed{P}_T$.
(Right) The difference of the $M_{3C,UB}$ and $M_{3C,LB}$ distributions with and without the cut $\slashed{P}_T > 20$ GeV.
The smallest bins of $M_{3C,LB}$ are the only bins to be
statistically significantly affected. }
\end{figure}

As described in \cite{Ghosh:1999ix,Bisset:2005rn,Ross:2007rm,Barr:2008ba}, the Standard Model four leptons with missing transverse momentum backgrounds are very strongly suppressed after a missing transverse momentum cut.
This requires an analysis of what type of an effect a $\slashed{P}_T > 20$ GeV cut will have on the distribution shape.
Fig.~\ref{FigM3CPtCutDependenceDifference} shows that the effect of this cut is dominantly on the smallest $M_{3C}$ bins.
On the left we see the $M_{3C,LB}$ result versus the $\slashed{P}_T$.
Unlike the $M_{2C}$ case in Fig~\ref{FigPTCutDependence}, the $M_{3C}$ solutions in Fig.~\ref{FigM3CPtCutDependenceDifference} do not correlate with the $\slashed{P}_T$.
The right side of the figure shows difference between the $M_{3C,UB}$ and $M_{3C,LB}$ distributions with and without the cut $\slashed{P}_T > 20$ GeV.
The smallest bins of $M_{3C,LB}$ are the only bins to be
statistically significantly affected.  The left-side suggests this lack of dependence on $\slashed{P}_T$ cuts is somewhat accidental and is due to the nearly uniform distribution of $M_{3C}$ solutions being removed by the cut.
The stronger dependence of the smallest $M_{3C}$ bins on the $\slashed{P}_T$ cut means we can either model the effect or exclude the first bins (about $10$ GeV worth) from the distribution used to predict the mass.  We will choose the latter because we will find that the background events also congregate in these first several bins.

\begin{itemize}
\item {\bf{Spin correlations}}
\end{itemize}

In our simulation to produce the ideal curves, we assumed each decay was uncorrelated with its spin in the rest frame of the decaying particle.
Spin correlations at production may affect this; however, such spin correlations are washed out when each branch of our assumed topology is at the end of longer decay chain. These upstream decays are the source of considerable UTM.

Some spin correlation information can be easily taken into account. The $m_{12}$ (or $m_{34}$) distribution's shape is sensitive to the spin correlations along the decay chain \cite{Athanasiou:2006ef}.  The observed $m_{12}$ (or $m_{34}$) distribution  can be used as an input to producing the ideal distribution shape.  In this way spin correlations along the decay chain can be taken into account in the simulations of the ideal distributions.

\begin{figure}
\centerline{\includegraphics[width=3.2in]{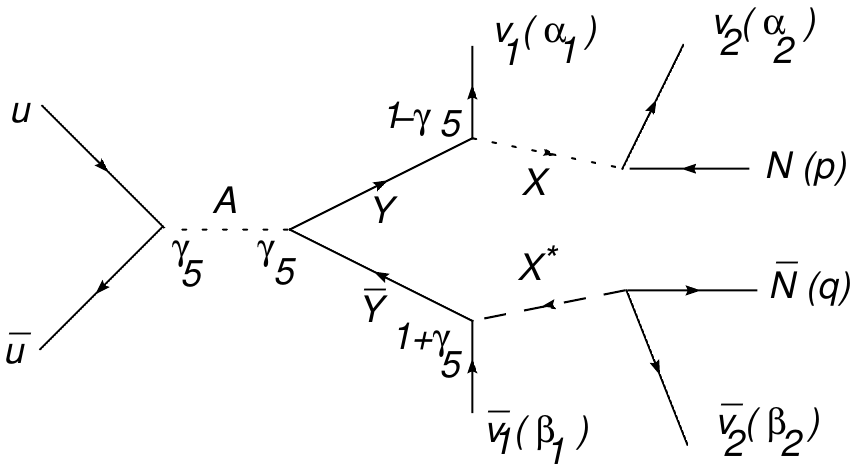}
\includegraphics[width=3.2in]{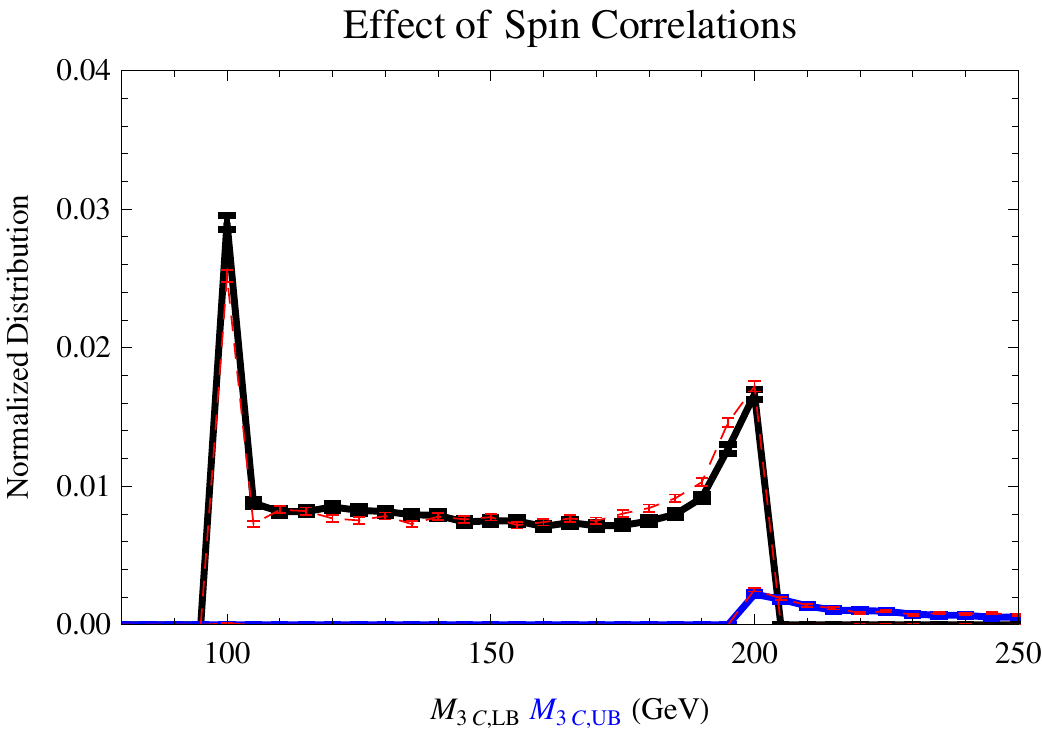}}
\caption{\label{FigM3CSpinCorrelations}Effect of a spin correlated process on the $M_{3C}$ distributions. Modeled masses are $M_Y=200$ GeV, $M_X=150$ GeV, and $M_N=100$ GeV.  The thick black and thick blue lines show the distributions of the uncorrelated lower bound and upper bound $M_{3C}$. The dotted red lines show the distributions of the spin correlated process.}
\end{figure}

Spin correlations between the two branches can also affect the distribution shape.
To demonstrate this, we modeled a strongly spin-correlated direct production process.
Fig \ref{FigM3CSpinCorrelations} (Left) shows the spin-correlated process that we consider.
Fig \ref{FigM3CSpinCorrelations} (Right) shows the $M_{3C}$ upper and lower bound distributions from this process compared to the $M_{3C}$ distribution from the same topology and masses but without spin correlations.  We compare distributions with perfect energy resolution, $m_{v_1}=m_{v_2}=0$ GeV, $M_Y=200$ GeV, $M_X=150$ GeV, and $M_N=100$ GeV.  Our maximally spin-correlated process involves pair production of $Y$ through a pseudoscalar $A$.  The fermion $Y$ in both branches decays to a complex scalar $X$ and visible fermion $v_1$ through a purely chiral interaction. The scalar $X$ then decays to the dark-matter particle $N$ and another visible particle $v_2$.
The production of the pseudoscalar ensures that the $Y$ and $\bar{Y}$ are in a configuration $\sqrt{2}^{\,-1}(|\uparrow\downarrow\rangle + |\downarrow\uparrow\rangle)$.
The particle $Y$ then decays with $X$ preferentially aligned with the spin. The $\bar{Y}$ decays with $X^*$ preferentially aligned against the spin.
Because $X$ is a scalar, the particle $N$ decays uniformly in all directions from the rest frame of $X$.
The correlated directions of $X$ causes the two sources of missing transverse momentum to be preferentially parallel.  The resulting greater magnitude of missing transverse momentum increases the cases where $M_{3C}$ has a solution closer to the endpoint.  For this reason the spin correlated distribution (red dotted distribution) is above the uncorrelated distribution (black thick lower bound distribution and blue thick upper bound distribution).  The upper bound distribution is statistically identical after 25000 events. The lower bound distribution clearly has been changed, but not as much as the $M_{2C}$ distribution in Fig.~\ref{FigM2CSpinCorrelations}.  This is due to the subsequent decay of the $X$ particle which lessens the likelihood that the two $N$s will be parallel.
For the remainder of the chapter we assume no such spin correlations are present.

\begin{itemize}
\item {\bf{Backgrounds}}
\end{itemize}

The Standard Model backgrounds for four-leptons and missing transverse momentum are studied in  \cite{Ghosh:1999ix,Bisset:2005rn}.
In the two previous chapters we
 summarized the SM backgrounds and the dominance of SUSY backgrounds
for this channel (see also \cite{Ross:2007rm}\cite{Barr:2008ba}).
As was mentioned earlier, the Standard Model backgrounds for four leptons with missing transverse momentum are very strongly suppressed after a missing transverse momentum cut.

To improve the quality of the fit, a model for backgrounds can be created based on assumptions about the origin of the events and wedge-box analysis like those described in Bisset \etal \cite{Bisset:2008hm} and references therein.
We preformed such a model in the previous chapter, and found the distribution shape isolated the correct mass of $\N{1}$ to within $1$ GeV with versus without the background model.
Although the quality of the fit without modeling the backgrounds is decreased, we find that the mass of the LSP associated with the best-fit is tolerant to unknown backgrounds.
In the SPS 1a example studied in the next section, SUSY background events
form about $12\%$ of the events.
We again see less than a $1$ GeV shift in $M_{\N{1}}$ with versus without the background events.
As such, we do not try to model the background in this $M_{3C}$ study.

\section{Estimated Performance}
\label{SecM3CPerformance}

With an understanding of the factors affecting the shapes of the $M_{3C}$ distributions, we combine all the influences together and consider the mass
determination performance.
We follow the same modeling and simulation procedures used in Sec.~\ref{M2CSecModeling}
except now we include an on-shell intermediate state and calculate $M_{3C}$.
We use \herwig\ \cite{Corcella:2002jc,Moretti:2002eu,Marchesini:1991ch} to generate events according to the SPS 1a model \cite{Allanach:2002nj}.
This is an {mSUGRA} model with $m_o=100$ GeV, $m_{1/2} = 250$ GeV, $A_o = -100$ GeV,  $\tan \beta = 10$, and sign$(\mu)=+$.
We initially assume the mass differences $\Delta M_{YN}=80.8$ GeV and $\Delta M_{XN}=47.0$ GeV
have been previously measured and take them as exact.
We later show how the distribution shape with the $m_{ll}$ endpoint also solves for the two mass differences.

Like $M_{2C}$, the $M_{3C}$ distributions is able to be well-predicted
from observations.
When we are determining masses based on distribution shapes, the larger the area difference between two distributions representing different masses, the more accurately and precisely we will be able to tell the difference.  Unfortunately, the $M_{3C}$ distribution is sensitive to the finite momentum resolution errors and combinatoric errors which have the effect to decrease the large area difference between  the distributions of two different masses shown in Figs.~\ref{FigM3CIdeal150100}.

Just as in Chapter \ref{ChapterM2CwUTM}, we model the distribution shape with a simple Mathematica Monte Carlo event generator, and compare the predicted distribution shapes
against the \herwig\ data modeling the benchmark point SPS 1a.
We again use the observed UTM as an input to the Mathematica simulated ideal distributions.
By modeling with the Mathematica, which does not use the SUSY cross sections, and comparing to more realistic \herwig\ generated data,
we hope to test that we understand the major
dependencies of the shape of the $M_{3C}$ distributions.  The Mathematica event generator produces events based on assumptions uniform angular distribution of the parents in the COM frame, the parent particles decay with a uniform angular distribution in the rest frame of the parent.  The particles are all taken to be on shell. $k_T>0$ is simulated by boosting the event in the transverse plane to compensate a specified $k_T$.

\begin{figure}
\centerline{
\includegraphics[width=3.1in]{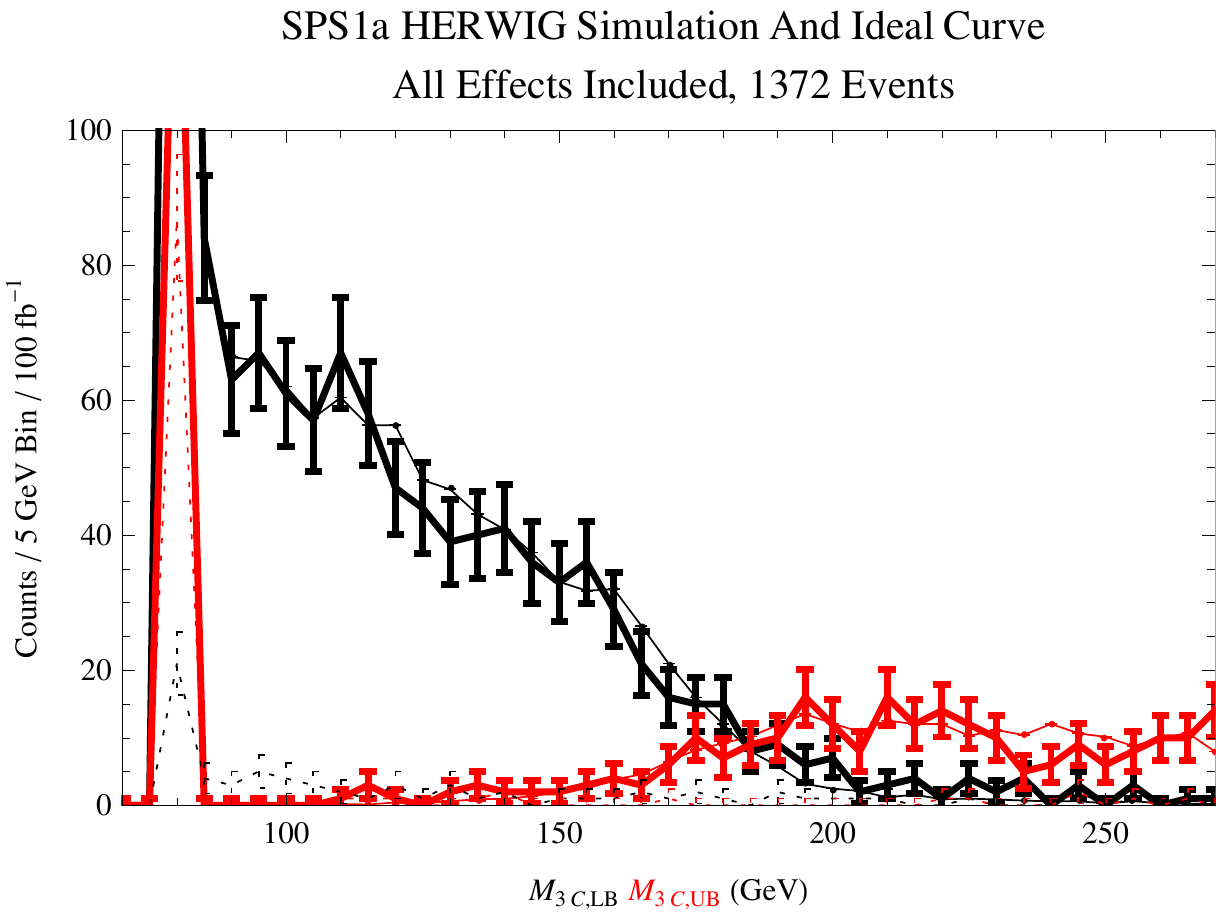}
\includegraphics[width=3.1in]{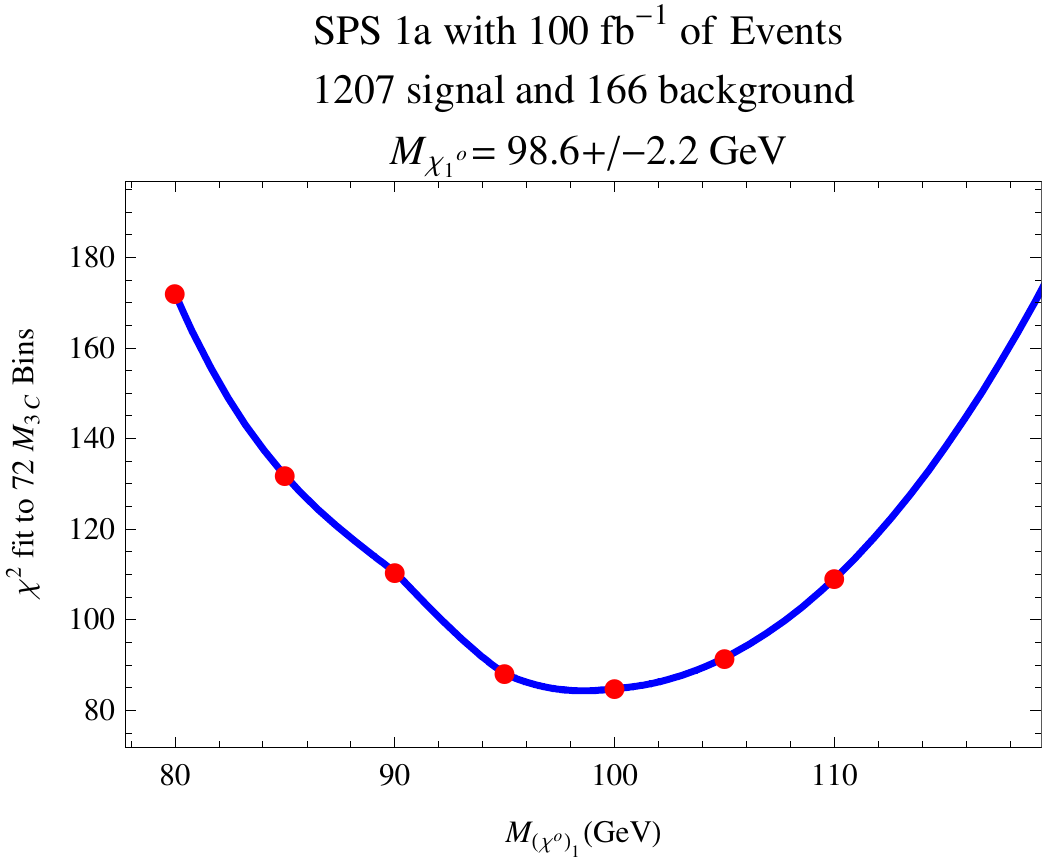}}
\caption{\label{FigM3CFitAllEffects} Fit of ideal $M_{3C}(M_{\N{1}})$ distributions to the \herwig\ generated $M_{3C}$ distributions. Includes combinatoric errors, backgrounds, energy resolution, and $\slashed{P}_T$ cuts. (Left) The observed \herwig\ counts versus the expected counts for ideal $M_{\N{1}}=95$ GeV.  (Right) The $\chi^2$ fit to ideal distributions of $M_{\N{1}}=80, 85, 90, 95, 100, 105, 110$ GeV. The correct mass is $M_{\N{1}}= 96.0$ GeV.}
\end{figure}

Figure~\ref{FigM3CFitAllEffects} shows the performance.
The left side of Fig.~\ref{FigM3CFitAllEffects} shows the $M_{3C}$ lower bound
and upper bound counts per $5$ GeV bin from the \herwig\ generated data, and it shows the predicted ideal counts calculated with Mathematica using the observed UTM distribution and assuming $M_{\N{1}}=95$ GeV.
The upper bound and lower bound show very close agreement.
The background events are shown in dotted lines and are seen accumulating in the first few bins.  These are the same bins dominantly affected by $\slashed{P}_T$ cuts. For this reason we excluded these first two bins from the distribution fit.
The right side of Fig~\ref{FigM3CFitAllEffects} shows the $\chi^2$ fit of the \herwig\ simulated data $M_{3C}$ distribution to the ideal $M_{3C}(M_{\N{1}})$ distribution with $M_{\N{1}}$ taken as the independent variable.
Ideal distribution shapes are calculated at values of $M_{\N{1}} = 80, 85, 90, 95, 100, 105, 110$ GeV.  The $\chi^2$ fitting procedure is described in more detail in Appendix \ref{AppendixChiSqFitting}.
All effects discussed in this chapter are included: combinatoric errors, SUSY backgrounds, energy resolution, and $\slashed{P}_T$ cuts.  Our ideal curves were based on the Mathematica simulations with $25000$ events per ideal curve.
Despite the presence of backgrounds, the $\chi^2$ is not much above $1$ per bin.

 The particular fit shown in Fig~\ref{FigM3CFitAllEffects} gives $M_{\N{1}}=98.6 \pm 2.2$ GeV where we measure uncertainty by using the positions at which $\chi^2$ is increased by one.
We repeat the fitting procedure on nine sets each with $\approx 100 \fb^{-1}$ of \herwig\ data ($\approx 1400$ events for each set).
The mean and standard deviation of these nine fits give
$M_{\N{1}} = 96.8 \pm 3.7$ GeV. After $300 \fb^{-1}$ one should expect a $\sqrt{3}$ improvement in the uncertainty giving $\pm 2.2 \GeV$.  The correct mass is $M_{\N{1}}= 96.05$ GeV.

\begin{figure}
\centerline{\includegraphics[width=3.3in]{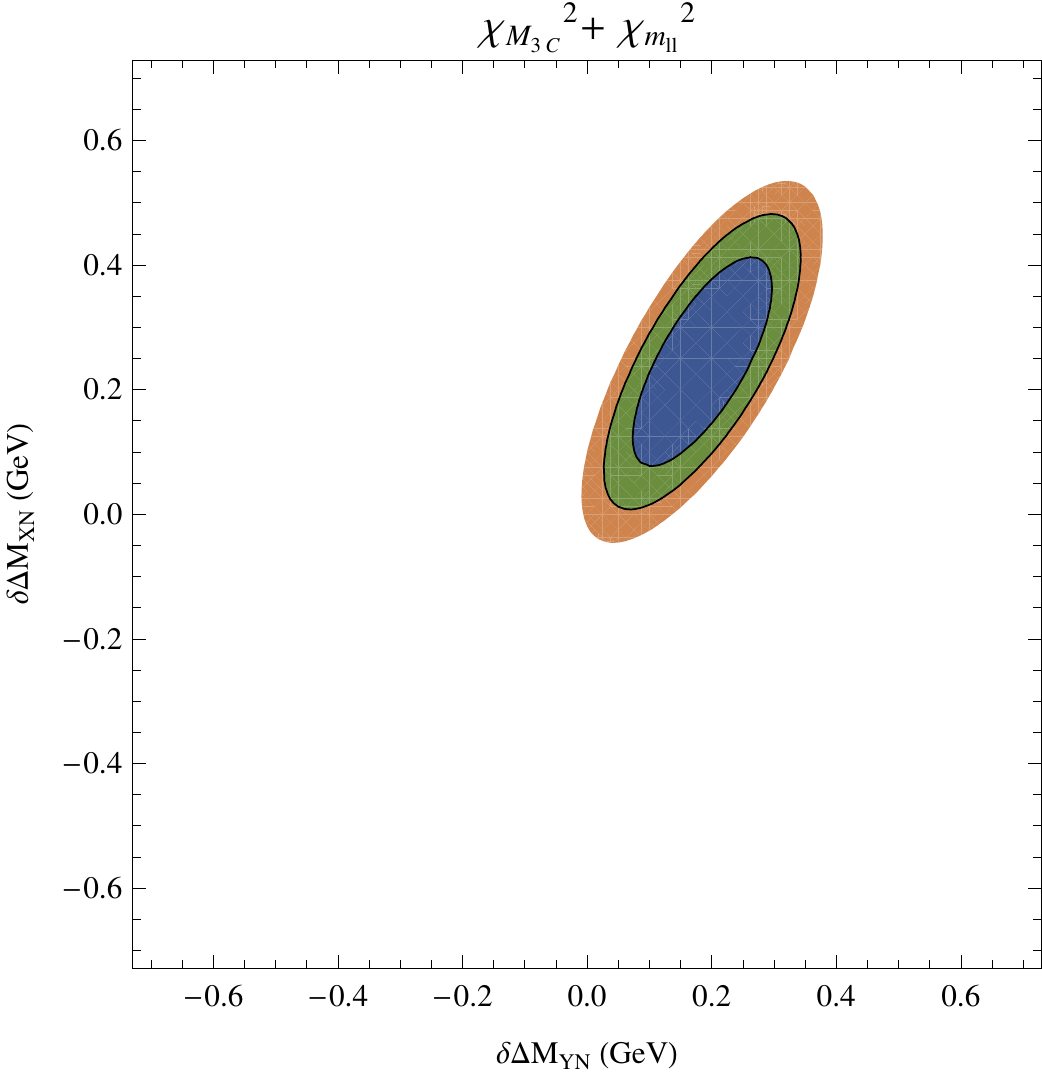}}
\caption{\label{FigM3CMllConstraint} Combined constraint from fitting both $\max m_{ll}$ and $M_{3C}$ with the mass difference as free parameters. We parameterized the difference from the true values in the model by $\Delta M_{YN}= 80.8 \GeV + \delta \Delta M_{YN}$ and $\Delta M_{XN}=47.0 \GeV + \delta \Delta M_{XN}$. We shown the $1, 2, 3 \sigma$ contours.}
\end{figure}

Our technique also enables a combined fit to both the mass differences and the mass scale. The $m_{ll}$ endpoint in Eq(\ref{EqTwoBodyDecayEdge}) constrains a relationship between the three masses.
Gjelsten, Miller, and Osland estimate this edge can be measured to better than $0.08 \GeV$  \cite{Gjelsten:2004ki,Gjelsten:2006tg}
using many different channels that
lead to the same edge, and after modeling energy resolution and background.
In the next several paragraphs we show that by combining this edge with the fits to the $M_{3C}$ upper bound and lower bound distribution shapes, we can constrain all three masses.

We first numerically calculated the effect of errors in the mass differences.
We used $300 \fb^{-1}$ of events (about $3600$ signal and $450$ background) including all the effects discussed.
We parameterize the error from the correct mass difference in the model by the variables $\delta \Delta M_{YN}$ and $\delta \Delta M_{XN}$ so that mass differences are given by $\Delta M_{YN}= 80.8 \GeV + \delta \Delta M_{YN}$ and $\Delta M_{XN}=47.0 \GeV + \delta \Delta M_{XN}$.
We calculated the $\chi^2_{M_{3C}}$ at $8$ points surrounding the correct mass difference by amounts $\delta \Delta M_{YN} = \pm 1 \GeV$ and $\delta \Delta M_{XN} = \pm 1 \GeV$.
The minimum $\chi^2_{M_{3C}}$ at each of the $9$ points gives the value of $M_{\N{1}}$ for each mass difference assumed.  The position of the minima can be parameterized by a quadratic near the true mass difference.  The resulting fit
 \begin{equation}
 M_{\N{1}}=96.4 + 1.9 \, (\delta \Delta M_{XN})^2 + 2.5  \,\delta \Delta M_{YN}\,
 \delta \Delta M_{XN}+3.2
   \delta \Delta M_{XN}-3.8 \,(\delta \Delta M_{YN})^2 - 8.3 \, \delta \Delta M_{YN}
   \label{EqMN1wDeltaMPropagated}
 \end{equation}
 shows in units of GeV how the mass $M_{\N{1}}$ is affected by small errors in the mass difference.

The $\chi^2_{M_{3C}}$ at these $9$ different values for the mass difference provides another constraint on the mass differences.
Fitting the $\chi^2_{M_{3C}}$ to a general quadratic near the true mass difference gives
\begin{equation}
\chi^2_{M_{3C}} = 162 + 38 \, (\delta \Delta M_{XN})^2 - 8 \, \delta \Delta M_{YN} \delta \Delta M_{XN}
- 5 \,   \delta \Delta M_{XN}  - 25 \, \delta \Delta M_{YN} \label{EqChiSqM3C}.
\end{equation}
The $\chi^2_{M_{3C}}$ described by Eq(\ref{EqChiSqM3C}) shows a sloping valley.
The sides of the valley constrain $\delta \Delta M_{XN}$ as seen by
the large positive coefficient of $(\delta \Delta M_{XN})^2$.
The valley slopes downward along $\delta \Delta M_{YN}$ as can be seen by the large negative coefficient of $\delta \Delta M_{YN}$ which leaves this axis unbounded within the region studied.

The unconstrained direction along $\Delta M_{YN}$ can by constrained by the mass relationships given by  the endpoint $\max m_{ll}$ or by $M_{T2}$ as described in Chapter \ref{ChapterMT2onCascadeDecays}.  Here we work with $\max m_{ll}$ to provide this constraint.
We calculate the $\chi^2_{\max m_{ll}}$ using $\delta \max m_{ll} = 0.08 \GeV$, and Eq(\ref{EqTwoBodyDecayEdge}) with $M_Y=\Delta M_{YN}+\delta \Delta M_{YN}+M_{\N{1}}$ and $M_X=\Delta M_{XN}+\delta \Delta M_{XN}+M_{\N{1}}$ where we use $M_{\N{1}}$ from Eq(\ref{EqMN1wDeltaMPropagated}).
This $\chi^2_{\max m_{ll}}$ constrains a diagonal path in $(\delta \Delta M_{YN},\delta \Delta M_{XN})$. The value of the $\chi^2_{\max m_{ll}}$ at the minimum is a constant along this path.
The combined constraint $\chi^2_{M_{3C}}$ to $\chi^2_{\max m_{ll}}$ leads to the
a minimum at $\delta \Delta M_{YN} = 0.18 \GeV$ and  $\delta \Delta M_{XN}= 0.25 \GeV$ where $M_{\N{1}}=95.7 \GeV$ as shown in Fig.~\ref{FigM3CMllConstraint}.
We have shown the contours where $\chi^2$ increases from its minimum by $1$,$2$ and $3$.
The uncertainty in the mass differences around this minimum is about $\pm 0.2 \GeV$.
The bias from the true mass differences is due to the unconstrained $\chi^2_{M_{3C}}$ along $\delta \Delta M_{YN}$.
We can use modeling to back out the unbiased mass differences.
Propagating the effects of uncertainty in the mass differences, we estimate a final performance of $M_{\N{1}}= 96.4 \pm 2.4$ GeV after $300 \fb^{-1}$ with about $3600$ signal events amid $450$ background events. We find the mass differences (without bias correction) of $ M_{\N{2}} - M_{\N{1}} = 81.0 \pm 0.2$ GeV and $M_{\tilde{l}_R} - M_{\N{1}}=44.3 \pm 0.2$ GeV.  This is to be compared to the  \herwig\ values of $M_{\N{1}}=96.0$ GeV, $M_{\N{2}} - M_{\N{1}} = 80.8$ GeV, and $M_{\tilde{l}_R} - M_{\N{1}}=44.3$ GeV.

How does this performance compare to other techniques?
Because SPS 1a is commonly used as a test case, we can approximately compare performance with two different groups.
The technique of \cite{Bachacou:1999zb,Gjelsten:2004ki,Lester:2006yw,Gjelsten:2006tg} which uses edges from cascade decays determines the LSP mass to $\pm 3.4 \GeV$ with about 500 thousand events from $300 \fb^{-1}$.  The approach of CEGHM \cite{Cheng:2008mg} assumes a pair of symmetric decay chains and assumes two events have the same structure.  They reach $\pm 2.8 \GeV$ using $700$ signal events after $300 \fb^{-1}$, but have a $2.5 \GeV$ systematic bias that needs modeling to remove.
Both techniques also constrain the mass differences.
By comparison we find $\pm 3.7$ GeV after $100 \fb^{-1}$ ($1200$ signal, $150$ background)
and estimate $\pm 2.4$ GeV after $300 \fb^{-1}$ ($3600$ signal, $450$ background) and propagating reasonable uncertainties in the mass differences.
The uncertainty calculations differ amongst groups. Some groups estimate uncertainty from repeated trials, and others use the amount one can change the mass before $\chi^2$ increases by one.
Without careful comparison under like circumstance by the same research group, the optimal method is not clear.  What is clear is that fitting the $M_{3C}$ and $M_{3C,UB}$ distributions determines the mass of
invisible particles as well if not better than the other known methods in both accuracy and precision.

\section*{Chapter Summary}

In this chapter, we have extended the constrained mass variable to the case with three new particle states.  We assume events with a symmetric, on-shell intermediate-state topology shown in Fig.~\ref{FigEventTopologyTovey}.
We can either assume that we have measured the mass difference between these new states through other techniques, or combine our technique with the $\max m_{ll}$ edge to find both mass differences and the mass scale.
The new constrained mass variables associated with events with these three new particle states are called
 $M_{3C,LB}$ and $M_{3C,UB}$, and they represent an event-by-event lower bound and upper bound (respectively) on the mass of the third lightest state possible while maintaining the constraints described in Eqs(\ref{EqM3Cc1}-\ref{EqM3Cc7}).
We have shown that most of the $M_{2C}$ distribution properties described in the previous chapter carry through to $M_{3C}$.
The additional particle state and mass difference enable a tighter event-by-event bound on the true mass.  The $M_{3C}$ distribution is more sensitive than the $M_{2C}$ distribution to the momentum and energy resolution errors.  Studying the performance on the SPS 1a benchmark point, we find that despite the energy resolution degradation, we are able to determine $M_{\N{1}}$ to at least the same level of precision and accuracy or possibly even better precision and accuracy than that found by using cascade decays or by using other MSTs.

\chapter{Discussions and Conclusions}
\label{ChapterConclusionsThesis}

In this thesis, we have started in Chapter \ref{ChapterPastCurrentMassPredictions} with a study of principles that in the past have successfully predicted the existence and the mass of new particle states.
We showed astrophysical evidence for dark matter and discussed properties that make supersymmetry an
attractive theory for the dark-matter candidate.
We discussed the grand unification of the three gauge couplings, and we introduced the mass-matrix unification suggested by Georgi and Jarlskog.
In Chapter \ref{ChapterUnificationAndFermionMassStructure} we showed how the mass unification is being quantitatively challenged by tighter constraints on the parameters, but that the mass unification hypothesis leads to a favored class of $\tan \beta$ enhanced threshold corrections where $\tan \beta$
is large and the sign of the gluino mass is opposite that of the wino mass.  Chapter \ref{ChapterMassDeterminationReview} turns to the task of measuring the masses of new particle states if they are produced with dark matter at hadron colliders.
We observed that determining the sign of $M_3$, determining spin, and determining the entire sparticle mass spectra are all facilitated by a model-independent  measure the mass of the dark-matter particle.
We discussed the current techniques to constrain the masses of the new particle states.
We developed ideas of how to use $M_{T2}$ in Chapter \ref{ChapterMT2onCascadeDecays} which enables us to find new constraints between particles in symmetric cascade decays, and we argued that $M_{T2}$ is a better variable to extract this information than $M_{CT}$.
Chapters \ref{ChapterM2Cdirect} and \ref{ChapterM2CwUTM} discussed the constrained mass variables $M_{2C}$ and $M_{2C,UB}$ which uses a previously measured mass difference to find the maximum and minimum mass of the dark-matter particle on an event-by-even basis.
This technique benefits from an $M_{2C}$ and $M_{2C,UB}$ distribution that can be determined from experimental observables and the unknown dark-matter mass.
The distribution shape can be fit to the bulk of the data to estimate the mass of the dark-matter particle more accurately than end-point techniques.
Chapter \ref{ChapterM3C} discussed the benefits and down-falls of the constraint mass variable $M_{3C}$.
Unlike the edges-from-cascade-decays method \cite{Bachacou:1999zb,Gjelsten:2004ki,Lester:2006yw,Gjelsten:2006tg} and the mass-shell technique of CEGHM\cite{Cheng:2008mg} each of which require the existence of four new particle states, the constrained mass variables only require two new particle states for $M_{2C}$ and three new particle states for $M_{3C}$.
Although proper comparisons to other techniques are difficult, it is clear that the constrained mass variables can determine the mass of
invisible particles at least as well if not better than the other known model-independent methods in both accuracy and precision.


%

From here what are the directions for continued research?
The origin of the Yukawa mass matrices remains unknown.
The Georgi-Jarlskog relations appear to hint at a solution, but only decrease the number of unknowns slightly.
The fits to the GUT-scale Yukawa parameters is a starting point for
more research on explaining this structure.
There are relatively few studies on how to measure the sign of the gluino mass parameter relative to the winos mass parameter; there are even fewer that do not make use of simplifying assumptions about the underlying supersymmetry model.
The results of Chapter \ref{ChapterMT2onCascadeDecays}  need to be extended to include the effects of combinatoric ambiguities.
In Chapters \ref{ChapterM2CwUTM}-\ref{ChapterM3C} there are several distribution properties which should be better studied. We would like to understand the physics of the invariance of the constrained mass variable distributions to changes in $\sqrt{s}$ and choice of $k^2$ when $k_T \neq 0$.  In Chapter \ref{ChapterM3C} it would also be nice to classify the conditions which determine how many islands of solutions exist for $M_{3C}$.  The $M_{3C}$ method could be tested on top-quark data to remeasure the masses of the $t$, $W$ and $\nu$.
In Chapters \ref{ChapterM2Cdirect}-\ref{ChapterM3C} we have considered two constrained mass variables $M_{2C}$ and $M_{3C}$ where the decay chains are symmetric, but we may find greater statistics for asymmetric events.
For example in many models one expects more events where $\N{2}$ is produced with $\N{1}$ or $\tilde{\chi}^\pm$.
The concepts of the constrained mass variable could be extended to deal with these asymmetric event topologies or to cases with both {LSP}s and neutrinos.

In conclusion, the thesis contributed to predictions of mass properties of new particle states that may be discovered at the LHC and developed the tools to make precision mass measurements that may help falsify or validate this prediction and many others.
Modern physics is rooted in an interplay between creating theoretical models,
developing experimental techniques, making observations, and falsifying theories.
The LHC's results will complete this long-awaited cycle making experimental observations to constrain the growing balloon of theories of which this thesis is a small part.

\appendix


\chapter{Renormalization Group Running}

\label{AppendixRGRunningDetails}
This appendix details our use of the Renormalization
Group Equations (RGEs) to relate the reported
values for observables at low-energy scales to their values
the GUT scale.
The one-loop and two-loop RG equations for
the gauge coupling and the Yukawa couplings
in the Standard Model and in the MSSM that we
use in this thesis come from a number of
sources
\cite{Chankowski:2001mx}\cite{Ramond:1999vh}\cite{Barger:1992ac}\cite{Fusaoka:1998vc}.
We solve the system using the internal
differential equation solver in Mathematica
6.0 on a basic Intel laptop.  We considered
using SoftSUSY or SuSpect to decrease the
likelihood of programming errors on our part, but neither
met our needs sufficiently.
SoftSUSY\cite{Allanach:2001kg}
 has only real Yukawa couplings, and
SuSpect \cite{Djouadi:2002ze} focuses
primarily on the third generation RG
evolution.

We assume a model for the neutrino spectrum
that does not impact the running of the quark
or charged lepton parameters below the GUT
scale.  This assumption is not trivial; there
are neutrino models where $Y^\nu$ will affect
the ratio $y_b / y_\tau$ by as much as 15\%
\cite{Bajc:2002iw}.  Furthermore due to
unitarity constraints, the UV completion of
the effective dimension-five neutrino mass
operators must take place before the GUT
scale \cite{Campbell:2006nv}. One way to have
a UV completion below the GUT scale while
making the RGE running independent of the
specific model is to set-up a non-degenerate
right-handed neutrino spectrum. One then
constructs the couplings such that large
Yukawa couplings in $Y^\nu$ are integrated
out with the heaviest right-handed neutrino
mass which is at or above the GUT scale. One
can then have the remaining right-handed
neutrinos considerably below the GUT scale
which couple to much smaller values $Y^\nu$
that do not affect the running of $Y^e$ or
$Y^u$. The rules for running the neutrino
parameters and decoupling the right-handed
neutrinos for the $U_{MNS}$ are
found in
Refs.~\cite{Chankowski:2001mx}\cite{Dighe:2006sr}
\cite{Vissani:1994fy} \cite{Antusch:2002rr}
\cite{Leontaris:1995be}.  For our work finding
the quark and lepton masses and mixings at the GUT scale,
we assume the effects of the neutrinos on the
RGE equations effectively decouple at or
above the GUT scale.

Our running
involves three models at different energy
scales.  Between $M_X$ to an effective SUSY
scale $M_S$, we assume the MSSM.
Between $M_S$
to $M_Z$, we assume the Standard Model. Below
$M_Z$, we assume $QCD$ and $QED$.
We do not
assume unification, rather we use the  low-energy values in Table \ref{TableParametersToFitTo} and Eq(\ref{EqLightQuarkMassConsistent})
provide our boundary condition. The
uncertainty in the final digit(s) of each
parameter is listed in parenthesis. For
example $123.4 \pm 0.2$ is written as
$123.4(2)$.


First, we use QCD and QED
low-energy effective theories to run the
quark and lepton masses from the scales at
which the particle data book reports their
values to $M_Z$. The $V_{CKM}$ mixing angles are set at this scale.
Next, we run using the Standard-Model
parameters, including the Higgs VEV and Higgs
self coupling from $M_Z$ to an effective SUSY scale
$M_S$.
At an effective SUSY scale $M_S$, we match
the  Standard
Model Yukawa couplings using the running
Higgs VEV $v(M_S)$ onto MSSM Yukawa couplings.
In the matching
procedure, we convert the gauge coupling
constants from $\overline{MS}$ to
$\overline{DR}$ as described in
\cite{Allanach:2001kg}, and we apply
approximate supersymmetric threshold
corrections as parameterized in Chapter \ref{ChapterUnificationAndFermionMassStructure}.
Details of the RG equations and the matchings are
described in the following sections.

\section{RGE Low-Energy $SU(3)_c \times U(1)_{EM}$ up to the Standard Model}
\label{SecLowEnergyRunning}

Below $M_Z$ we use QCD and QED RG Equations
to move the parameters to the scale reported
in the Particle Data Book.  We run the
light-quark masses between  $\mu=M_Z$ to
$\mu_L=2$ GeV
  with the following
factors:
 \begin{equation}
   \eta_{u,d,s} = (1.75 \pm 0.03 ) + 19.67 \left( \alpha_s^{(5)}(M_Z) - 0.118
   \right)
 \end{equation}
\begin{equation}
   \eta_{c} = (2.11 \pm 0.03 ) + 41.13 \left( \alpha_s^{(5)}(M_Z) - 0.118
   \right)
 \end{equation}
\begin{equation}
   \eta_{b} = (1.46 \pm 0.02 ) + 8.14 \left( \alpha_s^{(5)}(M_Z) - 0.118
   \right),
 \end{equation}
that are calculated with the 4-loop RGE
equations using Chetyrkin's RUNDEC software
for Mathematica \cite{Chetyrkin:2000yt}.
These factors are used as $m_b(m_b)\, / \eta_b
= m_b(M_Z)$. The uncertainty in the first
term is an estimate of the theoretical error
from neglecting the five-loop running give by
$\delta \eta / \eta \approx \exp(
{\mathcal{O}}(1) \langle \alpha_s \rangle^5
\log\frac{M_Z}{\mu_L})-1$. RUNDEC also
converts the top quark's pole mass to an
$\overline{MS}$ running mass, and applies the
small threshold corrections from decoupling
each of the quarks.  The different $\eta$
factors take the parameters from a six-quark
effective model at $\mu=M_Z$ to their
respective scales.

The leptons can be converted from their Pole mass
into running masses at $M_Z$ by a similar
running factor:
 \begin{equation}
  \eta_e = 1.046 \ \ \  \eta_\mu = 1.027  \ \
  \ \eta_\tau = 1.0167.
 \end{equation}
We did not incorporate the lepton masses as
error sources.

The input values in Table
\ref{TableParametersToFitTo} lead to the
values at $M_Z$ in Table \ref{TableMZvalues} where we have propagated
uncertainty from both the strong coupling
constant uncertainty and the uncertainty in
the mass at the starting scale.  These values for the parameters at $M_Z$ agree
with those calculated recently elsewhere in the literature \cite{Dorsner:2006hw}.
 \begin{table}
\centerline{\begin{tabular}{|c|c|c|}
  \hline
  Parameter & At Scale $\mu=M_Z$  \\ \hline
  $m_t$ & $169.6(2.3)$ GeV  \\
  $m_c$ & $599(50)$ MeV   \\
  $m_u$ & $1.54(50)$ MeV  \\
  $m_b$ & $2.87(6)$ GeV \\
  $m_s$ & $57(12)$ MeV \\
  $m_d$ & $3.1(5)$ MeV \\
  \hline
\end{tabular}}
\caption{\label{TableMZvalues} The $\overline{MS}$ values for the running quark masses at $M_Z$.}
 \end{table}

The gauge coupling constants $g_1$ and $g_2$
are determined from $e_{EM}$, $\sin^2 \theta_W$ at the scale $M_Z$ by
 \begin{equation}
   {g_1^2(M_Z)} = \frac{5}{3} \frac{{e^2_{EM}(M_Z)}
   }{\cos^2 \theta_W} \ \ \    {g^2_2(M_Z)} =  \frac{{e^2_{EM}(M_Z)}
   }{\sin^2 \theta_W}.
    \end{equation}

\section{RGE for Standard Model with Neutrinos up to MSSM}
\label{SecRGESM}

We found using the two-loop vs the one-loop
RG equations for the gauge-coupling produced
a shift in the low energy parameters of more
than $1 \sigma$. For this reason, we always
use two-loop RG equations for the gauge
couplings.  For the Yukawa couplings we find
the two-loop vs the one-loop RGE equations
shift the low-energy parameters much less
than the experimental uncertainty. Therefore,
we perform our minimizations using the
one-loop RG equations for the Yukawa and
checked the final fits against the two-loop
RG equations.

We reproduce here for reference the two-loop
gauge coupling RG equations and the one-loop
Yukawa coupling RG equations as this level of
detail is sufficient to reproduce our
results. We define $t=\log M$.  The equations
are compiled from \cite{Antusch:2002rr},
\cite{Chankowski:2001mx},
\cite{Fusaoka:1998vc}.
\begin{eqnarray}
  \frac{d}{dt} g_1 & = & \frac{1}{16 \pi^2} \frac{41}{10} g_1^3 \nonumber \\ & & +
    \frac{g_1^3}{(16 \pi^2)^2} \left( \frac{199}{50} g_1^2 + \frac{27}{15} g_2^2 +
    \frac{44}{5} g_3^2
      -\frac{17}{10}{\rm{Tr}}(Y^u Y^{u\,\dag})
      - \frac{1}{2} {\rm{Tr}}(Y^d Y^{d\,\dag})
      - \frac{3}{2} {\rm{Tr}}(Y^e Y^{e\,\dag})
      \right) \\
  \frac{d}{dt} g_2 & = & \frac{1}{16 \pi^2}  \frac{-19}{6} g_2^3 \nonumber \\ & & +
    \frac{g_2^3}{(16 \pi^2)^2} \left( \frac{9}{10}\, g_1^2 + \frac{35}{6}\, g_2^2 + 12\, g_3^2
      - \frac{3}{2}{\rm{Tr}}(Y^u Y^{u\,\dag})- \frac{3}{2} {\rm{Tr}}(Y^d Y^{d\,\dag})
      - \frac{1}{2} {\rm{Tr}}(Y^e Y^{e\,\dag}) \right)
      \\
  \frac{d}{dt} g_3 & = & \frac{-7}{16 \pi^2}  g_3^3 \nonumber \\ & & +
    \frac{g_3^3}{(16 \pi^2)^2} \left( \frac{11}{10}\, g_1^2 + \frac{9}{2}\, g_2^2  -26 g_3^2
      -2{\rm{Tr}}(Y^u Y^{u\,\dag})- 2 {\rm{Tr}}(Y^d Y^{d\,\dag})
      \right)
\end{eqnarray}
\begin{equation}
  T = {\rm{Tr}}( 3 Y^u Y^{u \dag} + 3 Y^d Y^{d \dag} + Y^e Y^{e \dag} + Y^\nu Y^{\nu
  \dag} )
\end{equation}
\begin{eqnarray}
 \frac{d}{dt} Y^u & = & \frac{1}{16 \pi^2}
                   \left(- \textbf{I}\, G^u
                   +       \textbf{I}\, T
                   + \frac{3}{2}\, Y^u\, Y^{u \dag}
                   - \frac{3}{2}\, Y^d \, Y^{d\,\dag}  \right) .  Y^u  \\
 \frac{d}{dt} Y^\nu & = & \frac{1}{16 \pi^2}
                   \left( - \textbf{I}\, G^\nu
                   +       \textbf{I}\,T
                   + \frac{3}{2} Y^\nu Y^{\nu \, \dag}
                   - \frac{3}{2}   Y^e \, Y^{e\,\dag}  \right) .  Y^\nu  \\
 \frac{d}{dt} Y^d & = & \frac{1}{16 \pi^2}
                   \left( - \textbf{I}\, G^d
                   +       \textbf{I}\, T
                    +   \frac{3}{2}\, Y^d \, Y^{d\,\dag}
                      - \frac{3}{2}\, Y^u\, Y^{u \dag} \right) .  Y^d  \\
 \frac{d}{dt} Y^e & = & \frac{1}{16 \pi^2}
                   \left( - \textbf{I}\, G^e
                   +       \textbf{I}\,T
                   +  \frac{3}{2}\, Y^e \, Y^{e\,\dag}
                   - \frac{3}{2}\, Y^\nu \, Y^{\nu\,\dag}  \right) .  Y^e
\end{eqnarray}
where
\begin{eqnarray}
 \begin{matrix}
   G^u = \frac{17}{20} \,g_1^2 + \frac{9}{4}\,g_2^2 + 8\,g_3^2 &
   G^d =  \frac{1}{4} \,g_1^2 + \frac{9}{4}\,g_2^2 + 8\,g_3^2 \cr
   G^e =    \frac{9}{4} \,g_1^2 + \frac{9}{4}\,g_2^2 &
   G^\nu =  \frac{9}{20}\, g_1^2 + \frac{9}{4}\,g_2^2.
  \end{matrix}
\end{eqnarray}

The Higgs self-interaction $\lambda$
(following the convention $-\lambda (H^\dag
H)^2 \subset {\mathcal{L}}$) obeys the RG
equation:
\begin{eqnarray}
 16 \pi^2  \frac{d}{dt} \lambda & = & 12
 \lambda^2 - ( \frac{9}{5} g_1^2 + 9 g_2^2)
 \lambda + \frac{9}{4} (\frac{3}{25} g_1^4 +
 \frac{2}{5} g_1^2 g_2^2 + g_2^4) + 4 T \,
 \lambda \\ & &
 - 4 {\rm{Tr}}( 3\, Y^u Y^{u \dag}  Y^u Y^{u
 \dag} + 3\, Y^d Y^{d \dag}  Y^d Y^{d \dag} +
 Y^e Y^{e \dag}  Y^e Y^{e \dag} ).
\end{eqnarray}

Between $M_Z$ and $M_S$, we run the Standard-Model VEV using \cite{Ramond:1999vh}:
\begin{equation}
16 \pi^2 \frac{d}{dt} v = v \left(
\frac{9}{4} ( \frac{1}{5} g_1^2 + g_2^2) - T
\right). \end{equation}
 \begin{eqnarray}
16 \pi^2  \frac{d}{dt} \kappa & = &
  \left( -\frac{3}{2} Y^e Y^{e\,\dag} \kappa -\frac{3}{2} \kappa Y^e
  Y^{e \, \dag} + \frac{1}{2} Y^\nu Y^{\nu\,\dag} \kappa
  + \frac{1}{2} \kappa Y^\nu Y^{\nu\,\dag} \right) \nonumber
  \\ & &
  +   \textbf{I} \left( {\rm{Tr}}(2 \,Y^\nu Y^{\nu \dag} + 2 \,Y^e Y^{e \dag} + 6 Y^u Y^{u\,\dag} + 6 Y^d Y^{d\,\dag} )
   -   3 g_2^2  + 4 \lambda \right) \kappa \nonumber
 \end{eqnarray}
 \begin{equation}
  \frac{d}{dt} M_{RR} =  \frac{1}{16 \pi^2}
  \left( (Y^{\nu \dag} \, Y^\nu)^T \, M_{RR} +
         M_{RR}\, Y^{\nu \dag} \, Y^\nu  \right).
  \end{equation}
We normalize $g_1$ to the $SU(5)$ convention:
$g_1^2 = \frac{3}{5} g_Y^2$.

\section{RGE for the MSSM with Neutrinos up to GUT Scale}
\label{SecRGEMSSM}

The MSSM RG equations are in the
$\overline{DR}$ scheme.  To convert
$\overline{MS}$ running quark masses to
$\overline{DR}$ masses at the same scale, we
use \cite{Baer:2002ek}
 \begin{equation}
  \delta m_{sch} =  \frac{ m_{\overline{DR}}}{  m_{\overline{MS}}} =
    \left(1 - \frac{1}{3} \frac{g_3^2}{4 \pi^2} - \frac{29}{72} (\frac{g_3^2}{4 \pi^2} )^2
    \right).
    \label{EqMSDRMass}
 \end{equation}
No corrections were used for switching lepton
running masses from one scheme to the other.
The gauge coupling constants are related via
\cite{Allanach:2001kg}
 \begin{eqnarray}
  \alpha_{3\, \overline{DR}}^{-1} =
  \alpha_{3\,
  \overline{MS}}^{-1} + \frac{3}{12 \pi} \\
    \alpha_{2\, \overline{DR}}^{-1} =
    \alpha_{2\,
  \overline{MS}}^{-1} + \frac{2}{12 \pi}
 \end{eqnarray}
where $\alpha_3 = g_3^2/ 4 \pi$ and $\alpha_2
= g_2^2 / 4 \pi$.  We define $t=\log M$. The
RG equations are:
\begin{eqnarray}
  \frac{d}{dt} g_1 & = & \frac{1}{16 \pi^2} \frac{33}{5} g_1^3 \nonumber \\ & & +
    \frac{g_1^3}{(16 \pi^2)^2} \left( \frac{199}{25} g_1^2 + \frac{27}{5} g_2^2 + \frac{88}{5} g_3^2
      -\frac{26}{5}{\rm{Tr}}(Y^u Y^{u\,\dag}) - \frac{14}{5} {\rm{Tr}}(Y^d Y^{d\,\dag})
      - \frac{18}{5} {\rm{Tr}}(Y^e Y^{e\,\dag})
      \right) \\
  \frac{d}{dt} g_2 & = & \frac{1}{16 \pi^2}  g_2^3 \nonumber \\ & & +
    \frac{g_2^3}{(16 \pi^2)^2} \left( \frac{9}{5}\, g_1^2 + 25\, g_2^2 + 24\, g_3^2
      -6{\rm{Tr}}(Y^u Y^{u\,\dag})- 6 {\rm{Tr}}(Y^d Y^{d\,\dag})
      - 2 {\rm{Tr}}(Y^e Y^{e\,\dag}) \right)
      \\
  \frac{d}{dt} g_3 & = & \frac{-3}{16 \pi^2}  g_3^3 \nonumber \\ & & +
    \frac{g_3^3}{(16 \pi^2)^2} \left( \frac{11}{5}\, g_1^2 + 9\, g_2^2 + 14\, g_3^2
      -4{\rm{Tr}}(Y^u Y^{u\,\dag})- 4 {\rm{Tr}}(Y^d Y^{d\,\dag})
      \right)
\end{eqnarray}
\begin{eqnarray}
 \frac{d}{dt} Y^u & = & \frac{1}{16 \pi^2}
                   \left(- \textbf{I}\, G^u
                   +       \textbf{I}\,3\, {\rm{Tr}}(Y^u Y^{u\,\dag})
                   +       \textbf{I}\, {\rm{Tr}}(Y^\nu Y^{\nu\,\dag})
                   + 3\, Y^u\, Y^{u \dag} + Y^d \, Y^{d\,\dag}  \right) .  Y^u  \label{EqRGMSSMYu} \\
 \frac{d}{dt} Y^\nu & = & \frac{1}{16 \pi^2}
                   \left( - \textbf{I}\, G^\nu
                   +       \textbf{I}\,3\, {\rm{Tr}}(Y^u Y^{u\,\dag})
                   +       \textbf{I}\, {\rm{Tr}}(Y^\nu Y^{\nu \,\dag})
                   + 3 Y^\nu Y^{\nu \, \dag} +   Y^e \, Y^{e\,\dag}  \right) .  Y^\nu  \label{EqRGMSSMYnu} \\
 \frac{d}{dt} Y^d & = & \frac{1}{16 \pi^2}
                   \left( - \textbf{I}\, G^d
                   +       \textbf{I}\, 3\, {\rm{Tr}}(Y^d Y^{d\,\dag})
                   +       \textbf{I}\, {\rm{Tr}}(Y^e Y^{e\,\dag})
                    +   3\, Y^d \, Y^{d\,\dag} +  Y^u\, Y^{u \dag} \right) .  Y^d  \label{EqRGMSSMYd}\\
 \frac{d}{dt} Y^e & = & \frac{1}{16 \pi^2}
                   \left( - \textbf{I}\, G^e
                   +       \textbf{I}\,3\, {\rm{Tr}}(Y^d Y^{d\,\dag})
                   +       \textbf{I}\, {\rm{Tr}}(Y^e Y^{e\,\dag})
                   +  3\, Y^e \, Y^{e\,\dag} + Y^\nu \, Y^{\nu\,\dag}  \right) .
                   Y^e \label{EqRGMSSMYe}
\end{eqnarray}
where
\begin{eqnarray}
 \begin{matrix}
   G^u = \frac{13}{15} \,g_1^2 + 3\,g_2^2 +
\frac{16}{3}\,g_3^2 &
   G^d =  \frac{17}{15} \,g_1^2 + 3\,g_2^2 +
   \frac{16}{3}\,g_3^2 \cr
   &
   \cr
   G^e =    \frac{9}{5} \,g_1^2 + 3\,g_2^2 &
   G^\nu =  \frac{3}{5}\, g_1^2 + 3\,g_2^2.
  \end{matrix}
\end{eqnarray}

 \begin{eqnarray}
 16 \pi^2  \frac{d}{dt} \kappa =
  \left(  Y^e Y^{e\,\dag} \kappa + \kappa Y^e
  Y^{e \, \dag} + Y^\nu Y^{\nu\,\dag} \kappa
  + \kappa Y^\nu Y^{\nu\,\dag} \right)  \nonumber \\
  +  \textbf{I} \left(\, 2 \, {\rm{Tr}}(Y^\nu Y^{\nu \dag} + 6 Y^u Y^{u\,\dag} )
  -   \frac{6}{5} g_1^2 - 6 g_2^2 \right) \kappa
 \end{eqnarray}

 \begin{equation}
  \frac{d}{dt} M_{RR} =  \frac{1}{16 \pi^2}
  \left( 2\,(Y^{\nu \dag} \, Y^\nu)^T \, M_{RR} +
         2\,M_{RR}\, Y^{\nu \dag} \, Y^\nu  \right)
  \end{equation}
 We normalize $g_1$ to the $SU(5)$
convention: $g_1^2 = \frac{3}{5} g_Y^2$.

\section{Approximate Running Rules of Thumb}

\begin{figure}
\centerline{
\includegraphics[width=6.5in]{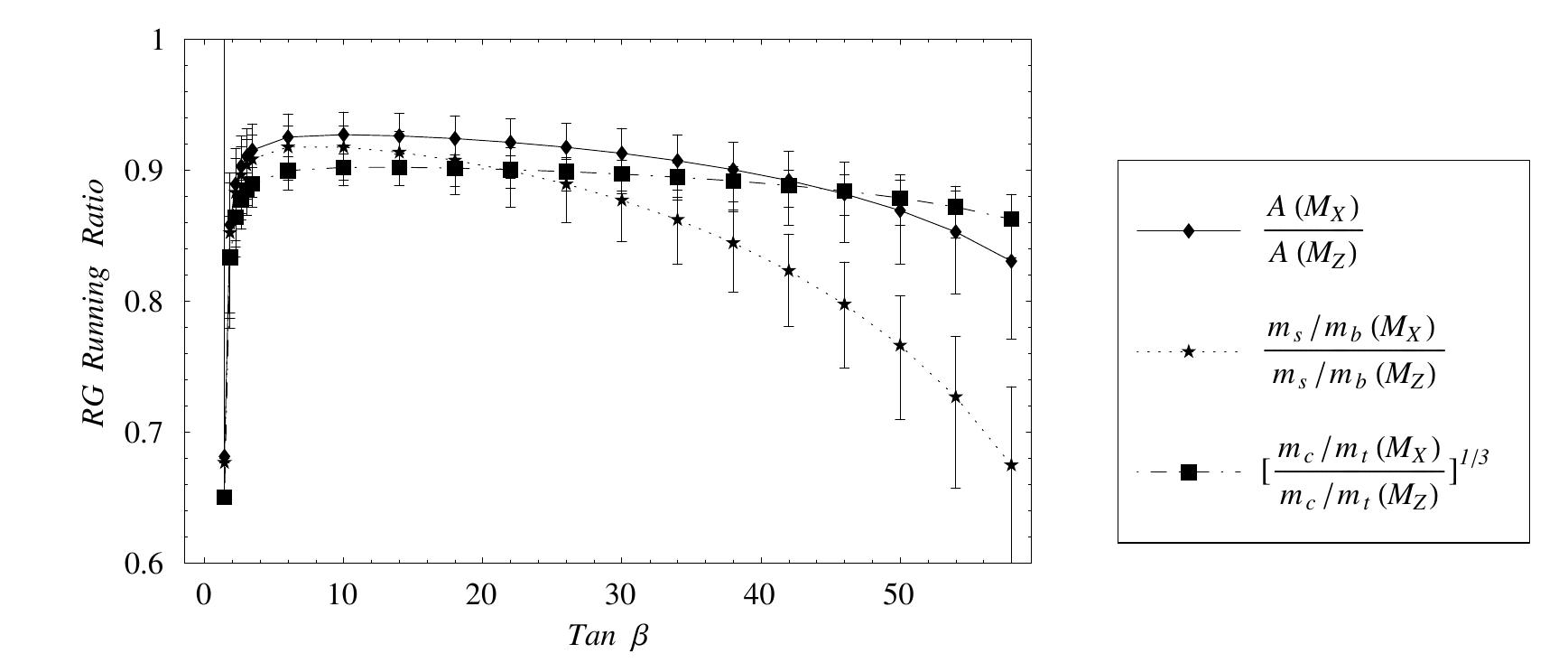}}
  \caption{\label{FigRGRatios}
 The impact of RG running parameter ratios with $M_S=500$ GeV. These
 ratios determine $\chi$ defined in Eq.~\ref{EqChiDefined}.
 If $M_S=M_Z$, all three are degenerate at small $\tan \beta$.}
 \end{figure}
At $\tan \beta < 10$ and when $M_S \sim M_Z$,
the running of parameters from $M_Z$ to $M_X$
obey the following relationships exploited in
the RRRV \cite{Roberts:2001zy} study:
 \begin{equation}
  \frac{ \bar{\eta}_o(M_X)}{\bar{\eta}_o(M_Z)} \approx  \frac{ \bar{\rho}_o(M_X)}{\bar{\rho}_o(M_Z)} \approx  \frac{ \lambda_o(M_X)}{\lambda_o(M_Z)}  \approx 1
  \ \ \ \ \frac{ A_o(M_X)}{A_o(M_Z)}  \approx  \chi
 \end{equation}
 \begin{equation}
  \frac{(m_u/m_c)_o(M_X)}{(m_u/m_c)_o(M_Z)} \approx  \frac{(m_d/m_s)_o(M_X)}{(m_d/m_s)_o(M_Z)} \approx
  1 \ \ \ \ \frac{(m_s/m_b)_o(M_X)}{(m_s/m_b)_o(M_Z)}
  \approx \chi \ \ \ \  \frac{(m_c/m_t)_o(M_X)}{(m_c/m_t)_o(M_Z)}
  \approx \chi^3
 \end{equation}
where
 \begin{equation}
   \chi \approx \exp \left( \int_{t_o}^{t_f} \frac{- y_t^2}{16 \pi^2}\, dt \right)
   \label{EqChiDefined}
 \end{equation}
and $t_o = \log M_Z$ and $t_f = \log M_X$.
Fig.~\ref{FigRGRatios} shows the two-loop
results reflecting how the running of these
ratios change when we consider  $M_S = 500 $
GeV and larger values of $\tan \beta$. If we
had plotted the $M_S=M_Z$ case, the curves in
Fig.~\ref{FigRGRatios} would have been
completely degenerate at small $\tan \beta$.
As a consistency check, we compare our
results to Fusaoka and Koide
\cite{Fusaoka:1998vc}. They find $\chi
\approx 0.851$ at $\tan \beta = 10$ with
their choice of $m_t(M_Z) = 180$. Our code
also gives $\chi=0.851$ at $\tan \beta=10$ if
we select $M_S=M_Z$ and omit the conversion
to $\overline{DR}$ scheme.

 \providecommand{\Mvariable}[1]{}
 \providecommand{\etal}{\textit{et.al.}}
 \providecommand{\foundit}{\setboolean{found}{true}}
 \providecommand{\MeV}{{\rm{MeV}}}
 \providecommand{\GeV}{{\rm{GeV}}}
 \providecommand{\eV}{{\rm{eV}}}
 \providecommand{\imag}[1]{\,i\,}
 \providecommand{\tanbloopfactor}{}

\renewcommand{\Mvariable}[1]{
 \setboolean{found}{false}
 \ifthenelse{\equal{#1}{\epsilon d}}{\epsilon_d \foundit}{}%
 \ifthenelse{\equal{#1}{\epsilon u}}{\epsilon_u \foundit}{}%
 \ifthenelse{\equal{#1}{\epsilon e}}{\epsilon_e \foundit}{}%
 \ifthenelse{\equal{#1}{\phi bp}}{\phi_{b'} \foundit}{}%
 \ifthenelse{\equal{#1}{fp}}{f' \foundit}{}%
 \ifthenelse{\equal{#1}{fpc}}{f'^* \foundit}{}%
 \ifthenelse{\equal{#1}{Vts}}{V_{ts} \foundit}{}%
 \ifthenelse{\equal{#1}{Vtd}}{V_{td} \foundit}{}%
 \ifthenelse{\equal{#1}{Vcb}}{V_{cb} \foundit}{}%
 \ifthenelse{\equal{#1}{Vub}}{V_{ub} \foundit}{}%
 \ifthenelse{\equal{#1}{f}}{f \foundit}{}%
 \ifthenelse{\equal{#1}{fc}}{f^* \foundit}{}%
 \ifthenelse{\equal{#1}{\phi c}}{\phi_{c} \foundit}{}%
 \ifthenelse{\equal{#1}{\epsilon 3}}{\epsilon_3 \foundit}{}%
 \ifthenelse{\equal{#1}{bp}}{b' \foundit}{}%
 \ifthenelse{\equal{#1}{bc}}{b^* \foundit}{}%
 \ifthenelse{\equal{#1}{cp}}{c' \foundit}{}%
 \ifthenelse{\equal{#1}{c}}{c \foundit}{}%
 \ifthenelse{\equal{#1}{a}}{a \foundit}{}%
 \ifthenelse{\equal{#1}{dc}}{d^* \foundit}{}%
 \ifthenelse{\equal{#1}{ac}}{a^* \foundit}{}%
 \ifthenelse{\equal{#1}{apc}}{a'^* \foundit}{}%
 \ifthenelse{\equal{#1}{ap}}{a' \foundit}{}%
 \ifthenelse{\equal{#1}{bpc}}{b'^* \foundit}{}%
 \ifthenelse{\equal{#1}{cpc}}{c'^* \foundit}{}%
 \ifthenelse{\equal{#1}{cc}}{c^* \foundit}{}%
 \ifthenelse{\equal{#1}{\epsilon \nu L}}{\epsilon_{\nu L} \foundit}{}%
 \ifthenelse{\equal{#1}{\epsilon \nu R}}{\epsilon_{\nu R} \foundit}{}%
 \ifthenelse{\equal{#1}{YAu}}{B_u \foundit}{}%
 \ifthenelse{\equal{#1}{YBu}}{C_u \foundit}{}%
 \ifthenelse{\equal{#1}{YCu}}{D_u \foundit}{}%
 \ifthenelse{\equal{#1}{YAd}}{B_d \foundit}{}%
 \ifthenelse{\equal{#1}{YBd}}{C_d \foundit}{}%
 \ifthenelse{\equal{#1}{YCd}}{D_d \foundit}{}%
 \ifthenelse{\equal{#1}{Y2323u}}{A_u \foundit}{}%
 \ifthenelse{\equal{#1}{Y2323d}}{A_d \foundit}{}%
 \ifthenelse{\equal{#1}{MeV}}{MeV \foundit}{}%
 \ifthenelse{\equal{#1}{v\theta}}{v_\theta \foundit}{}%
 \ifthenelse{\equal{#1}{MdR}}{M_{dR} \foundit}{}%
 \ifthenelse{\equal{#1}{Yu11}}{Y^u_{11} \foundit}{}%
 \ifthenelse{\equal{#1}{Yu12}}{Y^u_{12} \foundit}{}%
 \ifthenelse{\equal{#1}{Yu13}}{Y^u_{13} \foundit}{}%
 \ifthenelse{\equal{#1}{Yu21}}{Y^u_{21} \foundit}{}%
 \ifthenelse{\equal{#1}{Yu22}}{Y^u_{22} \foundit}{}%
 \ifthenelse{\equal{#1}{Yu23}}{Y^u_{23} \foundit}{}%
 \ifthenelse{\equal{#1}{Yu31}}{Y^u_{31} \foundit}{}%
 \ifthenelse{\equal{#1}{Yu32}}{Y^u_{32} \foundit}{}%
 \ifthenelse{\equal{#1}{Yu33}}{Y^u_{33} \foundit}{}%
 \ifthenelse{\equal{#1}{Yu11c}}{Y^{u*}_{11} \foundit}{}%
 \ifthenelse{\equal{#1}{Yu12c}}{Y^{u*}_{12} \foundit}{}%
 \ifthenelse{\equal{#1}{Yu13c}}{Y^{u*}_{13} \foundit}{}%
 \ifthenelse{\equal{#1}{Yu21c}}{Y^{u*}_{21} \foundit}{}%
 \ifthenelse{\equal{#1}{Yu22c}}{Y^{u*}_{22} \foundit}{}%
 \ifthenelse{\equal{#1}{Yu23c}}{Y^{u*}_{23} \foundit}{}%
 \ifthenelse{\equal{#1}{Yu31c}}{Y^{u*}_{31} \foundit}{}%
 \ifthenelse{\equal{#1}{Yu32c}}{Y^{u*}_{32} \foundit}{}%
 \ifthenelse{\equal{#1}{Yu33c}}{Y^{u*}_{33} \foundit}{}%
 \ifthenelse{\equal{#1}{Yd11}}{Y^d_{11} \foundit}{}%
 \ifthenelse{\equal{#1}{Yd12}}{Y^d_{12} \foundit}{}%
 \ifthenelse{\equal{#1}{Yd13}}{Y^d_{13} \foundit}{}%
 \ifthenelse{\equal{#1}{Yd21}}{Y^d_{21} \foundit}{}%
 \ifthenelse{\equal{#1}{Yd22}}{Y^d_{22} \foundit}{}%
 \ifthenelse{\equal{#1}{Yd23}}{Y^d_{23} \foundit}{}%
 \ifthenelse{\equal{#1}{Yd31}}{Y^d_{31} \foundit}{}%
 \ifthenelse{\equal{#1}{Yd32}}{Y^d_{32} \foundit}{}%
 \ifthenelse{\equal{#1}{Yd33}}{Y^d_{33} \foundit}{}%
 \ifthenelse{\equal{#1}{Yd11c}}{Y^{d*}_{11} \foundit}{}%
 \ifthenelse{\equal{#1}{Yd12c}}{Y^{d*}_{12} \foundit}{}%
 \ifthenelse{\equal{#1}{Yd13c}}{Y^{d*}_{13} \foundit}{}%
 \ifthenelse{\equal{#1}{Yd21c}}{Y^{d*}_{21} \foundit}{}%
 \ifthenelse{\equal{#1}{Yd22c}}{Y^{d*}_{22} \foundit}{}%
 \ifthenelse{\equal{#1}{Yd23c}}{Y^{d*}_{23} \foundit}{}%
 \ifthenelse{\equal{#1}{Yd31c}}{Y^{d*}_{31} \foundit}{}%
 \ifthenelse{\equal{#1}{Yd32c}}{Y^{d*}_{32} \foundit}{}%
 \ifthenelse{\equal{#1}{Yd33c}}{Y^{d*}_{33} \foundit}{}%
 \ifthenelse{\not{\boolean{found}}}{#1 Help}{}%
 }

 \chapter{Hierarchical Yukawa couplings and Observables}
 \label{SecCKMParams}

In this appendix, we find the masses and
unitary matrices associated with a general
hierarchical Yukawa coupling matrix.  We use
this expansion to find general expressions
for $V_{CKM}$ to the order needed for
comparison to current experiments.
Consider a general hierarchical Yukawa matrix
   \begin{equation}
       Y^u = y_t \left( \begin{matrix}
       Y^u_{11} &   Y^u_{12} &   Y^u_{13} \cr
       Y^u_{21} &   Y^u_{22} &   Y^u_{23} \cr
       Y^u_{31} &   Y^u_{32} &   1
       \end{matrix}
       \right)
\ \   \approx \ \
       Y^u = y_t \left( \begin{matrix}
      {\mathcal{O}}(\epsilon^4) &  {\mathcal{O}}(\epsilon^3) &   {\mathcal{O}}(\epsilon^3) \cr
      {\mathcal{O}}(\epsilon^3) &  {\mathcal{O}}( \epsilon^2) &  {\mathcal{O}}( \epsilon^2) \cr
      {\mathcal{O}}(\epsilon^3) &   {\mathcal{O}}(\epsilon^2) &   1
       \end{matrix}
       \right)
       \label{EqYukawaOrderOfMagnitude}
   \end{equation}
where $\epsilon$ is a small parameter.
Diagonalization of $Y^u \, Y^{u \dag}$ leads
 to
 the diagonal matrix from
 $U^u_L\, Y^u\, Y^{u \dag}\, U^{u \dag}_L = |D_u|^2$
 where
 $Y^u= U_L^{u \dag} D_u U^u_R$. The matrix $|D^u|^2$ gives
 the square of the mass eigenstates.
  \begin{eqnarray}
    |D^u_3 / y_t|^2 & = & 1 + \Mvariable{Yu23}\,\Mvariable{Yu23c} +
  \Mvariable{Yu32}\,\Mvariable{Yu32c} + {\mathcal{O}(\epsilon^5)} \\
  |D^u_2 / y_t|^2 & = &
  \Mvariable{Yu22}\,\Mvariable{Yu22c} + \Mvariable{Yu12}\,\Mvariable{Yu12c} +
  \Mvariable{Yu21}\,\Mvariable{Yu21c}  -
  \Mvariable{Yu22c}\,\Mvariable{Yu23}\,
   \Mvariable{Yu32} -
  \Mvariable{Yu22}\,\Mvariable{Yu23c}\,
   \Mvariable{Yu32c} + {\mathcal{O}(\epsilon^7)}  \\
    |D^u_3 / y_t |^2 & = & \Mvariable{Yu11}\,\Mvariable{Yu11c} -
  \frac{\Mvariable{Yu11c}\,\Mvariable{Yu12}\,
     \Mvariable{Yu21}}{\Mvariable{Yu22}} -
  \frac{\Mvariable{Yu11}\,\Mvariable{Yu12c}\,
     \Mvariable{Yu21c}}{\Mvariable{Yu22c}} +
  \frac{\Mvariable{Yu12}\,\Mvariable{Yu12c}\,
     \Mvariable{Yu21}\,\Mvariable{Yu21c}}{
     \Mvariable{Yu22}\,\Mvariable{Yu22c}} + {\mathcal{O}(\epsilon^{10})}
  \end{eqnarray}
  The resulting approximate
 expressions  for $U_L^{u \dag}$ are
 \begin{eqnarray}
   (U_L^{u \dag})^{13} = Y^u_{13}  +  Y^u_{12}\,  Y^{u*}_{32}  + {\mathcal{O}}(\epsilon^6)\\
   (U_L^{u \dag})^{23} = Y^u_{23}  +   Y^u_{22}\,  Y^{u*}_{32} + {\mathcal{O}}(\epsilon^6) \\
   (U_L^{u \dag})^{33} = 1 - \frac{1}{2} Y^u_{23} Y^{u*}_{23} + {\mathcal{O}}(\epsilon^6)
 \end{eqnarray}
 \begin{eqnarray}
   (U_L^{u \dag})^{12} &=&
\frac{\Mvariable{Yu12}}{\Mvariable{Yu22}} \\
& &  \nonumber -
  \frac{{\Mvariable{Yu12}}^2\,\Mvariable{Yu12c}}
   {2\,{\Mvariable{Yu22}}^2\,\Mvariable{Yu22c}} -
  \frac{\Mvariable{Yu12}\,\Mvariable{Yu21}\,
     \Mvariable{Yu21c}}{{\Mvariable{Yu22}}^2\,
     \Mvariable{Yu22c}} +
  \frac{\Mvariable{Yu11}\,\Mvariable{Yu21c}}
   {\Mvariable{Yu22}\,\Mvariable{Yu22c}} -
  \frac{\Mvariable{Yu13}\,\Mvariable{Yu32}}
   {\Mvariable{Yu22}} +
  \frac{\Mvariable{Yu12}\,\Mvariable{Yu23}\,
     \Mvariable{Yu32}}{{\Mvariable{Yu22}}^2}
  + {\mathcal{O}}(\epsilon^5)     \\
   (U_L^\dag)^{22} &=& 1 - \frac{1}{2} \left( |Y^u_{23}|^2 + \left|\frac{Y^u_{12}}{Y^u_{22}}\right|^2  \right) +  {\mathcal{O}}(\epsilon^4)\\
   (U_L^\dag)^{32} &=&  - \Mvariable{Yu23c} -
  \Mvariable{Yu22c}\,\Mvariable{Yu32}
   - \frac{\Mvariable{Yu12}\,\Mvariable{Yu13c}}
     {\Mvariable{Yu22}}   +
  \frac{\Mvariable{Yu12}\,\Mvariable{Yu12c}\,
     \Mvariable{Yu23c}}{2\,\Mvariable{Yu22}\,
     \Mvariable{Yu22c}} +  {\mathcal{O}}(\epsilon^5)
 \end{eqnarray}
 \begin{eqnarray}
   (U_L^{u \dag})^{11} &=&  1 - \frac{1}{2} \left|\frac{Y^u_{12}}{Y^u_{22}} \right|^2 +  {\mathcal{O}}(\epsilon^4)\\
   (U_L^{u \dag})^{21} &=&  -
  \frac{\Mvariable{Yu12c}}{\Mvariable{Yu22c}} \\
  & & + \frac{\Mvariable{Yu12}\,{\Mvariable{Yu12c}}^2}
   {2\,\Mvariable{Yu22}\,{\Mvariable{Yu22c}}^2} +
  \frac{\Mvariable{Yu12c}\,\Mvariable{Yu21}\,
     \Mvariable{Yu21c}}{\Mvariable{Yu22}\,
     {\Mvariable{Yu22c}}^2}  -
  \frac{\Mvariable{Yu11c}\,\Mvariable{Yu21}}
   {\Mvariable{Yu22}\,\Mvariable{Yu22c}} +
  \frac{\Mvariable{Yu13c}\,\Mvariable{Yu32c}}
   {\Mvariable{Yu22c}} -
  \frac{\Mvariable{Yu12c}\,\Mvariable{Yu23c}\,
     \Mvariable{Yu32c}}{{\Mvariable{Yu22c}}^2} +  {\mathcal{O}}(\epsilon^5)  \nonumber \\
   (U_L^\dag)^{31} &=&  -\Mvariable{Yu13c} + \frac{\Mvariable{Yu12c}\,
     \Mvariable{Yu23c}}{\Mvariable{Yu22c}} +  {\mathcal{O}}(\epsilon^5).
 \end{eqnarray}
\normalsize

\baselineskip=21pt plus1pt
\noindent
These results have been checked numerically
including the phase.

If we assume the $Y^d$ Yukawa matrices follow
the same form as
Eq.~\ref{EqYukawaOrderOfMagnitude}, we can
obtain parallel expressions for $U^d_L$. The
CKM matrix is then given by $V_{CKM} = U^u_L
U^{d \dag}_L$ and, we obtain these
expressions for the individual components:

\small
 \begin{eqnarray}
V_{us} & = &
\frac{\Mvariable{Yd12}}{\Mvariable{Yd22}} -
\frac{\Mvariable{Yu12}}{\Mvariable{Yu22}} +
\\ \nonumber
& & \frac{- {\Mvariable{Yd12}}^2\,
       \Mvariable{Yd12c}  }{2\,
     {\Mvariable{Yd22}}^2\,\Mvariable{Yd22c}} -
  \frac{\Mvariable{Yd12}\,\Mvariable{Yd21}\,
     \Mvariable{Yd21c}}{{\Mvariable{Yd22}}^2\,
     \Mvariable{Yd22c}} +
  \frac{\Mvariable{Yd11}\,\Mvariable{Yd21c}}
   {\Mvariable{Yd22}\,\Mvariable{Yd22c}} -
  \frac{\Mvariable{Yd13}\,\Mvariable{Yd32}}
   {\Mvariable{Yd22}} +
  \frac{\Mvariable{Yd12}\,\Mvariable{Yd23}\,
     \Mvariable{Yd32}}{{\Mvariable{Yd22}}^2} +
  \frac{\Mvariable{Yd12}\,\Mvariable{Yd12c}\,
     \Mvariable{Yu12}}{2\,\Mvariable{Yd22}\,
     \Mvariable{Yd22c}\,\Mvariable{Yu22}} \\ & & +
  \frac{{\Mvariable{Yu12}}^2\,\Mvariable{Yu12c}}
   {2\,{\Mvariable{Yu22}}^2\,\Mvariable{Yu22c}} +
  \frac{\Mvariable{Yu12}\,\Mvariable{Yu21}\,
     \Mvariable{Yu21c}}{{\Mvariable{Yu22}}^2\,
     \Mvariable{Yu22c}}  -
  \frac{\Mvariable{Yd12}\,\Mvariable{Yu12}\,
     \Mvariable{Yu12c}}{2\,\Mvariable{Yd22}\,
     \Mvariable{Yu22}\,\Mvariable{Yu22c}} -
  \frac{\Mvariable{Yu11}\,\Mvariable{Yu21c}}
   {\Mvariable{Yu22}\,\Mvariable{Yu22c}} +
  \frac{\Mvariable{Yu13}\,\Mvariable{Yu32}}
   {\Mvariable{Yu22}} -
  \frac{\Mvariable{Yu12}\,\Mvariable{Yu23}\,
     \Mvariable{Yu32}}{{\Mvariable{Yu22}}^2}
     +
{\mathcal{O}(\epsilon^{5})} \nonumber
\\
V_{cd} & = &  -
\frac{\Mvariable{Yd12c}}{\Mvariable{Yd22c}}+
\frac{\Mvariable{Yu12c}}{\Mvariable{Yu22c}}
\\ \nonumber & & -
  \frac{\Mvariable{Yu12}\,{\Mvariable{Yu12c}}^2}
   {2\,\Mvariable{Yu22}\,{\Mvariable{Yu22c}}^2} -
  \frac{\Mvariable{Yu12c}\,\Mvariable{Yu21}\,
     \Mvariable{Yu21c}}{\Mvariable{Yu22}\,
     {\Mvariable{Yu22c}}^2}  +
  \frac{\Mvariable{Yd12c}\,\Mvariable{Yu12}\,
     \Mvariable{Yu12c}}{2\,\Mvariable{Yd22c}\,
     \Mvariable{Yu22}\,\Mvariable{Yu22c}} +
  \frac{\Mvariable{Yu11c}\,\Mvariable{Yu21}}
   {\Mvariable{Yu22}\,\Mvariable{Yu22c}} -
  \frac{\Mvariable{Yu13c}\,\Mvariable{Yu32c}}
   {\Mvariable{Yu22c}} +
  \frac{\Mvariable{Yu12c}\,\Mvariable{Yu23c}\,
     \Mvariable{Yu32c}}{{\Mvariable{Yu22c}}^2}
\\ & &
\frac{\Mvariable{Yd12}\,{\Mvariable{Yd12c}}^2}
   {2\,\Mvariable{Yd22}\,{\Mvariable{Yd22c}}^2} +
  \frac{\Mvariable{Yd12c}\,\Mvariable{Yd21}\,
     \Mvariable{Yd21c}}{\Mvariable{Yd22}\,
     {\Mvariable{Yd22c}}^2} -
  \frac{\Mvariable{Yd11c}\,\Mvariable{Yd21}}
   {\Mvariable{Yd22}\,\Mvariable{Yd22c}} +
  \frac{\Mvariable{Yd13c}\,\Mvariable{Yd32c}}
   {\Mvariable{Yd22c}} -
  \frac{\Mvariable{Yd12c}\,\Mvariable{Yd23c}\,
     \Mvariable{Yd32c}}{{\Mvariable{Yd22c}}^2}-
  \frac{\Mvariable{Yd12}\,\Mvariable{Yd12c}\,
     \Mvariable{Yu12c}}{2\,\Mvariable{Yd22}\,
     \Mvariable{Yd22c}\,\Mvariable{Yu22c}}
     +
{\mathcal{O}(\epsilon^{5})} \nonumber
 \end{eqnarray}
 \begin{eqnarray}
V_{ub} & = & \Mvariable{Yd13} -
\Mvariable{Yu13} -
  \frac{\Mvariable{Yd23}\,\Mvariable{Yu12}}
   {\Mvariable{Yu22}} +
  \frac{\Mvariable{Yu12}\,\Mvariable{Yu23}}
   {\Mvariable{Yu22}} +
{\mathcal{O}(\epsilon^{5})}
\\
V_{td} & = & -\Mvariable{Yd13c} +
  \Mvariable{Yu13c} +
\frac{\Mvariable{Yd12c}\,
     \Mvariable{Yd23c}}{\Mvariable{Yd22c}}  - \frac{\Mvariable{Yd12c}\,
     \Mvariable{Yu23c}}{\Mvariable{Yd22c}} +
{\mathcal{O}(\epsilon^{5})}
 \end{eqnarray}
 \begin{eqnarray}
V_{cb} & = & - \Mvariable{Yu23}
+\Mvariable{Yd23} +
{\mathcal{O}(\epsilon^{4})}
\\
V_{ts} & = &  \Mvariable{Yu23c} -
\Mvariable{Yd23c} +
{\mathcal{O}(\epsilon^{4})}
 \end{eqnarray}
\normalsize
\baselineskip=21pt plus1pt

\noindent
In the above equations, we have included
enough terms to achieve experimental accuracy
for the Yukawa couplings.

For completeness, we include the diagonal
entries
\begin{eqnarray}
V_{ud} & = & 1 -
\frac{\Mvariable{Yd12}\,\Mvariable{Yd12c}}
   {2\,\Mvariable{Yd22}\,\Mvariable{Yd22c}} +
  \frac{\Mvariable{Yd12c}\,\Mvariable{Yu12}}
   {\Mvariable{Yd22c}\,\Mvariable{Yu22}} -
  \frac{\Mvariable{Yu12}\,\Mvariable{Yu12c}}
   {2\,\Mvariable{Yu22}\,\Mvariable{Yu22c}}
\\
V_{cs} & = & 1 -
\frac{\Mvariable{Yd12}\,\Mvariable{Yd12c}}
   {2\,\Mvariable{Yd22}\,\Mvariable{Yd22c}} +
  \frac{\Mvariable{Yd12}\,\Mvariable{Yu12c}}
   {\Mvariable{Yd22}\,\Mvariable{Yu22c}} -
  \frac{\Mvariable{Yu12}\,\Mvariable{Yu12c}}
   {2\,\Mvariable{Yu22}\,\Mvariable{Yu22c}} \\
V_{tb} & = & 1 +
\Mvariable{Yd23}\,\Mvariable{Yu23c} -
\frac{\Mvariable{Yd23}\,\Mvariable{Yd23c}}{2}
 -
  \frac{\Mvariable{Yu23}\,\Mvariable{Yu23c}}{2}
 \end{eqnarray}

To express the  Wolfenstein $A$ and $\lambda$
terms to experimental accuracy, one needs the
lengthy expression for $V_{us}$. Because the
Wolfenstein parameters are redundant with the
$V_{CKM}$ elements from which they are based,
we will omit explicit expressions for $A$ and
$\lambda$.  The Wolfenstein parameters $\bar
\rho + i\,\bar \eta$ represent a phase
convention independent expression for the CP
violation and can be expressed compactly:
 \begin{eqnarray}
 \bar \rho + i \, \bar \eta &= & - \frac{\Mvariable{Yd13c} -
      \Mvariable{Yu13c} -
      \frac{\Mvariable{Yd23c}\,\Mvariable{Yu12c}}
       {\Mvariable{Yu22c}} +
      \frac{\Mvariable{Yu12c}\,\Mvariable{Yu23c}}
       {\Mvariable{Yu22c}}}{\left( - \frac
             {\Mvariable{Yd12c}}{\Mvariable{Yd22c}}
             +
        \frac{\Mvariable{Yu12c}}{\Mvariable{Yu22c}}
        \right) \,\left( \Mvariable{Yd23c} -
        \Mvariable{Yu23c} \right) }.
 \end{eqnarray}
 The Jarkslog CP Invariant is given by
 \small
 \begin{eqnarray}
  J_{CP} & = & {\rm{Im}} \left(
  - \frac{\left( \frac{\Mvariable{Yd12c}}
            {\Mvariable{Yd22c}} -
           \frac{\Mvariable{Yu12c}}
            {\Mvariable{Yu22c}} \right) \,
         \left( -\left( \Mvariable{Yd23}\,
              \Mvariable{Yu12} \right)  +
           \Mvariable{Yd13}\,\Mvariable{Yu22} -
           \Mvariable{Yu13}\,\Mvariable{Yu22} +
           \Mvariable{Yu12}\,\Mvariable{Yu23}
           \right) \,
         \left( \Mvariable{Yd23c} -
           \Mvariable{Yu23c} \right) }{\Mvariable{Yu22}} \right).
 \end{eqnarray}
\normalsize
\baselineskip=21pt plus1pt





\chapter{Verifying $M_{T2}$ in Eq(5.4)}
\label{AppendixMT2Eq54}

We derived the $M_{T2}$ side of Eq(\ref{EqMCTMT2Equality2}) by following the analytic solution given by Barr and Lester in \cite{Lester:2007fq}.  In this appendix, we outline how to verify that $M_{T2}$ is is indeed given by
  \begin{equation}
    M_{T2} (\chi=0, \alpha , \beta , \slashed{P}_T = -\alpha_T - \beta_T) = 2 ( {\alpha}_T \cdot {\beta}_T + |{\alpha}_T |\,| {\beta}_T |) \label{EqAppendixEquality}
  \end{equation}
when $\alpha^2=\beta^2=0$ and $p^2=q^2=\chi^2=0$ and $g_T=0$.  To do this we note that $M_{T2}$ can also be defined as the minimum value of $(\alpha+p)^2$ minimized over $p$ and $q$ subject to the conditions $p^2=q^2=\chi^2$ (on-shell dark-matter particle state), and $(\alpha+p)^2 = (\beta+q)^2$ (equal on-shell parent-particle state), and  $(\alpha+\beta+p+q+g)_T=0$ (conservation of transverse momentum) \cite{Ross:2007rm}.

The solution which gives Eq(\ref{EqAppendixEquality}) has $p_T = -\beta_T$ and $q_T=-\alpha_T$ with the rapidity of $p$($q$) equal to the rapidity of $\alpha$ ($\beta$).  We now verify that this solution satisfies all these constraints.  Transverse momentum conservation is satisfied trivially: $(\alpha+\beta+p+q)_T=(\alpha+\beta-\alpha -\beta)_T =0$.  The constraint to have the parent particles on-shell can be verified with $2|\alpha_T||p_T| - 2\vec{p}_T  \cdot \vec{\alpha}_T =2|\beta_T||q_T| - 2\vec{q}_T \cdot \vec{\beta}_T  = 2|\beta_T||\beta_T| + 2\vec{\alpha}_T \cdot \vec{\beta}_T$.

Now all that remains is to show that the parent particle's mass is at a minimum with respect to ways in which one splits up $p$ and $q$ to satisfy $p_T+q_T=\slashed{P}_T$ while satisfying the above constraints.  We take $p$ and $q$ to be a small deviation from the stated solution ${p}_T = -{\beta}_T +{\delta}_T$ and ${q}_T = -\alpha_T - {\delta}_T$ where $\delta_T$ is the small deviation in the transverse plane.   We keep $p$ and $q$ on shell at $\chi=0$. The terms $p_o$, $p_z$, $q_o$, $q_z$ are maintained at their minimum by keeping the rapidity of $p$ and $q$ equal to $\alpha$ and $\beta$.  The condition $(p+\alpha)^2=(q+\beta)^2$ is satisfied for a curve of values for $\delta_T$.  The deviation tangent to this curve near $|\delta_T|=0$ is given by
$\delta_T(\lambda) = \lambda\, \hat{z} \times (\alpha_T |\beta_T| + \beta_T |\alpha_T|)$ where $\times$ is a cross product, $\hat{z}$ denotes the beam direction, and we
parameterized the magnitude by the scalar $\lambda$.  Finally, we can substitute $p$ and $q$ with the deviation $\delta_T(\lambda)$ back into the expression  $(\alpha+p)^2$  and verify that $2 ( {\alpha}_T \cdot {\beta}_T + |{\alpha}_T |\,| {\beta}_T |)$ at $\lambda=0$ is indeed the minimum.

\chapter{Fitting Distributions to Data}
\label{AppendixChiSqFitting}

In order to determine $M_N$(which is $M_{\N{1}}$ in our case studies) we perform a $\chi^{2}$
fit between ideal constrained mass variable distributions and the \herwig\ data or other low-statistics data.  Because the mass difference is a given, it does not matter whether we work with $M_N$ or $M_Y$ as the independent variable.

First we illustrate the procedure if there is only one distribution being fit as in Chapter \ref{ChapterM2Cdirect}.
We work with the constrained mass variable $M_{2C}$ using $M_Y$ as the independent variable.
To do this we define a $\chi^{2}$
distribution by computing the number of events, $C_{j},$ in a given range,
$j,$ (bin $j$) of $M_{2C}.$  The variable $N$ is the total number of events in bins that will be fit.
Assuming a Poisson distribution, we assign an
uncertainty, $\sigma_{j}$, to each bin $j$ given by
\begin{equation}
\sigma_{j}^{2}=\frac{1}{2}\left(  N\,f({M_{XC}}_{j},M_{Y}) + C_{j}\right)  .
\end{equation}
Here the normalized distribution of ideal events is $f(M_{2C},M_{Y})$, and the
second term has been added to ensure that the contribution of bins with very
few events, where Poisson statistics does not apply\footnote{By this we mean
that $N\,f({M_{2C}}_{j},M_{Y})$ has a large percent error when used as a
predictor of the number of counts $C_{j}$ when $N\,f({M_{2C}}_{j},M_{Y})$ is
less than about 5.}, have a reasonable weighting. Then $\chi^{2}$ is given by
\begin{equation}
\chi^{2}(M_{Y})=\sum_{\mathrm{bin}\ j}  \left(  \frac
{C_{j}-N\,f({M_{2C}}_{j},M_{Y})}{\sigma_{j}}\right)  ^{2} .
\end{equation}
The minimum $\chi^{2}(M_{Y})$ is our estimate of $M_{Y}$. The amount $M_{Y}$
changes for an increase of $\chi^{2}$ by one gives our $1\,\sigma$
uncertainty, $\delta M_{Y}$, for $M_{Y}$ \cite{Bevington}. As justification
for this we calculate ten different seed random numbers to generate ten
distinct groups of 250 events. We check that the $M_{Y}$ estimates for the ten
sets are distributed with about 2/3 within $\delta M_{Y}$ of the true $M_{Y}$
as one would expect for $1\,\sigma$ error bars. One might worry that with our
definition of $\chi^{2}$, the value of $\chi^{2}$ per degree of freedom is
less than one. However this is an artifact of the fact that the bins with very
few or zero events are not adequately described by Poisson statistics and if
we remove them we do get a reasonable $\chi^{2}$ per degree of freedom. The
determination of $M_{Y}$ using this reduced set gives similar results.

In Chapters \ref{ChapterM2CwUTM} and \ref{ChapterM3C} we determine the
mass with two distributions (the upper and the lower bound distributions).
We treat these two distributions separately and add the resulting $\chi^2$ to form a final $\chi^2$.  In these chapters we choose $M_{\N{1}}$ as the independent variable. We illustrate with $M_{2C,LB}$ and $M_{2C,UB}$, but the procedure is the same for $M_{3C}$.

First we update the definitions for this case.
We define $N_{LB}$ as the number of $M_{2C}$ events in the region to be fit, and likewise
$N_{UB}$ is the number of $M_{2C,UB}$ events in the region to be fit.
The $M_{2C}$ of the events are grouped into bins, $C_{j},$ in a given range,
$j$.  The variable $f_{LB}(M_{2C},M_{\N{1}})$ is the normalized $M_{2C}$ distribution of ideal events expected in bin $j$ as calculated with an assumed $M_{\N{1}}$, the measured $\Deltam$, the observed $m_{ll}$ distribution, the observed UTM distribution, and the appropriate detector simulator.  We likewise define the upper bound distribution to be $f_{UB}(M_{2C,UB},M_{\N{1}})$.
We also define the background distribution for lower-bound and
upper-bound distributions to be $f_{B,LB}(M_{2C})$  and $f_{B,UB}(M_{2C,UB})$ and the fraction of the total events we estimate are from background $\lambda$.

Again, we assign an
uncertainty, $\sigma_{j}$, to each bin $j$ given by
\begin{equation}
\sigma_{LB,j}^{2}(M_{\N{1}})=\frac{1}{2}\left(  N_{LB}\,((1-\lambda)f_{LB}({M_{2C}}_{j},M_{\N{1}})+\lambda f_{B,LB}({M_{2C}}_j))  + C_{j}\right),
\end{equation}
and likewise for the upper-bound distribution.
The second term has been added to ensure an appropriate weighting of bins with very
few events that does not bias the fit towards or away from this end-point. In bins with few counts, normal Poisson statistics does not apply\footnote{By this we mean
that $N\,f({M_{2C}}_{j},M_{\N{1}})$ has a large percent error when used as a
predictor of the number of counts $C_{j}$ when $N\,f({M_{2C}}_{j},M_{\N{1}})$ is
less than about 5.}.

The $\chi^{2}$ is given by
\begin{eqnarray}
\chi^{2}(M_{\N{1}}) & = &\sum_{\mathrm{bin}\ j} \left(
\frac{C_{j}-N_{LB}\,(1-\lambda)\,f_{LB}({M_{2C}}_{j},M_{\N{1}}) - N_{LB} \,\lambda\,f_{B,LB}({M_{2C}}_j,M_{\N{1}})}{\sigma_{LB,j}}\right)  ^{2} \\
 & & + \sum_{\mathrm{bin}\ j} \left(
\frac{C_{UB,j}-N_{UB}\,(1-\lambda)\,f_{UB}({M_{2C,UB}}_{j},M_{\N{1}}) - N_{UB} \,\lambda\,f_{B,UB}({M_{2C,UB}}_j,M_{\N{1}})}{\sigma_{UB,j}}\right) ^{2}. \nonumber
\end{eqnarray}
We calculate ideal distributions for $M_{\N{1}}=50, 60, 70, 80, 90 \GeV$. We fit quadratic interplant through the points.
The minimum $\chi^{2}(M_{\N{1}})$ of the interplant is our estimate of $M_{\N{1}}$.
The amount $M_{\N{1}}$
changes for an increase in $\chi^{2}$ by one gives our $1\,\sigma$
uncertainty, $\delta M_{\N{1}}$, for $M_{\N{1}}$ \cite{Bevington}.

The $M_{3C}$ fits in Chapter \ref{ChapterM3C} were performed following this same procedure.  In this case we choose $M_{\N{1}}=80, 85, 90, 95, 100, 105, 110$ GeV.  For the $M_{3C}$ studies we did not create a background model and therefore fixed the background parameter to $\lambda=0$.  The studies involving nine different choices of $(\Delta M_{YN},\Delta M_{XN})$ were done using $M_{\N{1}}=90, 100, 110$ GeV.

Some subtleties for which we should check.
If the bin size is not large enough, the artificially large variations in the distribution sometimes bias the fits to place the endpoint near a fluctuation.
The bin size should be made large enough so that this does not happen.
This can be checked for by testing if the results are invariant with respect to changing the bin size by a small amount.
Because of the larger number of lower bound events, the optimal bin size may be different for the upper-bound and the lower-bound distributions.

\chapter{Uniqueness of Event Reconstruction}
\label{AppendixUniquenessOfReconstruction}


In the Chapters \ref{ChapterM2Cdirect} and \ref{ChapterM3C} we claim that the events near an endpoint of $M_{2C}$ and $M_{3C}$ distributions (events that nearly saturate the bound) are nearly reconstructed.  This appendix offers a proof of the claim.  To prove uniqueness, we need to establish that as $M_{3C}$ or $M_{2C}$ of an event (lower bound or upper bound) approach the endpoint of the distributions, the solutions with different values of $q$ and $p$ approach a common solution.

\begin{figure}
\centerline{\includegraphics{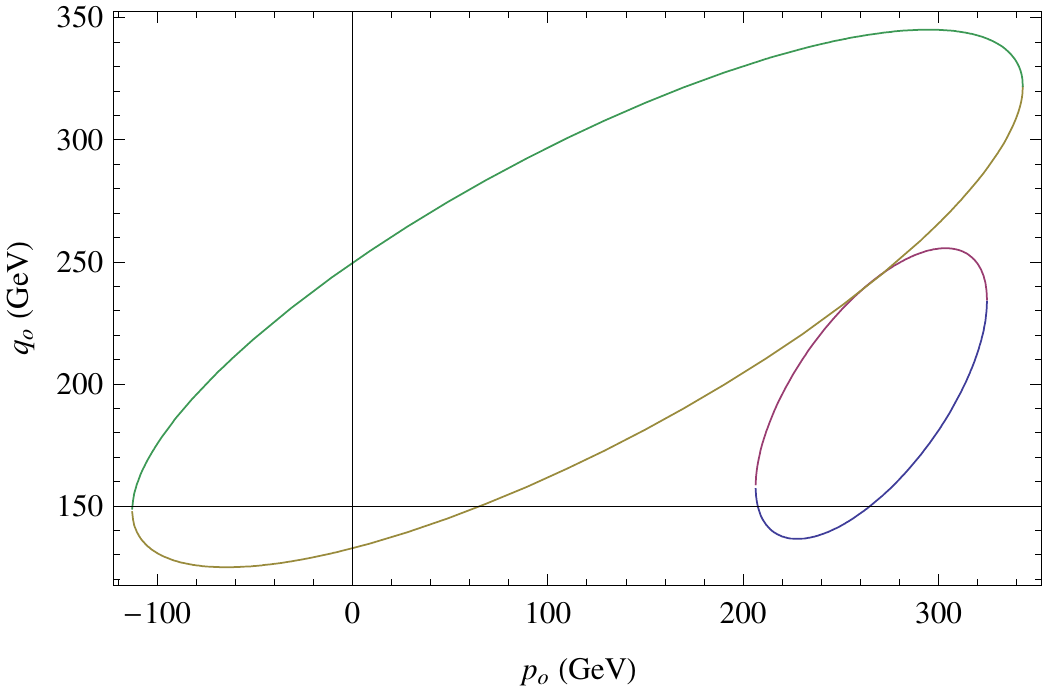}}
\caption{\label{FigpoqoEllipse} Shows the ellipses defined for $p_o$ and $q_o$ in  Eqs(\ref{Eqpo}-\ref{Eqqo}) using the correct mass scale for an event that nearly saturates the $M_{3C}$ endpoint.  For this event, the $M_{3C}$ lies within $1\%$  of the endpoint and reconstructs $p$ and $q$ to within $4\%$. Perfect error resolution and combinatorics are assumed.}
\end{figure}
We begin with $M_{3C}$.  Sec.~\ref{SecNumericallyCalculatingM3C} shows that there are at most four solutions given $M_N$, $M_X$ and $M_Y$ formed by the intersection of two ellipses in $(p_o, q_o)$ defined by Eqs(\ref{Eqpo}-\ref{Eqqo}) as shown in Fig~\ref{FigpoqoEllipse}.  Consider the case that an event has a lower bound $M_{3C}$ near $M_Y$.  We are guaranteed that a solution occurs at the true mass scale when we choose the correct combinatoric assignments.  As one varies the mass scale downward, the two ellipses drift and change shape and size so that four solutions become two solutions and eventually, at the value of $M_{3C}$ for the event, become one single solution.
When the disconnection of the two ellipses occurs near the true mass scale, the value of $M_{3C}$ will be near the endpoint.  The unique solutions for $p$ and $q$ given at $M_{3C}$  are nearly degenerate with the true values of $p$ and $q$ found when one uses the true masses to solve for $p$ and $q$.  The closer $M_{3C}$ is to the endpoint the closer the two ellipses are to intersecting at a single point when the true masses are used and to giving a unique reconstruction.  The example pictured in Fig.~\ref{FigpoqoEllipse} show an event with $M_{3C}$ within $1\%$ of the endpoint and where the $p$ and $q$ are reconstructed to within $4\%$.  This shows that for $M_{3C}$ events that are near the endpoint allowing for any choice of combinatorics then nearly reconstruct the true values for $p$ and $q$.   If there are combinatoric ambiguities, one need to test all combinatoric possibilities.  If the minimum combinatoric option has a lower bound at the end-point, the above arguments follow unchanged.
The above arguments can be repeated to show $M_{3C,UB}$ near the endpoint also reconstructs the correct $p$ and $q$.

Next we turn to $M_{2C}$. For every event the lower-bounds satisfy $M_{2C}(M_Y-M_N) \leq M_{3C}(M_Y-M_N,M_X-M_N)$.  With $M_{2C}$ the propagator $(p+\alpha_2)^2$, which we can equate with $\chi_X^2$, is not fixed.  The kinematically allowed values for $\chi_X$ are $M_N^2 < \chi^2_X < M_Y^2$ assuming the visible states $\alpha_1$ and $\alpha_2$ are massless.
Eq(\ref{EqMatrixSolForpqVec}) shows that $\vec{p}$ and $\vec{q}$ solutions are linear in $\chi^2_X$ with no terms dependent on $\chi_X$ alone or other powers of $\chi_X$.  Including $\chi_X^2$ as a free parameter in Eqs(\ref{Eqpo}-\ref{Eqqo}) leads to two ellipsoids (or hyperboloids) in the space $(p_o,q_o,\chi_X^2)$.  We will assume without loss of generality that these are ellipsoids.  Again, at the true mass scale we are guaranteed the two ellipsoids intersect at an ellipse.  Now as one varies the mass scale the two ellipsoids drift and change shape and size.  The $M_{2C}$ value then corresponds to the mass scale where the two ellipsoids are in contact at one point.
As we select events with a value of $M_{2C}$ that approaches the true mass scale the intersection of the two ellipsoids shrink to a point giving a unique reconstruction of $p$ and $q$.
The combinatoric ambiguities for $M_{2C}$ are avoided by selecting events with two distinct OSSF pairs.
Events that saturate the upper bound of $M_{2C}$ also reconstruct $p$ and $q$ by the same logic as above.

\normalsize

\chapter{Acronyms List}

\begin{itemize}
\item {$\Lambda$CDM} $\Lambda$(Cosmological Constant) Cold Dark Matter
\item {CKM} Cabibbo-Kobayashi-Maskawa
\item {CMB} Cosmic Microwave Background
\item {$\overline{DR}$} Dimensional Reduction (Renormalization Scheme)
\item {GIM}  Glashow-Iliopoulos-Maiani
\item {GJ} Georgi-Jarlskog
\item {GUT} Grand Unified Theory
\item {ISR}  Initial State Radiation
\item {KK} Kaluza Klein
\item {LHC} Large Hadron Collider
\item {LKP} Lightest Kaluza Klein particle
\item {LSP} Lightest supersymmetric particle
\item {$\overline{MS}$} Minimal Subtraction (Renormalization Scheme)
\item {MACHO} MAssive Compact Halo Objects
\item {OSSF} Opposite-Sign Same Flavor
\item {QFT} Quantum Field Theory
\item {REWSB} Radiative Electroweak Symmetry Breaking
\item {RG} Renormalization Group 
\item {RGE} Renormalization Group Equations
\item {SM} Standard Model
\item {SSB} Spontaneous Symmetry Breaking
\item {SUSY} Supersymmetry
\item {UED} Universal Extra Dimensions
\item {UTM} Upstream Transverse Momentum
\item {VEV} Vacuum Expectation Value
\end{itemize}





\newcommand{\noopsort}[1]{} \newcommand{\printfirst}[2]{#1}
  \newcommand{\singleletter}[1]{#1} \newcommand{\switchargs}[2]{#2#1}
\providecommand{\href}[2]{#2}\begingroup\raggedright\endgroup


\begin{thebibliography}{10%
0}

\bibitem{Roberts:2001zy}
R.~G. Roberts, A.~Romanino, G.~G. Ross, and L.~Velasco-Sevilla, {\it Precision
  test of a fermion mass texture},  {\em Nucl. Phys.} {\bf B615} (2001)
  358--384, [\href{http://xxx.lanl.gov/abs/hep-ph/0104088}{{\tt
  hep-ph/0104088}}].

\bibitem{VandelliTesiPhD}
W.~Vandelli, {\em Prospects for the detection of chargino-neutralino direct
  production with ATLAS detector at the LHC}.
\newblock {PhD} dissertation, Universita Degli Studi Di Pavia, Dipartimento Di
  Fisica, 2006.

\bibitem{Allanach:2002nj}
B.~C. Allanach {\em et.~al.}, {\it The snowmass points and slopes: Benchmarks
  for susy searches},  \href{http://xxx.lanl.gov/abs/hep-ph/0202233}{{\tt
  hep-ph/0202233}}.

\bibitem{Boyle:1686ab}
R.~Boyle, {\em A Free Enquiry into the Vulgarly Received Notion of Nature}.
\newblock Cambridge University Press, 1685.
\newblock Introduction of 1996 edition.

\bibitem{wilczek:2004428ab}
F.~Wilczek, {\it From ``not wrong" to (maybe) right},  {\em Nature} {\bf 428}
  (2004) 261.

\bibitem{dagostini:2004ab}
G.~D'Agostini, {\it From observations to hypotheses: Probabilistic reasoning
  versus falsificationism and its statistical variations},
  \href{http://xxx.lanl.gov/abs/physics/0412148}{{\tt physics/0412148}}.

\bibitem{FermiLuminocity}
{CDF Collaboration}, {\it Luminocity web page},  {\em
  {http://www-cdf.fnal.gov/$\sim$konigsb/lum$\_$official$\_$page.html}} (June,
  2008).

\bibitem{PDBook2008}
W.-M. {Yao}, {\it {Review of Particle Physics}},  {\em {Journal of Physics G}}
  {\bf 33} (2008) 1+.

\bibitem{AtlasTDR}
A.~Collaboration, {\it {ATLAS} computing : Technical design report},  {\em
  {CERN}, {ATLAS-TDR-017},{CERN-LHCC-2005-022}} (2005).

\bibitem{Ross:2007az}
G.~Ross and M.~Serna, {\it {Unification and Fermion Mass Structure}},  {\em
  Phys. Lett.} {\bf B664} (2008) 97--102,
  [\href{http://xxx.lanl.gov/abs/0704.1248}{{\tt 0704.1248}}].

\bibitem{Georgi:1979df}
H.~Georgi and C.~Jarlskog, {\it A new lepton - quark mass relation in a unified
  theory},  {\em Phys. Lett.} {\bf B86} (1979) 297--300.

\bibitem{Lester:1999tx}
C.~G. Lester and D.~J. Summers, {\it Measuring masses of semi-invisibly
  decaying particles pair produced at hadron colliders},  {\em Phys. Lett.}
  {\bf B463} (1999) 99--103,
  [\href{http://xxx.lanl.gov/abs/hep-ph/9906349}{{\tt hep-ph/9906349}}].

\bibitem{Barr:2003rg}
A.~Barr, C.~Lester, and P.~Stephens, {\it {m(T2): The truth behind the
  glamour}},  {\em J. Phys.} {\bf G29} (2003) 2343--2363,
  [\href{http://xxx.lanl.gov/abs/hep-ph/0304226}{{\tt hep-ph/0304226}}].

\bibitem{Serna:2008zk}
M.~Serna, {\it {A short comparison between $m_{T2}$ and $m_{CT}$}},  {\em JHEP}
  {\bf 06} (2008) 004, [\href{http://xxx.lanl.gov/abs/0804.3344}{{\tt
  0804.3344}}].

\bibitem{Ross:2007rm}
G.~G. Ross and M.~Serna, {\it {Mass Determination of New States at Hadron
  Colliders}},  {\em Phys. Lett.} {\bf B665} (2008) 212--218,
  [\href{http://xxx.lanl.gov/abs/0712.0943}{{\tt 0712.0943}}].

\bibitem{Barr:2008ba}
A.~J. Barr, G.~G. Ross, and M.~Serna, {\it {The Precision Determination of
  Invisible-Particle Masses at the LHC}},  {\em Phys. Rev.} {\bf D78} (2008)
  056006, [\href{http://xxx.lanl.gov/abs/0806.3224}{{\tt 0806.3224}}].

\bibitem{Corcella:2002jc}
G.~Corcella {\em et.~al.}, {\it {HERWIG 6.5 release note}},
  \href{http://xxx.lanl.gov/abs/hep-ph/0210213}{{\tt hep-ph/0210213}}.

\bibitem{Moretti:2002eu}
S.~Moretti, K.~Odagiri, P.~Richardson, M.~H. Seymour, and B.~R. Webber, {\it
  {Implementation of supersymmetric processes in the HERWIG event generator}},
  {\em JHEP} {\bf 04} (2002) 028,
  [\href{http://xxx.lanl.gov/abs/hep-ph/0204123}{{\tt hep-ph/0204123}}].

\bibitem{Marchesini:1991ch}
G.~Marchesini {\em et.~al.}, {\it {HERWIG: A Monte Carlo event generator for
  simulating hadron emission reactions with interfering gluons. Version 5.1 -
  April 1991}},  {\em Comput. Phys. Commun.} {\bf 67} (1992) 465--508.

\bibitem{Barr:2008hv}
A.~J. Barr, A.~Pinder, and M.~Serna, {\it {Precision Determination of
  Invisible-Particle Masses at the CERN LHC: II}},  {\em Phys. Rev.} {\bf D79}
  (2009) 074005, [\href{http://xxx.lanl.gov/abs/0811.2138}{{\tt 0811.2138}}].

\bibitem{Dirac:1928hu}
P.~A.~M. Dirac, {\it {The Quantum theory of electron}},  {\em Proc. Roy. Soc.
  Lond.} {\bf A117} (1928) 610--624.

\bibitem{PhysRev.43.491}
C.~D. Anderson, {\it The positive electron},  {\em Phys. Rev.} {\bf 43} (Mar,
  1933) 491--494.

\bibitem{NimaFineTune}
E.~Lane, {\it {Harvard's Nima Arkani-Hamed Ponders New Universes, Different
  Dimensions}},  {\em {AAAS}: Advancing Science Serving Society} (May, 2005).

\bibitem{CERNCourierWeisskopfSpecial}
J.~D. Jackson, {\it Weisskopf tribute},  {\em Cern Courier} (Dec, 2002).

\bibitem{PhysRev.56.72}
V.~F. Weisskopf, {\it On the self-energy and the electromagnetic field of the
  electron},  {\em Phys. Rev.} {\bf 56} (Jul, 1939) 72--85.

\bibitem{GellMann:1961ky}
M.~Gell-Mann, {\it {The Eightfold Way: A Theory of strong interaction
  symmetry}}, . CTSL-20.

\bibitem{Ne'eman:1961cd}
Y.~Ne'eman, {\it {Derivation of strong interactions from a gauge invariance}},
  {\em Nucl. Phys.} {\bf 26} (1961) 222--229.

\bibitem{NeemanRembered}
T.~Ne'eman, {\it In rememberance of yuval ne'eman (1925-2006)},  {\em
  Physicsaplus} (2006). {http://physicsaplus.org.il}.

\bibitem{Fraser:1998ih}
e.~. Fraser, G., {\it {The particle century}}, . Bristol, UK: IOP (1998) 232 p.

\bibitem{PhysRevLett.12.204}
V.~E. Barnes, P.~L. Connolly, D.~J. Crennell, B.~B. Culwick, W.~C. Delaney,
  W.~B. Fowler, P.~E. Hagerty, E.~L. Hart, N.~Horwitz, P.~V.~C. Hough, J.~E.
  Jensen, J.~K. Kopp, K.~W. Lai, J.~Leitner, J.~L. Lloyd, G.~W. London, T.~W.
  Morris, Y.~Oren, R.~B. Palmer, A.~G. Prodell, D.~Radoji\ifmmode
  \check{c}\else \v{c}\fi{}i\ifmmode~\acute{c}\else \'{c}\fi{}, D.~C. Rahm,
  C.~R. Richardson, N.~P. Samios, J.~R. Sanford, R.~P. Shutt, and J.~R. Smith,
  {\it Observation of a hyperon with strangeness minus three},  {\em Phys. Rev.
  Lett.} {\bf 12} (Feb, 1964) 204--206.

\bibitem{Aubert:2006dc}
{\bf BABAR} Collaboration, B.~Aubert {\em et.~al.}, {\it {Measurement of the
  spin of the Omega- hyperon at BABAR}},  {\em Phys. Rev. Lett.} {\bf 97}
  (2006) 112001, [\href{http://xxx.lanl.gov/abs/hep-ex/0606039}{{\tt
  hep-ex/0606039}}].

\bibitem{Okubo:1961jc}
S.~Okubo, {\it {Note on unitary symmetry in strong interactions}},  {\em Prog.
  Theor. Phys.} {\bf 27} (1962) 949--966.

\bibitem{WeinbergQFT}
S.~Weinberg, {\em Quantum Theory of Fields}.
\newblock Cambridge University Press, 1996.

\bibitem{ChengLi}
T.-P. Cheng and L.-F. Li, {\em Gauge theory of elementary particle physics}.
\newblock Oxford University Press, 1984.

\bibitem{Gaillard:1974hs}
M.~K. Gaillard and B.~W. Lee, {\it {Rare Decay Modes of the K-Mesons in Gauge
  Theories}},  {\em Phys. Rev.} {\bf D10} (1974) 897.

\bibitem{Lee:1968ks}
T.~D. Lee, {\it {ANALYSIS OF DIVERGENCES IN A NEUTRAL SPIN 1 MESON THEORY WITH
  PARITY NONCONSERVING INTERACTIONS}},  {\em Nuovo Cim.} {\bf A59} (1969)
  579--598.

\bibitem{Glashow:1970gm}
S.~L. Glashow, J.~Iliopoulos, and L.~Maiani, {\it {Weak Interactions with
  Lepton-Hadron Symmetry}},  {\em Phys. Rev.} {\bf D2} (1970) 1285--1292.

\bibitem{Borodulin:1995xd}
V.~I. Borodulin, R.~N. Rogalev, and S.~R. Slabospitsky, {\it {CORE: COmpendium
  of RElations: Version 2.1}},
  \href{http://xxx.lanl.gov/abs/hep-ph/9507456}{{\tt hep-ph/9507456}}.

\bibitem{Srednicki:2007qs}
M.~Srednicki, {\it {Quantum field theory}}, . Cambridge, UK: Univ. Pr. (2007)
  641 p.

\bibitem{Ramond:1999vh}
P.~Ramond, {\it {Journeys beyond the standard model}}, . Reading, Mass.,
  Perseus Books, 1999.

\bibitem{PDBook2006}
W.-M. {Yao}, {\it {Review of Particle Physics}},  {\em {Journal of Physics G}}
  {\bf 33} (2006) 1+.

\bibitem{Weinberg:1967tq}
S.~Weinberg, {\it {A Model of Leptons}},  {\em Phys. Rev. Lett.} {\bf 19}
  (1967) 1264--1266.

\bibitem{Greenberg:1964pe}
O.~W. Greenberg, {\it {Spin and Unitary Spin Independence in a Paraquark Model
  of Baryons and Mesons}},  {\em Phys. Rev. Lett.} {\bf 13} (1964) 598--602.

\bibitem{Greenberg:2008fs}
O.~W. Greenberg, {\it {The color charge degree of freedom in particle
  physics}},  \href{http://xxx.lanl.gov/abs/0805.0289}{{\tt 0805.0289}}.

\bibitem{Gross:1973ju}
D.~J. Gross and F.~Wilczek, {\it {Asymptotically Free Gauge Theories. 1}},
  {\em Phys. Rev.} {\bf D8} (1973) 3633--3652.

\bibitem{Politzer:1973fx}
H.~D. Politzer, {\it {RELIABLE PERTURBATIVE RESULTS FOR STRONG INTERACTIONS?}},
   {\em Phys. Rev. Lett.} {\bf 30} (1973) 1346--1349.

\bibitem{LlewellynSmith:1981yk}
C.~H. Llewellyn~Smith and J.~F. Wheater, {\it {Electroweak Radiative
  Corrections and the Value of sin**2- Theta-W}},  {\em Phys. Lett.} {\bf B105}
  (1981) 486.

\bibitem{Arnison:1983rp}
{\bf UA1} Collaboration, G.~Arnison {\em et.~al.}, {\it {Experimental
  observation of isolated large transverse energy electrons with associated
  missing energy at s**(1/2) = 540-GeV}},  {\em Phys. Lett.} {\bf B122} (1983)
  103--116.

\bibitem{Arnison:1983mk}
{\bf UA1} Collaboration, G.~Arnison {\em et.~al.}, {\it {Experimental
  observation of lepton pairs of invariant mass around 95-GeV/c**2 at the CERN
  SPS collider}},  {\em Phys. Lett.} {\bf B126} (1983) 398--410.

\bibitem{Bertone:2004pz}
G.~Bertone, D.~Hooper, and J.~Silk, {\it {Particle dark matter: Evidence,
  candidates and constraints}},  {\em Phys. Rept.} {\bf 405} (2005) 279--390,
  [\href{http://xxx.lanl.gov/abs/hep-ph/0404175}{{\tt hep-ph/0404175}}].

\bibitem{Baer:2008uu}
H.~Baer and X.~Tata, {\it {Dark matter and the LHC}},
  \href{http://xxx.lanl.gov/abs/0805.1905}{{\tt 0805.1905}}.

\bibitem{Oortz1932}
J.~Oortz, {\it The force exerted by the stellar system in the direction
  perpendicular to the galactic plane and some related problems},  {\em
  Bulletin of the Astronomical Institutes of the Netherlands} {\bf VI} (1932)
  249.

\bibitem{2001ARA&A..39..137S}
Y.~{Sofue} and V.~{Rubin}, {\it {Rotation Curves of Spiral Galaxies}},  {\em
  Annual Reviews Astronomy and Astrophysics} {\bf 39} (2001) 137--174,
  [\href{http://xxx.lanl.gov/abs/arXiv:astro-ph/0010594}{{\tt
  arXiv:astro-ph/0010594}}].

\bibitem{2004IAUS..220..211R}
D.~{Russeil}, O.~{Garrido}, P.~{Amram}, and M.~{Marcelin}, {\it {Rotation curve
  of our Galaxy and field galaxies}},  in {\em Dark Matter in Galaxies}
  (S.~{Ryder}, D.~{Pisano}, M.~{Walker}, and K.~{Freeman}, eds.), vol.~220 of
  {\em IAU Symposium}, pp.~211--+, July, 2004.

\bibitem{Zwicky1937}
F.~Swicky, {\it On the masses of nebulae and of clusers of nebulae},  {\em The
  Astrophysical Journal} {\bf 86} (1937) 217.

\bibitem{Bottema:2002zv}
R.~Bottema, J.~L.~G. Pestana, B.~Rothberg, and R.~H. Sanders, {\it {MOND
  rotation curves for spiral galaxies with Cepheid- based distances}},  {\em
  Astron. Astrophys.} {\bf 393} (2002) 453--460,
  [\href{http://xxx.lanl.gov/abs/astro-ph/0207469}{{\tt astro-ph/0207469}}].

\bibitem{Clowe:2006eq}
D.~Clowe {\em et.~al.}, {\it {A direct empirical proof of the existence of dark
  matter}},  {\em Astrophys. J.} {\bf 648} (2006) L109--L113,
  [\href{http://xxx.lanl.gov/abs/astro-ph/0608407}{{\tt astro-ph/0608407}}].

\bibitem{Spergel:1996hw}
D.~Spergel, {\it {Particle Dark Matter}},
  \href{http://xxx.lanl.gov/abs/astro-ph/9603026}{{\tt astro-ph/9603026}}.

\bibitem{Hartle:2003yu}
J.~B. Hartle, {\it {An introduction to Einstein's general relativity}}, . San
  Francisco, USA: Addison-Wesley (2003) 582 p.

\bibitem{Frieman:2008sn}
J.~Frieman, M.~Turner, and D.~Huterer, {\it {Dark Energy and the Accelerating
  Universe}},  \href{http://xxx.lanl.gov/abs/0803.0982}{{\tt 0803.0982}}.

\bibitem{Ahmed:2008eu}
{\bf CDMS} Collaboration, Z.~Ahmed {\em et.~al.}, {\it {A Search for WIMPs with
  the First Five-Tower Data from CDMS}},
  \href{http://xxx.lanl.gov/abs/0802.3530}{{\tt 0802.3530}}.

\bibitem{Angle:2007uj}
{\bf XENON} Collaboration, J.~Angle {\em et.~al.}, {\it {First Results from the
  XENON10 Dark Matter Experiment at the Gran Sasso National Laboratory}},  {\em
  Phys. Rev. Lett.} {\bf 100} (2008) 021303,
  [\href{http://xxx.lanl.gov/abs/0706.0039}{{\tt 0706.0039}}].

\bibitem{Bernabei:1998td}
{\bf DAMA} Collaboration, R.~Bernabei {\em et.~al.}, {\it {On a further search
  for a yearly modulation of the rate in particle dark matter direct search}},
  {\em Phys. Lett.} {\bf B450} (1999) 448--455.

\bibitem{Bernabei:2008yi}
{\bf DAMA} Collaboration, R.~Bernabei {\em et.~al.}, {\it {First results from
  DAMA/LIBRA and the combined results with DAMA/NaI}},
  \href{http://xxx.lanl.gov/abs/0804.2741}{{\tt 0804.2741}}.

\bibitem{Baer:2006te}
H.~Baer, A.~Mustafayev, E.-K. Park, and X.~Tata, {\it {Target dark matter
  detection rates in models with a well- tempered neutralino}},  {\em JCAP}
  {\bf 0701} (2007) 017, [\href{http://xxx.lanl.gov/abs/hep-ph/0611387}{{\tt
  hep-ph/0611387}}].

\bibitem{Ellis:2005mb}
J.~R. Ellis, K.~A. Olive, Y.~Santoso, and V.~C. Spanos, {\it {Update on the
  direct detection of supersymmetric dark matter}},  {\em Phys. Rev.} {\bf D71}
  (2005) 095007, [\href{http://xxx.lanl.gov/abs/hep-ph/0502001}{{\tt
  hep-ph/0502001}}].

\bibitem{Arrenberg:2008wy}
S.~Arrenberg, L.~Baudis, K.~Kong, K.~T. Matchev, and J.~Yoo, {\it {Kaluza-Klein
  Dark Matter: Direct Detection vis-a-vis LHC}},
  \href{http://xxx.lanl.gov/abs/0805.4210}{{\tt 0805.4210}}.

\bibitem{deBoer:2005tm}
W.~de~Boer, C.~Sander, V.~Zhukov, A.~V. Gladyshev, and D.~I. Kazakov, {\it
  {EGRET excess of diffuse galactic gamma rays as tracer of dark matter}},
  {\em Astron. Astrophys.} {\bf 444} (2005) 51,
  [\href{http://xxx.lanl.gov/abs/astro-ph/0508617}{{\tt astro-ph/0508617}}].

\bibitem{Akerib:2004fq}
{\bf CDMS} Collaboration, D.~S. Akerib {\em et.~al.}, {\it {First results from
  the cryogenic dark matter search in the Soudan Underground Lab}},  {\em Phys.
  Rev. Lett.} {\bf 93} (2004) 211301,
  [\href{http://xxx.lanl.gov/abs/astro-ph/0405033}{{\tt astro-ph/0405033}}].

\bibitem{Tegmark:2005cy}
M.~Tegmark, {\it {Cosmological neutrino bounds for non-cosmologists}},  {\em
  Phys. Scripta} {\bf T121} (2005) 153--155,
  [\href{http://xxx.lanl.gov/abs/hep-ph/0503257}{{\tt hep-ph/0503257}}].

\bibitem{ArkaniHamed:1998rs}
N.~Arkani-Hamed, S.~Dimopoulos, and G.~R. Dvali, {\it {The hierarchy problem
  and new dimensions at a millimeter}},  {\em Phys. Lett.} {\bf B429} (1998)
  263--272, [\href{http://xxx.lanl.gov/abs/hep-ph/9803315}{{\tt
  hep-ph/9803315}}].

\bibitem{Appelquist:2000nn}
T.~Appelquist, H.-C. Cheng, and B.~A. Dobrescu, {\it {Bounds on universal extra
  dimensions}},  {\em Phys. Rev.} {\bf D64} (2001) 035002,
  [\href{http://xxx.lanl.gov/abs/hep-ph/0012100}{{\tt hep-ph/0012100}}].

\bibitem{Servant:2002aq}
G.~Servant and T.~M.~P. Tait, {\it {Is the lightest Kaluza-Klein particle a
  viable dark matter candidate?}},  {\em Nucl. Phys.} {\bf B650} (2003)
  391--419, [\href{http://xxx.lanl.gov/abs/hep-ph/0206071}{{\tt
  hep-ph/0206071}}].

\bibitem{Battaglia:2005zf}
M.~Battaglia, A.~Datta, A.~De~Roeck, K.~Kong, and K.~T. Matchev, {\it
  {Contrasting supersymmetry and universal extra dimensions at the CLIC
  multi-TeV e+ e- collider}},  {\em JHEP} {\bf 07} (2005) 033,
  [\href{http://xxx.lanl.gov/abs/hep-ph/0502041}{{\tt hep-ph/0502041}}].

\bibitem{Haag:1974qh}
R.~Haag, J.~T. Lopuszanski, and M.~Sohnius, {\it {All Possible Generators of
  Supersymmetries of the s Matrix}},  {\em Nucl. Phys.} {\bf B88} (1975) 257.

\bibitem{Gliozzi:1976qd}
F.~Gliozzi, J.~Scherk, and D.~I. Olive, {\it {Supersymmetry, Supergravity
  Theories and the Dual Spinor Model}},  {\em Nucl. Phys.} {\bf B122} (1977)
  253--290.

\bibitem{Stockinger:2006zn}
D.~Stockinger, {\it The muon magnetic moment and supersymmetry},
  \href{http://xxx.lanl.gov/abs/hep-ph/0609168}{{\tt hep-ph/0609168}}.

\bibitem{Martin:1997ns}
S.~P. Martin, {\it A supersymmetry primer},
  \href{http://xxx.lanl.gov/abs/hep-ph/9709356}{{\tt hep-ph/9709356}}.

\bibitem{Baer:2006rs}
H.~Baer and X.~Tata, {\it {Weak scale supersymmetry: From superfields to
  scattering events}}, . Cambridge, UK: Univ. Pr. (2006) 537 p.

\bibitem{Gates:1983nr}
S.~J. Gates, M.~T. Grisaru, M.~Rocek, and W.~Siegel, {\it Superspace, or one
  thousand and one lessons in supersymmetry},  {\em Front. Phys.} {\bf 58}
  (1983) 1--548, [\href{http://xxx.lanl.gov/abs/hep-th/0108200}{{\tt
  hep-th/0108200}}].

\bibitem{Wess:1992cp}
J.~Wess and J.~Bagger, {\it {Supersymmetry and supergravity}}, . Princeton,
  USA: Univ. Pr. (1992) 259 p.

\bibitem{Cahill:1999aq}
K.~E. Cahill, {\it {Elements of supersymmetry}},
  \href{http://xxx.lanl.gov/abs/hep-ph/9907295}{{\tt hep-ph/9907295}}.

\bibitem{Aitchison:2005cf}
I.~J.~R. Aitchison, {\it {Supersymmetry and the MSSM: An elementary
  introduction}},  \href{http://xxx.lanl.gov/abs/hep-ph/0505105}{{\tt
  hep-ph/0505105}}.

\bibitem{Ibanez:2007pf}
L.~E. Ibanez and G.~G. Ross, {\it {Supersymmetric Higgs and radiative
  electroweak breaking}},  {\em Comptes Rendus Physique} {\bf 8} (2007)
  1013--1028, [\href{http://xxx.lanl.gov/abs/hep-ph/0702046}{{\tt
  hep-ph/0702046}}].

\bibitem{Ibanez:1982fr}
L.~E. Ibanez and G.~G. Ross, {\it {$SU(2)_L \times U(1)$} symmetry breaking as
  a radiative effect of supersymmetry breaking in guts},  {\em Phys. Lett.}
  {\bf B110} (1982) 215--220.

\bibitem{Inoue:1982pi}
K.~Inoue, A.~Kakuto, H.~Komatsu, and S.~Takeshita, {\it {Aspects of Grand
  Unified Models with Softly Broken Supersymmetry}},  {\em Prog. Theor. Phys.}
  {\bf 68} (1982) 927.

\bibitem{AlvarezGaume:1983gj}
L.~Alvarez-Gaume, J.~Polchinski, and M.~B. Wise, {\it {Minimal Low-Energy
  Supergravity}},  {\em Nucl. Phys.} {\bf B221} (1983) 495.

\bibitem{Raby:1980uk}
S.~Raby, {\it {SOME THOUGHTS ON THE MASS OF THE TOP QUARK}},  {\em Nucl. Phys.}
  {\bf B187} (1981) 446.

\bibitem{Glashow:1980xq}
S.~L. Glashow, {\it {WHERE IS THE TOP QUARK?}},  {\em Phys. Rev. Lett.} {\bf
  45} (1980) 1914.

\bibitem{Mahanthappa:1979ek}
K.~T. Mahanthappa and M.~A. Sher, {\it {THE MASS OF THE TOP QUARK IN SU(5)}},
  {\em Phys. Lett.} {\bf B86} (1979) 294.

\bibitem{Yanagida:1979gs}
T.~Yanagida, {\it {HORIZONTAL SYMMETRY AND MASS OF THE TOP QUARK}},  {\em Phys.
  Rev.} {\bf D20} (1979) 2986.

\bibitem{Pendleton:1980as}
B.~Pendleton and G.~G. Ross, {\it Mass and mixing angle predictions from
  infrared fixed points},  {\em Phys. Lett.} {\bf B98} (1981) 291.

\bibitem{Bardeen:1993rv}
W.~A. Bardeen, M.~Carena, S.~Pokorski, and C.~E.~M. Wagner, {\it Infrared fixed
  point solution for the top quark mass and unification of couplings in the
  mssm},  {\em Phys. Lett.} {\bf B320} (1994) 110--116,
  [\href{http://xxx.lanl.gov/abs/hep-ph/9309293}{{\tt hep-ph/9309293}}].

\bibitem{PhysRevD.24.1681}
S.~Dimopoulos, S.~Raby, and F.~Wilczek, {\it Supersymmetry and the scale of
  unification},  {\em Phys. Rev. D} {\bf 24} (Sep, 1981) 1681--1683.

\bibitem{Ibanez:1981yh}
L.~E. Ibanez and G.~G. Ross, {\it {Low-Energy Predictions in Supersymmetric
  Grand Unified Theories}},  {\em Phys. Lett.} {\bf B105} (1981) 439.

\bibitem{Dimopoulos:1981zb}
S.~Dimopoulos and H.~Georgi, {\it {Softly Broken Supersymmetry and SU(5)}},
  {\em Nucl. Phys.} {\bf B193} (1981) 150.

\bibitem{Sakai:1981gr}
N.~Sakai, {\it {Naturalness in Supersymmetric Guts}},  {\em Zeit. Phys.} {\bf
  C11} (1981) 153.

\bibitem{deBoer:2003xm}
W.~de~Boer and C.~Sander, {\it {Global electroweak fits and gauge coupling
  unification}},  {\em Phys. Lett.} {\bf B585} (2004) 276--286,
  [\href{http://xxx.lanl.gov/abs/hep-ph/0307049}{{\tt hep-ph/0307049}}].

\bibitem{Willenbrock:2003ca}
S.~Willenbrock, {\it {Triplicated trinification}},  {\em Phys. Lett.} {\bf
  B561} (2003) 130--134, [\href{http://xxx.lanl.gov/abs/hep-ph/0302168}{{\tt
  hep-ph/0302168}}].

\bibitem{Georgi:1974sy}
H.~Georgi and S.~L. Glashow, {\it Unity of all elementary particle forces},
  {\em Phys. Rev. Lett.} {\bf 32} (1974) 438--441.

\bibitem{Pati:1974yy}
J.~C. Pati and A.~Salam, {\it Lepton number as the fourth color},  {\em Phys.
  Rev.} {\bf D10} (1974) 275--289.

\bibitem{Ross:1985ai}
G.~G. Ross, {\it {GRAND UNIFIED THEORIES}}, . Reading, Usa: Benjamin/cummings (
  1984) 497 P. ( Frontiers In Physics, 60).

\bibitem{Buras:1977yy}
A.~J. Buras, J.~R. Ellis, M.~K. Gaillard, and D.~V. Nanopoulos, {\it {Aspects
  of the Grand Unification of Strong, Weak and Electromagnetic Interactions}},
  {\em Nucl. Phys.} {\bf B135} (1978) 66--92.

\bibitem{deMedeirosVarzielas:2005Earlier}
S.~F. de~Medeiros~Varzielas and G.~G. Ross, {\it {SU(3)} family symmetry and
  bi-tri maximial mixing},  {\em Nuclear Physics B} {\bf 733} (2006) 31 -- 47,
  [\href{http://xxx.lanl.gov/abs/hep-ph/0507176}{{\tt hep-ph/0507176}}].

\bibitem{Appelquist:1974tg}
T.~Appelquist and J.~Carazzone, {\it {Infrared Singularities and Massive
  Fields}},  {\em Phys. Rev.} {\bf D11} (1975) 2856.

\bibitem{Pich:1998xt}
A.~Pich, {\it Effective field theory},
  \href{http://xxx.lanl.gov/abs/hep-ph/9806303}{{\tt hep-ph/9806303}}.

\bibitem{Martemyanov:2005bt}
B.~V. Martemyanov and V.~S. Sopov, {\it Light quark mass ratio from dalitz plot
  of $\eta \to \pi^+\pi^-\pi^0$ decay},  {\em Phys. Rev.} {\bf D71} (2005)
  017501, [\href{http://xxx.lanl.gov/abs/hep-ph/0502023}{{\tt
  hep-ph/0502023}}].

\bibitem{Group:2007bx}
T.~E.~W. Group, {\it A combination of cdf and d0 results on the mass of the top
  quark},  \href{http://xxx.lanl.gov/abs/hep-ex/0703034}{{\tt hep-ex/0703034}}.

\bibitem{PDBook2000}
D.~{Groom}, {\it {Review of Particle Physics}},  {\em {The European Physical
  Journal}} {\bf C15} (2000) 1+.

\bibitem{Fusaoka:1998vc}
H.~Fusaoka and Y.~Koide, {\it Updated estimate of running quark masses},  {\em
  Phys. Rev.} {\bf D57} (1998) 3986--4001,
  [\href{http://xxx.lanl.gov/abs/hep-ph/9712201}{{\tt hep-ph/9712201}}].

\bibitem{Baer:2002ek}
H.~Baer, J.~Ferrandis, K.~Melnikov, and X.~Tata, {\it Relating bottom quark
  mass in dr-bar and ms-bar regularization schemes},  {\em Phys. Rev.} {\bf
  D66} (2002) 074007, [\href{http://xxx.lanl.gov/abs/hep-ph/0207126}{{\tt
  hep-ph/0207126}}].

\bibitem{Lubicz:2007zv}
V.~Lubicz, {\it Lattice qcd, flavor physics and the unitarity triangle
  analysis},  \href{http://xxx.lanl.gov/abs/hep-ph/0702204}{{\tt
  hep-ph/0702204}}.

\bibitem{Chankowski:2001mx}
P.~H. Chankowski and S.~Pokorski, {\it Quantum corrections to neutrino masses
  and mixing angles},  {\em Int. J. Mod. Phys.} {\bf A17} (2002) 575--614,
  [\href{http://xxx.lanl.gov/abs/hep-ph/0110249}{{\tt hep-ph/0110249}}].

\bibitem{Barger:1992ac}
V.~D. Barger, M.~S. Berger, and P.~Ohmann, {\it Supersymmetric grand unified
  theories: Two loop evolution of gauge and yukawa couplings},  {\em Phys.
  Rev.} {\bf D47} (1993) 1093--1113,
  [\href{http://xxx.lanl.gov/abs/hep-ph/9209232}{{\tt hep-ph/9209232}}].

\bibitem{Barr:2002mw}
S.~M. Barr and I.~Dorsner, {\it Atmospheric neutrino mixing and b - tau
  unification},  {\em Phys. Lett.} {\bf B556} (2003) 185--191,
  [\href{http://xxx.lanl.gov/abs/hep-ph/0211346}{{\tt hep-ph/0211346}}].

\bibitem{Diaz-Cruz:2000mn}
J.~L. Diaz-Cruz, H.~Murayama, and A.~Pierce, {\it Can supersymmetric loops
  correct the fermion mass relations in su(5)?},  {\em Phys. Rev.} {\bf D65}
  (2002) 075011, [\href{http://xxx.lanl.gov/abs/hep-ph/0012275}{{\tt
  hep-ph/0012275}}].

\bibitem{Pierce:1996zz}
D.~M. Pierce, J.~A. Bagger, K.~T. Matchev, and R.-j. Zhang, {\it Precision
  corrections in the minimal supersymmetric standard model},  {\em Nucl. Phys.}
  {\bf B491} (1997) 3--67, [\href{http://xxx.lanl.gov/abs/hep-ph/9606211}{{\tt
  hep-ph/9606211}}].

\bibitem{Blazek:1995nv}
T.~Blazek, S.~Raby, and S.~Pokorski, {\it Finite supersymmetric threshold
  corrections to ckm matrix elements in the large tan beta regime},  {\em Phys.
  Rev.} {\bf D52} (1995) 4151--4158,
  [\href{http://xxx.lanl.gov/abs/hep-ph/9504364}{{\tt hep-ph/9504364}}].

\bibitem{Carena:1999py}
M.~Carena, D.~Garcia, U.~Nierste, and C.~E.~M. Wagner, {\it Effective
  lagrangian for the anti-t b h+ interaction in the mssm and charged higgs
  phenomenology},  {\em Nucl. Phys.} {\bf B577} (2000) 88--120,
  [\href{http://xxx.lanl.gov/abs/hep-ph/9912516}{{\tt hep-ph/9912516}}].

\bibitem{Carena:2002es}
M.~Carena and H.~E. Haber, {\it Higgs boson theory and phenomenology. ((v))},
  {\em Prog. Part. Nucl. Phys.} {\bf 50} (2003) 63--152,
  [\href{http://xxx.lanl.gov/abs/hep-ph/0208209}{{\tt hep-ph/0208209}}].

\bibitem{Tobe:2003bc}
K.~Tobe and J.~D. Wells, {\it Revisiting top-bottom-tau yukawa unification in
  supersymmetric grand unified theories},  {\em Nucl. Phys.} {\bf B663} (2003)
  123--140, [\href{http://xxx.lanl.gov/abs/hep-ph/0301015}{{\tt
  hep-ph/0301015}}].

\bibitem{Hagiwara:2006jt}
K.~Hagiwara, A.~D. Martin, D.~Nomura, and T.~Teubner, {\it Improved predictions
  for g-2 of the muon and alpha(qed)(m(z)**2)},
  \href{http://xxx.lanl.gov/abs/hep-ph/0611102}{{\tt hep-ph/0611102}}.

\bibitem{Everett:2001tq}
L.~L. Everett, G.~L. Kane, S.~Rigolin, and L.-T. Wang, {\it Implications of
  muon g-2 for supersymmetry and for discovering superpartners directly},  {\em
  Phys. Rev. Lett.} {\bf 86} (2001) 3484--3487,
  [\href{http://xxx.lanl.gov/abs/hep-ph/0102145}{{\tt hep-ph/0102145}}].

\bibitem{Randall:1998uk}
L.~Randall and R.~Sundrum, {\it Out of this world supersymmetry breaking},
  {\em Nucl. Phys.} {\bf B557} (1999) 79--118,
  [\href{http://xxx.lanl.gov/abs/hep-th/9810155}{{\tt hep-th/9810155}}].

\bibitem{Hall:1993gn}
L.~J. Hall, R.~Rattazzi, and U.~Sarid, {\it The top quark mass in
  supersymmetric so(10) unification},  {\em Phys. Rev.} {\bf D50} (1994)
  7048--7065, [\href{http://xxx.lanl.gov/abs/hep-ph/9306309}{{\tt
  hep-ph/9306309}}].

\bibitem{Komine:2001rm}
S.~Komine and M.~Yamaguchi, {\it Bottom-tau unification in susy su(5) gut and
  constraints from b --> s gamma and muon g-2},  {\em Phys. Rev.} {\bf D65}
  (2002) 075013, [\href{http://xxx.lanl.gov/abs/hep-ph/0110032}{{\tt
  hep-ph/0110032}}].

\bibitem{Pallis:2003aw}
C.~Pallis, {\it b - tau unification and sfermion mass non-universality},  {\em
  Nucl. Phys.} {\bf B678} (2004) 398--426,
  [\href{http://xxx.lanl.gov/abs/hep-ph/0304047}{{\tt hep-ph/0304047}}].

\bibitem{Ramage:2003pf}
M.~R. Ramage and G.~G. Ross, {\it Soft susy breaking and family symmetry},
  {\em JHEP} {\bf 08} (2005) 031,
  [\href{http://xxx.lanl.gov/abs/hep-ph/0307389}{{\tt hep-ph/0307389}}].

\bibitem{King:2000vp}
S.~F. King and M.~Oliveira, {\it Yukawa unification as a window into the soft
  supersymmetry breaking lagrangian},  {\em Phys. Rev.} {\bf D63} (2001)
  015010, [\href{http://xxx.lanl.gov/abs/hep-ph/0008183}{{\tt
  hep-ph/0008183}}].

\bibitem{Blazek:2001sb}
T.~Blazek, R.~Dermisek, and S.~Raby, {\it Predictions for higgs and susy
  spectra from so(10) yukawa unification with mu > 0},  {\em Phys. Rev. Lett.}
  {\bf 88} (2002) 111804, [\href{http://xxx.lanl.gov/abs/hep-ph/0107097}{{\tt
  hep-ph/0107097}}].

\bibitem{Blazek:2002ta}
T.~Blazek, R.~Dermisek, and S.~Raby, {\it Yukawa unification in so(10)},  {\em
  Phys. Rev.} {\bf D65} (2002) 115004,
  [\href{http://xxx.lanl.gov/abs/hep-ph/0201081}{{\tt hep-ph/0201081}}].

\bibitem{Auto:2003ys}
D.~Auto, H.~Baer, C.~Balazs, A.~Belyaev, J.~Ferrandis, and X.~Tata, {\it Yukawa
  coupling unification in supersymmetric models},  {\em JHEP} {\bf 06} (2003)
  023, [\href{http://xxx.lanl.gov/abs/hep-ph/0302155}{{\tt hep-ph/0302155}}].

\bibitem{Balazs:2003mm}
C.~Balazs and R.~Dermisek, {\it Yukawa coupling unification and non-universal
  gaugino mediation of supersymmetry breaking},  {\em JHEP} {\bf 06} (2003)
  024, [\href{http://xxx.lanl.gov/abs/hep-ph/0303161}{{\tt hep-ph/0303161}}].

\bibitem{Gatto:1968ss}
R.~Gatto, G.~Sartori, and M.~Tonin, {\it Weak selfmasses, cabibbo angle, and
  broken su(2) x su(2)},  {\em Phys. Lett.} {\bf B28} (1968) 128--130.

\bibitem{Matsuda:2006xa}
K.~Matsuda and H.~Nishiura, {\it Can four-zero-texture mass matrix model
  reproduce the quark and lepton mixing angles and cp violating phases?},  {\em
  Phys. Rev.} {\bf D74} (2006) 033014,
  [\href{http://xxx.lanl.gov/abs/hep-ph/0606142}{{\tt hep-ph/0606142}}].

\bibitem{Hill:1983xh}
C.~T. Hill, {\it {Are There Significant Gravitational Corrections to the
  Unification Scale?}},  {\em Phys. Lett.} {\bf B135} (1984) 47.

\bibitem{Ellis:1979fg}
J.~R. Ellis and M.~K. Gaillard, {\it {Fermion Masses and Higgs Representations
  in SU(5)}},  {\em Phys. Lett.} {\bf B88} (1979) 315.

\bibitem{Antusch:2008tf}
S.~Antusch and M.~Spinrath, {\it {Quark and lepton masses at the GUT scale
  including SUSY threshold corrections}},
  \href{http://xxx.lanl.gov/abs/0804.0717}{{\tt 0804.0717}}.

\bibitem{Mrenna:1999ai}
S.~Mrenna, G.~L. Kane, and L.-T. Wang, {\it {Measuring gaugino soft phases and
  the LSP mass at Fermilab}},  {\em Phys. Lett.} {\bf B483} (2000) 175--183,
  [\href{http://xxx.lanl.gov/abs/hep-ph/9910477}{{\tt hep-ph/9910477}}].

\bibitem{BaerTata:2006}
Baer and Tata, {\em Supersymmetry Phenomoenology}.
\newblock Cambridge University Press, 1996.

\bibitem{Abbott:2000dt}
{\bf D0} Collaboration, B.~Abbott {\em et.~al.}, {\it {Spin correlation in
  $t\bar{t}$ production from $p\bar{p}$ collisions at $\sqrt{s} = 1.8$ TeV}},
  {\em Phys. Rev. Lett.} {\bf 85} (2000) 256--261,
  [\href{http://xxx.lanl.gov/abs/hep-ex/0002058}{{\tt hep-ex/0002058}}].

\bibitem{Devenish:2004pb}
R.~Devenish and A.~Cooper-Sarkar, {\it {Deep inelastic scattering}}, . Oxford,
  UK: Univ. Pr. (2004) 403 p.

\bibitem{Maltoni:2007tc}
F.~Maltoni, T.~McElmurry, R.~Putman, and S.~Willenbrock, {\it {Choosing the
  factorization scale in perturbative QCD}},
  \href{http://xxx.lanl.gov/abs/hep-ph/0703156}{{\tt hep-ph/0703156}}.

\bibitem{Alwall:2007st}
J.~Alwall {\em et.~al.}, {\it Madgraph/madevent v4: The new web generation},
  {\em JHEP} {\bf 09} (2007) 028,
  [\href{http://xxx.lanl.gov/abs/arXiv:0706.2334 [hep-ph]}{{\tt arXiv:0706.2334
  [hep-ph]}}].

\bibitem{Boos:2004kh}
{\bf CompHEP} Collaboration, E.~Boos {\em et.~al.}, {\it Comphep 4.4: Automatic
  computations from lagrangians to events},  {\em Nucl. Instrum. Meth.} {\bf
  A534} (2004) 250--259, [\href{http://xxx.lanl.gov/abs/hep-ph/0403113}{{\tt
  hep-ph/0403113}}].

\bibitem{DurhamPDFFarm}
{The Durham HEP Databases}, {\it {Parton Distribution Functions}}, .
  {http://durpdg.dur.ac.uk/HEPDATA/PDF}.

\bibitem{Feng:1999fu}
J.~L. Feng, T.~Moroi, L.~Randall, M.~Strassler, and S.-f. Su, {\it {Discovering
  supersymmetry at the Tevatron in Wino LSP scenarios}},  {\em Phys. Rev.
  Lett.} {\bf 83} (1999) 1731--1734,
  [\href{http://xxx.lanl.gov/abs/hep-ph/9904250}{{\tt hep-ph/9904250}}].

\bibitem{Krasnikov:1996fu}
N.~V. Krasnikov, {\it {SUSY model with R-parity violation, longlived charged
  slepton and quasistable matter}},  {\em JETP Lett.} {\bf 63} (1996) 503--509,
  [\href{http://xxx.lanl.gov/abs/hep-ph/9602270}{{\tt hep-ph/9602270}}].

\bibitem{Jedamzik:2007cp}
K.~Jedamzik, {\it {The cosmic 6Li and 7Li problems and BBN with long-lived
  charged massive particles}},  {\em Phys. Rev.} {\bf D77} (2008) 063524,
  [\href{http://xxx.lanl.gov/abs/0707.2070}{{\tt 0707.2070}}].

\bibitem{Athanasiou:2006ef}
C.~Athanasiou, C.~G. Lester, J.~M. Smillie, and B.~R. Webber, {\it
  {Distinguishing spins in decay chains at the Large Hadron Collider}},  {\em
  JHEP} {\bf 08} (2006) 055,
  [\href{http://xxx.lanl.gov/abs/hep-ph/0605286}{{\tt hep-ph/0605286}}].

\bibitem{Phalen:2007te}
D.~J. Phalen and A.~Pierce, {\it {Sfermion interference in neutralino decays at
  the LHC}},  {\em Phys. Rev.} {\bf D76} (2007) 075002,
  [\href{http://xxx.lanl.gov/abs/arXiv:0705.1366 [hep-ph]}{{\tt arXiv:0705.1366
  [hep-ph]}}].

\bibitem{Drees:2006um}
M.~Drees, W.~Hollik, and Q.~Xu, {\it One-loop calculations of the decay of the
  next-to-lightest neutralino in the mssm},  {\em JHEP} {\bf 02} (2007) 032,
  [\href{http://xxx.lanl.gov/abs/hep-ph/0610267}{{\tt hep-ph/0610267}}].

\bibitem{Allanach:2000kt}
B.~C. Allanach, C.~G. Lester, M.~A. Parker, and B.~R. Webber, {\it {Measuring
  sparticle masses in non-universal string inspired models at the LHC}},  {\em
  JHEP} {\bf 09} (2000) 004,
  [\href{http://xxx.lanl.gov/abs/hep-ph/0007009}{{\tt hep-ph/0007009}}].

\bibitem{Allanach:2008ib}
B.~C. Allanach, J.~P. Conlon, and C.~G. Lester, {\it {Measuring Smuon-Selectron
  Mass Splitting at the CERN LHC and Patterns of Supersymmetry Breaking}},
  {\em Phys. Rev.} {\bf D77} (2008) 076006,
  [\href{http://xxx.lanl.gov/abs/0801.3666}{{\tt 0801.3666}}].

\bibitem{phdthesis-lester}
C.~Lester, {\em Model independent sparticle mass measurements at {ATLAS}}.
\newblock {PhD} dissertation, University of Cambridge, Department of Physics,
  December, 2001.
\newblock {CERN-THESIS-2004-003}.

\bibitem{Bachacou:1999zb}
H.~Bachacou, I.~Hinchliffe, and F.~E. Paige, {\it Measurements of masses in
  sugra models at lhc},  {\em Phys. Rev.} {\bf D62} (2000) 015009,
  [\href{http://xxx.lanl.gov/abs/hep-ph/9907518}{{\tt hep-ph/9907518}}].

\bibitem{Gjelsten:2004ki}
B.~K. Gjelsten, D.~J. Miller, and P.~Osland, {\it {Measurement of SUSY masses
  via cascade decays for SPS 1a}},  {\em JHEP} {\bf 12} (2004) 003,
  [\href{http://xxx.lanl.gov/abs/hep-ph/0410303}{{\tt hep-ph/0410303}}].

\bibitem{Lester:2006yw}
C.~G. Lester, {\it Constrained invariant mass distributions in cascade decays:
  The shape of the 'm(qll)-threshold' and similar distributions},  {\em Phys.
  Lett.} {\bf B655} (2007) 39--44,
  [\href{http://xxx.lanl.gov/abs/hep-ph/0603171}{{\tt hep-ph/0603171}}].

\bibitem{Gjelsten:2006tg}
B.~K. Gjelsten, D.~J. Miller, P.~Osland, and A.~R. Raklev, {\it Mass
  determination in cascade decays using shape formulas},  {\em AIP Conf. Proc.}
  {\bf 903} (2007) 257--260,
  [\href{http://xxx.lanl.gov/abs/hep-ph/0611259}{{\tt hep-ph/0611259}}].

\bibitem{Bisset:2008hm}
M.~Bisset, N.~Kersting, and R.~Lu, {\it {Improving SUSY Spectrum Determinations
  at the LHC with Wedgebox and Hidden Threshold Techniques}},
  \href{http://xxx.lanl.gov/abs/0806.2492}{{\tt 0806.2492}}.

\bibitem{Goldstein:1993mj}
G.~R. Goldstein, K.~Sliwa, and R.~H. Dalitz, {\it {Observing top-quark
  production at the Fermilab Tevatron}},  {\em Phys. Rev.} {\bf 47} (1993)
  967--972.

\bibitem{Kondo:1993in}
K.~Kondo, T.~Chikamatsu, and S.~H. Kim, {\it {Dynamical likelihood method for
  reconstruction of events with missing momentum. 3: Analysis of a CDF high
  p(T) e mu event as t anti-t production}},  {\em J. Phys. Soc. Jap.} {\bf 62}
  (1993) 1177--1182.

\bibitem{Raja:1996vz}
R.~Raja, {\it {On measuring the top quark mass using the dilepton decay
  modes}},  {\em ECONF} {\bf C960625} (1996) STC122,
  [\href{http://xxx.lanl.gov/abs/hep-ex/9609016}{{\tt hep-ex/9609016}}].

\bibitem{Raja:1997qs}
R.~Raja, {\it {Remark on the errors associated with the Dalitz-Goldstein
  method}},  {\em Phys. Rev.} {\bf D56} (1997) 7465--7465.

\bibitem{Brandt:2006uc}
O.~Brandt, {\it {Measurement of the mass of the top quark in dilepton final
  states with the D0 detector}}, . FERMILAB-MASTERS-2006-03.

\bibitem{Cheng:2007xv}
H.-C. Cheng, J.~F. Gunion, Z.~Han, G.~Marandella, and B.~McElrath, {\it {Mass
  Determination in SUSY-like Events with Missing Energy}},  {\em JHEP} {\bf 12}
  (2007) 076, [\href{http://xxx.lanl.gov/abs/0707.0030}{{\tt 0707.0030}}].

\bibitem{Cheng:2008mg}
H.-C. Cheng, D.~Engelhardt, J.~F. Gunion, Z.~Han, and B.~McElrath, {\it
  {Accurate Mass Determinations in Decay Chains with Missing Energy}},  {\em
  Phys. Rev. Lett.} {\bf 100} (2008) 252001,
  [\href{http://xxx.lanl.gov/abs/0802.4290}{{\tt 0802.4290}}].

\bibitem{Nojiri:2007pq}
M.~M. Nojiri, G.~Polesello, and D.~R. Tovey, {\it {A hybrid method for
  determining SUSY particle masses at the LHC with fully identified cascade
  decays}},  {\em JHEP} {\bf 05} (2008) 014,
  [\href{http://xxx.lanl.gov/abs/0712.2718}{{\tt 0712.2718}}].

\bibitem{Abazov:2002bu}
{\bf D0} Collaboration, V.~M. Abazov {\em et.~al.}, {\it {Improved $W$ boson
  mass measurement with the D\O\ detector}},  {\em Phys. Rev.} {\bf D66} (2002)
  012001, [\href{http://xxx.lanl.gov/abs/hep-ex/0204014}{{\tt
  hep-ex/0204014}}].

\bibitem{Lester:2007fq}
C.~Lester and A.~Barr, {\it {MTGEN : Mass scale measurements in pair-production
  at colliders}},  {\em JHEP} {\bf 12} (2007) 102,
  [\href{http://xxx.lanl.gov/abs/0708.1028}{{\tt 0708.1028}}].

\bibitem{Cho:2007dh}
W.~S. Cho, K.~Choi, Y.~G. Kim, and C.~B. Park, {\it {Measuring superparticle
  masses at hadron collider using the transverse mass kink}},  {\em JHEP} {\bf
  02} (2008) 035, [\href{http://xxx.lanl.gov/abs/0711.4526}{{\tt 0711.4526}}].

\bibitem{Cho:2008cu}
W.~S. Cho, K.~Choi, Y.~G. Kim, and C.~B. Park, {\it {Measuring the top quark
  mass with $m_{T2}$ at the LHC}},
  \href{http://xxx.lanl.gov/abs/0804.2185}{{\tt 0804.2185}}.

\bibitem{Hamaguchi:2008hy}
K.~Hamaguchi, E.~Nakamura, and S.~Shirai, {\it {A Measurement of Neutralino
  Mass at the LHC in Light Gravitino Scenarios}},
  \href{http://xxx.lanl.gov/abs/0805.2502}{{\tt 0805.2502}}.

\bibitem{Cho:2007qv}
W.~S. Cho, K.~Choi, Y.~G. Kim, and C.~B. Park, {\it {Gluino Stransverse Mass}},
   {\em Phys. Rev. Lett.} {\bf 100} (2008) 171801,
  [\href{http://xxx.lanl.gov/abs/0709.0288}{{\tt 0709.0288}}].

\bibitem{Gripaios:2007is}
B.~Gripaios, {\it Transverse observables and mass determination at hadron
  colliders},  \href{http://xxx.lanl.gov/abs/arXiv:0709.2740 [hep-ph]}{{\tt
  arXiv:0709.2740 [hep-ph]}}.

\bibitem{Barr:2007hy}
A.~J. Barr, B.~Gripaios, and C.~G. Lester, {\it {Weighing Wimps with Kinks at
  Colliders: Invisible Particle Mass Measurements from Endpoints}},  {\em JHEP}
  {\bf 02} (2008) 014, [\href{http://xxx.lanl.gov/abs/0711.4008}{{\tt
  0711.4008}}].

\bibitem{Nojiri:2008hy}
M.~M. Nojiri, Y.~Shimizu, S.~Okada, and K.~Kawagoe, {\it {Inclusive transverse
  mass analysis for squark and gluino mass determination}},
  \href{http://xxx.lanl.gov/abs/0802.2412}{{\tt 0802.2412}}.

\bibitem{Tovey:2008ui}
D.~R. Tovey, {\it {On measuring the masses of pair-produced semi-invisibly
  decaying particles at hadron colliders}},  {\em JHEP} {\bf 04} (2008) 034,
  [\href{http://xxx.lanl.gov/abs/0802.2879}{{\tt 0802.2879}}].

\bibitem{AtlasMT2Wiki}
{\it Atlas $m_{T2}$ wiki c++ library},  {\em
  {https://twiki.cern.ch/twiki//bin/view/Atlas/StransverseMassLibrary}}.

\bibitem{MoortgatPick:1999di}
G.~A. Moortgat-Pick, H.~Fraas, A.~Bartl, and W.~Majerotto, {\it Polarization
  and spin effects in neutralino production and decay},  {\em Eur. Phys. J.}
  {\bf C9} (1999) 521--534, [\href{http://xxx.lanl.gov/abs/hep-ph/9903220}{{\tt
  hep-ph/9903220}}].

\bibitem{MoortgatPick:2000db}
G.~A. Moortgat-Pick and H.~Fraas, {\it Implications of cp and cpt for
  production and decay of majorana fermions},
  \href{http://xxx.lanl.gov/abs/hep-ph/0012229}{{\tt hep-ph/0012229}}.

\bibitem{Choi:2005gt}
S.~Y. Choi, B.~C. Chung, J.~Kalinowski, Y.~G. Kim, and K.~Rolbiecki, {\it
  Analysis of the neutralino system in three-body leptonic decays of
  neutralinos},  {\em Eur. Phys. J.} {\bf C46} (2006) 511--520,
  [\href{http://xxx.lanl.gov/abs/hep-ph/0504122}{{\tt hep-ph/0504122}}].

\bibitem{Djouadi:2002ze}
A.~Djouadi, J.-L. Kneur, and G.~Moultaka, {\it Suspect: A fortran code for the
  supersymmetric and higgs particle spectrum in the mssm},
  \href{http://xxx.lanl.gov/abs/hep-ph/0211331}{{\tt hep-ph/0211331}}.

\bibitem{CMSTDR}
C.~Collaboration, {\em CMS physics : Technical Design Report}.
\newblock CERN Report No: CERN-LHCC-2006-001 ; CMS-TDR-008-1, 2006.

\bibitem{DeRoeck:2005bw}
A.~De~Roeck {\em et.~al.}, {\it Supersymmetric benchmarks with non-universal
  scalar masses or gravitino dark matter},  {\em Eur. Phys. J.} {\bf C49}
  (2007) 1041--1066, [\href{http://xxx.lanl.gov/abs/hep-ph/0508198}{{\tt
  hep-ph/0508198}}].

\bibitem{Bisset:2005rn}
M.~Bisset, N.~Kersting, J.~Li, F.~Moortgat, and Q.~Xie, {\it Pair-produced
  heavy particle topologies: Mssm neutralino properties at the lhc from gluino
  / squark cascade decays},  {\em Eur. Phys. J.} {\bf C45} (2006) 477--492,
  [\href{http://xxx.lanl.gov/abs/hep-ph/0501157}{{\tt hep-ph/0501157}}].

\bibitem{Ghosh:1999ix}
D.~K. Ghosh, R.~M. Godbole, and S.~Raychaudhuri, {\it Signals for
  r-parity-violating supersymmetry at a 500-gev e+ e- collider},
  \href{http://xxx.lanl.gov/abs/hep-ph/9904233}{{\tt hep-ph/9904233}}.

\bibitem{2005NuPhS.144..341T}
J.~{Tanaka}, {\it {Discovery potential of the Standard Model Higgs at the
  LHC}},  {\em Nuclear Physics B Proceedings Supplements} {\bf 144} (July,
  2005) 341--348.

\bibitem{Paige:2003mg}
F.~E. Paige, S.~D. Protopopescu, H.~Baer, and X.~Tata, {\it {ISAJET 7.69: A
  Monte Carlo event generator for p p, anti-p p, and e+ e- reactions}},
  \href{http://xxx.lanl.gov/abs/hep-ph/0312045}{{\tt hep-ph/0312045}}.

\bibitem{Akhmadalev:2001ar}
{\bf ATLAS} Collaboration, S.~Akhmadalev {\em et.~al.}, {\it {Hadron energy
  reconstruction for the ATLAS calorimetry in the framework of the
  non-parametrical method}},  {\em Nucl. Instrum. Meth.} {\bf A480} (2002)
  508--523, [\href{http://xxx.lanl.gov/abs/hep-ex/0104002}{{\tt
  hep-ex/0104002}}].

\bibitem{Allanach:2001kg}
B.~C. Allanach, {\it Softsusy: A c++ program for calculating supersymmetric
  spectra},  {\em Comput. Phys. Commun.} {\bf 143} (2002) 305--331,
  [\href{http://xxx.lanl.gov/abs/hep-ph/0104145}{{\tt hep-ph/0104145}}].

\bibitem{Bajc:2002iw}
B.~Bajc, G.~Senjanovic, and F.~Vissani, {\it b - tau unification and large
  atmospheric mixing: A case for non-canonical see-saw},  {\em Phys. Rev.
  Lett.} {\bf 90} (2003) 051802,
  [\href{http://xxx.lanl.gov/abs/hep-ph/0210207}{{\tt hep-ph/0210207}}].

\bibitem{Campbell:2006nv}
B.~A. Campbell and D.~W. Maybury, {\it Triviality and the supersymmetric
  see-saw},  \href{http://xxx.lanl.gov/abs/hep-ph/0605144}{{\tt
  hep-ph/0605144}}.

\bibitem{Dighe:2006sr}
A.~Dighe, S.~Goswami, and W.~Rodejohann, {\it Corrections to tri-bimaximal
  neutrino mixing: Renormalization and planck scale effects},
  \href{http://xxx.lanl.gov/abs/hep-ph/0612328}{{\tt hep-ph/0612328}}.

\bibitem{Vissani:1994fy}
F.~Vissani and A.~Y. Smirnov, {\it Neutrino masses and b - tau unification in
  the supersymmetric standard model},  {\em Phys. Lett.} {\bf B341} (1994)
  173--180, [\href{http://xxx.lanl.gov/abs/hep-ph/9405399}{{\tt
  hep-ph/9405399}}].

\bibitem{Antusch:2002rr}
S.~Antusch, J.~Kersten, M.~Lindner, and M.~Ratz, {\it Neutrino mass matrix
  running for non-degenerate see-saw scales},  {\em Phys. Lett.} {\bf B538}
  (2002) 87--95, [\href{http://xxx.lanl.gov/abs/hep-ph/0203233}{{\tt
  hep-ph/0203233}}].

\bibitem{Leontaris:1995be}
G.~K. Leontaris, S.~Lola, and G.~G. Ross, {\it Heavy neutrino threshold effects
  in low-energy phenomenology},  {\em Nucl. Phys.} {\bf B454} (1995) 25--44,
  [\href{http://xxx.lanl.gov/abs/hep-ph/9505402}{{\tt hep-ph/9505402}}].

\bibitem{Chetyrkin:2000yt}
K.~G. Chetyrkin, J.~H. Kuhn, and M.~Steinhauser, {\it Rundec: A mathematica
  package for running and decoupling of the strong coupling and quark masses},
  {\em Comput. Phys. Commun.} {\bf 133} (2000) 43--65,
  [\href{http://xxx.lanl.gov/abs/hep-ph/0004189}{{\tt hep-ph/0004189}}].

\bibitem{Dorsner:2006hw}
I.~Dorsner, P.~F. Perez, and G.~Rodrigo, {\it Fermion masses and the uv cutoff
  of the minimal realistic su(5)},
  \href{http://xxx.lanl.gov/abs/hep-ph/0607208}{{\tt hep-ph/0607208}}.

\bibitem{Bevington}
P.~Bevington and K.~Robinson, {\em Data Reduction and Error Analysis in the
  Physics Sciences}.
\newblock McGraw Hill, second edition~ed., 1992.

\end{thebibliography}
\end{document}